\documentclass[review]{elsarticlenew}

\usepackage{lineno}  % For line numbers during review
\usepackage{graphicx}
\usepackage{amssymb,amsmath}
\usepackage{hyperref}
\usepackage{float}
\usepackage{array} 
\usepackage{subcaption}
\journal{Computers and Geosciences}
\usepackage{graphicx} % Required for inserting images
\usepackage{amsmath}
\usepackage{amssymb}
\usepackage{xcolor}
\usepackage{appendix}
\usepackage{tcolorbox}  % Add to preamble

\newcommand{\Enc}{\operatorname{Enc}}
\newcommand{\Dec}{\operatorname{Dec}}
\newcommand{\diag}{\operatorname{diag}}

\title{Controlled Latent Diffusion Models for $3$D Porous Media Reconstruction \footnote{The final version of this article has been published in Computers and Geosciences, Volume 206, January 2026, 106038. DOI: \href{https://doi.org/10.1016/j.cageo.2025.106038}{10.1016/j.cageo.2025.106038}. \textbf{Readers should refer to the published version for the most up-to-date content.}}}
\author[a]{Danilo Naiff}
\address[a]{Department of Mechanical Engineering, Coppe, Federal University of Rio de Janeiro}
\ead{dfnaiff@mecanica.ufrj.br} 
\author[b]{Bernardo P. Schaeffer}
\address[b]{Institute of Mathematics, Federal University of Rio de Janeiro}
\author[c]{Gustavo Pires}
\address[c]{Department of Geology, Institute of Geosciences, Federal University of Rio de Janeiro}
\author[d]{Dragan Stojkovic}
\address[d]{ExxonMobil Technology and Engineering Company}
\author[e]{Thomas Rapstine}
\address[e]{ExxonMobil Upstream Company}
\author[b]{Fabio Ramos}

\date{March 2025}

\begin{document}

\begin{abstract}
\textbf{Note}: The final version of this article was published in Computers and Geosciences, Volume 206, January 2026, 106038. DOI: \href{https://doi.org/10.1016/j.cageo.2025.106038}{10.1016/j.cageo.2025.106038}. \textbf{Readers should refer to the published version for the most up-to-date content.} Three-dimensional digital reconstruction of porous media presents a fundamental challenge in geoscience, requiring simultaneous resolution of fine-scale pore structures while capturing representative elementary volumes. We introduce a computational framework that addresses this challenge through latent diffusion models operating within the EDM framework. Our approach reduces dimensionality via a custom Variational Autoencoder trained in binary geological volumes, improving efficiency and also enabling the generation of larger volumes than previously possible with diffusion models. A key innovation is our controlled unconditional sampling methodology, which enhances distribution coverage by first sampling target statistics from their empirical distributions, then generating samples conditioned on these values. Extensive testing on four distinct rock types demonstrates that conditioning on porosity—a readily computable statistic—is sufficient to ensure a consistent representation of multiple complex properties, including permeability, two-point correlation functions, and pore size distributions. The framework achieves better generation quality than pixel-space diffusion while enabling significantly larger volume reconstruction ($256^3$ voxels) with substantially reduced computational requirements, establishing a new state-of-the-art for digital rock physics applications.

\end{abstract}


\maketitle

\section{Introduction}

Three-dimensional imaging technologies, particularly X-ray computed tomography, have revolutionized our understanding of porous media by enabling direct visualization of both pore structures and fluid distributions within otherwise opaque materials \citep{andra2013digital1, andra2013digital2, cao2020dynamic}. This breakthrough has found applications in water resources management, fuel cell development, carbon capture/storage, underground hydrogen storage, and hydrocarbon recovery. However, imaging techniques face an inherent trade-off: high resolution is crucial for accurately resolving pore structures, while a sufficiently large field of view is essential for capturing representative volumes that exhibit meaningful macroscopic properties \citep{saxena2018imaging, zhu2019challenges, wang2021deep, zhu2022key}.

Statistical generation of porous space images using numerical methods, particularly generative models, presents a promising solution to this resolution-versus-field-of-view challenge \citep{liu2022multiscale, luo2024multi, mosser2017reconstruction, zhu2024generation}. Within this domain, diffusion models have emerged as powerful alternatives demonstrating superior resistance to mode collapse and more stable training dynamics \citep{luo2024multi}, while showing exceptional promise in capturing multi-scale features inherent in heterogeneous porous media and enabling integration of multiple physical constraints.

This work advances the application of diffusion models to porous media generation through two complementary approaches. First, we leverage Latent Diffusion \citep{rombach2022high,DIFEDERICO2025105755}, a computationally efficient pipeline that reduces processing requirements while maintaining generation quality. Second, we integrate this with the state-of-the-art EDM framework \citep{karras2022elucidating}, enhancing generation stability and quality. Our approach utilizes a pre-trained Variational Autoencoder (VAE) to encode input data into a compressed latent space where diffusion operates efficiently, with the VAE's decoder transforming generated latent representations back into the original data space.

We extend these frameworks to geological binary volume generation in two key ways. First, we developed a specialized VAE for 3D binary pore-scale geological images—crucial given the scarcity of pre-trained 3D VAEs—optimizing compression efficiency while minimizing information loss. Second, we enhanced our diffusion model with a Transformer conditional layer to process complex statistical inputs such as two-point correlation functions, which can be readily obtained through experimental techniques like small-angle X-ray scattering (SAXS) or nuclear magnetic resonance (NMR).

We also introduce a novel sampling methodology, termed controlled unconditional sampling, to enhance the coverage of the target distribution. This approach leverages a conditional generative model by first sampling a target statistic $y$ from its empirical distribution $r(y)$ observed in training data, then generating samples conditioned on this value using our trained conditional model $p(x|y)$. The marginalization over $y$, given by $\int p(x|y)r(y)dy$, approximates the desired unconditional distribution $p(x)$, ensuring comprehensive coverage provided that the conditional model accurately captures underlying relationships.
% A detailed mathematical formulation is presented in Section~\ref{methodology}.

\begin{figure}[H]
   \centering
   % First row - Porosity controlled (generated) samples
   \begin{subfigure}[t]{0.23\textwidth}
       \centering
       \includegraphics[width=\textwidth]{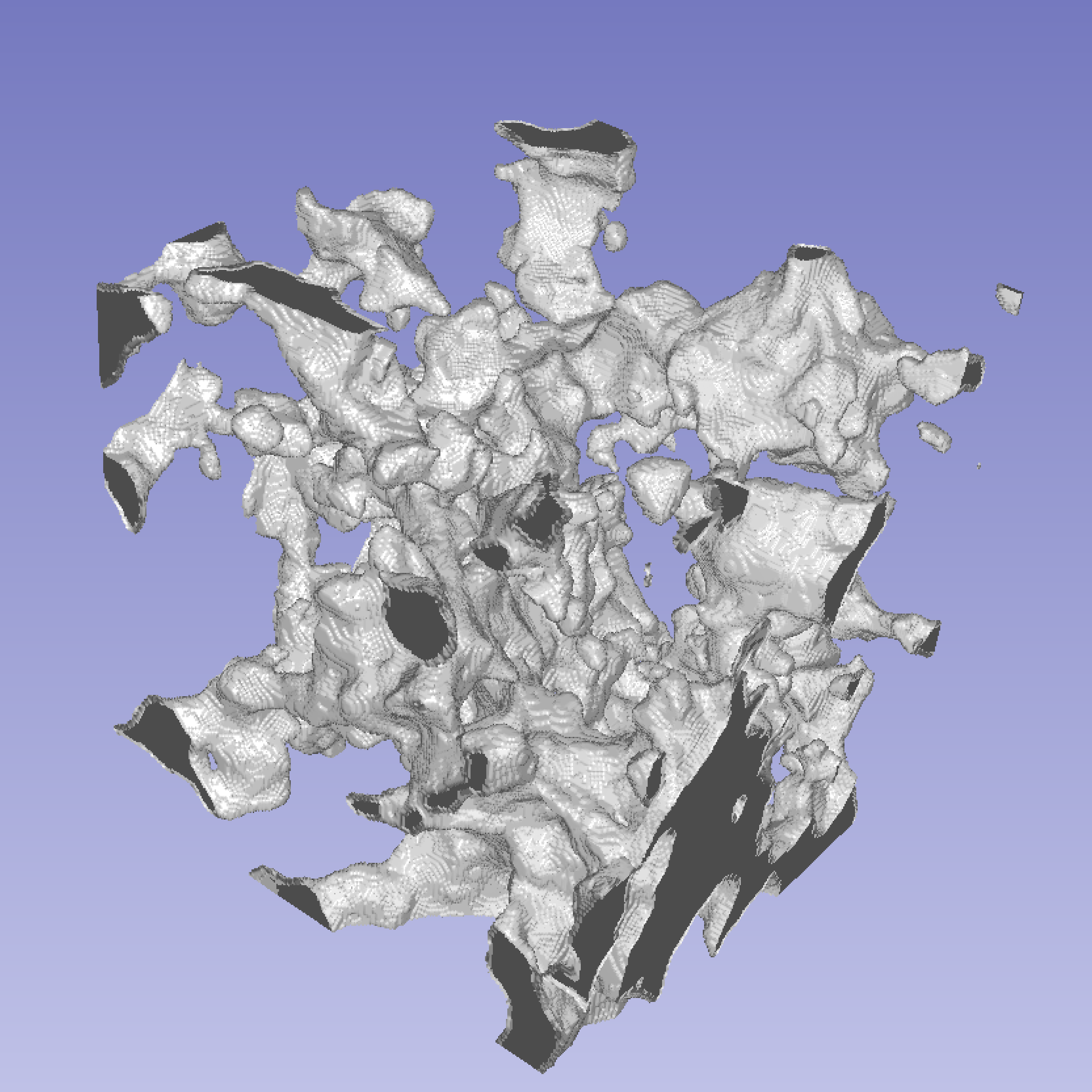}
       \caption{Doddington\\(generated)}
       \label{fig:dodd_pcond}
   \end{subfigure}%
   \hfill
   \begin{subfigure}[t]{0.23\textwidth}
       \centering
       \includegraphics[width=\textwidth]{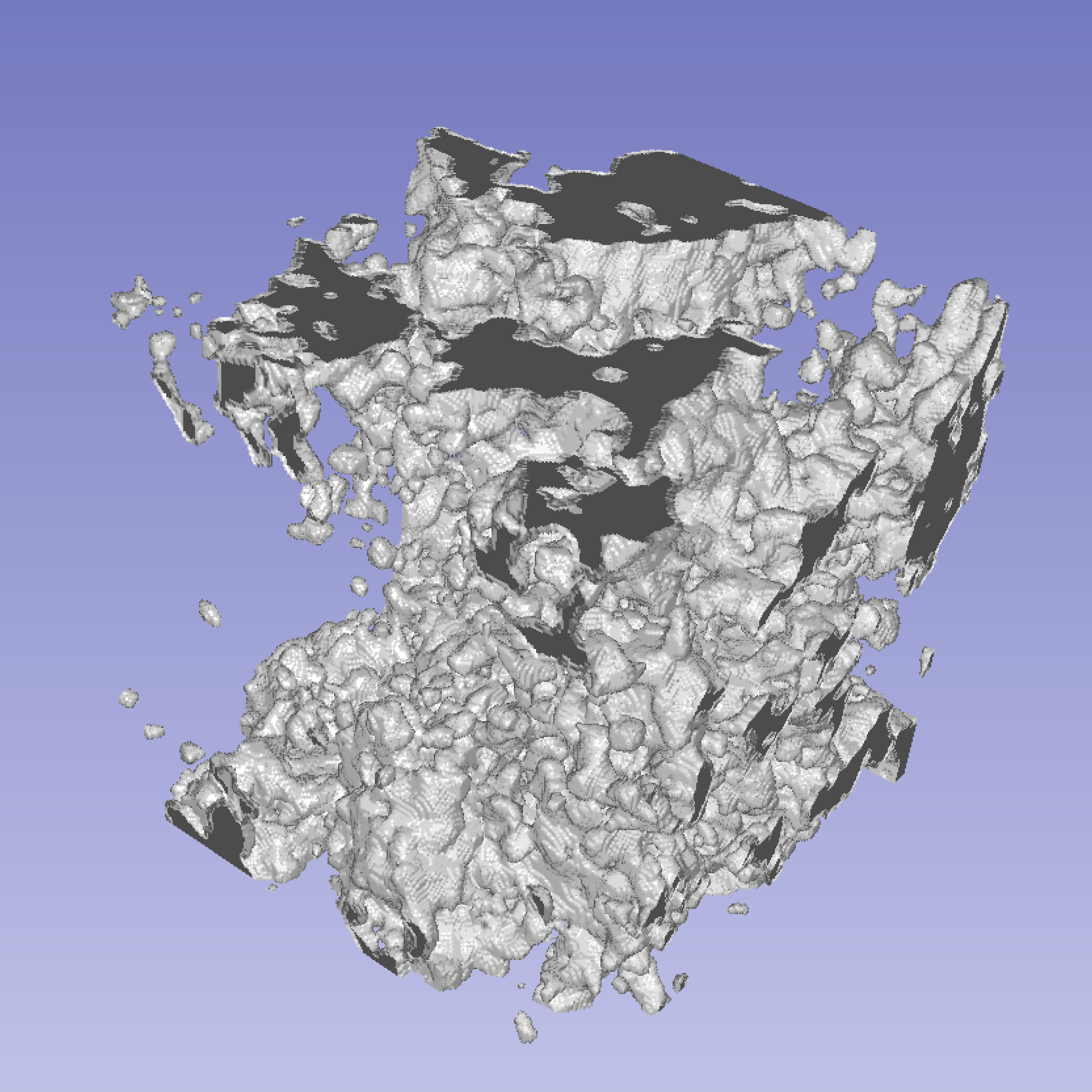}
       \caption{Estaillades\\(generated)}
       \label{fig:est_pcond}
   \end{subfigure}%
   \hfill
   \begin{subfigure}[t]{0.23\textwidth}
       \centering
       \includegraphics[width=\textwidth]{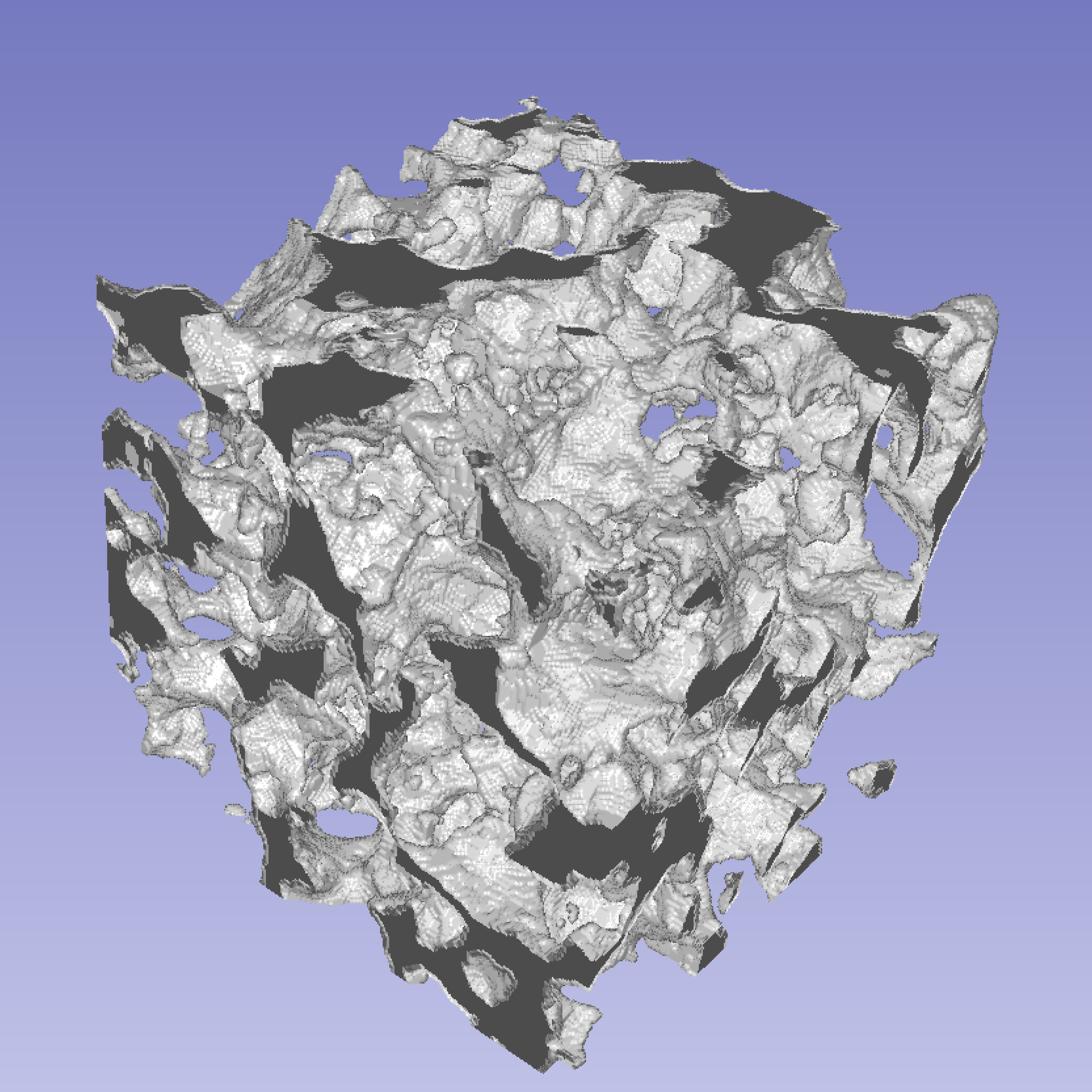}
       \caption{Bentheimer\\(generated)}
       \label{fig:bent_pcond}
   \end{subfigure}%
   \hfill
   \begin{subfigure}[t]{0.23\textwidth}
       \centering
       \includegraphics[width=\textwidth]{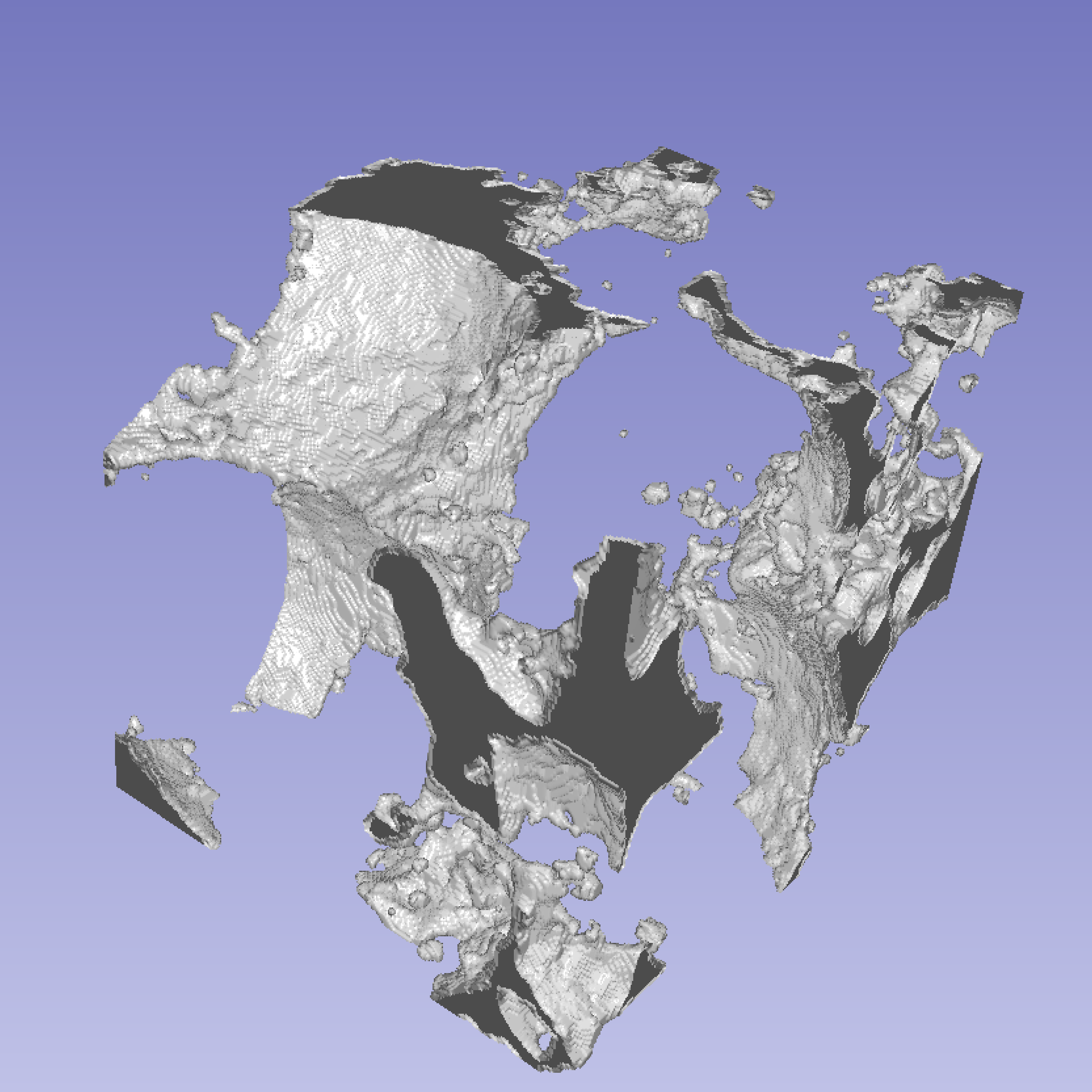}
       \caption{Ketton\\ (generated)}
       \label{fig:ket_pcond}
   \end{subfigure}
   
   % Second row - Validation samples
   \begin{subfigure}[t]{0.23\textwidth}
       \centering
       \includegraphics[width=\textwidth]{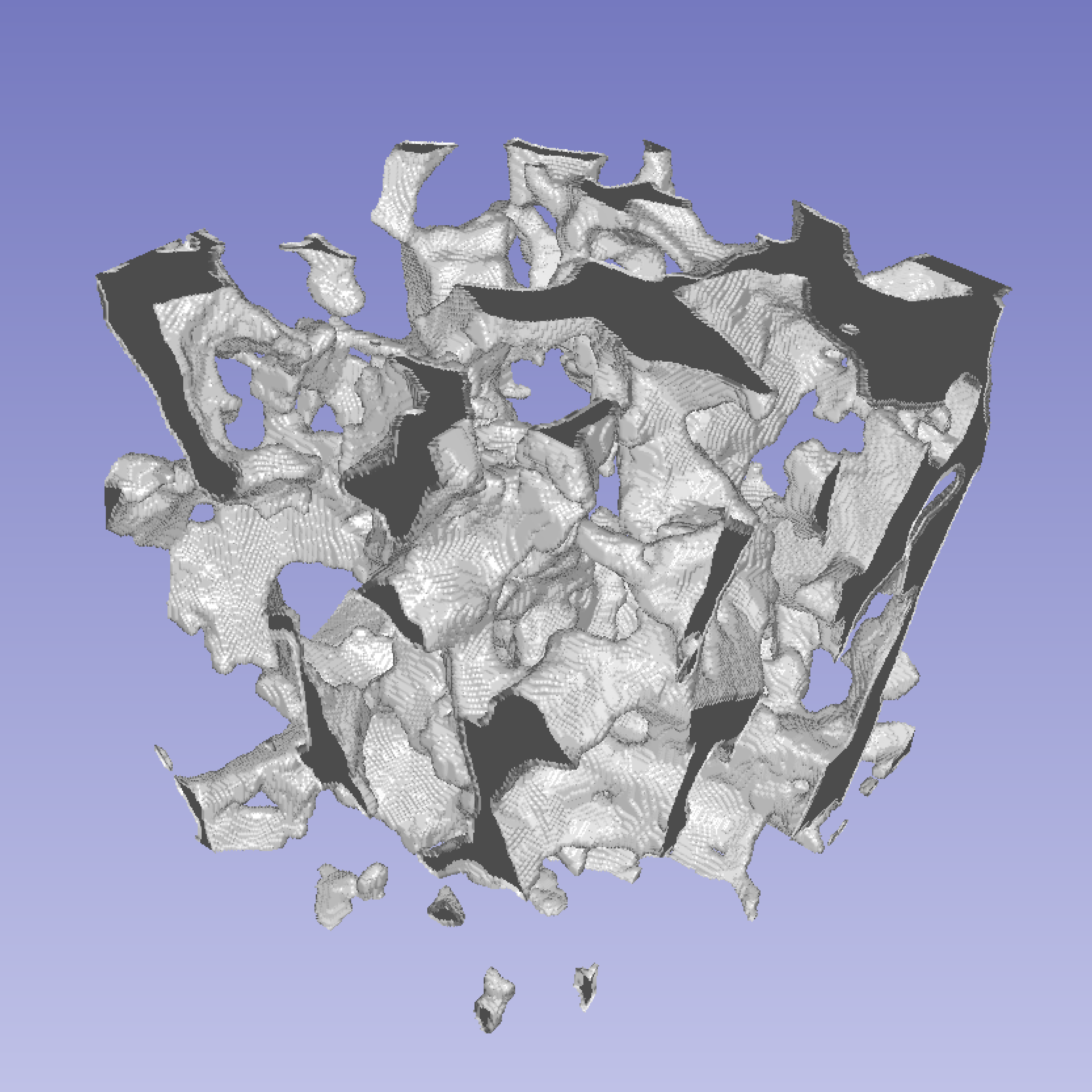}
       \caption{Doddington\\ (validation)}
       \label{fig:dodd_valid}
   \end{subfigure}%
   \hfill
   \begin{subfigure}[t]{0.23\textwidth}
       \centering
       \includegraphics[width=\textwidth]{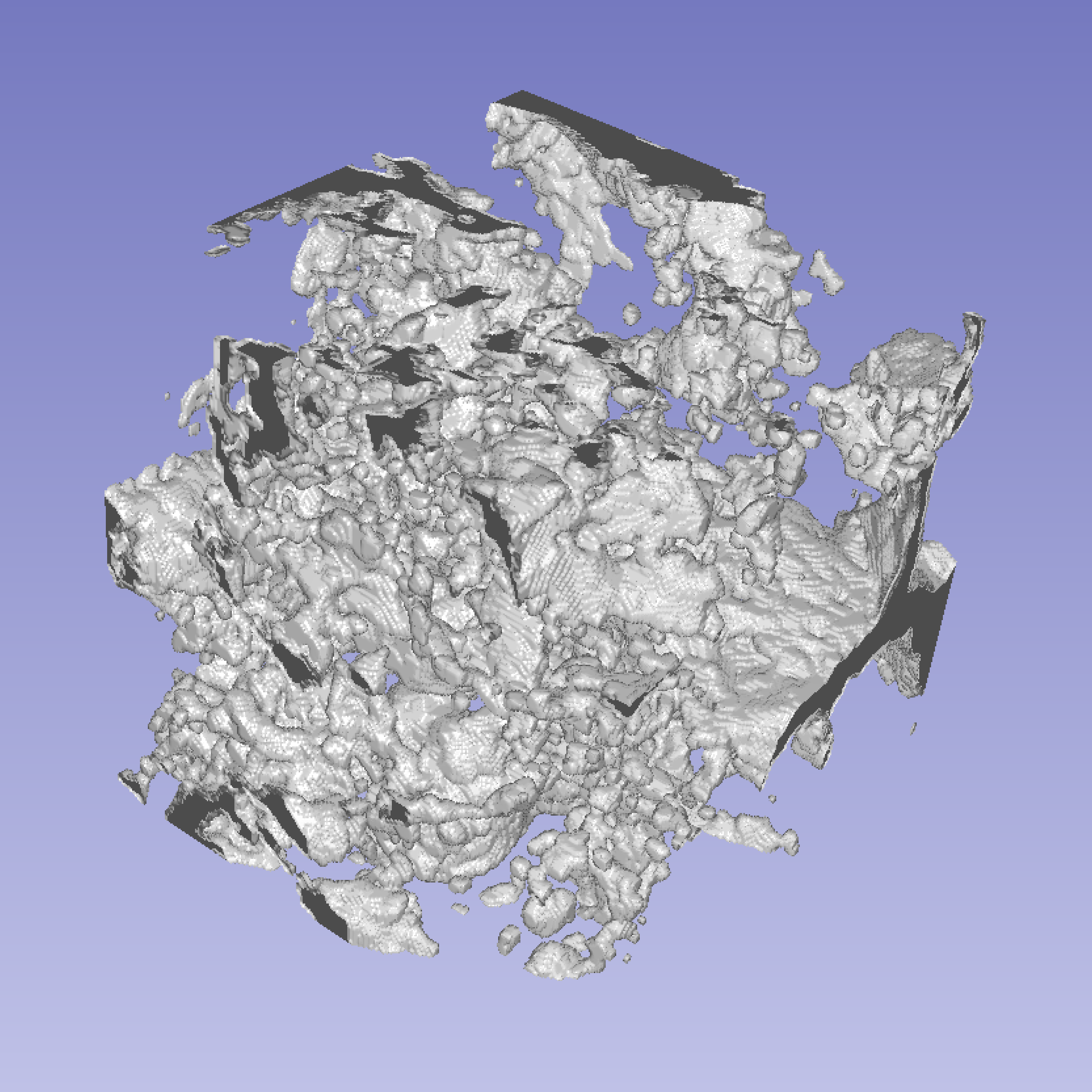}
       \caption{Estaillades\\ (validation)}
       \label{fig:est_valid}
   \end{subfigure}%
   \hfill
   \begin{subfigure}[t]{0.23\textwidth}
       \centering
       \includegraphics[width=\textwidth]{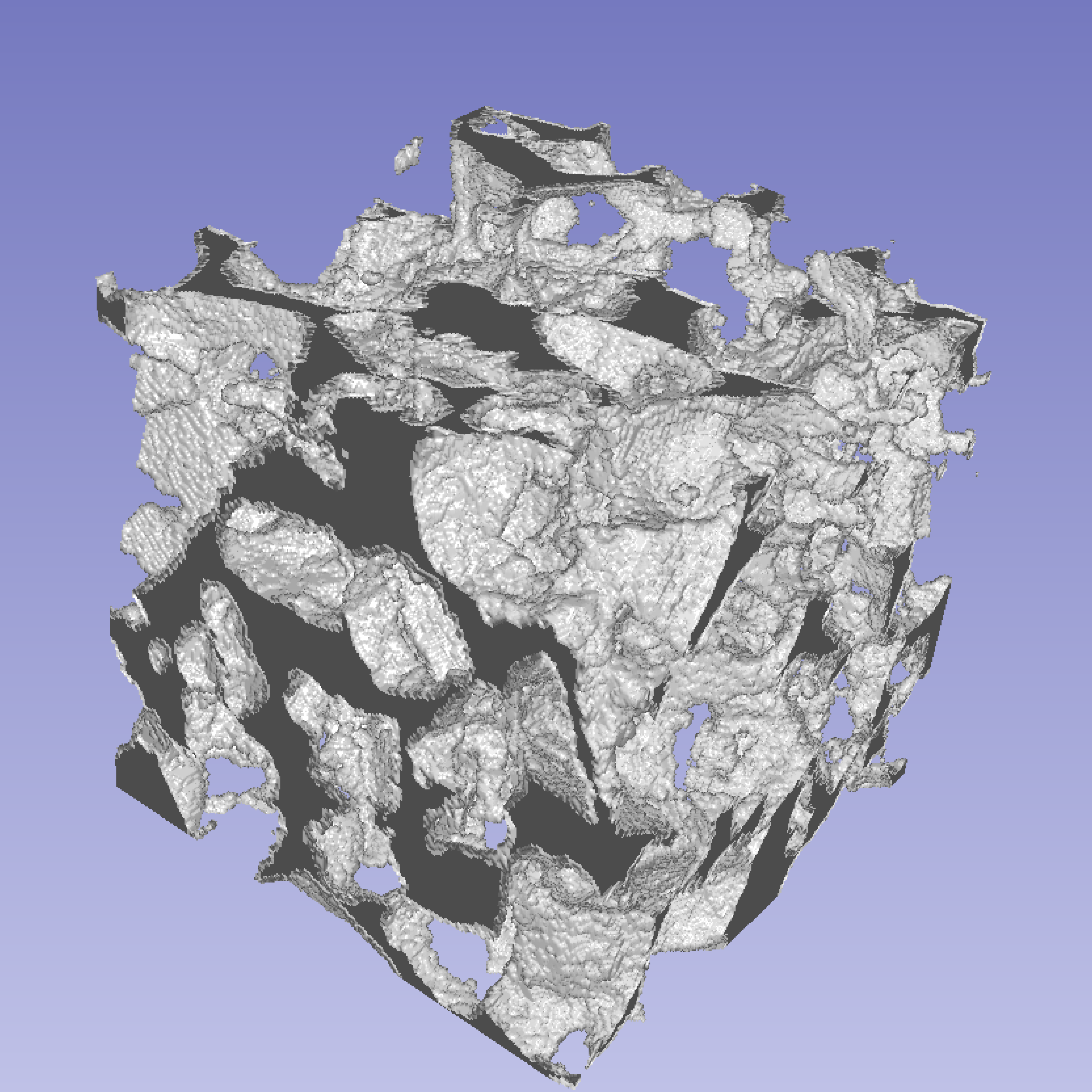}
       \caption{Bentheimer\\(validation)}
       \label{fig:bent_valid}
   \end{subfigure}%
   \hfill
   \begin{subfigure}[t]{0.23\textwidth}
       \centering
       \includegraphics[width=\textwidth]{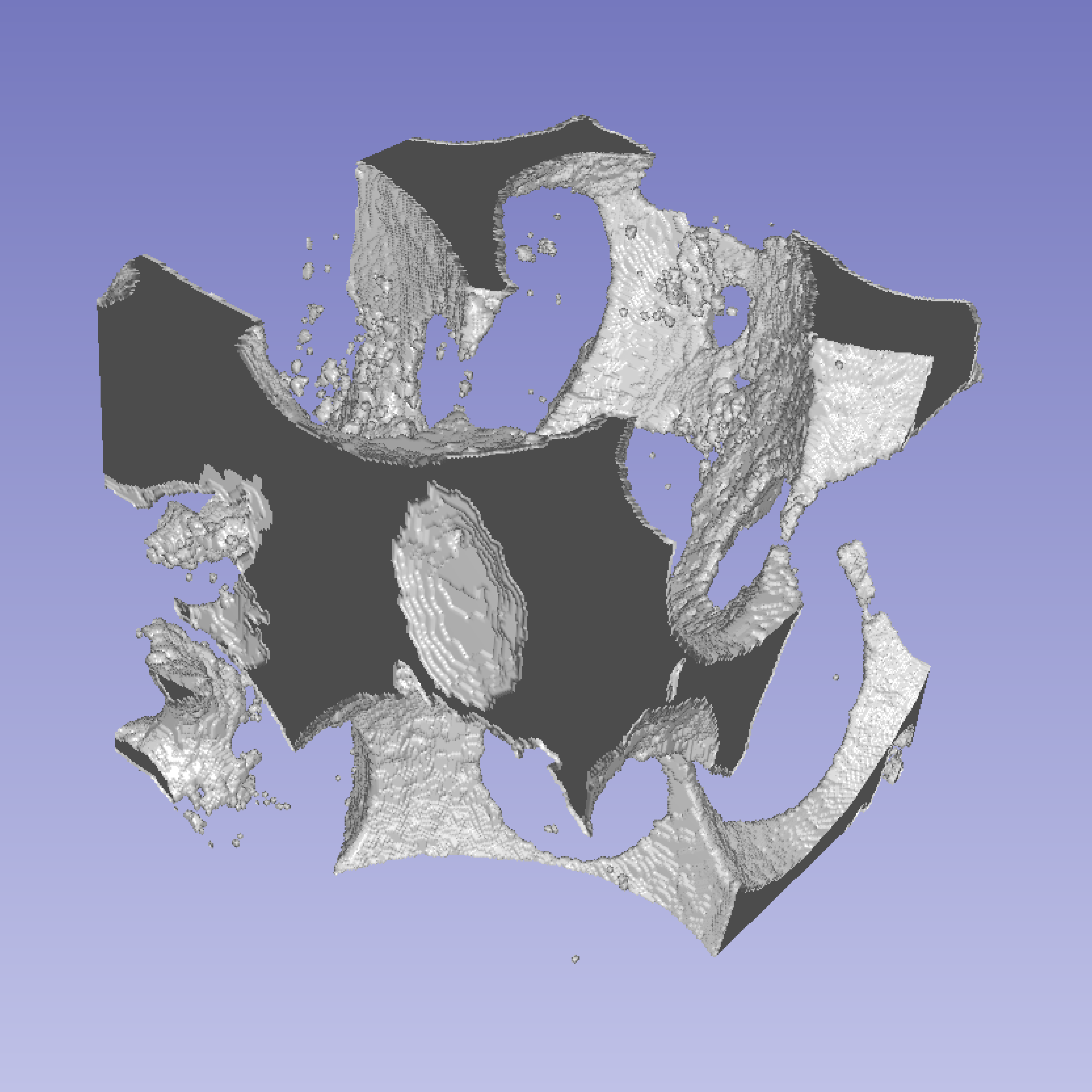}
       \caption{Ketton\\ (validation)}
       \label{fig:ket_valid}
   \end{subfigure}
   \caption{3D visualization of the pore-spaces of $256^3$ volumes. Top row: porosity-controlled generated samples. Bottom row: validation samples.}
   \label{fig:pore_spaces}
\end{figure}

Remarkably, conditioning on a single, easily computed statistic such as porosity achieves satisfactory coverage across multiple complex characteristics, including permeability, specific surface area, two-point correlation functions, and pore size distributions. When combined with data augmentation through symmetry operations, this technique proves particularly effective where limited training data would typically preclude successful unconditional generation. The efficacy of this approach is visually demonstrated in Figure~\ref{fig:pore_spaces}, which presents generated pore-space structures for four distinct rock types: Bentheimer and Doddington sandstones, and Estaillades and Ketton limestones.
% Pore spaces for 256 volumes

The key contributions of this work include: (1) A modern latent diffusion framework based on EDM for generating binary 3D porous media volumes, offering state-of-the-art generation quality; (2) Novel strategies including data augmentation and controlled unconditional sampling to overcome limited dataset limitations; (3) Extended conditioning capabilities to handle functional inputs through a Transformer embedding layer; and (4) A specialized 3D autoencoder optimized for binary porous media volumes, featuring low-regularization tuning for diffusion model applications.

Our framework represents a significant advancement in porous media characterization, introducing a computationally efficient pipeline for generating high-resolution, physically realistic samples at scales applicable to industrial and research applications. It establishes robust connections between statistical descriptors and physical structures, enabling unprecedented accuracy in representing complex geological features. Its ability to incorporate diverse conditioning data makes it an invaluable tool for uncertainty quantification in reservoir characterization studies \citep{li2023a}.

\section{State-of-the-art and previous approaches}

The field of porous media generation has evolved from traditional methods to modern machine-learning approaches. Early methodologies included process-based techniques \citep{coelho1997geometrical}, Gaussian field methods \citep{torquato1993chord}, simulated annealing \citep{hazlett1997statistical}, methods capturing spatial statistics or simulating sedimentary processes \citep{joshi1974class, blair1996using, yeong1998reconstructing, oren2002process}, and Sequential Indicator Simulation \citep{keehm2004permeability}. Multiple-point statistics, the previous state-of-the-art, demonstrated three-dimensional pore-space image generation \citep{okabe2004prediction}, though these conventional approaches often struggled to produce structurally diverse samples and fully capture complex porous structures \citep{sahimi2021physics}.

Recent breakthroughs in machine learning, particularly generative models, coupled with increased availability of high-quality three-dimensional training data, have transformed the landscape of porous media generation \citep{mosser2017reconstruction, zhu2019challenges, liu2022multiscale, li2023a, li2023b, zhang_3dpmrnn}. These modern approaches offer substantial advantages in characterizing complex pore spaces, generating images rapidly, producing diverse outputs, and incorporating physical constraints.

The introduction of Generative Adversarial Networks (GANs) marked a significant milestone, with implementations such as DCGAN and WGAN demonstrating the potential of deep learning for porous media generation \citep{mosser2017reconstruction, zha2020reconstruction}. However, these early GAN models faced inherent challenges in training stability, mode collapse, and particularly in capturing features across multiple spatial scales—critical for complex rocks with wide pore size distributions.

A significant advancement was achieved by \citep{zhu2024generation} through the Improved Pyramid Wasserstein GAN (IPWGAN), which introduced a Laplacian pyramid generator creating pore-space features across spatial scales, while feature statistics mixing regularization enhanced diversity and realism. The IPWGAN improved the reproduction of key physical properties, including two-point correlation functions, porosity, permeability, and Euler characteristic, with significant error reduction compared to previous approaches. However, IPWGAN remains strictly an \emph{unconditional generation model} and does not fully learn all image features, resulting in a parameter range smaller than that of real images.

Another notable application of GANs to the generation of porous media volumes can be found in \citep{nguyen2022_controlledGAN}, where the natural difficulty in performing conditional generation is overcome by the use of an actor-critic reinforcement learning module. This module learns the inputs that, when passed to the generator, produce an output with the desired morphological features, resulting in a pipeline capable of producing samples that reliably adhere to the targeted conditions.

Diffusion models have emerged as powerful alternatives to GANs, showing greater capability to avoid mode collapse and offering more stable training dynamics \citep{luo2024multi}, with particular promise in capturing complex multi-scale features while incorporating multiple physical constraints. A fundamental challenge has been the computational complexity in processing high-dimensional 3D geological data, particularly for heterogeneous formations like carbonate reservoirs. Among diffusion models formulations  (\citep{ho2020denoising}, \citep{song2021score}, \citep{ho2022classifier}), the EDM framework \citep{karras2022elucidating} provided an important simplification of diffusion model design-space, offering several advantages over discrete frameworks such as DDPM, used in \citep{BENTAMOU2025113596,DIFEDERICO2025105755,luo2024multi}. Its continuous description of the generation process allows explicit control of discretization errors, leading to more accurate sampling with fewer steps. Additionally, EDM's neural network preconditioning provides faster, more stable training dynamics, and its generality allows extensive parameter exploration, which is valuable for domain-specific applications where ideal parameters may lie outside general application ranges.

Recent research has made significant strides in applying diffusion models to geological modeling, although current approaches face limitations. The framework in \citep{luo2024multi} demonstrates multiconditional generation using DDPM in pixel-space but is constrained to volumes of $64^3$, insufficient for many digital rock analysis applications requiring larger representative elementary volumes. A complementary approach in \citep{DIFEDERICO2025105755} combines latent diffusion with DDIM to parameterize facies-based geomodels, but both studies rely on discrete-time diffusion formulations.

To address these limitations, we combine the EDM framework with latent diffusion, enabling the generation of substantially larger porous media volumes while reducing computational costs. Our compression factor makes training and inference for binary volumes of size $(256, 256, 256)$ more computationally efficient than processing volumes of $(64, 64, 64)$ in the pixel space, without compromising quality or diversity. Latent diffusion effectively addresses challenges in training diffusion models on limited datasets, particularly relevant in geophysical applications where high-quality three-dimensional imaging data is often scarce due to acquisition costs and technical constraints.

\section{Methodology}\label{methodology}

\subsection{Reconstruction with Unconditional and Controlled Generative Models}\label{reconstruction methods}

Generative models aim to learn complex data distributions to generate samples resembling real data. The primary goal is to approximate the true data distribution \( q(\mathbf{x}) \) with a model distribution \( p(\mathbf{x}) \) such that \( p(\mathbf{x}) \approx q(\mathbf{x}) \). We refer to this as an \emph{unconditional generative model} as it models \( \mathbf{x} \) without conditioning on external variables.

For a dataset \( \mathcal{D} = \{ \mathbf{x}_i \} \) with associated features \( \mathbf{y}_i = f(\mathbf{x}_i) \), we may instead model the conditional distribution \( q(\mathbf{x} \mid \mathbf{y}) \). Our objective becomes finding \( p(\mathbf{x} \mid \mathbf{y}) \) that approximates \( q(\mathbf{x} \mid \mathbf{y}) \) for values of \( \mathbf{y} \) that are typical under the feature distribution \( r(\mathbf{y}) \).

By modeling \( p(\mathbf{x} \mid \mathbf{y}) \) and integrating over \( \mathbf{y} \), we approximate the unconditional data distribution:

\begin{equation}
\label{eq:marginal}
p(\mathbf{x}) = \int p(\mathbf{x} \mid \mathbf{y}) \, r(\mathbf{y}) \, d\mathbf{y} \approx q(\mathbf{x}).
\end{equation}

In practice, we use an approximation \( \hat{r}(\mathbf{y}) \) of the true distribution \( r(\mathbf{y}) \) when direct access to the true feature distribution is unavailable or computationally intractable. This simplified approach maintains control over generative outcomes while enhancing practical utility.

For \( \mathbf{x} \in \mathbb{R}^N \) (with \( N \) typically large), neural networks accomplish this task by:

\begin{enumerate}
    \item Defining a parameterized distribution \( p_\theta(\mathbf{x} \mid \mathbf{y}) \) with sampling capabilities.
    \item Creating a differentiable loss function \( \mathcal{L}(\mathbf{x}, \mathbf{y}; \theta) \) such that minimizing the expected loss:
    \[
    L(\theta) := \mathbb{E}_{(\mathbf{x}, \mathbf{y}) \sim q(\mathbf{x}, \mathbf{y})} \, \mathcal{L}(\mathbf{x}, \mathbf{y}; \theta)
    \]
    leads \( p_\theta(\mathbf{x}, \mathbf{y}) := p_\theta(\mathbf{x} \mid \mathbf{y}) r(\mathbf{y}) \) to approximate \( q(\mathbf{x}, \mathbf{y}) \).
    \item Minimizing \( L(\theta) \) using stochastic gradient descent with minibatches from \( \mathcal{D} \).
\end{enumerate}

In digital rock reconstruction, \( \mathbf{x} \) represents a volumetric image of the rock's pore space, and \( \mathbf{y} \) represents features such as porosity or two-point correlation.

We call this process \emph{unconditional \( \mathbf{y} \)-controlled reconstruction}. It is unconditional because \( \mathbf{y} \) has been marginalized over, yet the guidance from \( \mathbf{y} \) during training significantly improves the recovery of the original distribution. For example, permeability statistics improve greatly when controlling on porosity (Section \ref{subsection: conditional results}), suggesting that the complex geometrical patterns of connected porosity and inter-cavity channels are learned in porosity-conditional training.

% The fully unconditional case occurs when \( \mathbf{y} = \emptyset \), simplifying to \( q(\mathbf{x}, \mathbf{y}) = q(\mathbf{x}) \) and \( p(\mathbf{x} \mid \mathbf{y}) = p(\mathbf{x}) \), eliminating the need for marginalization in Equation~\eqref{eq:marginal} since \( r(\mathbf{y}) \equiv 1 \).

% Beyond diffusion models (our focus), other generative approaches include GANs~\citep{goodfellow2014generative}, Normalizing Flows~\citep{papamakarios2021normalizing}, and VAEs~\citep{kingma2014auto}. For a comprehensive review, see Bond-Taylor et al.~\citep{bond-taylor2021deep}.

\subsection{Diffusion-based generative models}\label{subsection: diffusion}

Diffusion models \citep{sohl-dickstein2015deep, ho2020denoising, song2021score, karras2022elucidating} are a class of generative models in which the samples $x \sim p(x)$, with $x \in \mathbb{R}^N$ are generated through the following process.

\begin{itemize}
    \item First, we sample from a random normal distribution $x_{max} \sim \mathcal{N}(0, \sigma_{max}^2 I)$, which we refer to as \textit{noise}.
    \item Then, through a learned iterative process, we transform the sampled noise $x_{max}$ to a denoised sample $x$ such that $x \sim p(x)$.
\end{itemize}

While we present diffusion models in the unconditional generation framework above, extending to conditional generation is straightforward by allowing distributions to depend on an additional label $y$. In this work, we focus on the EDM framework \citep{karras2022elucidating}, a particular class of score-based diffusion models \citep{song2021score}.

The EDM framework bases its denoising process around the Probability Flow Ordinary Differential Equation (ODE). Defining $p(x_\sigma; \sigma)$ as
\begin{equation}\label{p_sigma definition}
    p(x_\sigma;\sigma) := \int \mathcal{N}(x_\sigma|x, \sigma^2 I) p(x) dx,
\end{equation}
a diffused version of $p(x)$, the Probability Flow ODE
\begin{equation}\label{flow ode}
    \frac{d x_\sigma}{d \sigma} = -\sigma \nabla\log p(x_\sigma, \sigma)
\end{equation}
preserves the densities $p(x_\sigma, \sigma)$. For any $0<\tau, \sigma$, if $X_{\tau}$ is distributed according to $p(x;\tau)$, then integrating \eqref{flow ode} from $\tau$ to $\sigma$ yields a random variable distributed according to $p(x;\sigma)$. This gives a time-reversible rule for trajectories that preserves probability distributions of a Gaussian diffusion.

For suitable $\sigma_{max}$, we can sample from $p(x)$ through the following process:
\begin{enumerate}
    \item First, sample noise $x_{\max} \sim \mathcal{N}(0, \sigma_{max}^2 I)$, with $\sim\mathcal{N}(0, \sigma_{max}^2 I)$ being a approximation of $p(x;\sigma_{max})$.
    \item Then, integrate equation \eqref{flow ode} backward in time from $\sigma_{\max}$ to $\sigma_{\min}\approx 0$.
\end{enumerate}

The score function $\nabla\log p(x_\sigma, \sigma)$ can be learned from the forward diffusion process through a neural network using denoising score-matching \citep{karras2022elucidating, vincent2011connection}. This involves training a denoiser function $D_{\theta}(x_\sigma, \sigma)$ with the loss
\begin{equation}\label{DSM loss}
    \mathcal{L}_{\text{DSM}}(\theta) = \mathbb{E}_{\sigma, X \sim q(x), X_\sigma \sim \mathcal{N}(X, \sigma^2 I)} \left[\lambda(\sigma) \lVert{D_\theta(X_\sigma, \sigma) - X}\rVert^2\right],
\end{equation}
where $\lambda(\sigma)$ is a loss weighting function. The function $s_{\theta}(x_\sigma, \sigma) := \frac{D_\theta(x_\sigma, \sigma) - x_\sigma}{\sigma^2}$ can be shown to approximate the true score function $\nabla\log p(x_\sigma,\sigma)$.

In summary, deploying a diffusion model in the EDM framework involves:
\begin{enumerate}
    \item A training phase, in which the score function is learned by a neural network with the loss function \eqref{DSM loss}, resulting in a trained score model.
    \item A sampling phase, where the ODE \eqref{flow ode} with the trained score function $s_\theta\approx\nabla\log p$ is solved from $\sigma_{\max}$ to $\sigma_{\min}$ using a numerical integrator.
\end{enumerate}

The computational cost of training significantly exceeds that of sampling, which requires only evaluating the score model at each discretization step. Typically, 30-50 discretization steps are used.

\subsection{Latent diffusion models}

\begin{figure}
    \centering
    \includegraphics[width=0.8\textwidth]{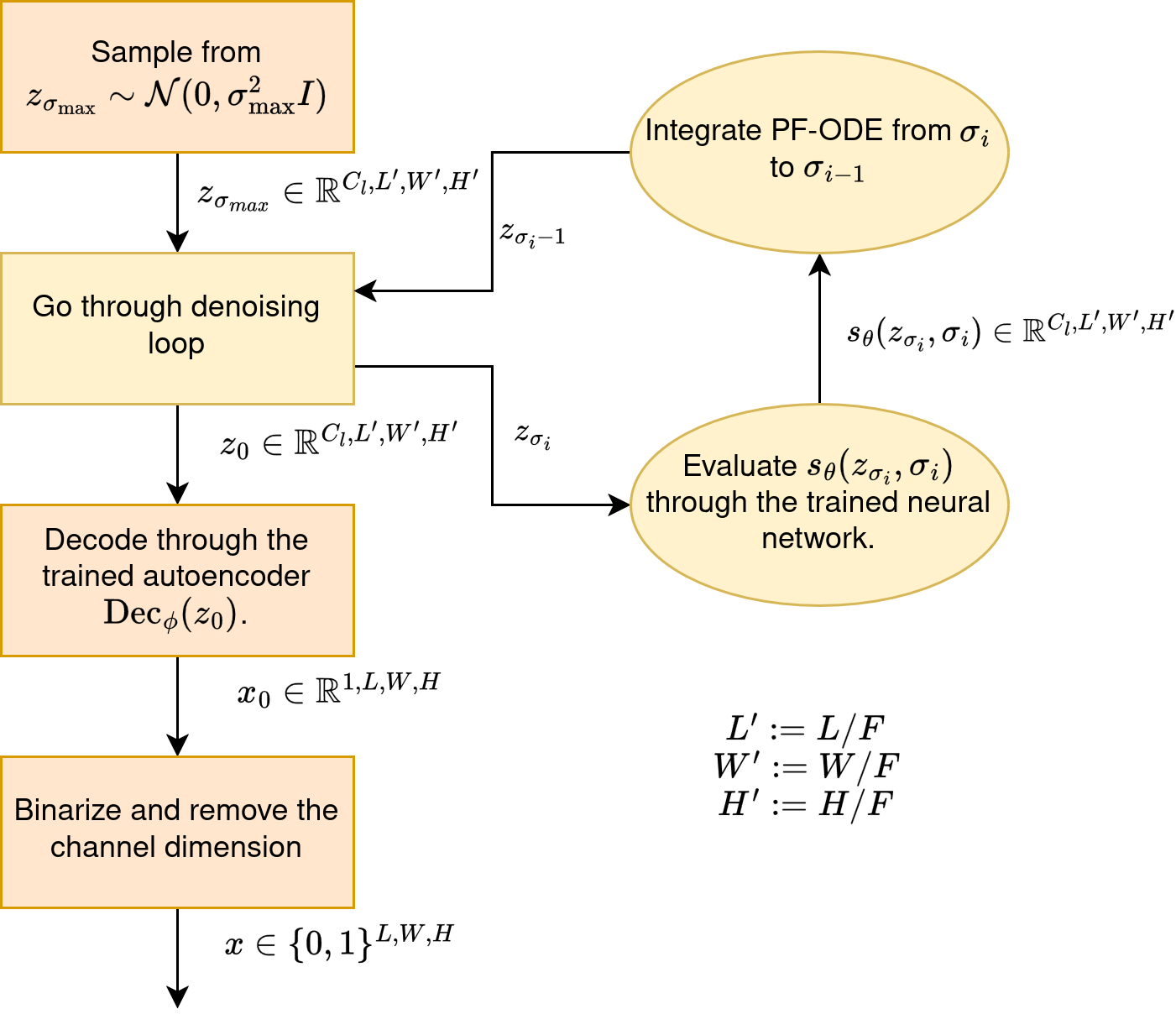}
    \caption{\label{fig:latent_diffusion_inference} Complete generation pipeline for binary volumes with a latent diffusion model. Here, $L, H, W$ are the desired volume dimensions, $C_l$ is the number of latent channels, and $F$ is the autoencoder reduction factor}
\end{figure}

Latent diffusion models (LDM) \citep{rombach2022high} are a state-of-the-art formulation of diffusion \citep{openai2024sora} that combines an autoencoder with the diffusion model discussed above. The core idea is to explicitly separate the tasks of data compression and distribution learning. First, an autoencoder is trained to map the high-dimensional data to a lower-dimensional latent space. Then, the autoencoder weights are fixed, and a diffusion model is trained on the latent space representations. Finally, at sampling, the diffusion model generates a sample in the latent space, which is decoded back to the original space. The complete pipeline is illustrated in Figure \ref{fig:latent_diffusion_inference}. This compression is associated with a substantial reduction in computation costs, which also allows diffusion models to be deployed on higher-dimensional data.

More precisely, the autoencoder consists of an encoder $\Enc_\phi(x)$ and a decoder $\Dec_\phi(z)$, both parameterized by neural networks. The latent diffusion framework uses the variational autoencoder (VAE) \citep{kingma2014auto}, in which the encoder maps each input $x$ to a Gaussian distribution:
\begin{equation*}
    x\mapsto p(z;x) := \mathcal{N}(\mu_z(x), \diag{\sigma_z}(x)),
\end{equation*}
and the decoder is a deterministic map $z\mapsto\Dec_{\phi}(z)$ from the latent space to the original space. Minimizing the log-likelihood of the model \citep{kingma2014auto} results in the following loss function
\begin{equation}\label{total loss}
    \mathcal{L}(\phi) = L_{\text{rec}}(\phi) + \lambda_{\text{KL}} \mathbb{E}_{x \sim q(x)} D_{KL}(p(z;\Enc_\phi(x))||\mathcal{N}(0, I)),
\end{equation}
where $D_{\text{KL}}$ is the Kullback-Leibler divergence, and $L_{\text{rec}}(\phi)$ is the $L^2$ reconstruction loss, given by
\begin{equation}\label{l2 recloss}
    \mathcal{L}_{\text{rec}}(\phi) = \mathbb{E}_{x \sim q(x)} \mathbb{E}_{z \sim p(z;\Enc_\phi(x))} \left[ \lVert x - \Dec_\phi(z) \rVert_2^2 \right].
\end{equation}

The second term can be seen as a regularization term, controlled by the hyperparameter $\lambda_{\text{KL}}$. Encoding $x$ into a distribution instead of a single value enforces regularity in the latent space, since the reconstruction loss \eqref{l2 recloss} enforces that small perturbations of a point $\mu_z(x)$ are reconstructed in a $\hat{x}\approx x$. This is a key advantage of using VAE for dimension reduction in latent diffusion, since it makes the latent distribution more tractable and better suited to generalization beyond the training dataset.

Latent diffusion models \citep{rombach2022high} originally aims at the generation of natural images, and uses a combination of a perceptual loss  (LPIPS) and an adversarial term \citep{rombach2022high, esser2021taming, zhang2018unreasonable}, instead of the \(L^2\) reconstruction loss in \(\eqref{l2 recloss}\). In principle, it is possible to use a generic pre-trained VAE to train a diffusion model on a specific dataset, but we found it to provide suboptimal compression and lead to poorly generated rock statistics. Thus, we opt to train our own VAE on binary porous media volumes with \eqref{l2 recloss}. Although the \(L^2\) loss is often avoided in natural image applications due to its tendency to produce softer contours, this issue is irrelevant in our setting since we perform a binarization step at the end. For generating grayscale images, it may be necessary to introduce an additional adversarial component in the reconstruction loss, but we leave that for future work.

\section{Data preparation and model details}

\subsection{Data}\label{datasection}

The data used in this work comprise micro-CT 3D volumes of size $1000^3$ of four well-known rock types: Bentheimer and Doddington sandstones (Table \ref{tab:sandstone-properties}), and Estaillades and Ketton limestones (Table \ref{tab:limestone-properties}). See \ref{datacode availability} for data availability.

\begin{table}[htbp]
\centering
\small
\caption{Properties of sandstone samples}
\begin{tabular}{>{\raggedright\arraybackslash}p{3.5cm}>{\raggedright\arraybackslash}p{3.5cm}>{\raggedright\arraybackslash}p{3.5cm}}
\hline
Property & Bentheimer & Doddington \\
\hline
Rock type & Quartzose sandstone & Quartzose sandstone \\
Geological group & Bentheim Sandstone & Fell Sandstone Formation \\
Place of origin & Bad Bentheim, Germany & Doddington, UK \\
Age (Million years) & 133--140 & 343--339 \\
Effective porosity & 0.20 & 0.192 \\
Permeability (m$^2$) & 1.875$\times$10$^{-12}$ & 1.038$\times$10$^{-12}$ \\
\hline
\end{tabular}
\label{tab:sandstone-properties}
\end{table}

\begin{table}[htbp]
\centering
\small
\caption{Properties of limestone samples}
\begin{tabular}{>{\raggedright\arraybackslash}p{3.5cm}>{\raggedright\arraybackslash}p{3.5cm}>{\raggedright\arraybackslash}p{3.5cm}}
\hline
Property & Estaillades & Ketton \\
\hline
Rock type & Bioclastic limestone & Ooidal limestone \\
Geological group & Estaillade Formation & Upper Lincolnshire \\
Place of origin & Oppède, France & Ketton, Rutland, UK \\
Age (Million years) & 22 & 169--176 \\
Effective porosity & 0.295 & 0.2337 \\
Permeability (m$^2$) & 1.490$\times$10$^{-13}$ & 2.807$\times$10$^{-12}$ \\
\hline
\end{tabular}
\label{tab:limestone-properties}
\end{table}

\begin{figure}
    \centering
    \includegraphics[width=1.0\textwidth]{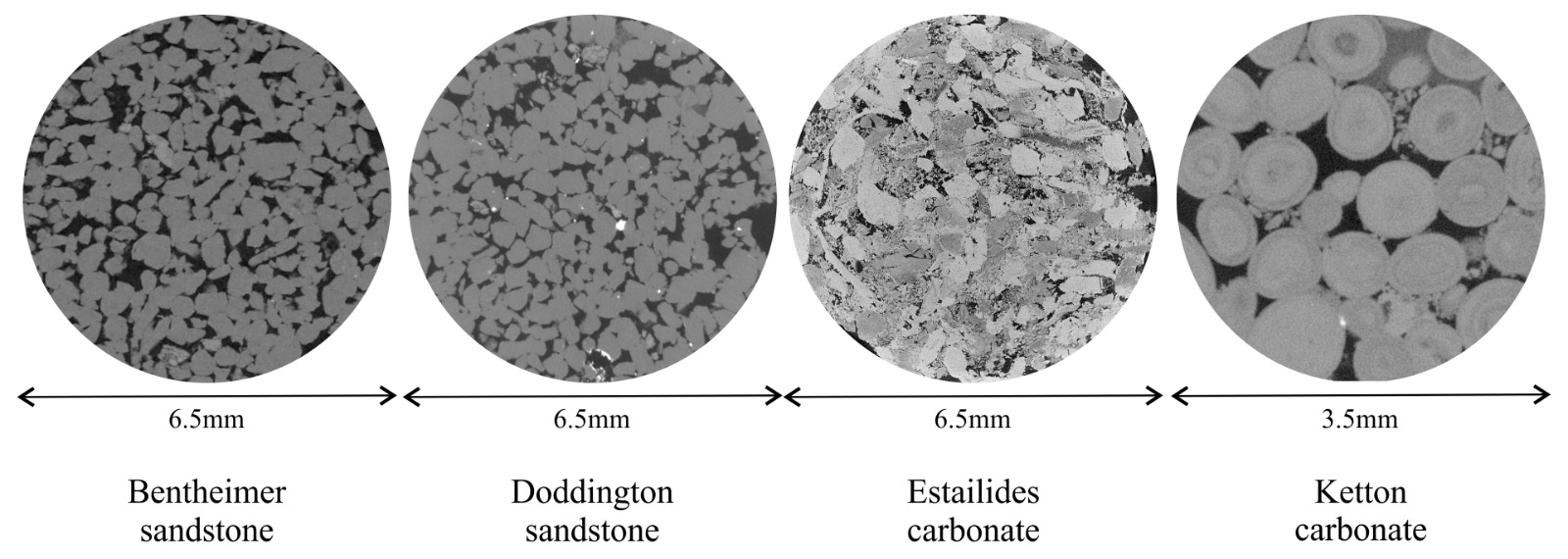}
\caption{\label{fig:slices} Grayscale 2D Slices of the Four Rock Types Used in This Study}
\end{figure}

Data used in this work satisfies the following properties: (i) rocks typically found as reservoirs; (ii) diverse compositional variations (quartz- and carbonate-rich rocks), degrees of heterogeneity and pore system complexity, including varying pore sizes, shapes and morphologies; (iii) well-known rocks with established porosity and permeability parameters. These properties enable the modeling of typical reservoir pore systems and investigation of how different mineralogical compositions and pore characteristics influence the system while providing reliable quantitative data to validate the generated models. Each $1000^3$ volume was partitioned into two datasets: one of size $(800, 1000, 1000)$ for training and another of size $(200, 1000, 1000)$ for validation, following standard machine learning practices to ensure the learned features generalize to unseen data. Grayscale 2D images of all four rock types can be seen in Figure \ref{fig:slices}.

\subsection{Network architectures}

\subsubsection{Autoencoder} \label{sunbsection: autoencoder architecture}
Our autoencoder architecture closely follows that in \citep{rombach2022high}, with the crucial modification of making the convolutional layers 3D instead of 2D. Also, the self-attention component was removed to improve efficiency, since we did not observe significant reconstruction improvements by its use. We attribute this phenomenon to the local character, i.e. the rapid decay of correlation statistics of pore-scale images, which can thus be fully captured by convolutional layers\footnote{One of the main advantages of attention layers is precisely the ability to efficiently capture long-range correlations \citep{attn_is_all}.}.

The dimension reduction performed by the autoencoder is linear in each dimension, and does not alter the tensorial dimension of the data in order to retain spatial structure in the latent space. That is, for a chosen reduction factor $F$, an input of dimension $[B, C_{in}, L, W, H]$ is encoded to an output $[B, C_l, \frac{L}{F}, \frac{W}{F}, \frac{H}{F}]$, where $B$ is the batch size, $C_{in}$ the number of input channels, $C_l$ the number of latent channels, and $L, W, H$ the spatial dimensions. For a binary image, we have $C_{in}=1$, and following \citep{rombach2022high} we take $C_l=4$. We choose $F=8$ for our reduction factor, as larger $F$ were observed to produce visually perceptible reconstruction errors. The autoencoder architecture is basically the same as in \citep{rombach2022high}, and we also take a very small regularization hyperparameter $\lambda_{\text{KL}}$, setting it to $\lambda_{\text{KL}}=10^{-4}$.

\subsubsection{Diffusion model}\label{diffusion network}
\begin{figure}
    \centering
    \includegraphics[width=\textwidth]{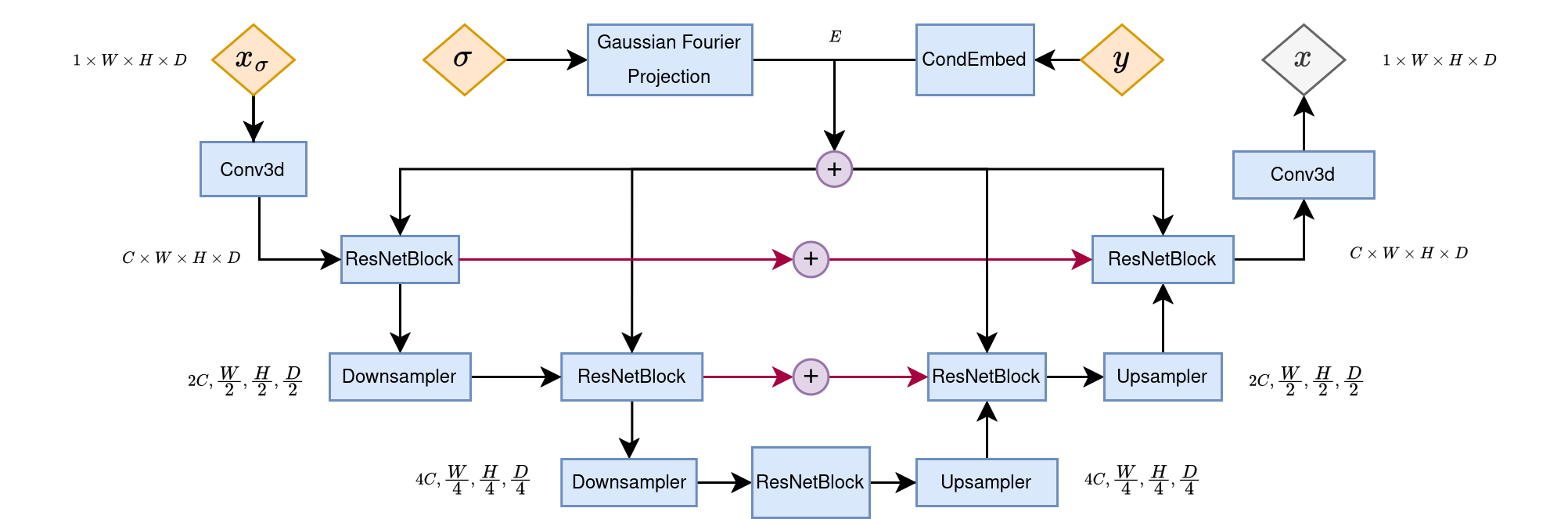}
    \caption{\label{fig:diffmodelouter} Outer part of the PUNet architecture, designed for 1-channel data. In our work, we let $C=64$, resulting in a neural network with 29.6 million parameters.}
\end{figure}

Our diffusion network architecture, named PUNet, is a 3D UNet \citep{ronneberger2015u} architecture, as it is commonly used for volume-based diffusion models. Its general outline is shown in Figure \ref{fig:diffmodelouter}. We point out that it also does not deploy self-attention, greatly improving computational performance while maintaining good performance. The architecture details can be found in \ref{architecturaldetails}.

\subsection{Training}

 In this work, following \citep{karras2022elucidating} we set the loss weighting function of equation \eqref{DSM loss} to $\lambda(\sigma) = (\sigma^2 + \sigma_{data}^2)/(\sigma \sigma_{data})^2$. Moreover, the distribution of $\sigma$ in the expectation is set to $\log \sigma \sim \mathcal{N}(-1.2, (1.2)^2 I)$ for pixel-space diffusion and to $\log \sigma \sim \mathcal{N}(-0.4, I)$ for latent diffusion, following \citep{karras2022elucidating, karras2024analyzingimprovingtrainingdynamics}. More specific details are left to \ref{app: training details}

\subsection{Sampling}

All samples shown in this work are generated with the Probability Flow ODE and $50$ discretization steps, following the remaining choices of sampling parameters of \citep{karras2022elucidating}. We also experimented with Song's reverse-time SDE and Karras's stochastic sampler, using $256$ steps, with generally less reliable generated statistics, especially for models less accurately trained. However, we do not discard further investigation of these sampling methods in future work.

\section{Results and discussion}\label{sec:results}

\subsection{Latent vs pixel-space diffusion}\label{subsection: pixel vs latent}

\begin{figure}[H]
    \centering
    
    % Bentheimer
    \begin{subfigure}{0.8\textwidth}
        \centering
        \includegraphics[width=\textwidth]{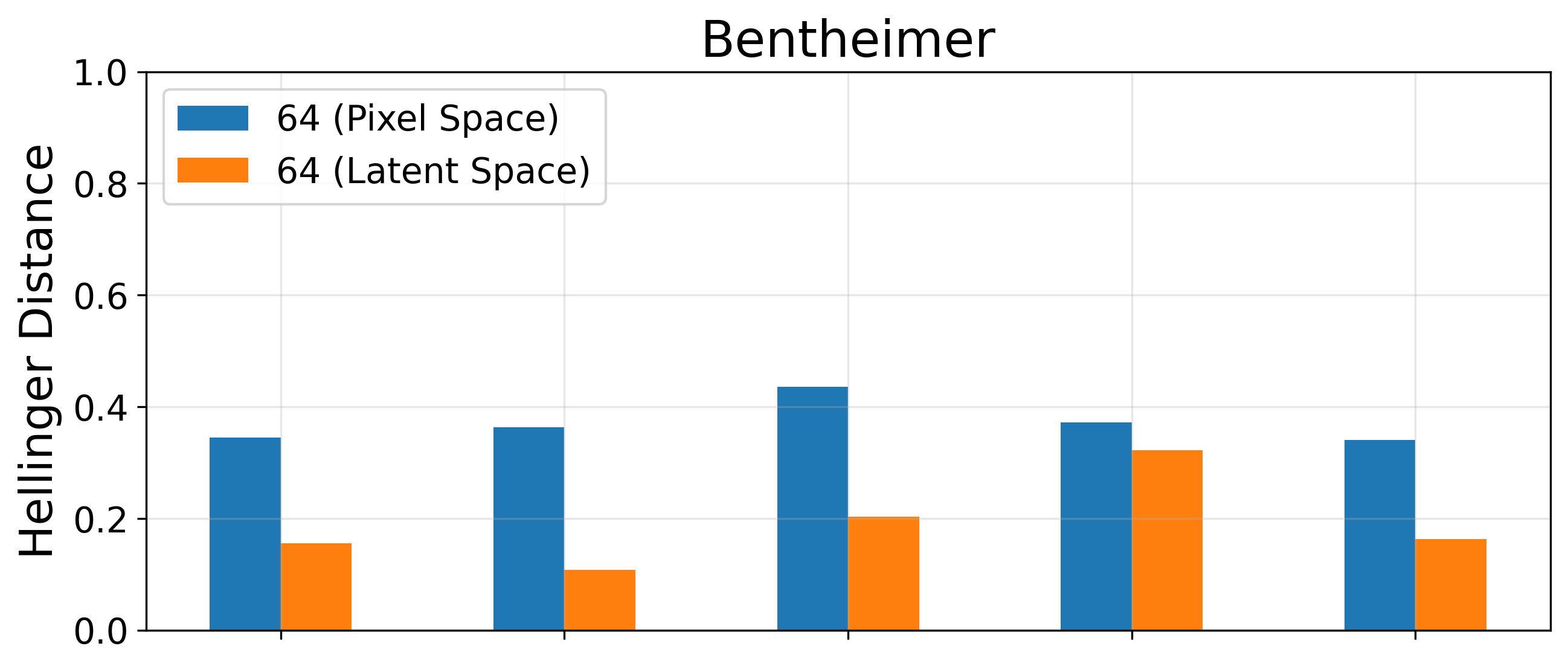}
    \end{subfigure}
    
    % Doddington
    \begin{subfigure}{0.8\textwidth}
        \centering
        \includegraphics[width=\textwidth]{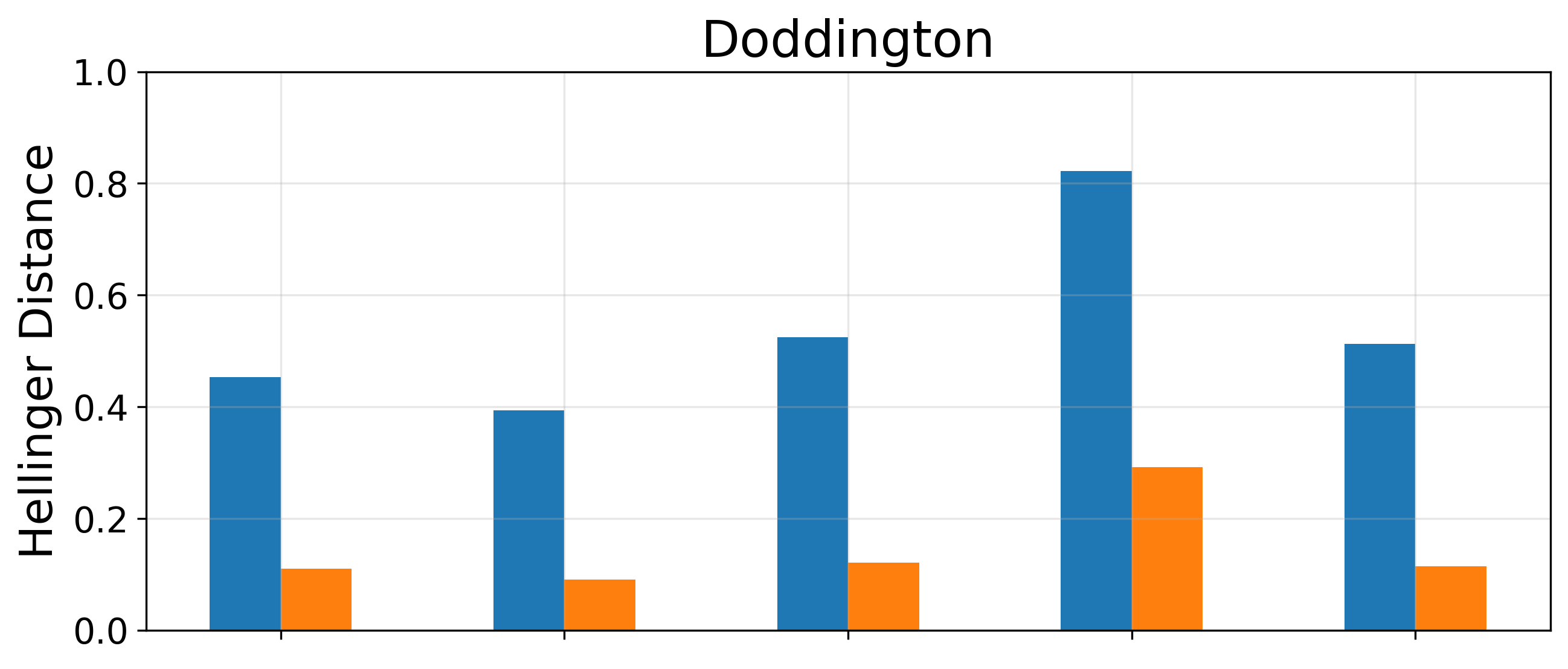}
    \end{subfigure}
    
    % Estaillades
    \begin{subfigure}{0.8\textwidth}
        \centering
        \includegraphics[width=\textwidth]{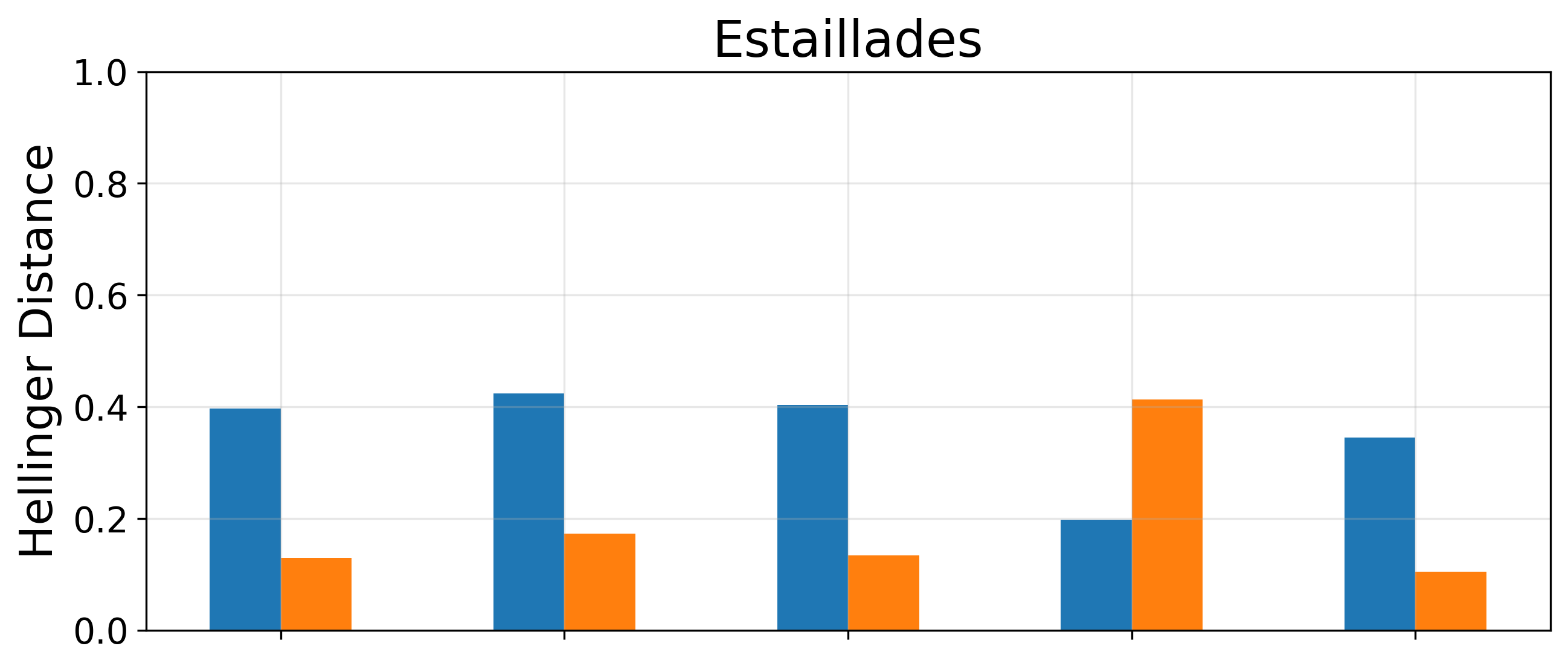}
    \end{subfigure}
    
    % Ketton
    \begin{subfigure}{0.8\textwidth}
        \centering
        \includegraphics[width=\textwidth]{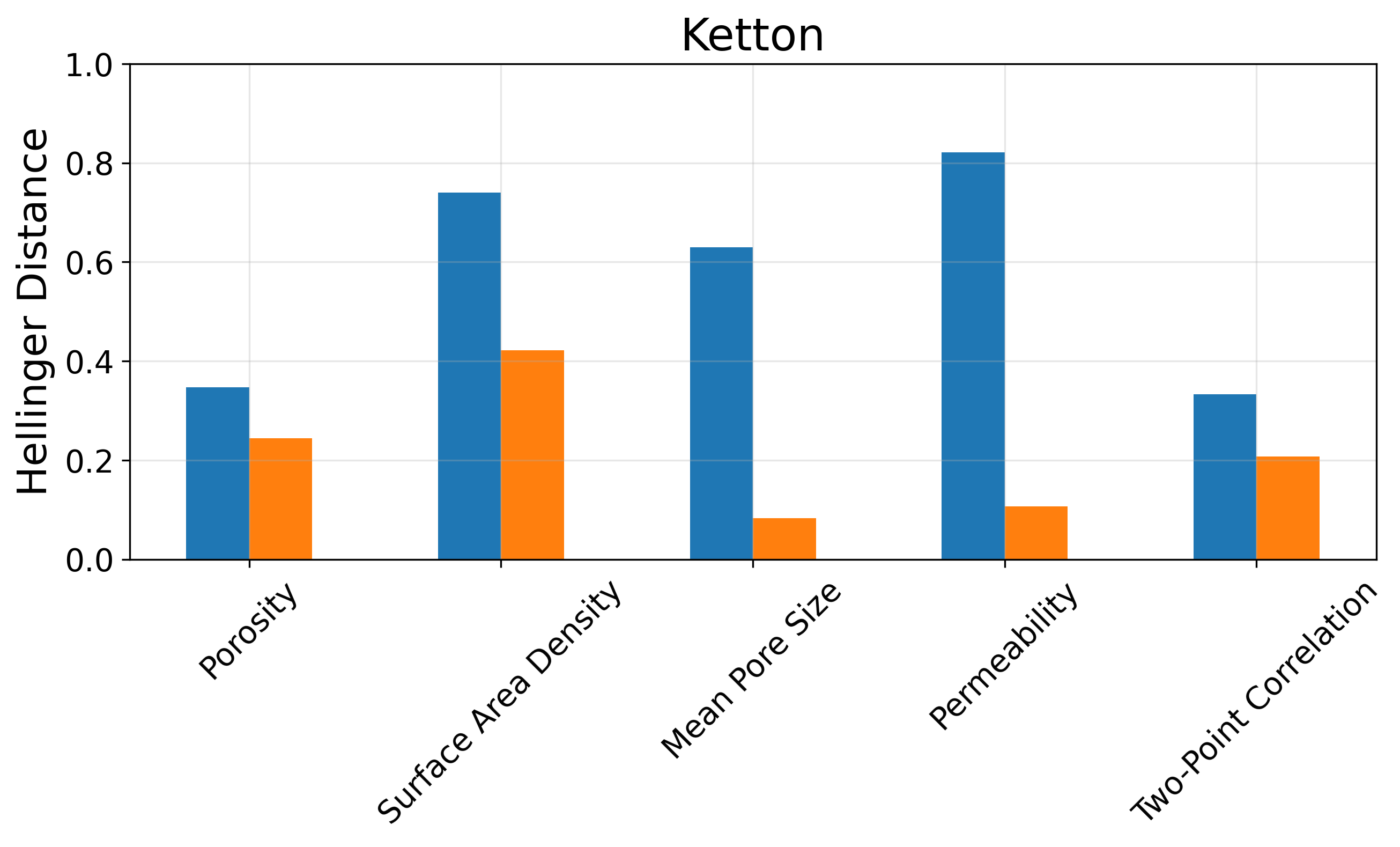}
    \end{subfigure}
    
    \caption{Hellinger distance of the statistics for $64^3$ volumes generated by pixel-space and latent diffusion, for different rock types.}
    \label{fig:hellinger_mre_lower}
\end{figure}

\begin{figure}[H]
    \centering
    % Top row
    \begin{subfigure}[t]{0.48\textwidth}
        \includegraphics[width=\textwidth]{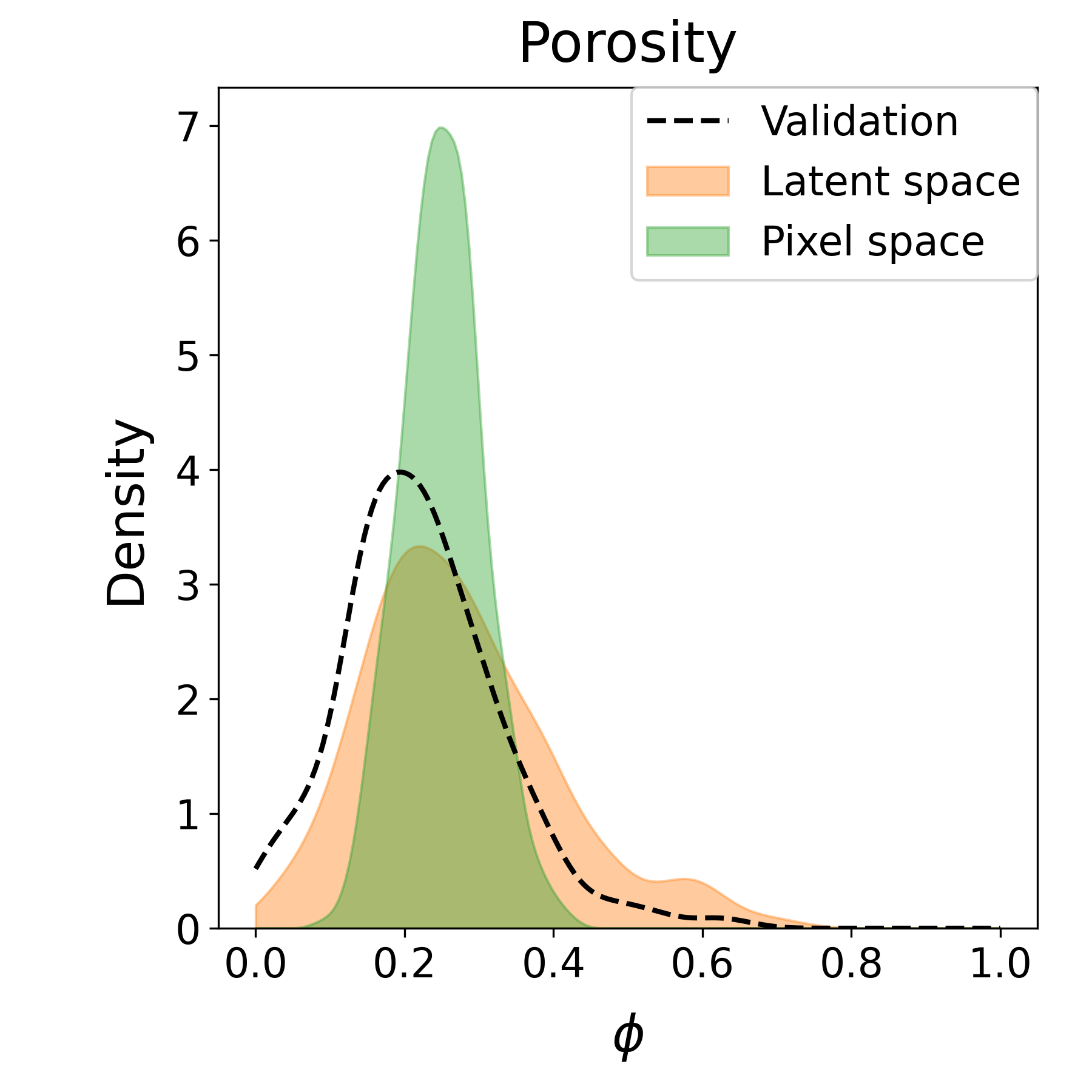}
        \caption{}
        \label{fig:64_bentheimer_porosity}
    \end{subfigure}
    \begin{subfigure}[t]{0.48\textwidth}
        \includegraphics[width=\textwidth]{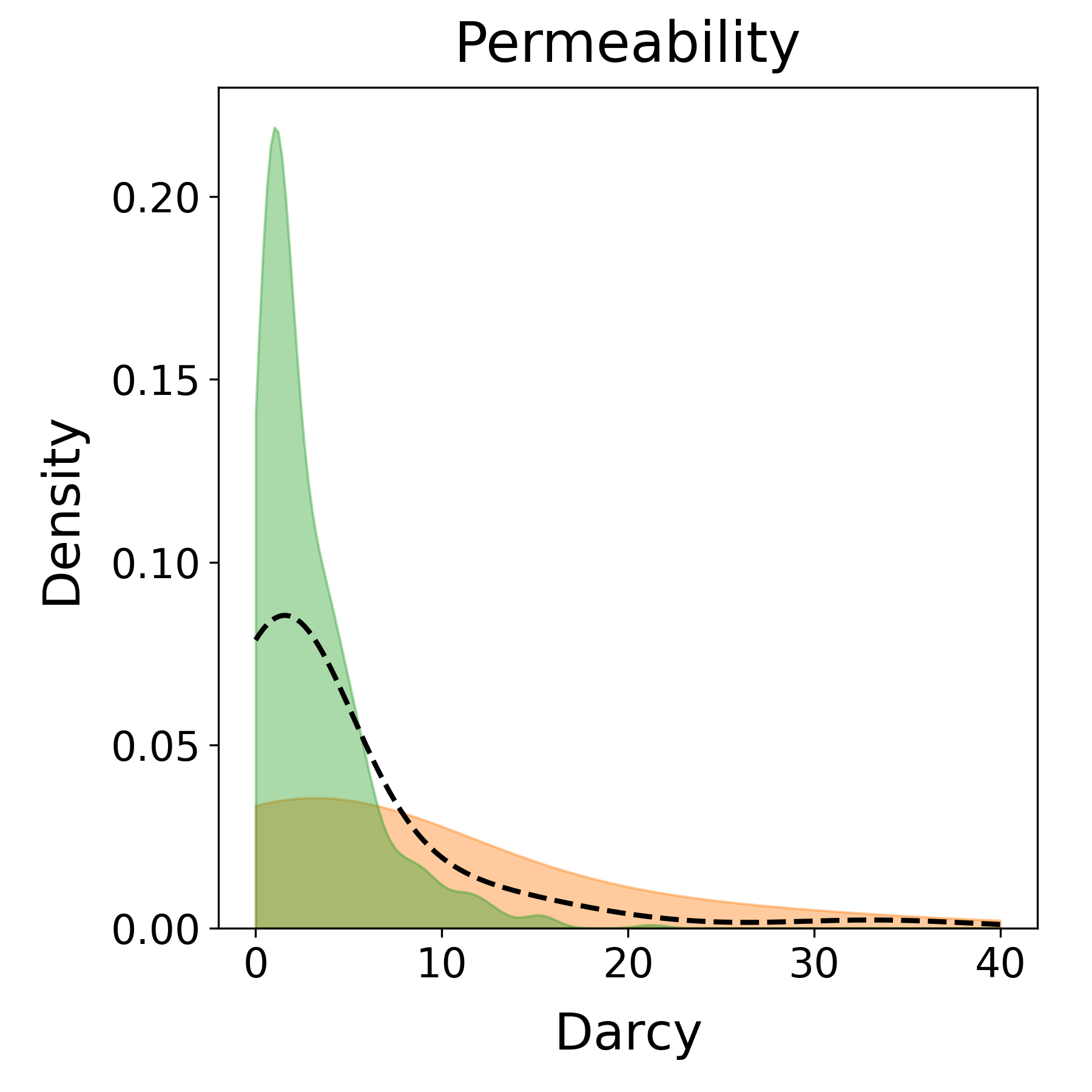}
        \caption{}
        \label{fig:64_bentheimer_perm}
    \end{subfigure}
    
    % Bottom row
    \begin{subfigure}[t]{0.48\textwidth}
        \includegraphics[width=\textwidth]{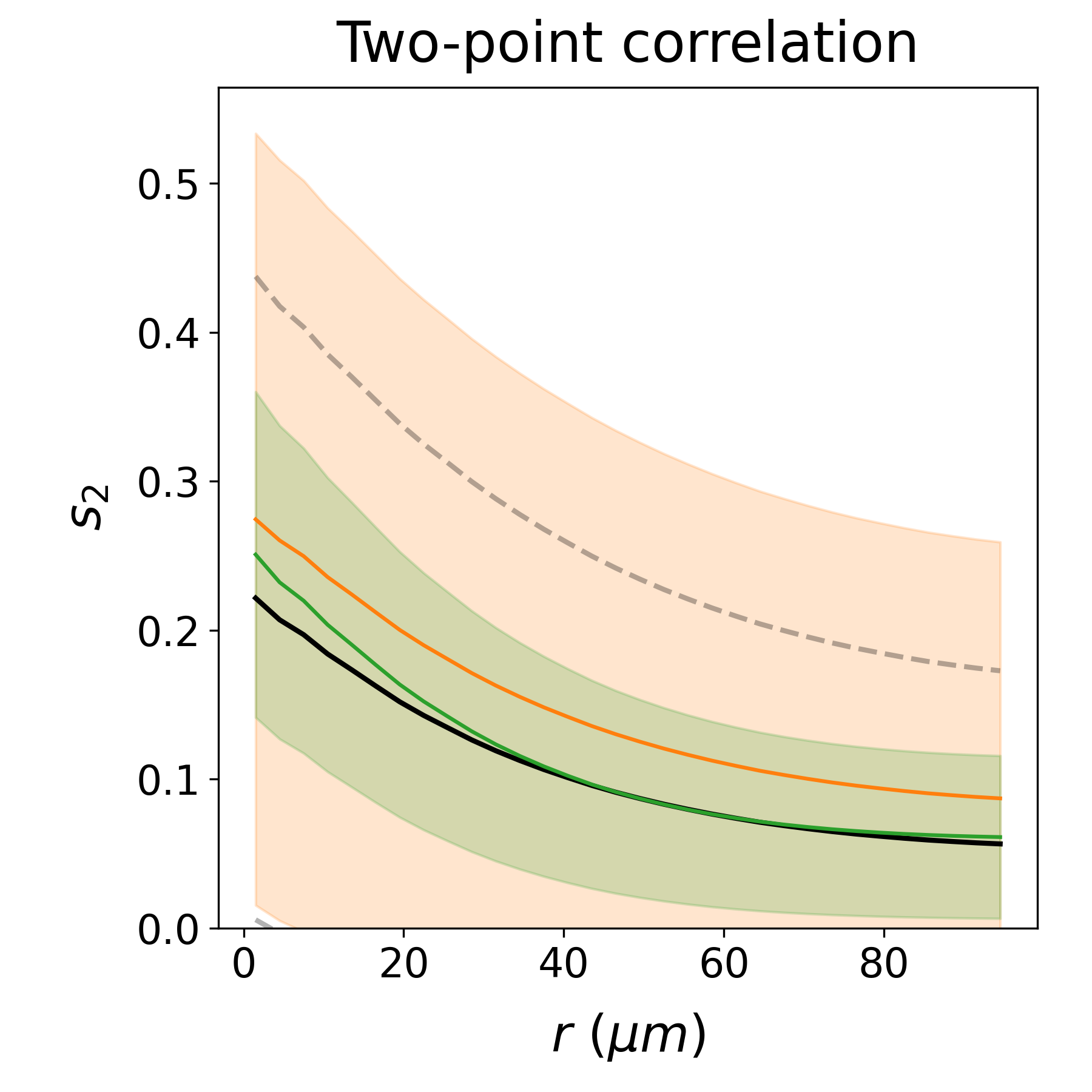}
        \caption{}
        \label{fig:64_bentheimer_tpc}
    \end{subfigure}
    \begin{subfigure}[t]{0.48\textwidth}
        \includegraphics[width=\textwidth]{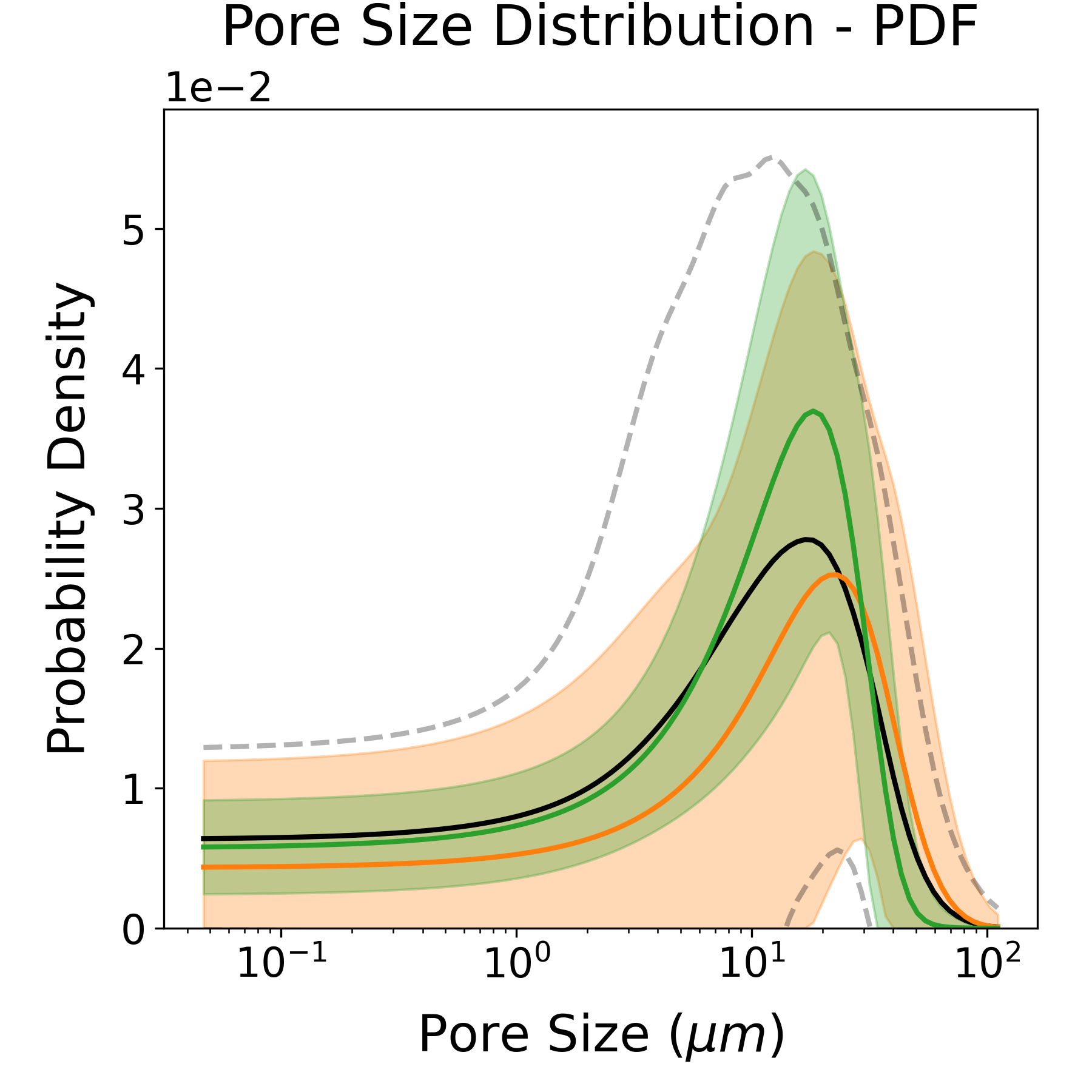}
        \caption{}
        \label{fig:64_bentheimer_psd}
    \end{subfigure}
    
    \caption{Statistical properties for Bentheimer sandstone at $64^3$ size: (a) porosity distribution, (b) permeability distribution, (c) two-point correlation function, and (d) pore size distribution.}
    \label{fig:bentheimer_64}
\end{figure}

For volumes up to $64^3$, it is possible to train a diffusion model directly on pixel space, as done in \citep{luo2024multi}, without the dimension reduction provided by the autoencoder. We compare the statistics of samples generated with pixel-space diffusion and latent diffusion to assess the impact of dimension reduction in the pipeline. Interestingly, latent diffusion yields significantly better results than pixel-space diffusion, as shown in Figure \ref{fig:hellinger_mre_lower}. The only negative impact is on the permeability for Estaillades, which might not be very reliable for $64^3$ volumes. We show the complete distributions for Bentheimer samples in Figure \ref{fig:bentheimer_64}, and similar figures for the remaining rocks can be found in \ref{app: all experiments}.

We will use the Hellinger distance, a standard statistical measure of the distance between two probability distributions, as our main numerical metric to evaluate the quality of the generated samples. Other numerical metrics are often used by previous work, such as the mean relative error (MRE) in \citep{zhu2024generation}. We observe that a mean-based metric is incapable of measuring a generative model`s coverage of all properties of the data distribution, and in particular, it completely misses the phenomenon of mode collapse, which consists of the concentration of a probability distribution near its mode, a major concern in this work. For completeness, we also calculate the MRE for our distributions, which can be found in \ref{fig:mre_stats}. The statistics are described in detail in \ref{section:evaluation_statistics}, and the definitions of the numerical metrics can be found in \ref{section:statistical_comparison}.

% One possible explanation is data scarcity. The lower-dimensional encoded distribution might need fewer data points to be sufficiently specified than the original one, which would give latent diffusion an important advantage when dealing with scarce data. This phenomenon may not have been explored in previous ML literature because data scarcity is rarely a concern in standard ML benchmarks.

%%\footnote{We briefly speculate here why this may be so. A well-observed phenomenon in ML literature\citep{manifold-hypothesis} is that real-world data probability distributions usually reside on lower dimensional manifolds inside the total configuration space. This is indeed one of the original motivations behind diffusion models, since diffusion makes the support of the distribution be the entire space, thus allowing to take $\nabla\log p$ of any distribution $p$. However, }.

%% What happens if we increase model size together with an adaptive loss weighting?

\subsection{Autoencoder}

\begin{figure}[htbp]
    \centering
    \begin{subfigure}[t]{0.95\textwidth}
        \centering
        \includegraphics[width=\textwidth]{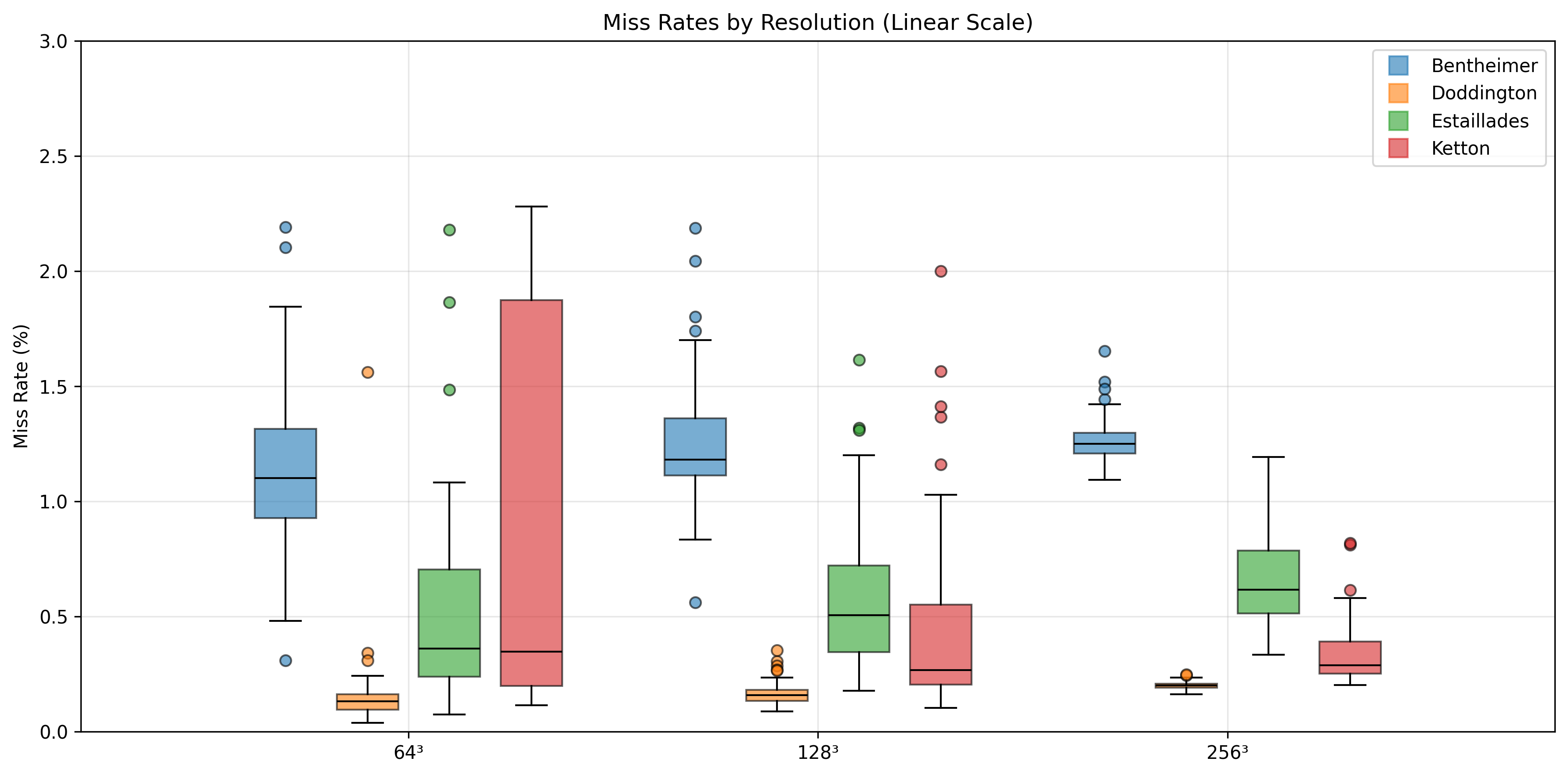}
        \caption{Boxplots of reconstruction error up to 3\% for different rock types and volume sizes.}
        \label{fig:miss_rates}
    \end{subfigure}
    
    \begin{subfigure}[t]{0.95\textwidth}
        \centering
        \includegraphics[width=\textwidth]{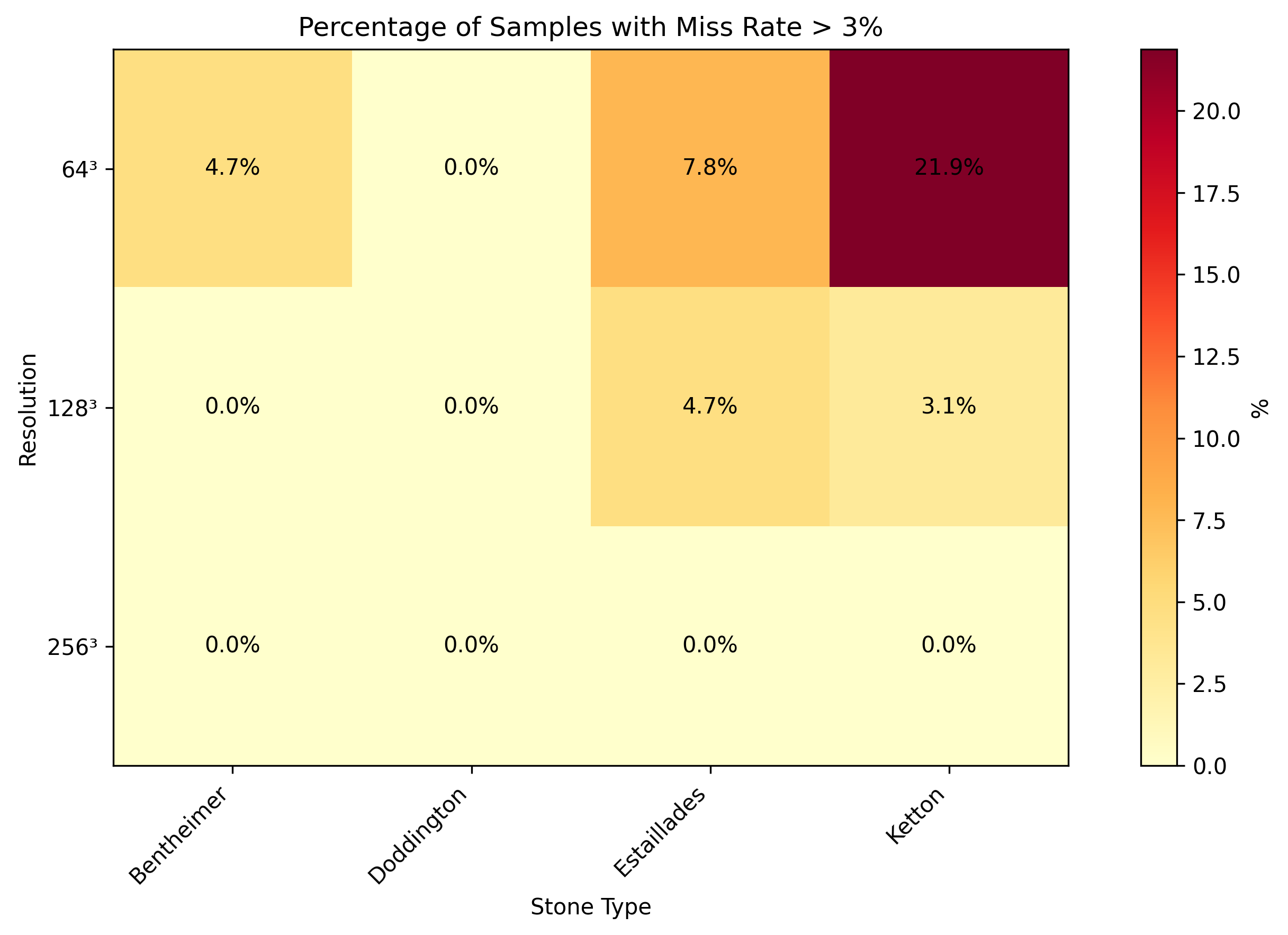}
        \caption{Percentage of reconstructions with error rates exceeding 3\%.}
        \label{fig:miss_table}
    \end{subfigure}
    \caption{Analysis of autoencoder reconstruction error for different rocks.}
    \label{fig:reconstruction_analysis}
\end{figure}

We trained a single autoencoder on all four rock types using subvolumes of size \( 64^3 \). As detailed in Section \ref{sunbsection: autoencoder architecture}, our autoencoder employs a fully convolutional architecture, enabling it to accurately reconstruct volumes of arbitrary size. Figure \ref{fig:reconstruction_analysis} presents the reconstruction errors for different datasets at volumes of \( 64^3 \), \( 128^3 \), and \( 256^3 \). Notably, the error percentage decreases as the volume size increases. This trend can be attributed to the reduced impact of rare patterns in larger volumes. Since these patterns occur infrequently in the training set, the model does not learn their encoding and reconstruction as effectively, leading to higher reconstruction errors in smaller volumes.  

As a side note, we report that the autoencoder was found to have an unexpected capacity to reconstruct rock types outside the training set, as can be seen in \ref{app: all experiments}, Figures \ref{fig:miss_rates} and \ref{fig:miss_table}. This suggests that a single general 3D autoencoder for porous media might perform successfully for any rock type, even when its training set is not sufficiently diverse.

% General autoencoder errors

\subsection{Unconditional generation of $128^3$ volumes}\label{subsection: uncond 128}

In Section \ref{subsection: pixel vs latent}, we presented results for volumes of size $64^3$ to assess performance differences between pixel- and latent-space diffusion, but it is clear that this volume size is too small for metrics such as pore-size-distribution and permeability to be meaningful for the rock data considered. Here, we show results for the unconditional generation of volumes of size $128^3$, for all four rock types. This volume size is large enough for the computation of these metrics and, on the other hand, small enough to allow successful training for an unconditional diffusion model, since our dataset consists of a single volume of size $1000^3$ for each rock. These results also highlight the advantages of latent diffusion, since training pixel-space diffusion on $128^3$ volumes is not feasible on our hardware due to memory constraints, and, to our knowledge, has not been done in the literature.

\begin{figure}[H]
    \centering
    % Top row
    \begin{subfigure}[t]{0.48\textwidth}
        \includegraphics[width=\textwidth]{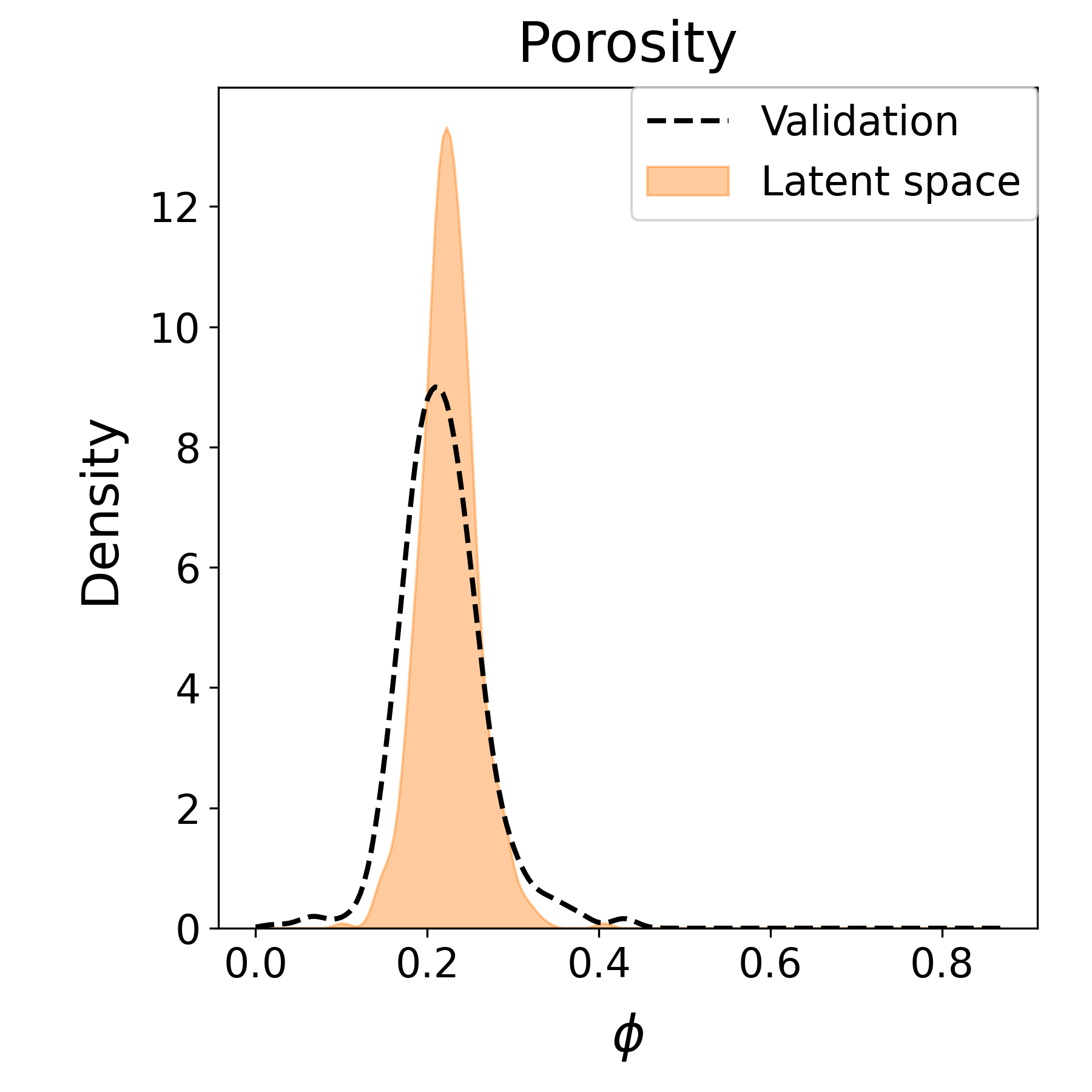}
        \caption{}
        \label{fig:128_bentheimer_porosity}
    \end{subfigure}
    \begin{subfigure}[t]{0.48\textwidth}
        \includegraphics[width=\textwidth]{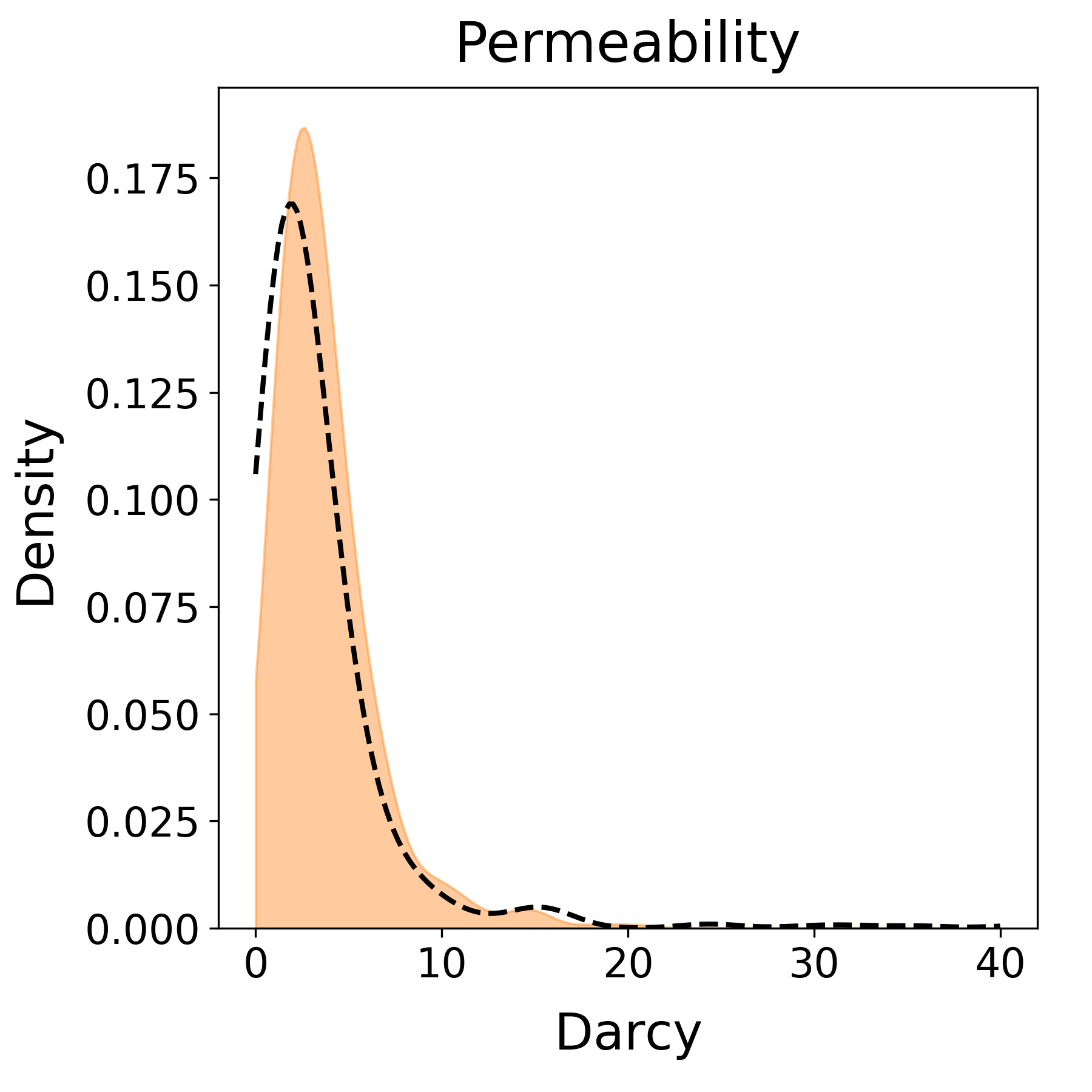}
        \caption{}
        \label{fig:128_bentheimer_perm}
    \end{subfigure}
    
    % Bottom row
    \begin{subfigure}[t]{0.48\textwidth}
        \includegraphics[width=\textwidth]{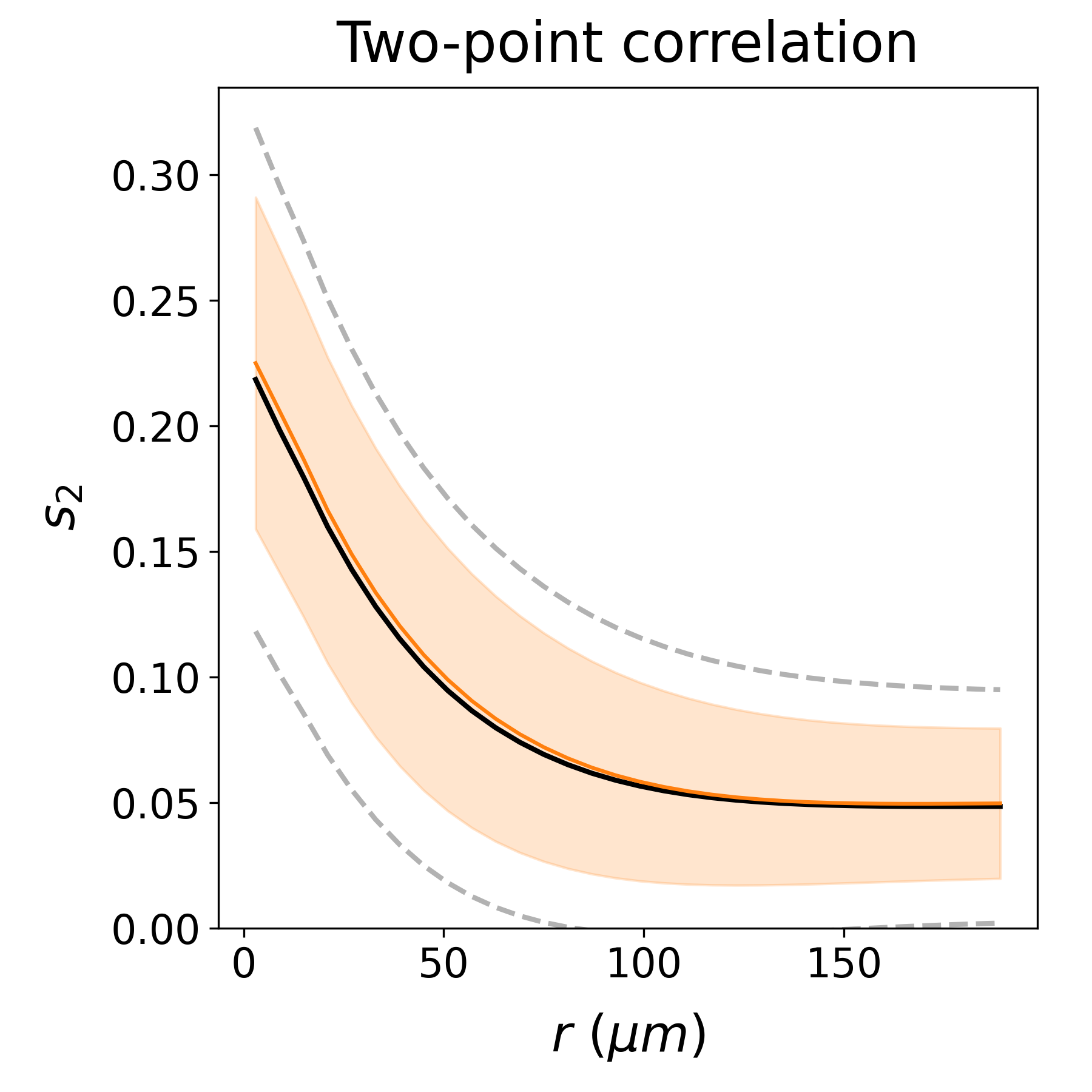}
        \caption{}
        \label{fig:128_bentheimer_tpc}
    \end{subfigure}
    \begin{subfigure}[t]{0.48\textwidth}
        \includegraphics[width=\textwidth]{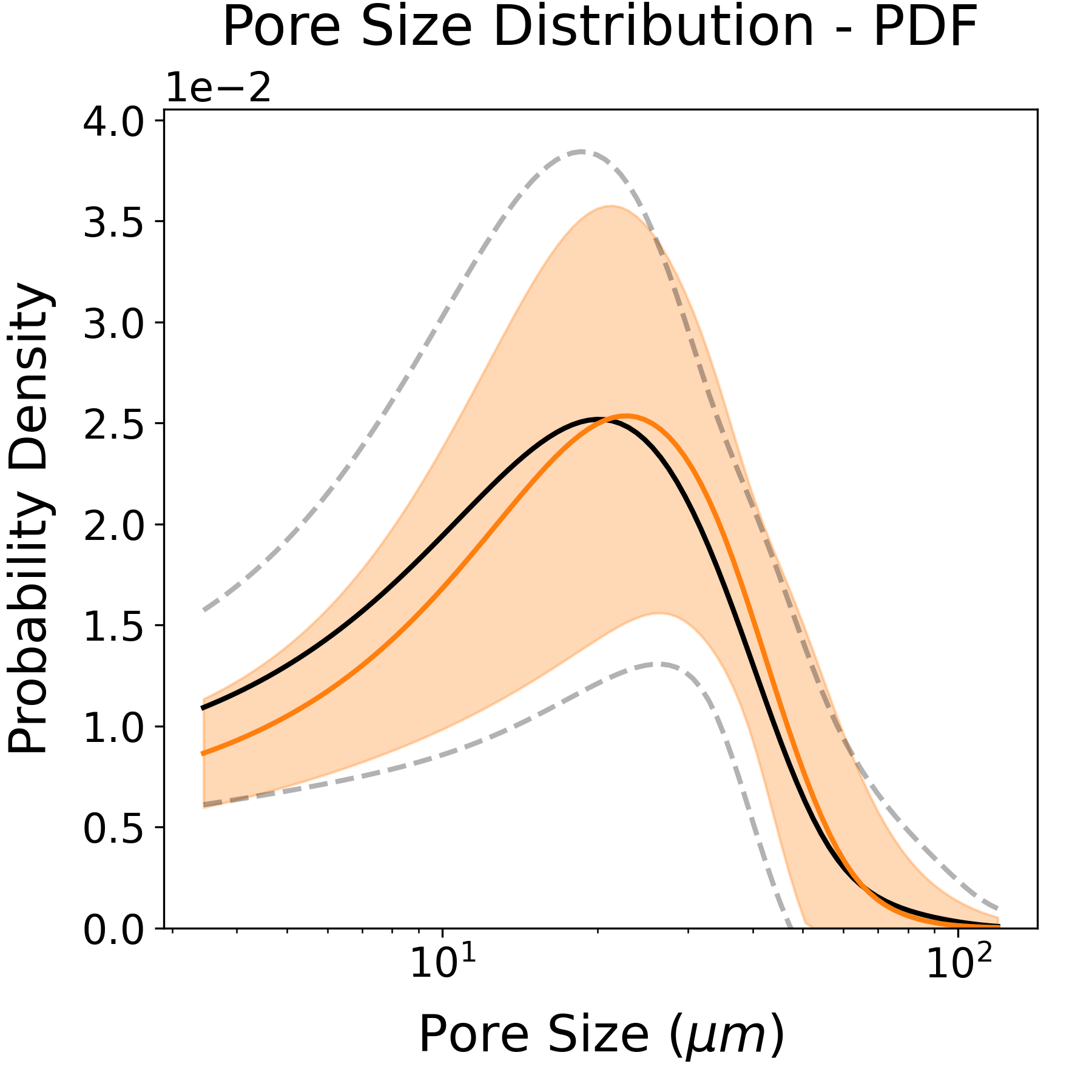}
        \caption{}
        \label{fig:128_bentheimer_psd}
    \end{subfigure}
    
    \caption{Statistical properties for Bentheimer sandstone at $128^3$ size: (a) porosity distribution, (b) permeability distribution, (c) two-point correlation function, and (d) pore size distribution.}
    \label{fig:bentheimer_128}
\end{figure}

\begin{figure}[H]
    \centering
    \includegraphics[width=\textwidth]{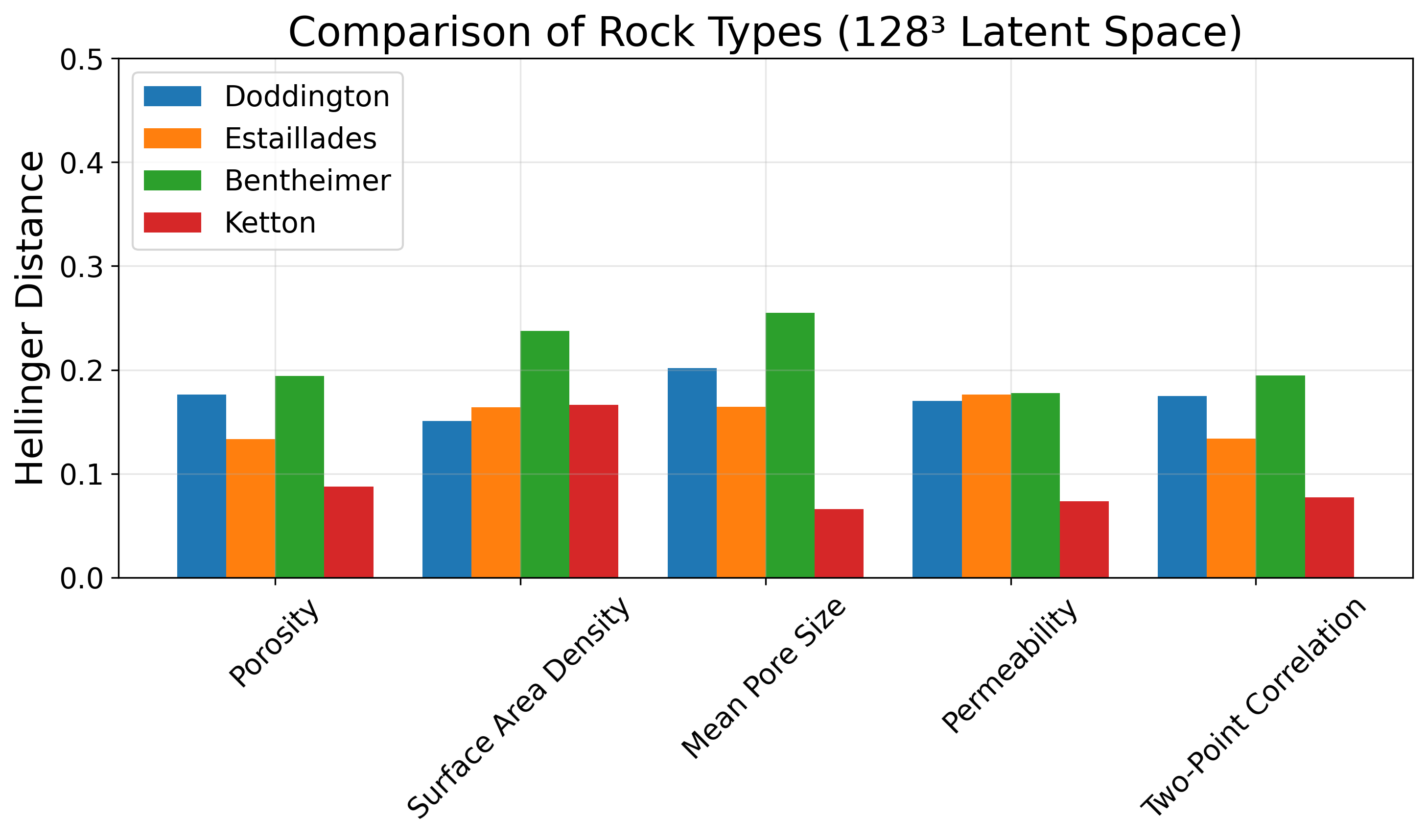}
    \caption{Hellinger distance of the statistics for $128^3$ samples generated by latent diffusion.}
    \label{fig:hellinger_mre_128}
\end{figure}

Figure \ref{fig:bentheimer_128} shows the distribution of the main statistics for Bentheimer $128^3$ volumes, calculated from 500 generated samples and 500 validation samples. In Figure \ref{fig:hellinger_mre_128}, we can further see that the generated samples of other rocks also have properties consistent with the data samples in all statistics considered: porosity distribution, two-point correlation, pore size distribution, pore surface ratio, and permeability. The distribution of statistics for the other rock types is shown in \ref{app: all experiments}.

\subsection{Conditional generation}\label{subsection: conditional results}

Volume generation conditioned on incomplete information about the data is another application of interest. As discussed in Section \ref{subsection: diffusion}, diffusion models are automatically suited for this task. A conditional model can be trained by passing to the network the data inputs together with the desired information, as described in \ref{diffusion network}.

\begin{figure}[ht]
    \centering
    \includegraphics[width=\textwidth]{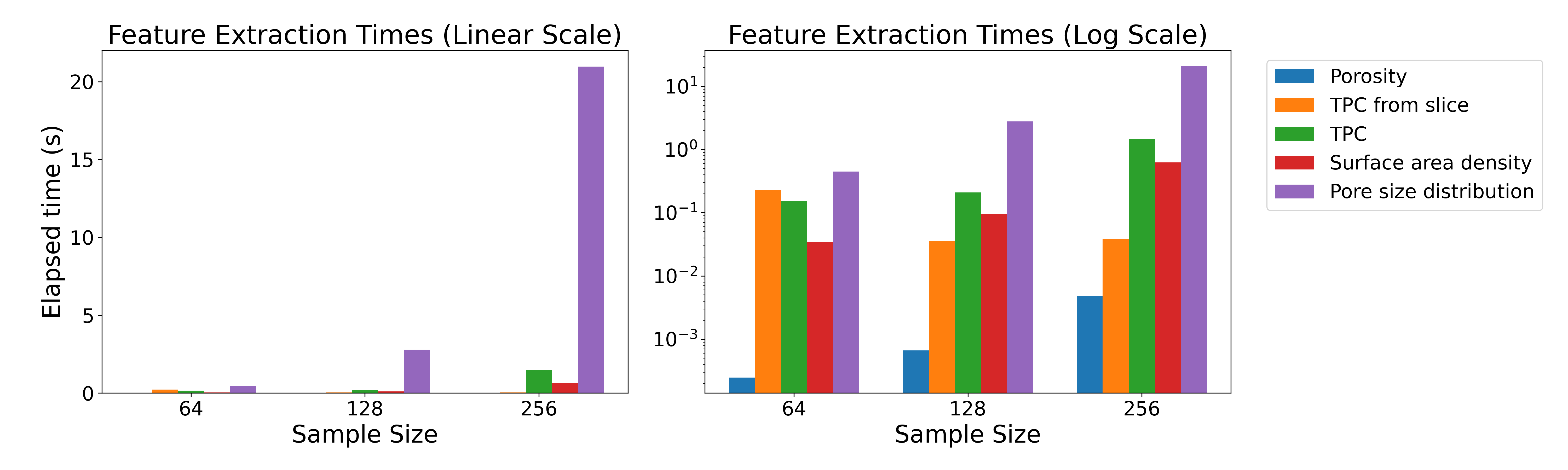}
    \caption{\label{fig:feature_extraction_time} Feature extraction time (using PoreSpy\citep{gostick2019porespy} in a 13th Gen Intel(R) Core(TM) i7-13650HX), as averaged from a random cubic volume of a binarized Bentheimer sandstone. Note that the figure in the right is in log scale.}
\end{figure}

We present results for generation conditioned on porosity and two-point correlation (TPC), as
estimated from a random slice of the rock volume. These two statistics can be computed at a relatively low computational cost, since the TPC from a slice gives an unbiased estimator for the TPC of the total volume, provided that the rock is isotropic. In \ref{appendix:isotropy_proof}, we show this is the case for the rocks studied in this work. Figure \ref{fig:feature_extraction_time} shows that extracting the surface area, pore size distribution, and two-point correlation from the full volume is unfeasible at training, especially for larger volumes\footnote{The TPC extraction from a slice for $256^3$ volumes already places a significant computational burden, essentially doubling training time.}.

\begin{figure}[H]
    \centering
    \includegraphics[width=0.95\textwidth]{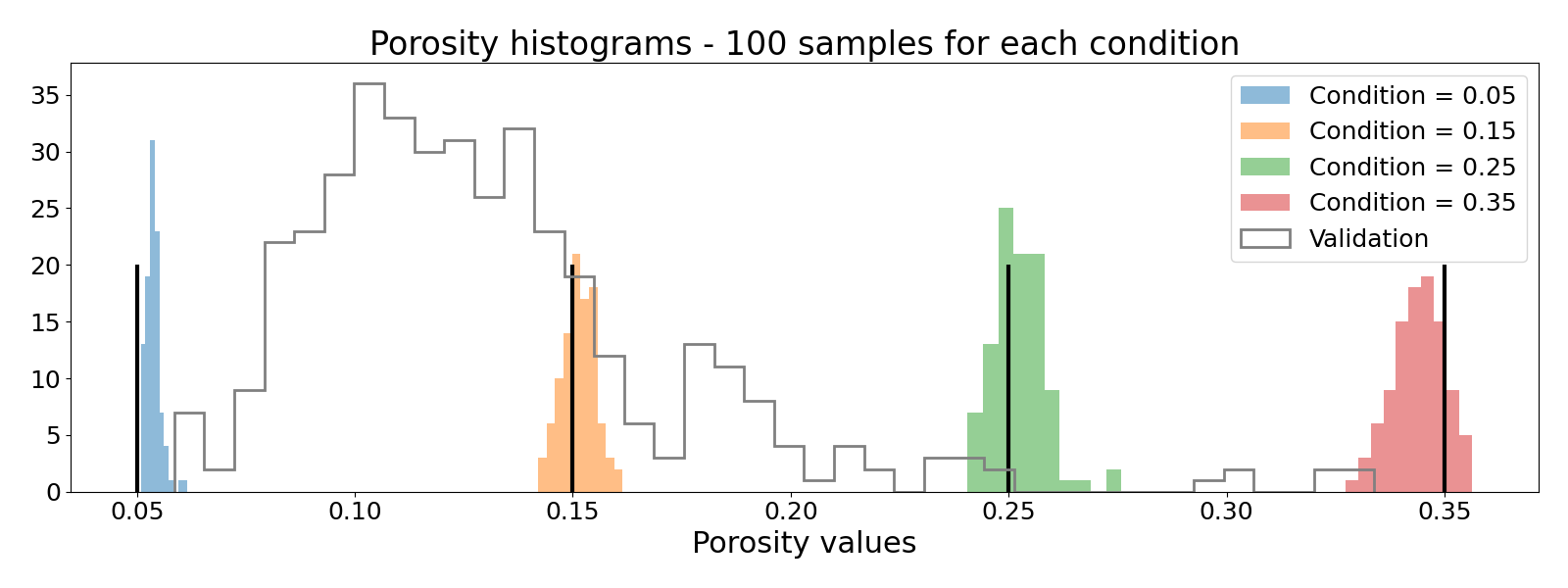}
    \includegraphics[width=0.95\textwidth]{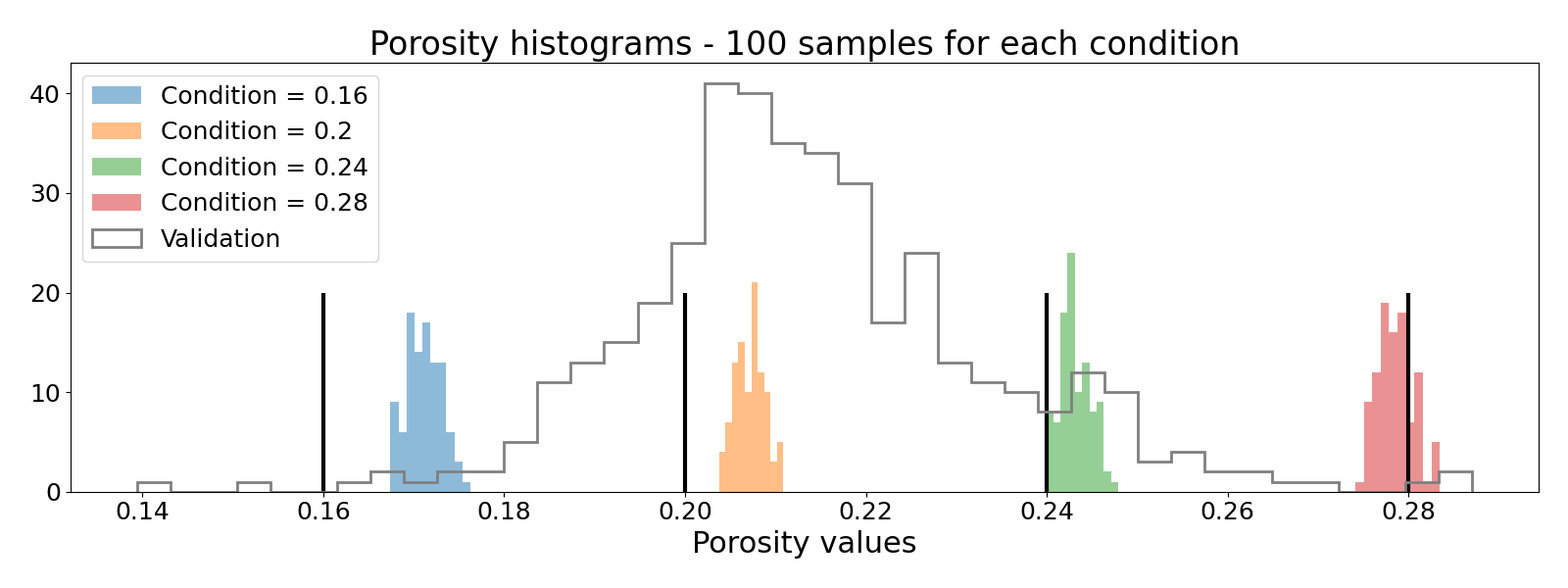}
    \caption{Porosity-conditioned samples for volumes of size $256^3$ of Estaillades(above) and Bentheimer (below), compared against the validation set. Black vertical lines indicate the conditioned value.}
    \label{fig:porosity conditional}
\end{figure}

Figure \ref{fig:porosity conditional} shows histograms for porosity-conditional generation, compared against the porosity histogram of the (unconditional) validation set. We show results for models trained on Bentheimer and Estaillades volumes of size $256^3$, which will be the conditional models used in \ref{sec: controlled} for controlled generation. As can be seen, the model generates samples with porosity near the conditioned value, with a slight bias for Bentheimer samples.

\begin{figure}[H]
    \centering
    \begin{subfigure}[t]{\textwidth}
        \includegraphics[width=\textwidth]{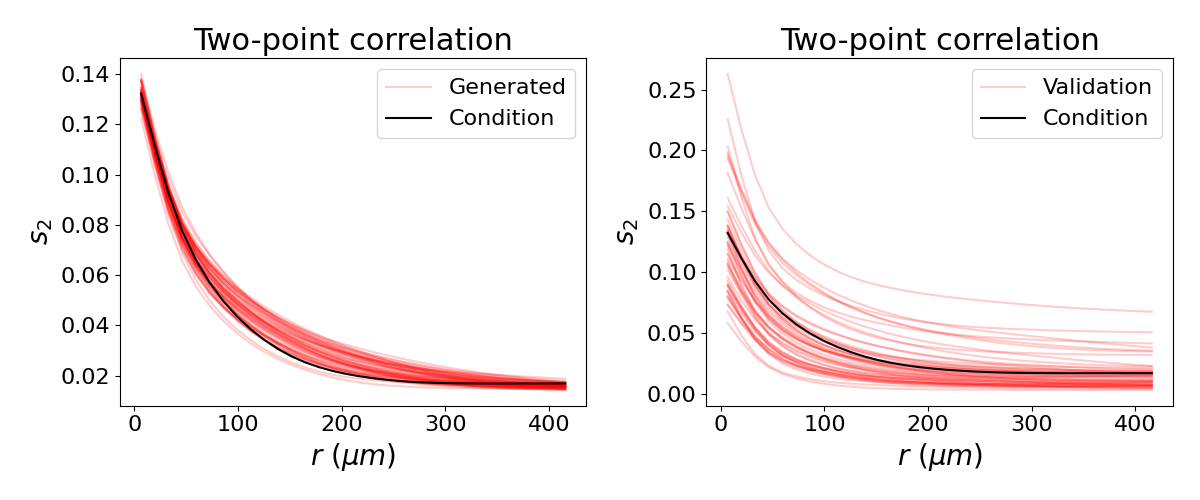}
        \caption{Unnormalized TPC curves from TPC-conditional generation (left) vs validation (right), compared against the considered TPC condition.}
    \end{subfigure}
    
    \begin{subfigure}[t]{\textwidth}
        \centering
        \includegraphics[width=0.4\textwidth]{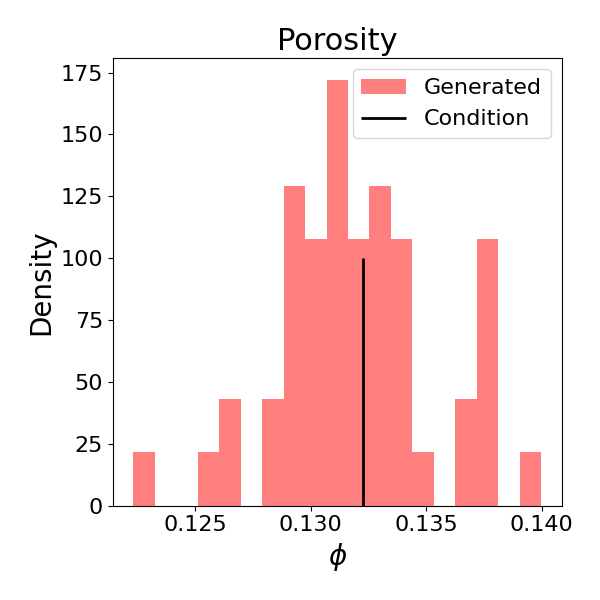}
        \caption{Porosity values from TPC-conditional generation.}
    \end{subfigure}
    \caption{\label{fig:tpc_conditional}Effect of conditioning on porosity and TPC curves for Estaillades \( 256^3 \) volumes, showcasing the model's multi-conditioning capability.}
\end{figure}

Figure \ref{fig:tpc_conditional} shows the effect of conditioning on a single pair of porosity and two-point-correlation curve, for Estaillades $256^3$ volumes, against a baseline of random samples from the validation. The conditions passed to the model are the porosity and the unnormalized TPC curve extracted from a random sample of the validation set. This highlights our model’s ability to incorporate multiple conditioning factors. In principle, we could also condition on other statistical measures, such as the mean and variance of the PSD, as done in \citep{luo2024multi} for \( 64^3 \) volumes. However, as previously discussed, computing the PSD becomes prohibitively expensive for \( 256^3 \) volumes. Furthermore, our transformer layer facilitates conditioning on more advanced statistical representations, such as a TPC curve, which, to the best of our knowledge, represents a novel approach in the geophysical generative modeling literature.

\subsection{Pushing the limits of size in image generation: \texorpdfstring{$256^3$}{256³} volumes}\label{256 generation}

Although the generation of $128^3$ volumes represents an important step in synthetic rock modeling, this size is not large enough to capture all relevant structure in the case of Ketton volumes, for example, due to its larger grain size. Motivated by this, we explore here the generation of larger images, which is made possible by the computational advantages of latent diffusion. For example, the latent representation of a volume of size $[1, 256, 256, 256]$ has dimensions $[4, 32, 32, 32]$, half of the total size of a $64^3$ volume in pixel-space.

However, since our dataset consists of a single $1000^3$ volume for each rock type, training a model on $256^3$ subvolumes becomes challenging due to data scarcity. In this section, we first present the problems found with direct unconditional generation, and then the solution given by controlled unconditional generation.

\subsubsection{Unconditional generation}

\begin{figure}[H]
    \centering
    \begin{subfigure}[t]{0.48\textwidth}
        \includegraphics[width=\textwidth]{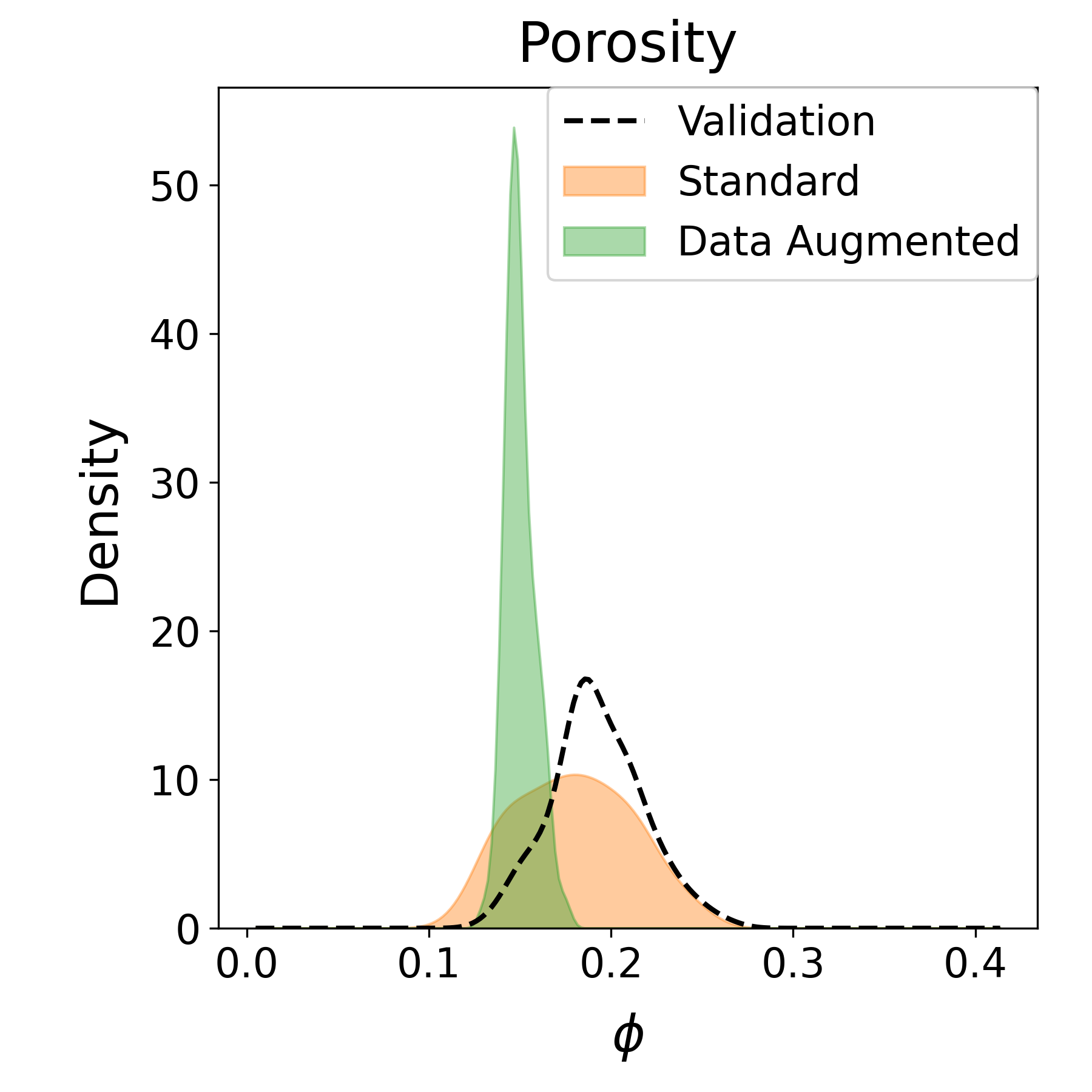}
        \caption{}
        \label{fig:doddington_256nc_a}
    \end{subfigure}
    \begin{subfigure}[t]{0.48\textwidth}
        \includegraphics[width=\textwidth]{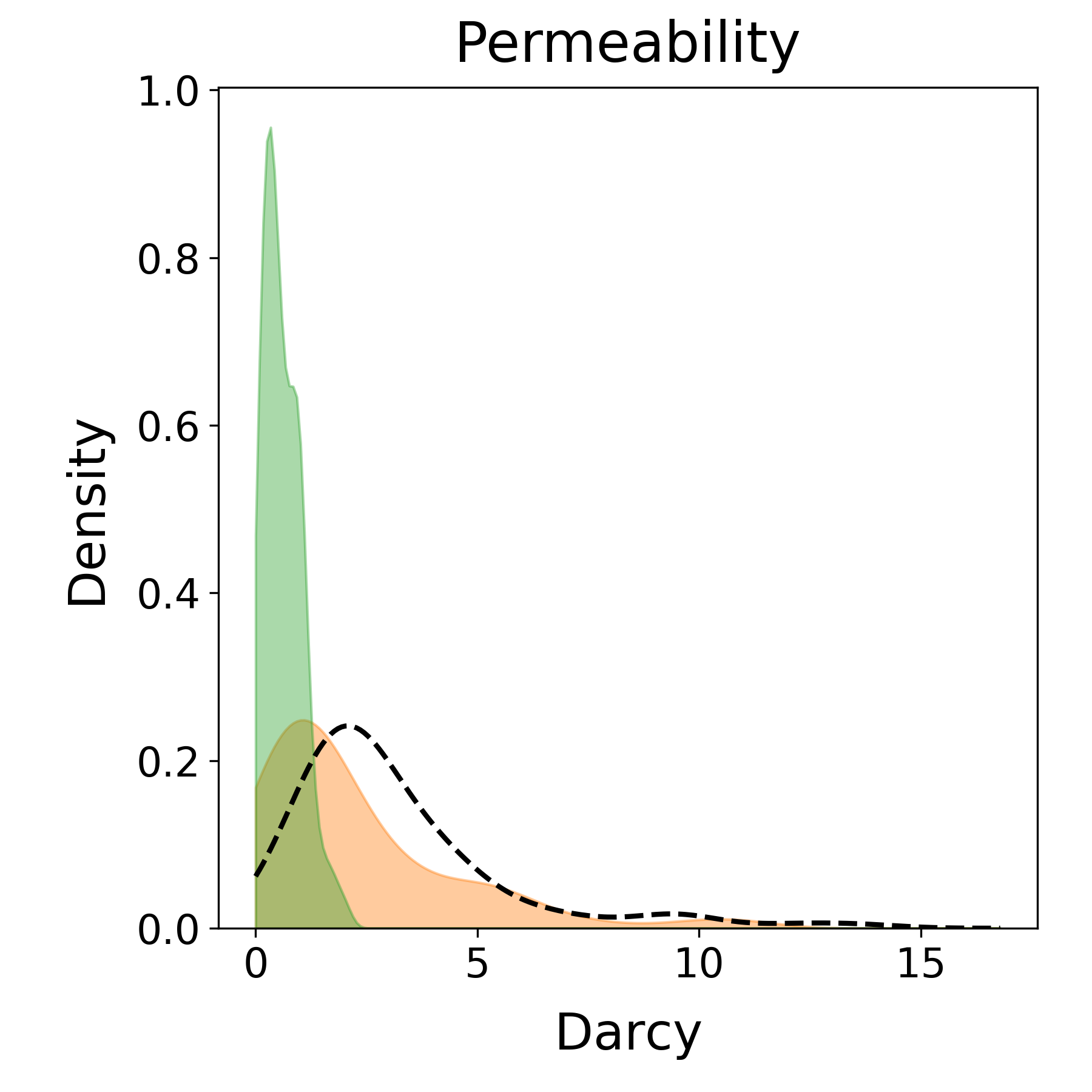}
        \caption{}
        \label{fig:doddington_256nc_b}
    \end{subfigure}
    
    \begin{subfigure}[t]{0.48\textwidth}
        \includegraphics[width=\textwidth]{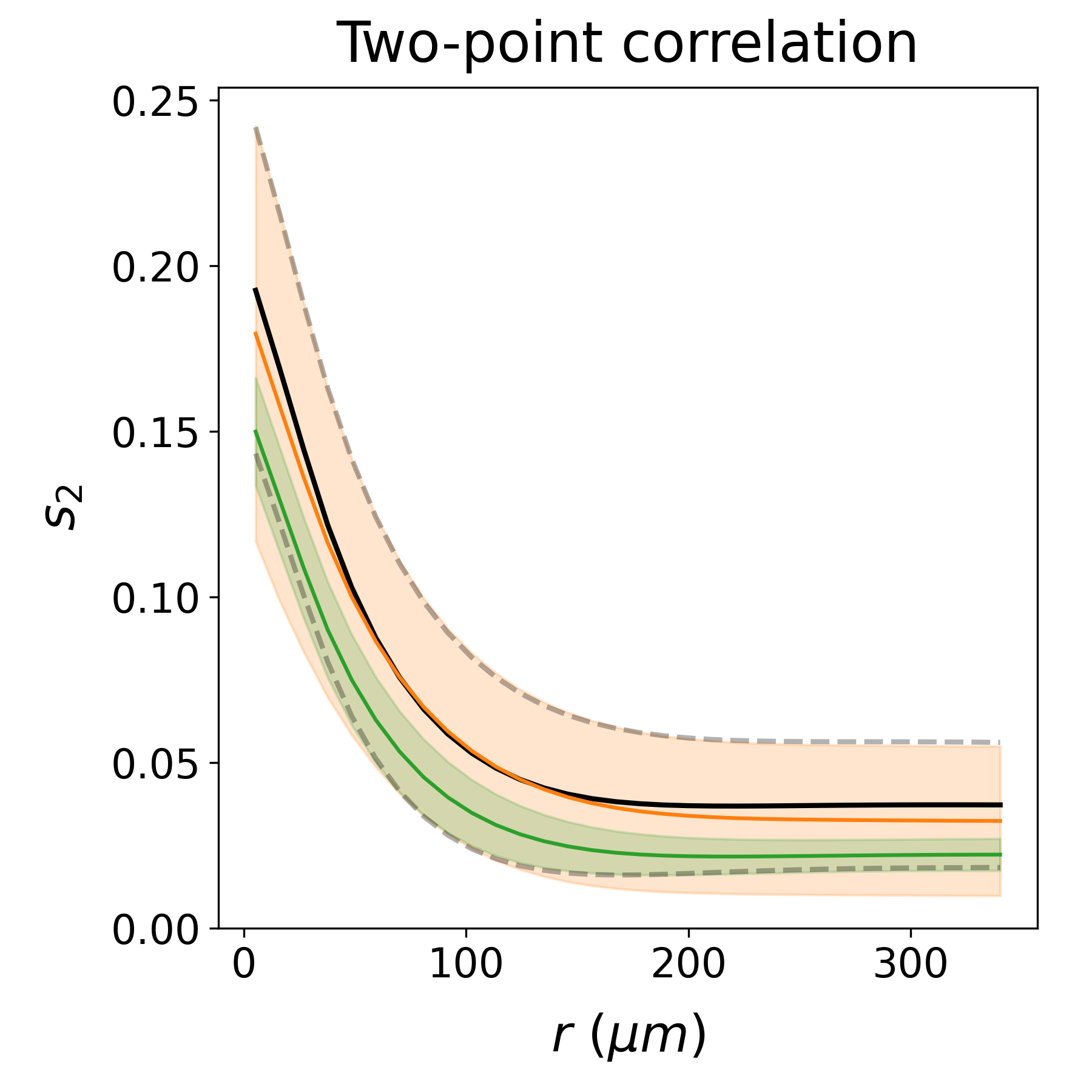}
        \caption{}
        \label{fig:doddington_256nc_c}
    \end{subfigure}
    \begin{subfigure}[t]{0.48\textwidth}
        \includegraphics[width=\textwidth]{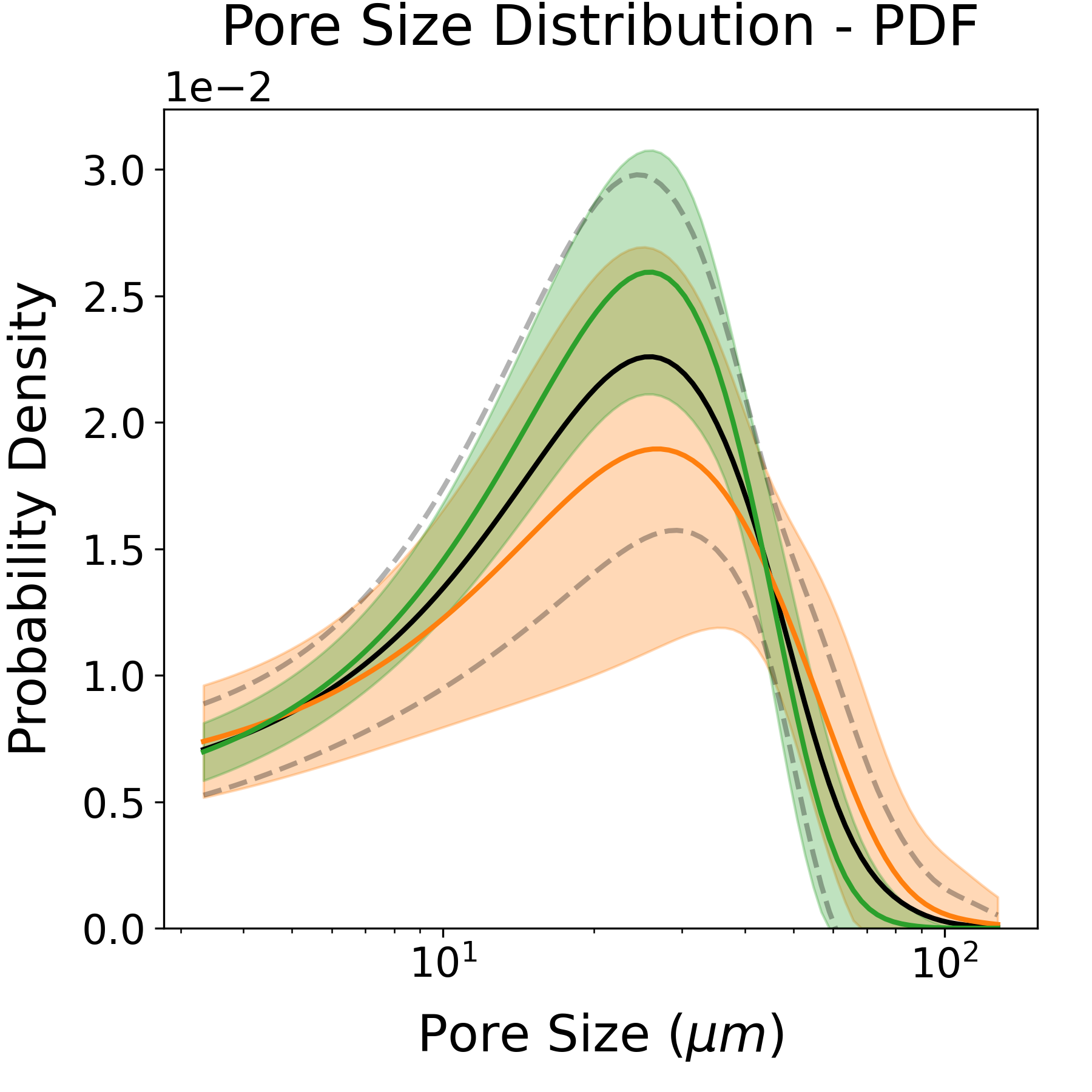}
        \caption{}
        \label{fig:doddington_256nc_d}
    \end{subfigure}
    \caption{Statistical properties for unconditional Doddington sandstone at 256$^3$ voxels: (a) porosity distribution, (b) permeability distribution, (c) two-point correlation function, and (d) pore size distribution.}
    \label{fig:doddington_256nc}
\end{figure}

\begin{figure}[H]
    \centering
    \begin{subfigure}[t]{0.48\textwidth}
        \includegraphics[width=\textwidth]{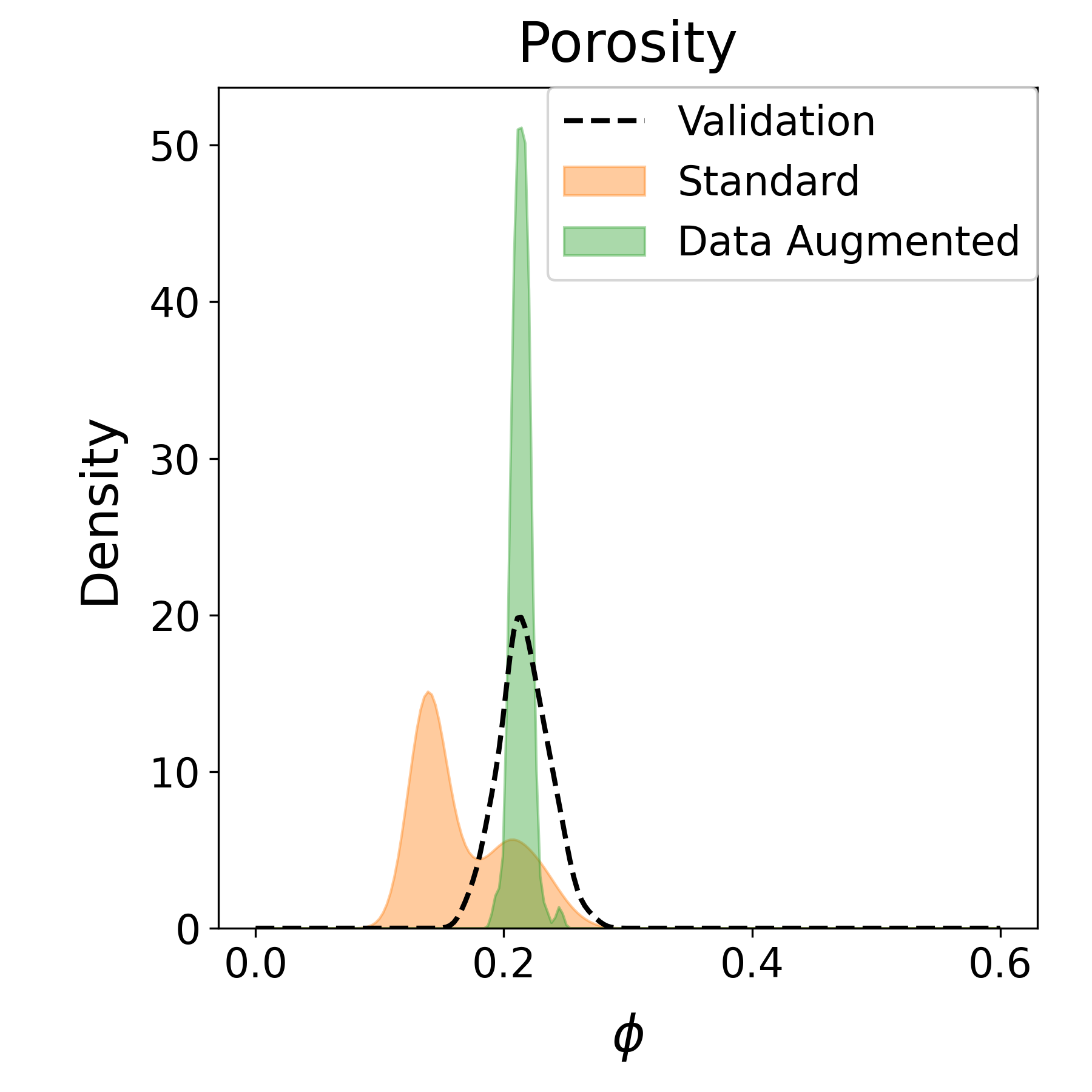}
        \caption{}
        \label{fig:bentheimer_256nc_a}
    \end{subfigure}
    \begin{subfigure}[t]{0.48\textwidth}
        \includegraphics[width=\textwidth]{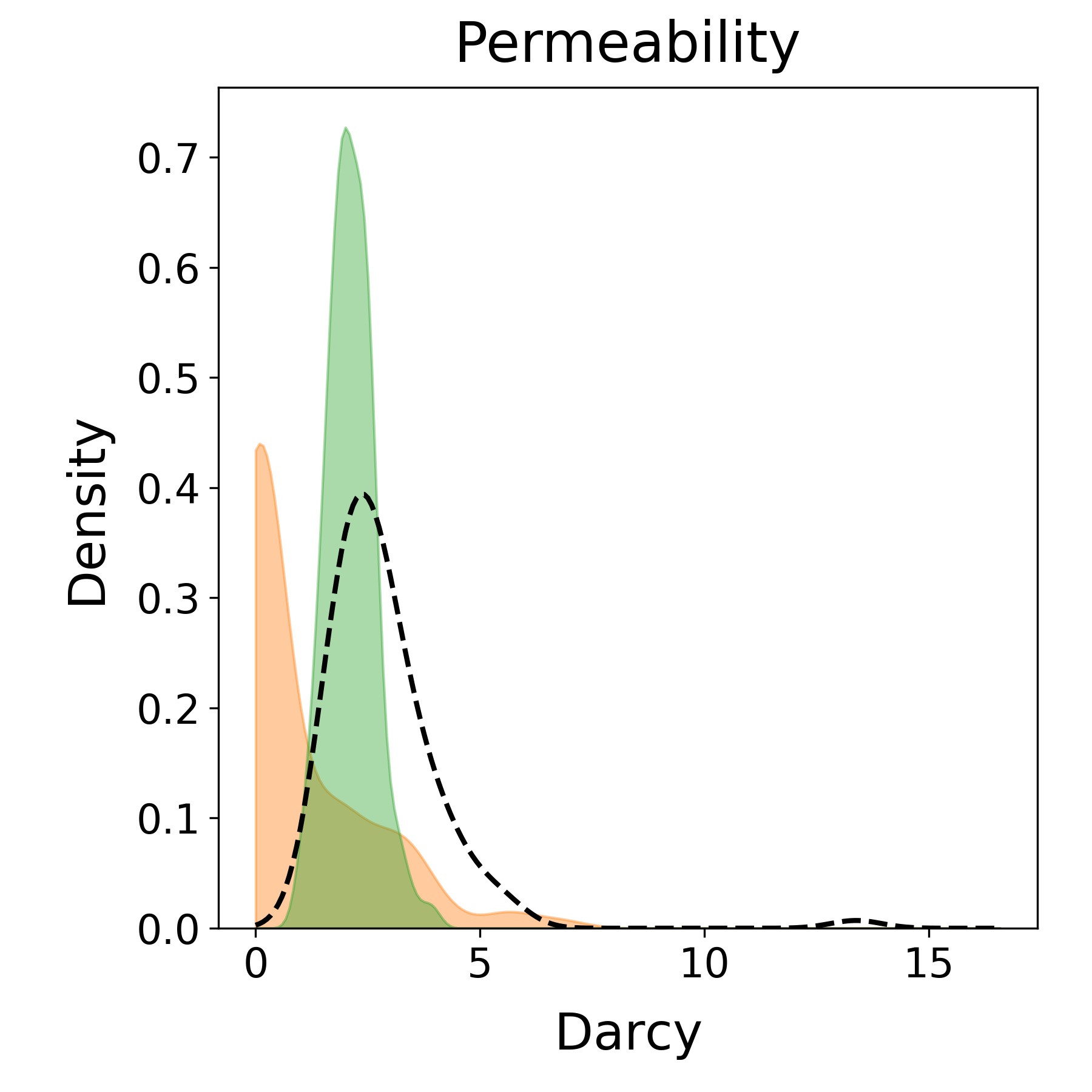}
        \caption{}
        \label{fig:bentheimer_256nc_b}
    \end{subfigure}
    
    \begin{subfigure}[t]{0.48\textwidth}
        \includegraphics[width=\textwidth]{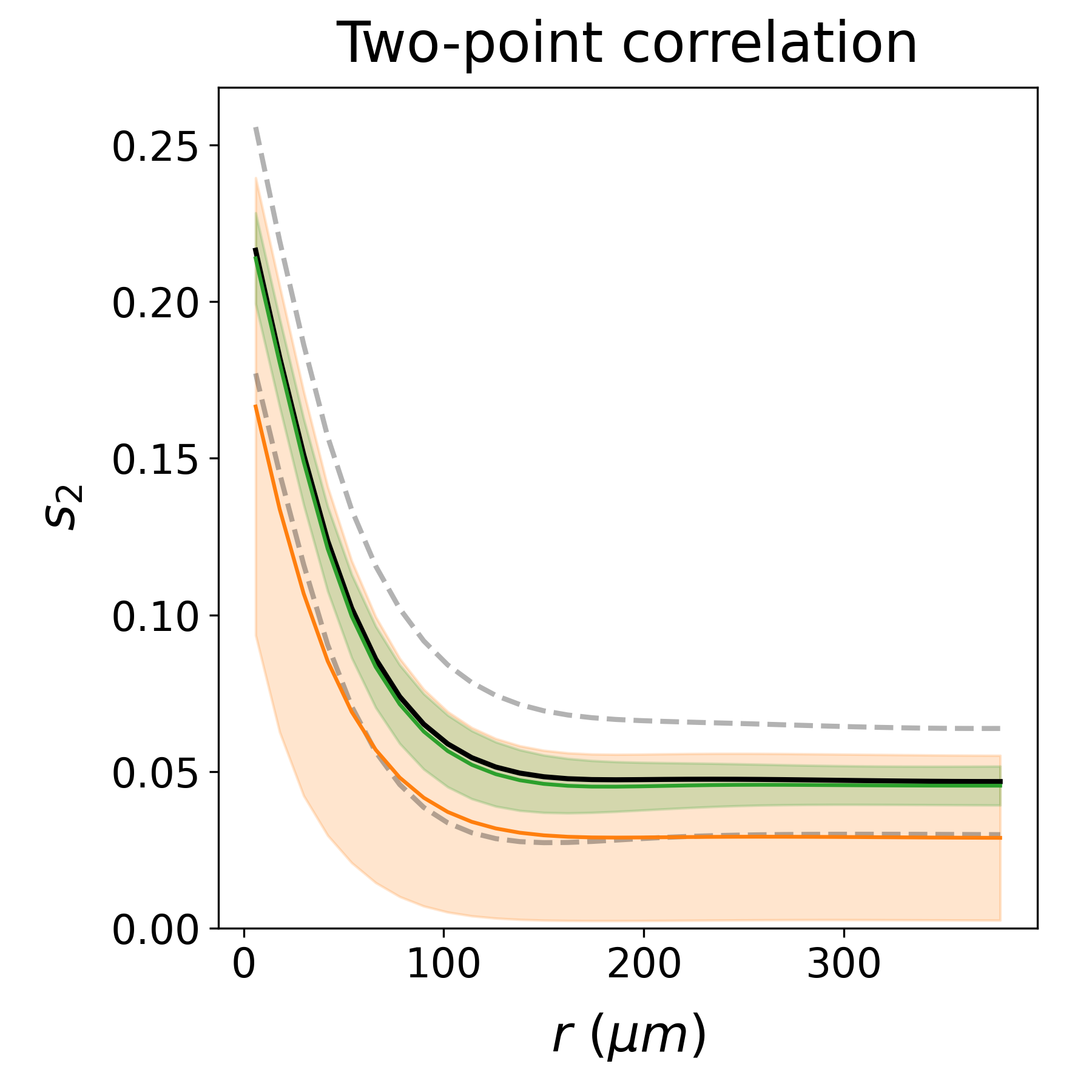}
        \caption{}
        \label{fig:bentheimer_256nc_c}
    \end{subfigure}
    \begin{subfigure}[t]{0.48\textwidth}
        \includegraphics[width=\textwidth]{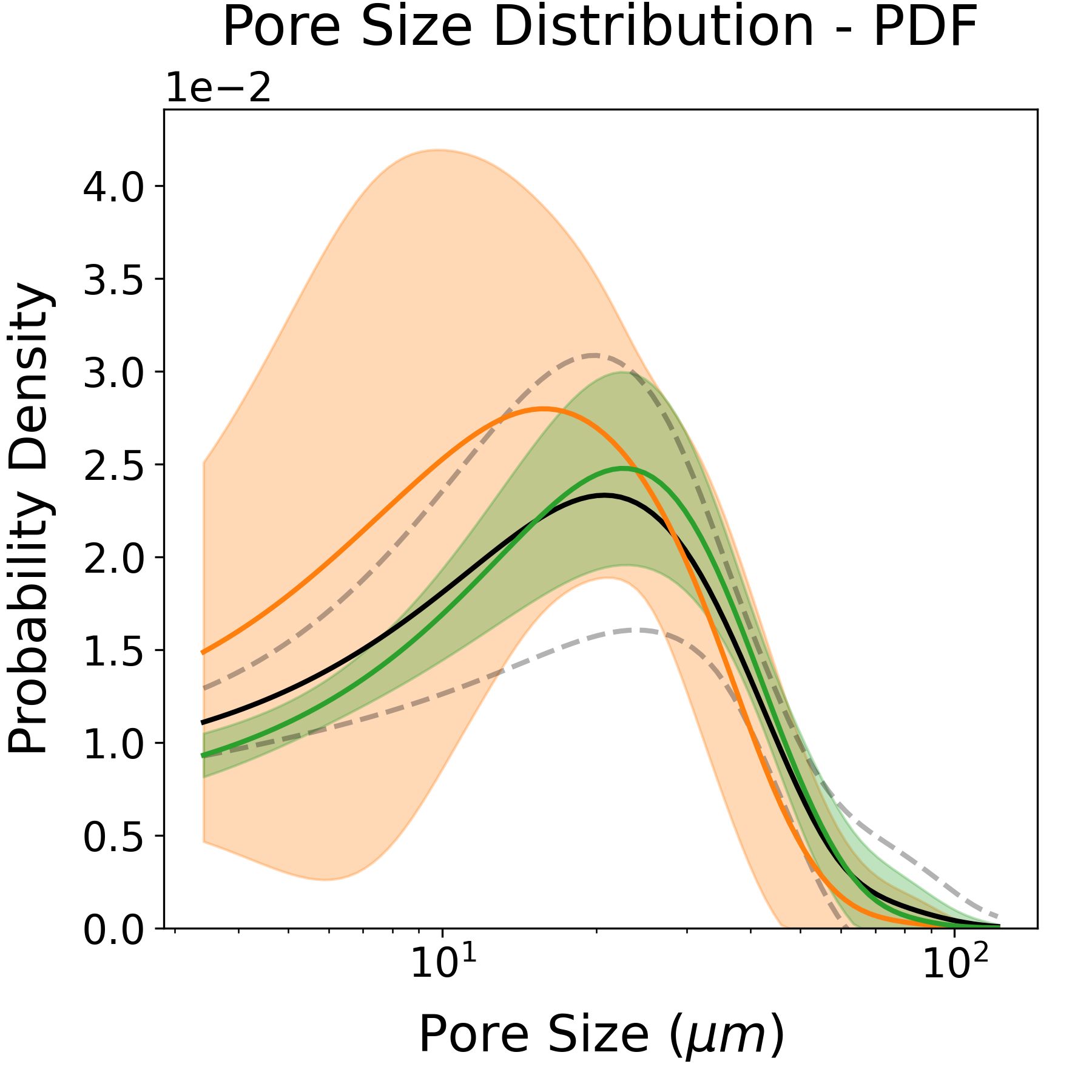}
        \caption{}
        \label{fig:bentheimer_256nc_d}
    \end{subfigure}
    \caption{Statistical properties for unconditional Bentheimer sandstone at 256$^3$ voxels: (a) porosity distribution, (b) permeability distribution, (c) two-point correlation function, and (d) pore size distribution.}
    \label{fig:bentheimer_256nc}
\end{figure}

First, we show results for samples generated with the same configurations as in Section \ref{subsection: uncond 128}, but with a model trained on $256^3$ volumes. In Figure \ref{fig:doddington_256nc}, it is possible to see that while the generated statistics for Doddington are consistent with the validation, the same is not true for Bentheimer (Figure \ref{fig:bentheimer_256nc}). In particular, a significant fraction of the generated samples have extremely low porosity values.

% Doddington 256 uncond

% Bentheimer 256 uncond

The relatively high structural complexity and heterogeneity of Bentheimer sandstone and Estailides carbonate require a larger and more diverse training set to capture their full range of pore geometries. When only a small number of examples are available, the model may fail to learn the complete porosity distribution, leading to artificially low-porosity outputs. By contrast, Doddington sandstone and Ketton carbonate exhibit more uniform pore architectures, enabling generative models to capture their variability even with fewer training samples. Consequently, the same modeling framework that struggles under limited data conditions for Bentheimer and Estailides can still perform reliably for Doddington and Ketton.

Therefore, in this work, we focus on a more robust pipeline that combines data augmentation with controlled sampling to ensure high-quality generation. In our first step toward this pipeline, we introduce a data augmentation procedure consisting of applying combinations of reflections to the training data\footnote{The set of reflections in all axes generates the symmetry group of the cuboid. Although the symmetry group of the cube also includes $90^{\circ}$ rotations, our main focus is to obtain a pipeline that is directly applicable to any rectangular shape.} reduced the out-of-distribution samples but produced a substantial mode collapse, as suggested by the acute concentration of the generated statistics distribution near its mode, which can be seen in Figures \ref{fig:bentheimer_256nc} for the case of Bentheimer sandstone. In rocks where there was no out-of-distribution problem, this considerably worsens the quality of the inference, as we see in Figure \ref{fig:doddington_256nc} for the case of the Doddington sandstone. Notice that these results show some similarity to the mode collapse observed with pixel-space diffusion, in Figure \ref{fig:bentheimer_64}. A natural way to prevent this type of collapse is given by controlled generation, whose results are presented in the next section.

\begin{figure}[H]
   \centering
   % Top row
   \begin{subfigure}[t]{0.48\textwidth}
       \centering
       \includegraphics[width=\textwidth]{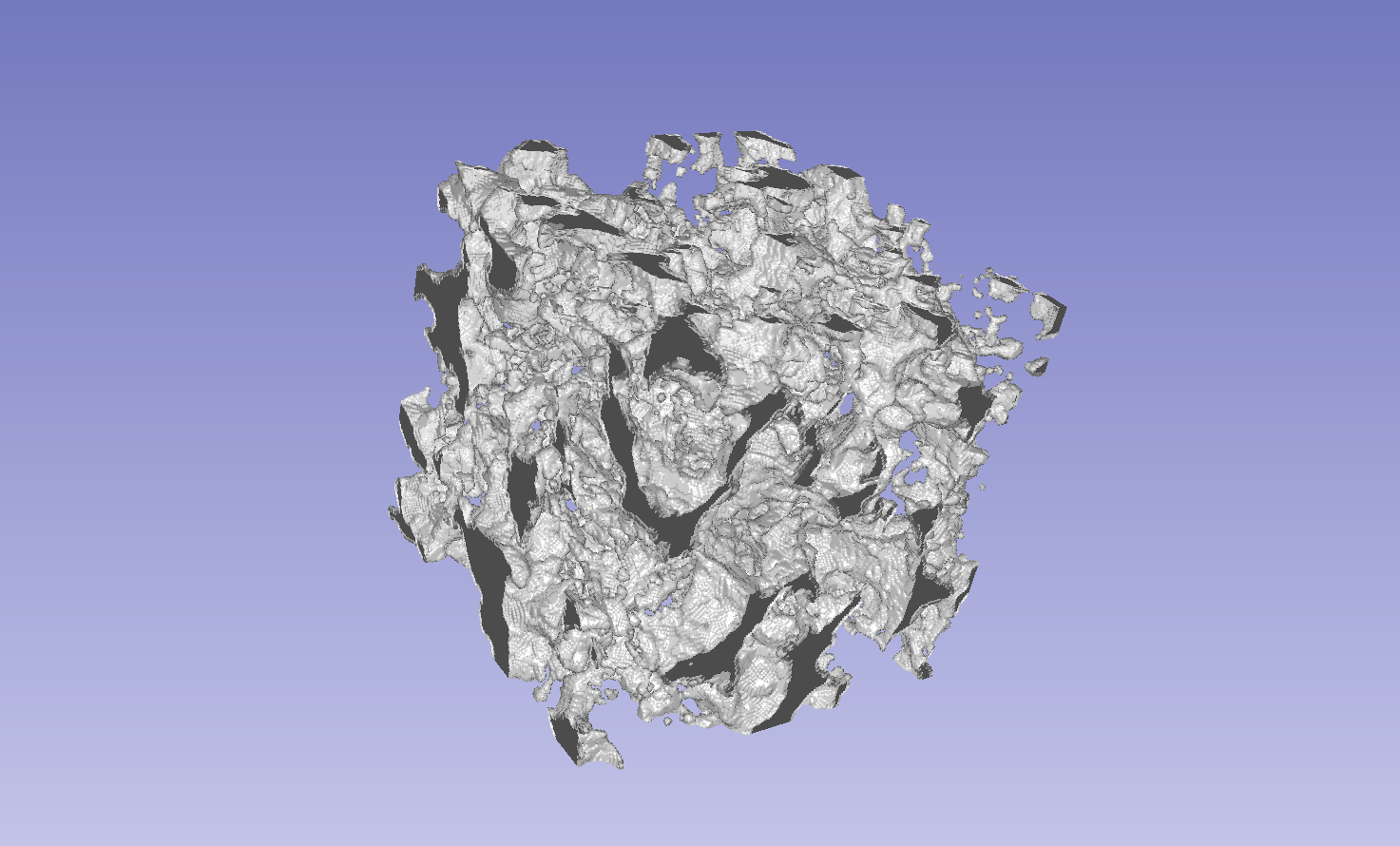}
       \caption{Pore space of a correctly generated Bentheimer rock}
       \label{fig:accept}
   \end{subfigure}%
   \hfill
   \begin{subfigure}[t]{0.48\textwidth}
       \centering
       \includegraphics[width=\textwidth]{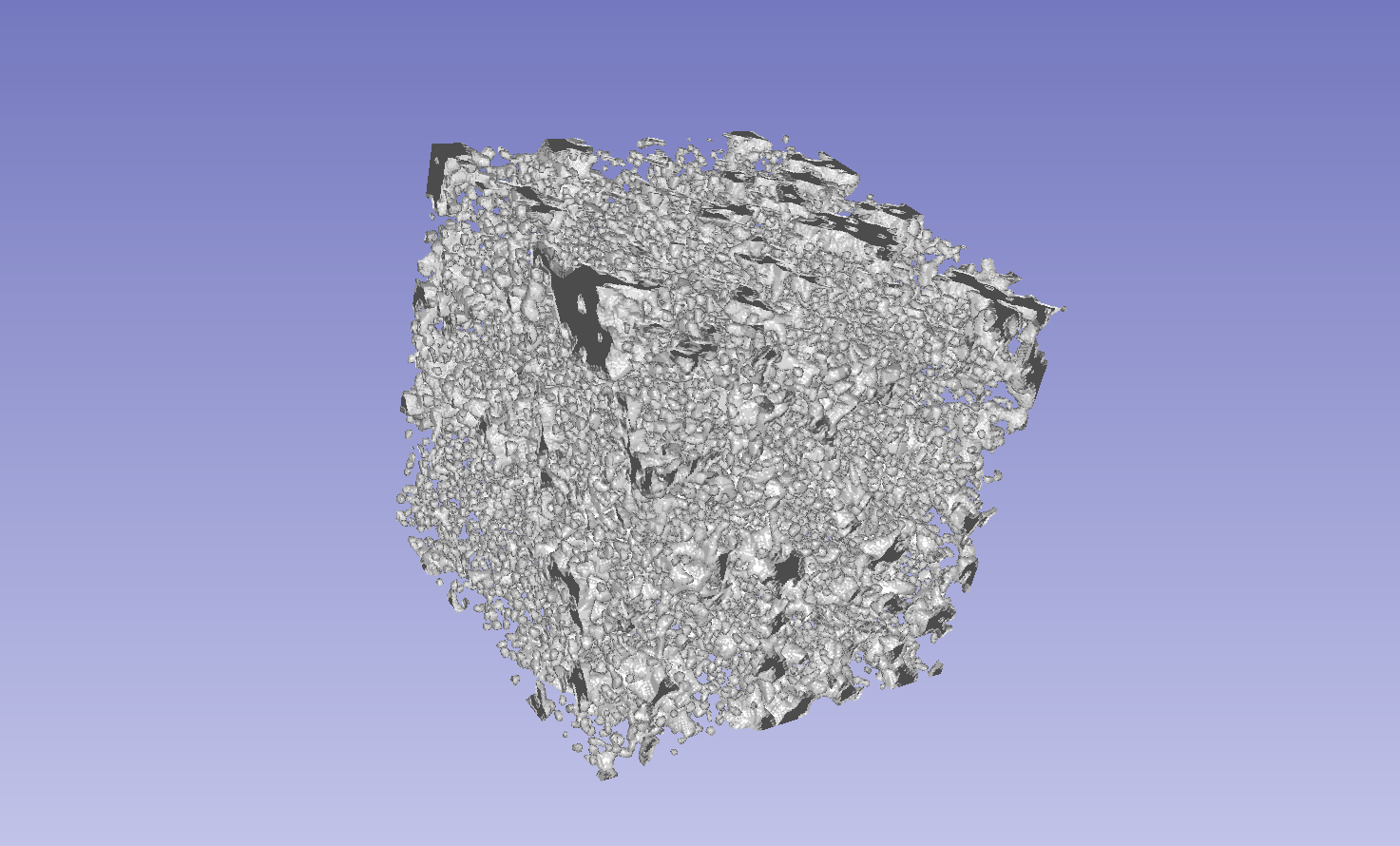}
       \caption{Pore space of an incorrectly generated Bentheimer rock}
       \label{fig:reject}
   \end{subfigure}
   
   % Bottom row
   \begin{subfigure}[t]{0.48\textwidth}
       \centering
       \includegraphics[width=\textwidth]{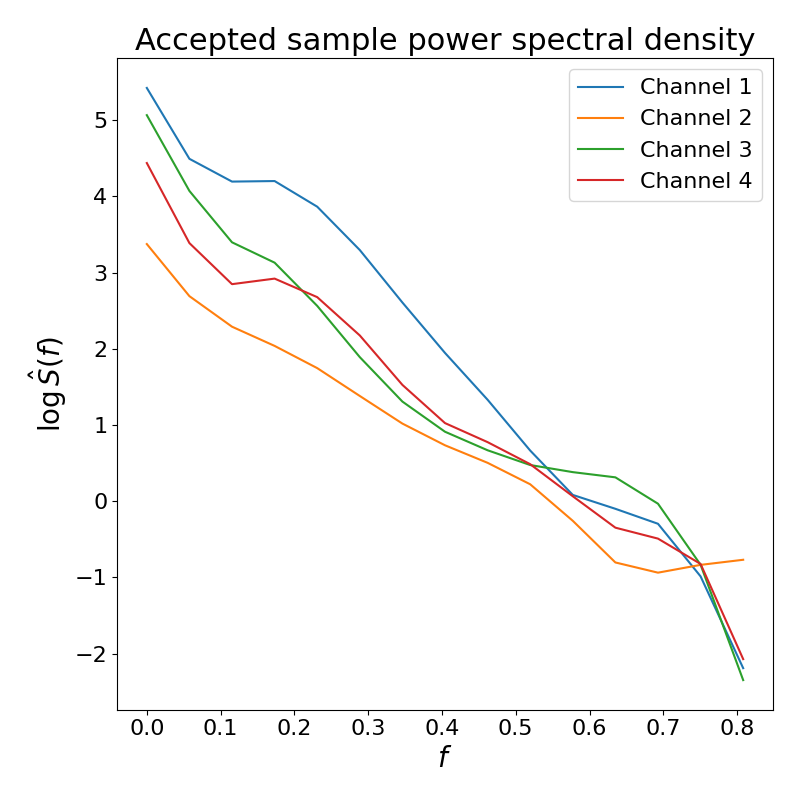}
       \caption{Power spectral density of the correctly generated sample}
       \label{fig:accept_spectra}
   \end{subfigure}
   \hfill
   \begin{subfigure}[t]{0.48\textwidth}
       \centering
       \includegraphics[width=\textwidth]{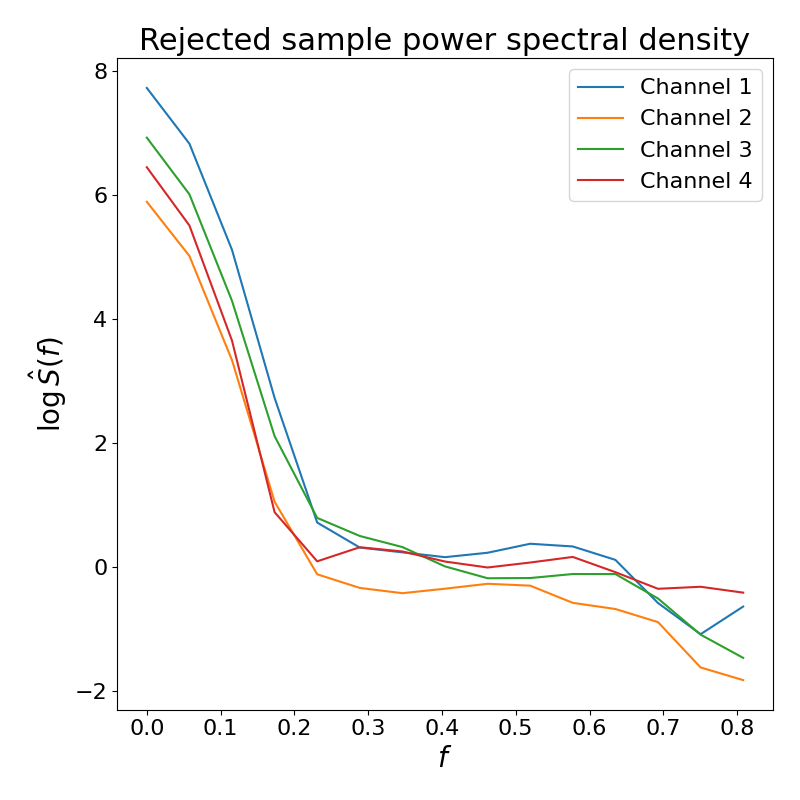}
       \caption{Power spectral density of the incorrectly generated sample}
       \label{fig:reject_spectra}
   \end{subfigure}
   \caption{Comparison of correctly and incorrectly generated Bentheimer rock samples with their corresponding power spectral densities.}
   \label{fig:comparisonspectra}
\end{figure}

As a final note, we examined the anomalous generated samples through power spectral density analysis of their latent space representations, drawing from image analysis techniques \citep{vanderSchaaf1996, Tolhurst1992}. For each generated sample $X \in \mathbb{R}^{C, H, W, D}$, we calculated the power spectral density across individual channels $X_c \in \mathbb{R}^{H, W, D}$ using Fourier analysis, applying a Gaussian window to reduce edge effects before computing the radially averaged spectrum $\hat{S}(f)$. Anomalous generations typically showed a high-frequency plateau characteristic of white noise, and successful ones exhibited continuous spectral decay, as can be seen in Figure \ref{fig:comparisonspectra}. However, this analysis had limitations, particularly for samples with very low porosity values, suggesting that more nuanced approaches would be needed to reliably distinguish between valid and spurious generated samples. We are currently investigating whether a refined version of this spectral analysis could serve as an effective filter for identifying anomalous generations.

% Spectra comparison

\subsubsection{Controlled unconditional generation}\label{sec: controlled}

One strategy to mitigate the problem of reduced variance in unconditional sampling is to use a conditional model conditioned on the distribution of some statistic from the training data. This corresponds to the marginalization
\begin{equation}
    p(x) = \int p(x|y) r(y) dy
\end{equation}
where $y$ is the considered statistic, with distribution $r(y)$. Therefore, unconditional sampling can be performed by first sampling the condition $y$ from the distribution $r(y)$ of some statistic, and then sampling from a conditional model with the condition $y$. This procedure is extensively discussed in Section \ref{reconstruction methods}, and referred as controlled unconditional sampling.

\begin{figure}[H]
    \centering
    \begin{subfigure}[t]{0.24\textwidth}
        \includegraphics[width=\textwidth]{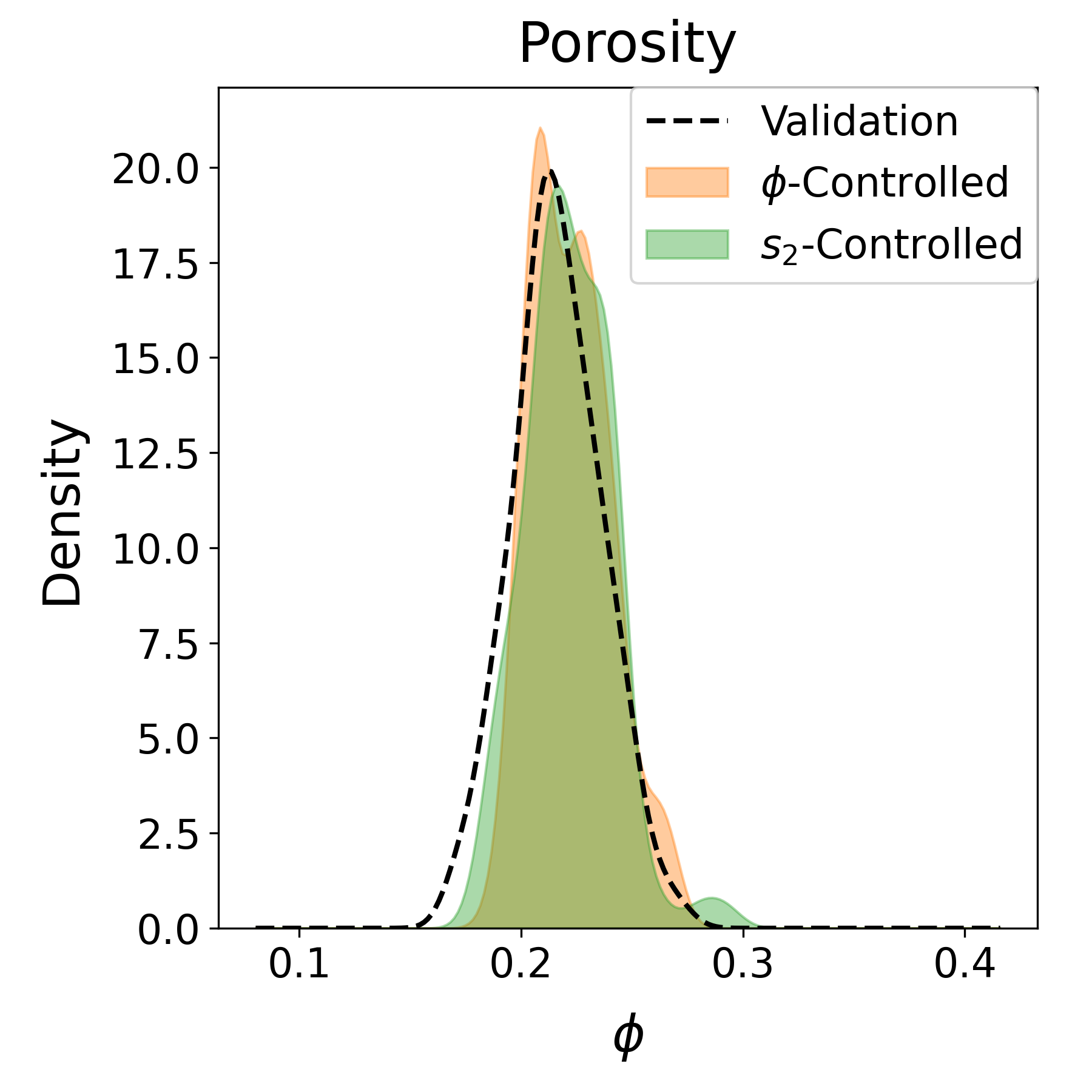}
        \caption{}
        \label{fig:bentheimer_256c_a}
    \end{subfigure}
    \begin{subfigure}[t]{0.24\textwidth}
        \includegraphics[width=\textwidth]{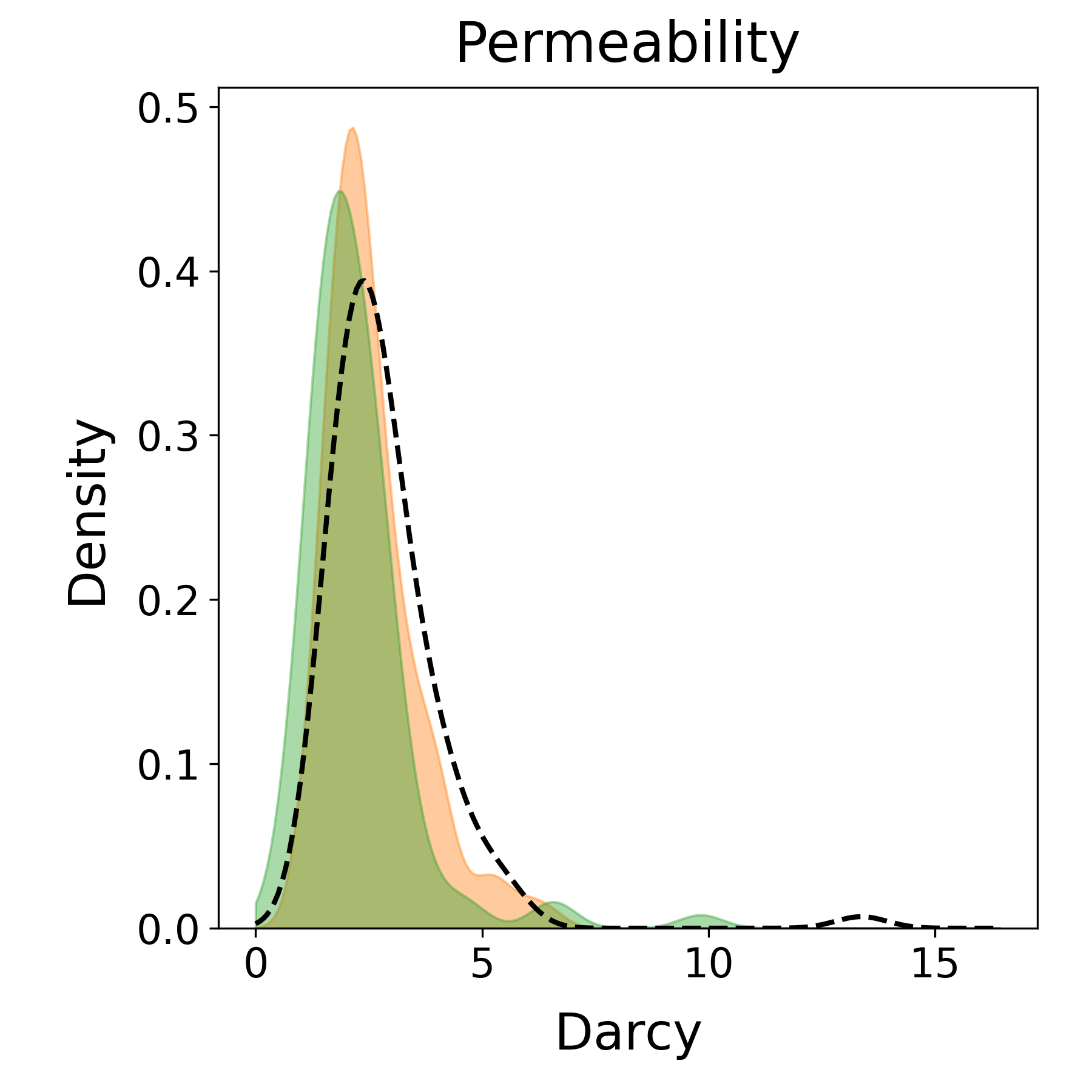}
        \caption{}
        \label{fig:bentheimer_256c_b}
    \end{subfigure}
    \begin{subfigure}[t]{0.24\textwidth}
        \includegraphics[width=\textwidth]{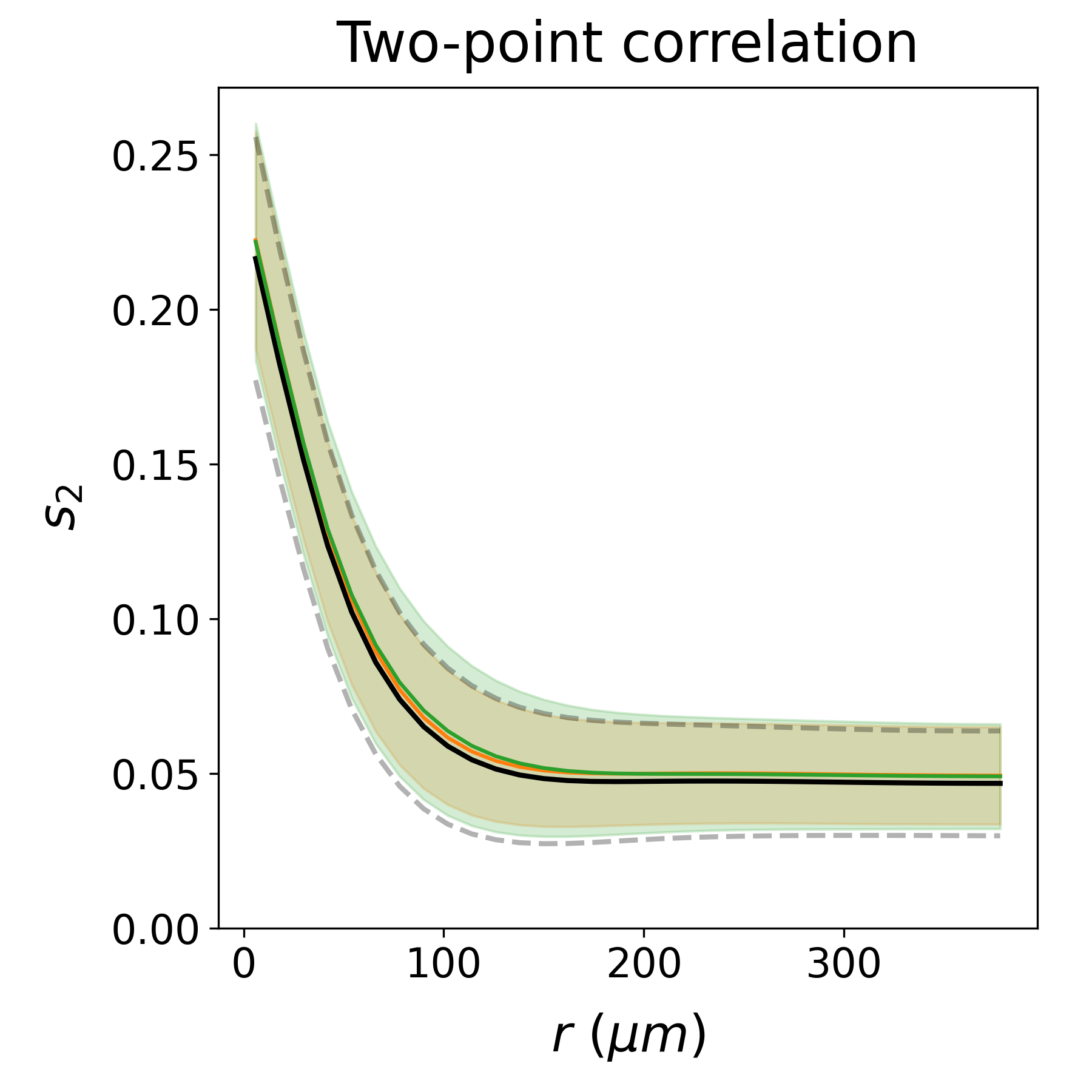}
        \caption{}
        \label{fig:bentheimer_256c_c}
    \end{subfigure}
    \begin{subfigure}[t]{0.24\textwidth}
        \includegraphics[width=\textwidth]{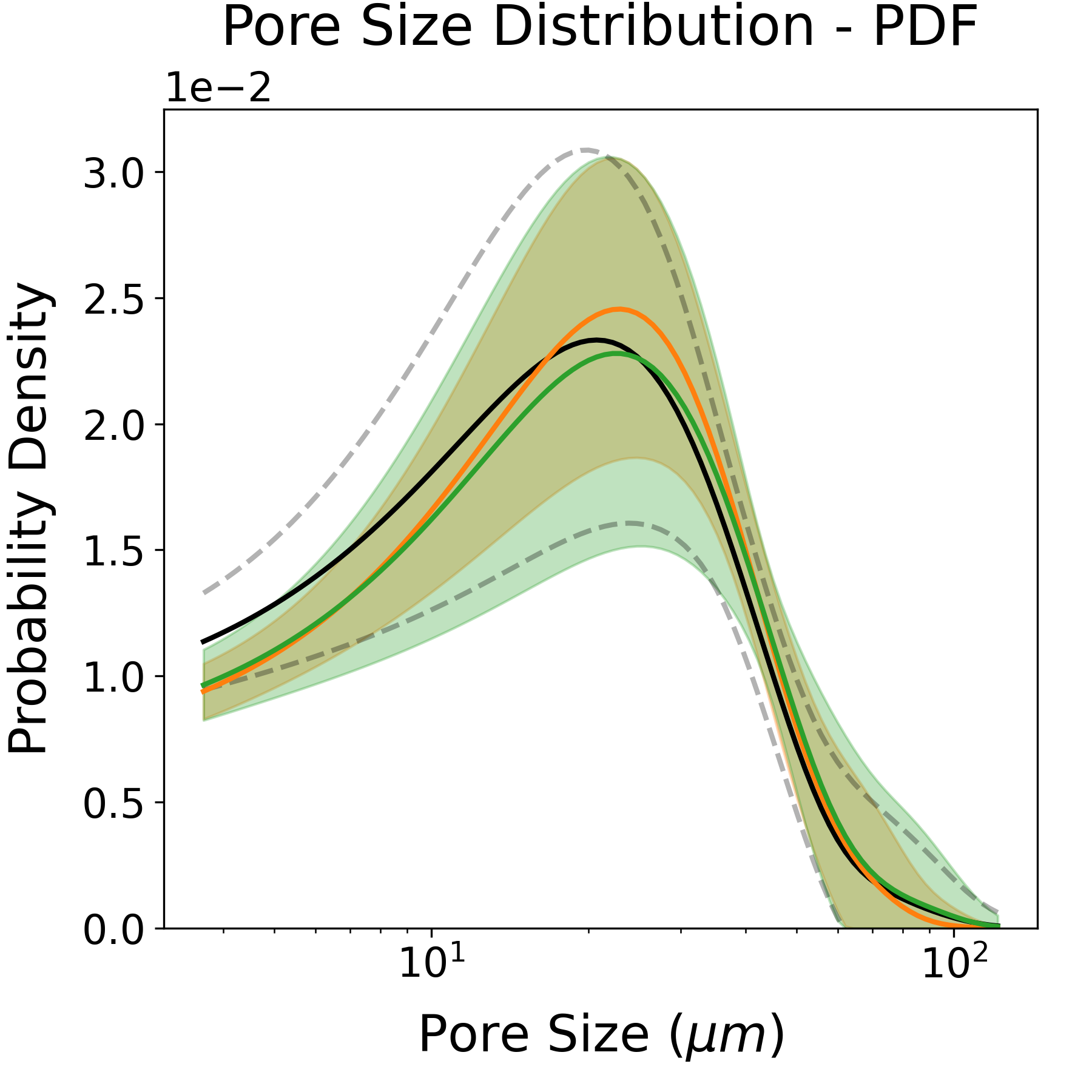}
        \caption{}
        \label{fig:bentheimer_256c_d}
    \end{subfigure}
    
    \begin{subfigure}[t]{0.24\textwidth}
        \includegraphics[width=\textwidth]{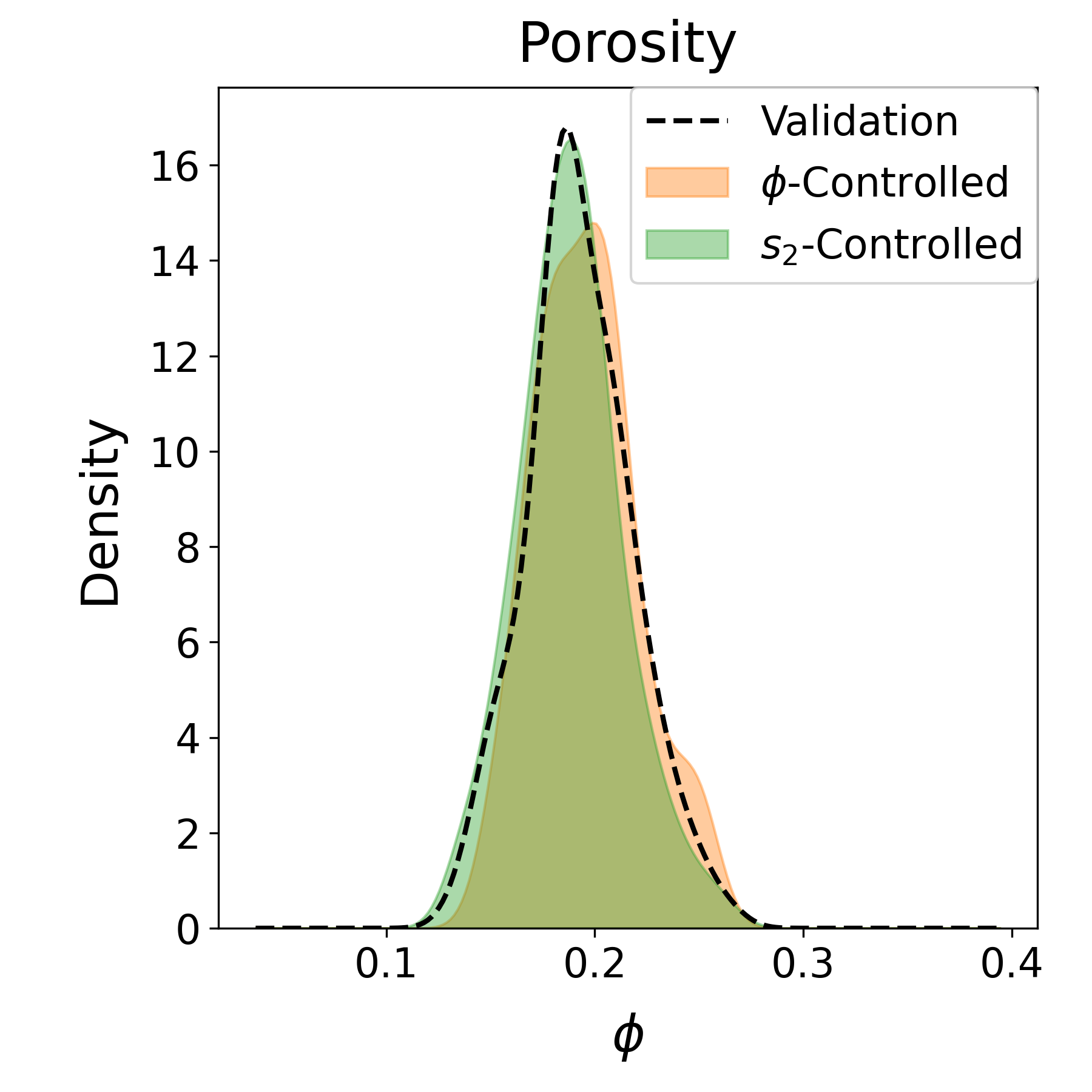}
        \caption{}
        \label{fig:doddington_256c_a}
    \end{subfigure}
    \begin{subfigure}[t]{0.24\textwidth}
        \includegraphics[width=\textwidth]{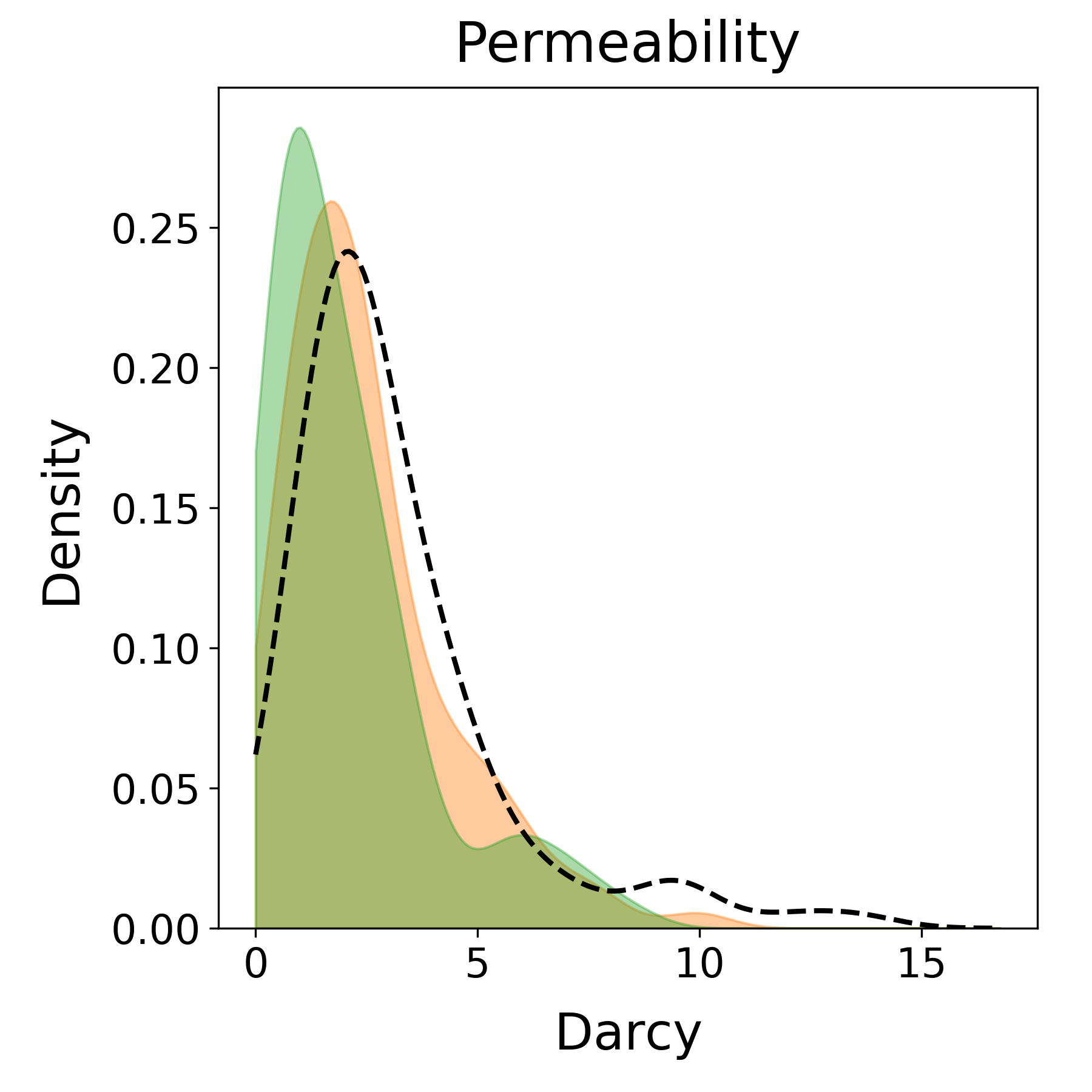}
        \caption{}
        \label{fig:doddington_256c_b}
    \end{subfigure}
    \begin{subfigure}[t]{0.24\textwidth}
        \includegraphics[width=\textwidth]{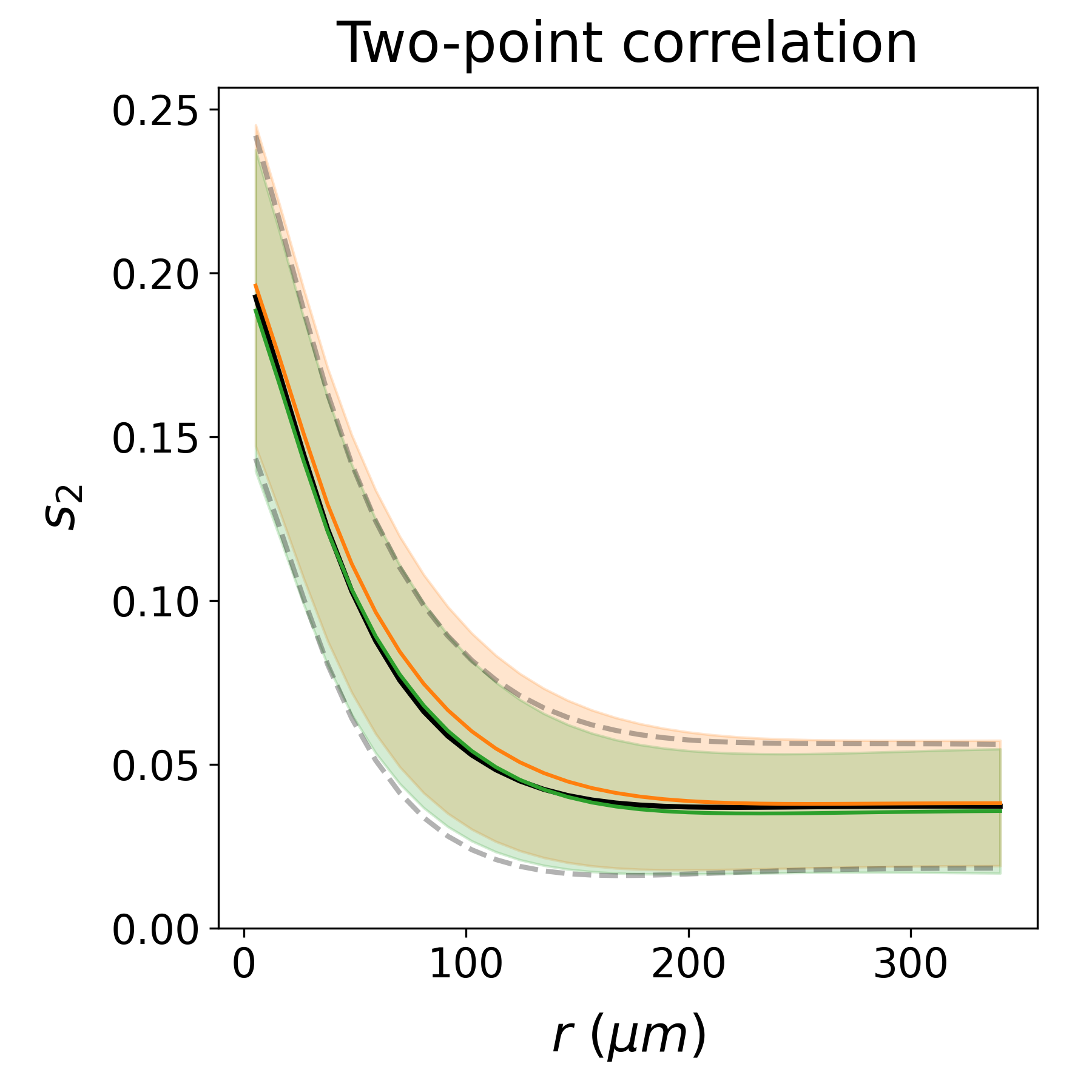}
        \caption{}
        \label{fig:doddington_256c_c}
    \end{subfigure}
    \begin{subfigure}[t]{0.24\textwidth}
        \includegraphics[width=\textwidth]{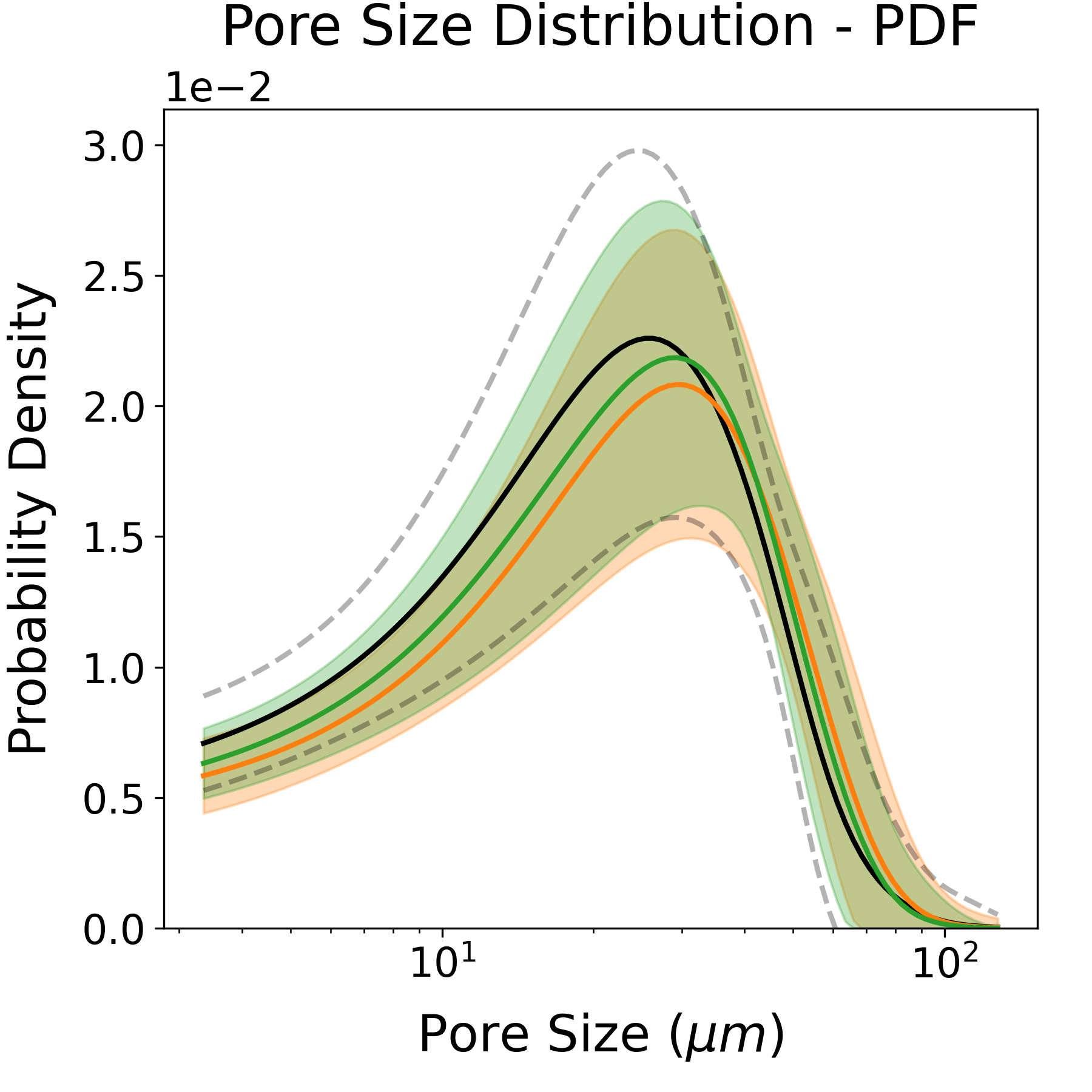}
        \caption{}
        \label{fig:doddington_256c_d}
    \end{subfigure}
    
    \begin{subfigure}[t]{0.24\textwidth}
        \includegraphics[width=\textwidth]{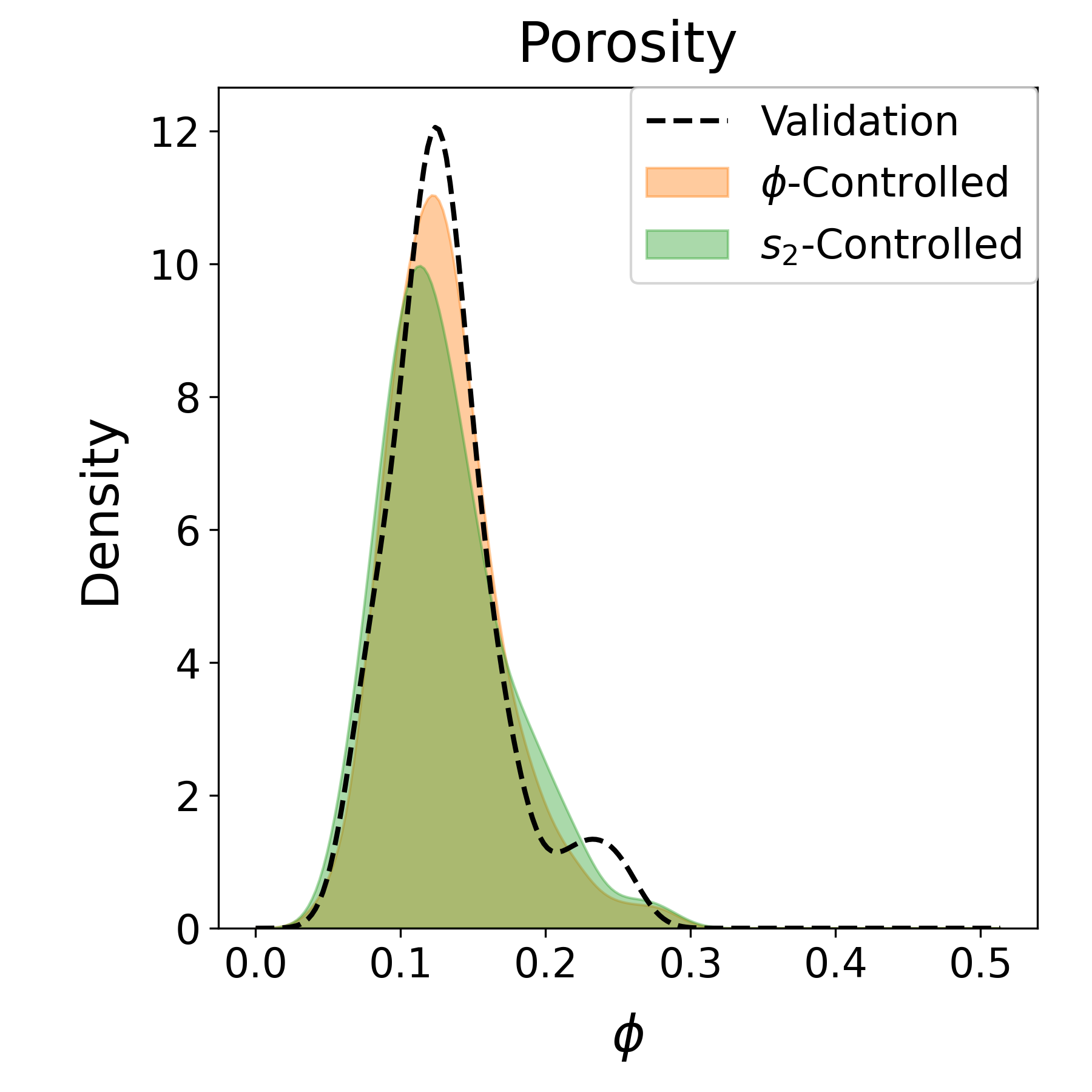}
        \caption{}
        \label{fig:estaillades_256c_a}
    \end{subfigure}
    \begin{subfigure}[t]{0.24\textwidth}
        \includegraphics[width=\textwidth]{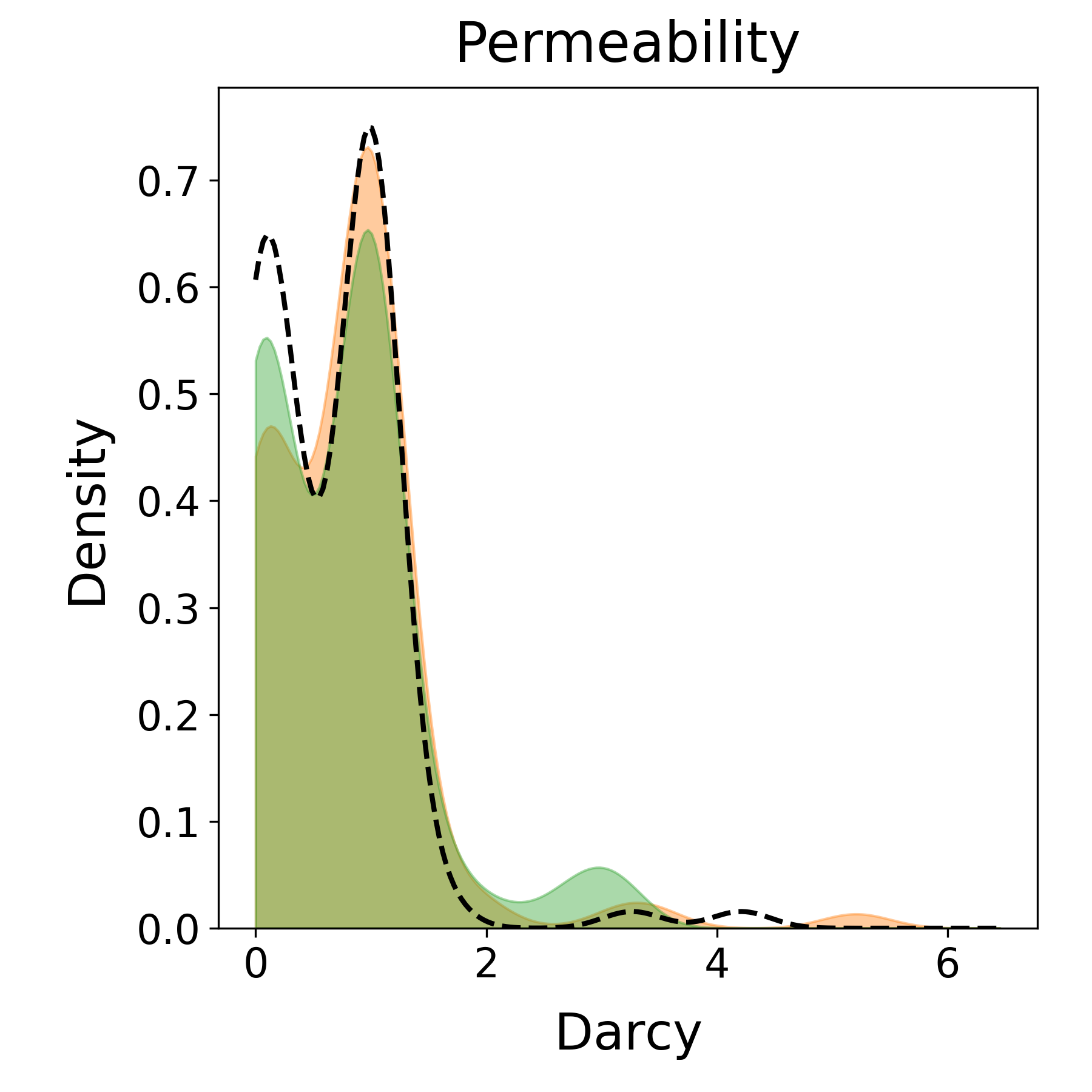}
        \caption{}
        \label{fig:estaillades_256c_b}
    \end{subfigure}
    \begin{subfigure}[t]{0.24\textwidth}
        \includegraphics[width=\textwidth]{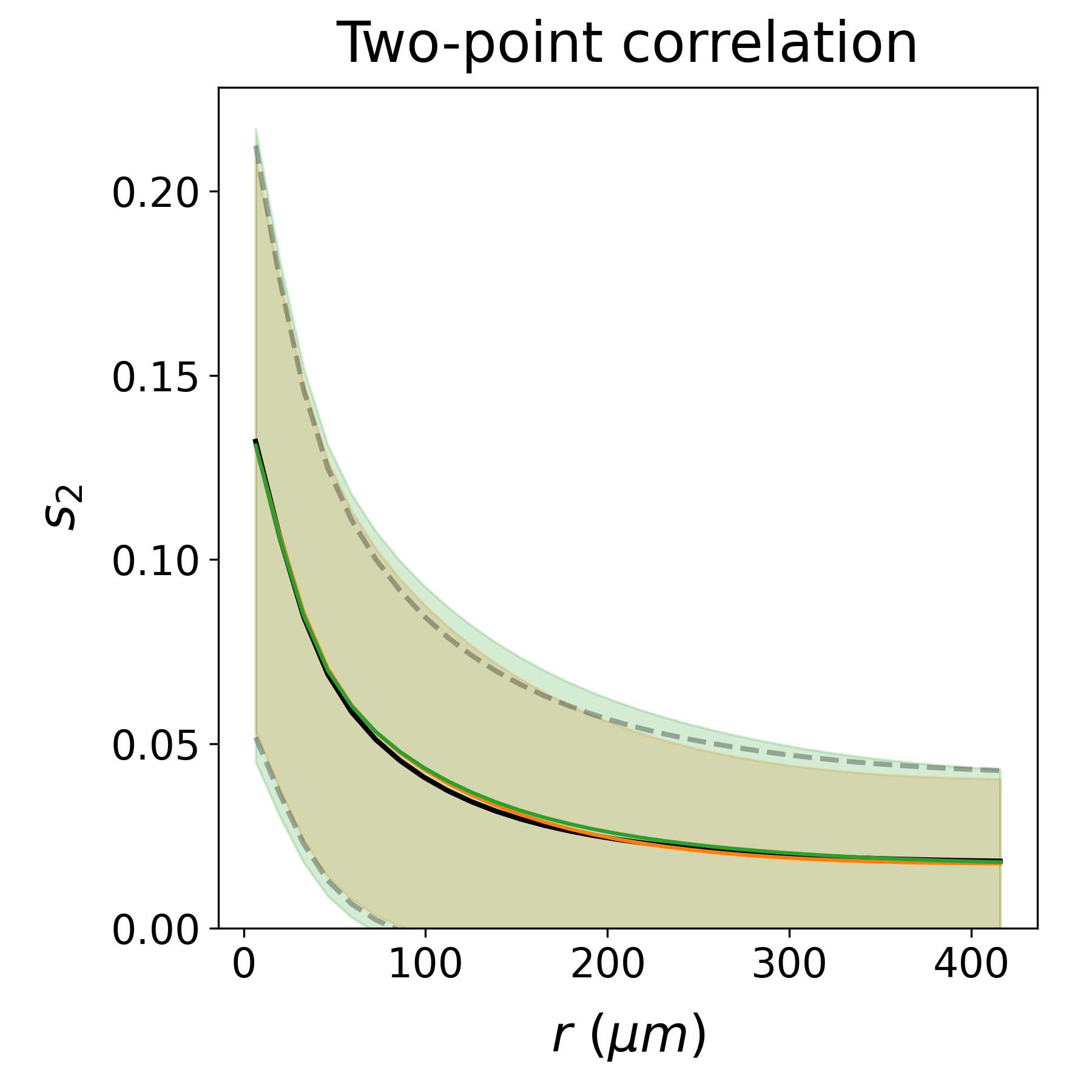}
        \caption{}
        \label{fig:estaillades_256c_c}
    \end{subfigure}
    \begin{subfigure}[t]{0.24\textwidth}
        \includegraphics[width=\textwidth]{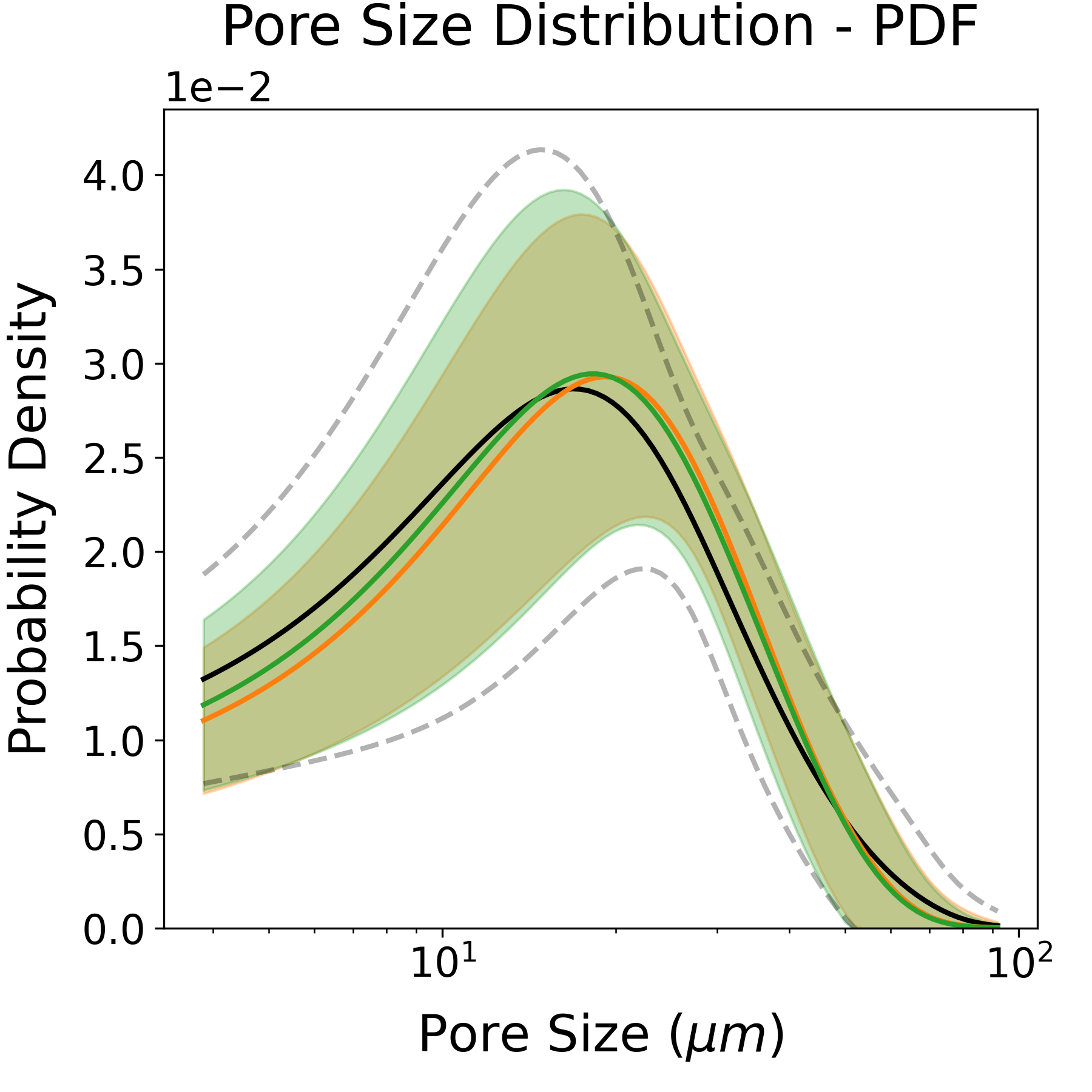}
        \caption{}
        \label{fig:estaillades_256c_d}
    \end{subfigure}
    
    \begin{subfigure}[t]{0.24\textwidth}
        \includegraphics[width=\textwidth]{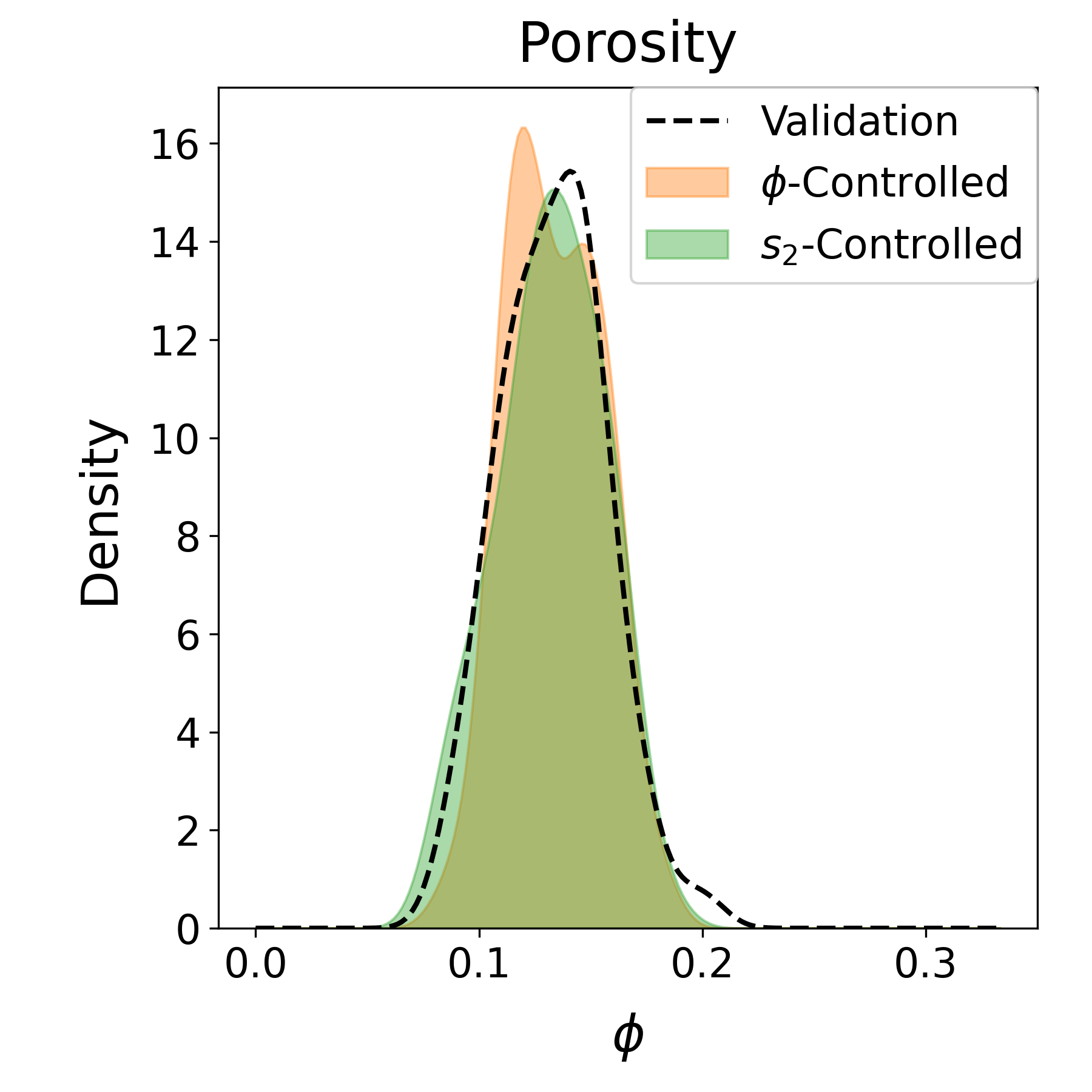}
        \caption{}
        \label{fig:ketton_256c_a}
    \end{subfigure}
    \begin{subfigure}[t]{0.24\textwidth}
        \includegraphics[width=\textwidth]{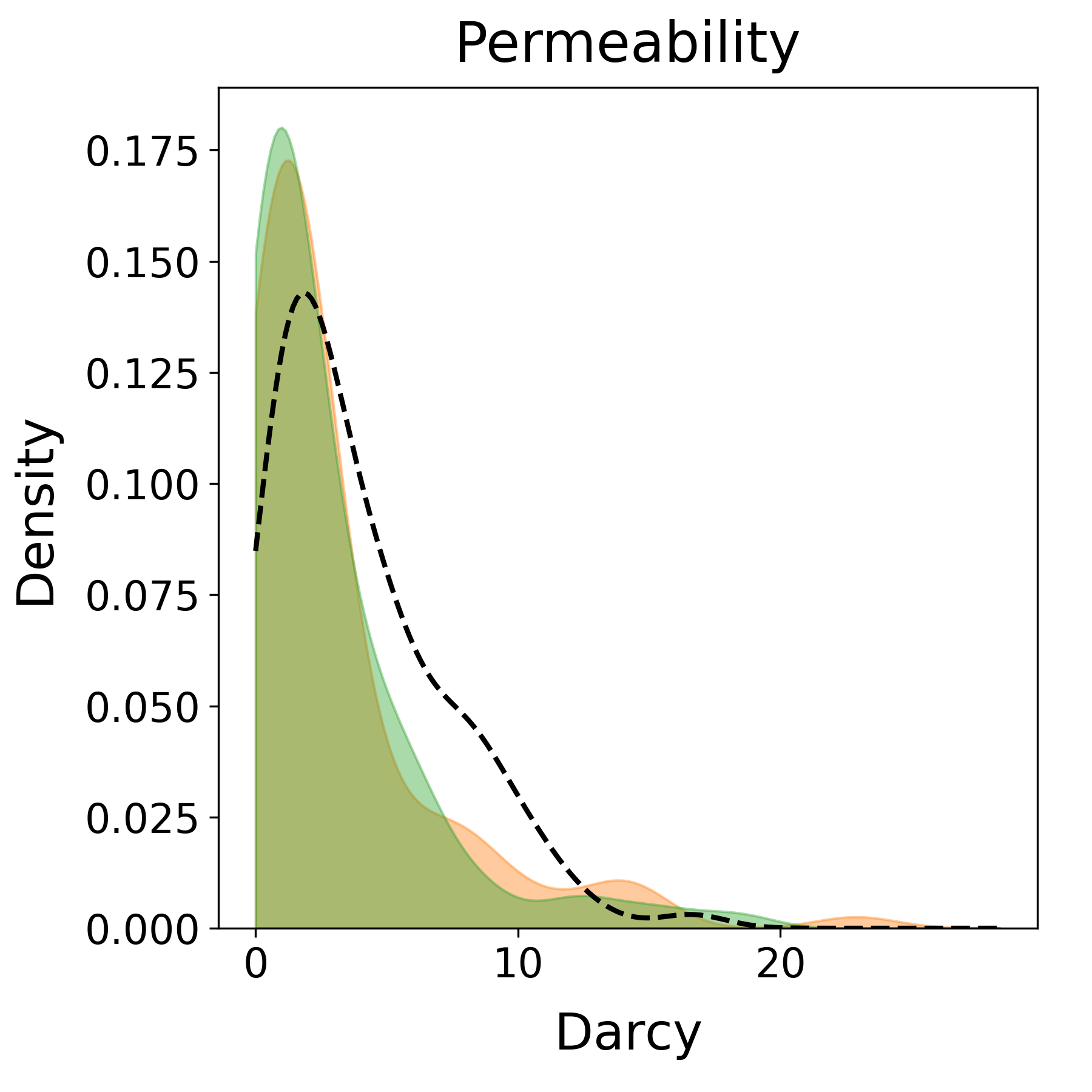}
        \caption{}
        \label{fig:ketton_256c_b}
    \end{subfigure}
    \begin{subfigure}[t]{0.24\textwidth}
        \includegraphics[width=\textwidth]{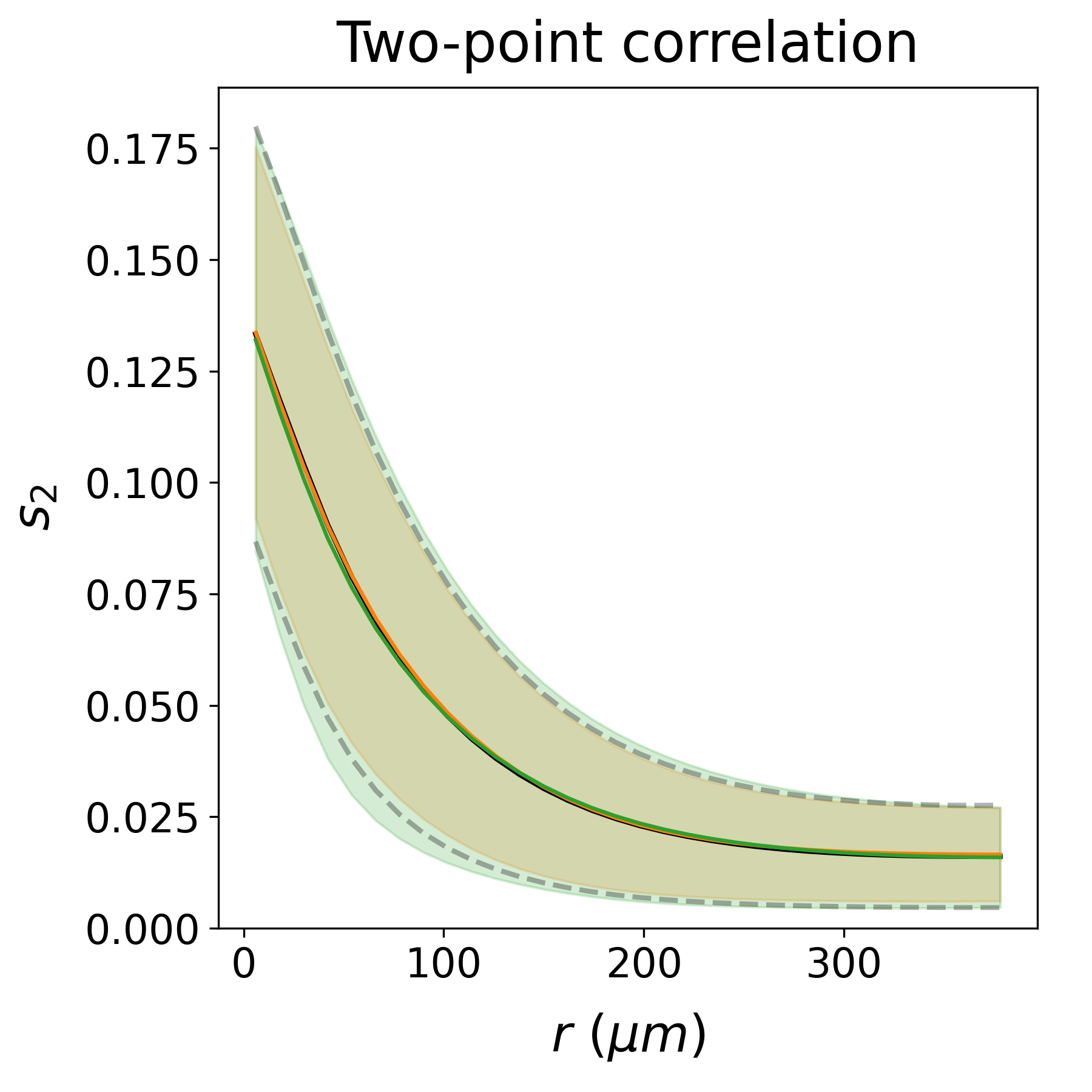}
        \caption{}
        \label{fig:ketton_256c_c}
    \end{subfigure}
    \begin{subfigure}[t]{0.24\textwidth}
        \includegraphics[width=\textwidth]{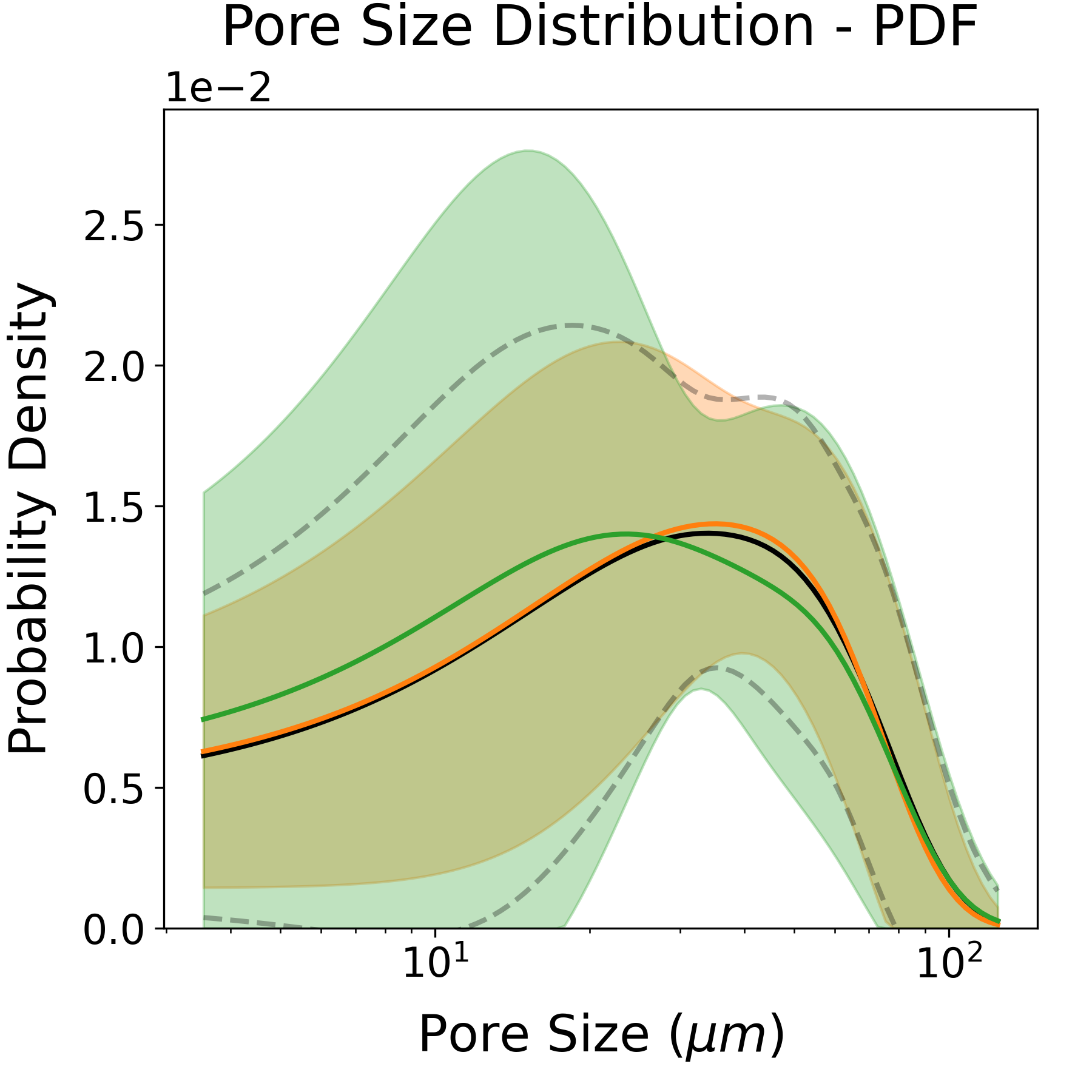}
        \caption{}
        \label{fig:ketton_256c_d}
    \end{subfigure}
    \caption{Comparison of statistical properties for controlled 256$^3$ volume samples across different rock types. Top to bottom: (a-d) Bentheimer, (e-h) Doddington, (i-l) Estaillades, and (m-p) Ketton sandstones. Each row shows (from left to right): porosity distribution, permeability distribution, two-point correlation function, and pore size distribution.}
    \label{fig:all_rocks_256c_comparison}
\end{figure}

\begin{figure}[H]
    \centering
    \begin{subfigure}[t]{0.48\textwidth}
        \includegraphics[width=\textwidth]{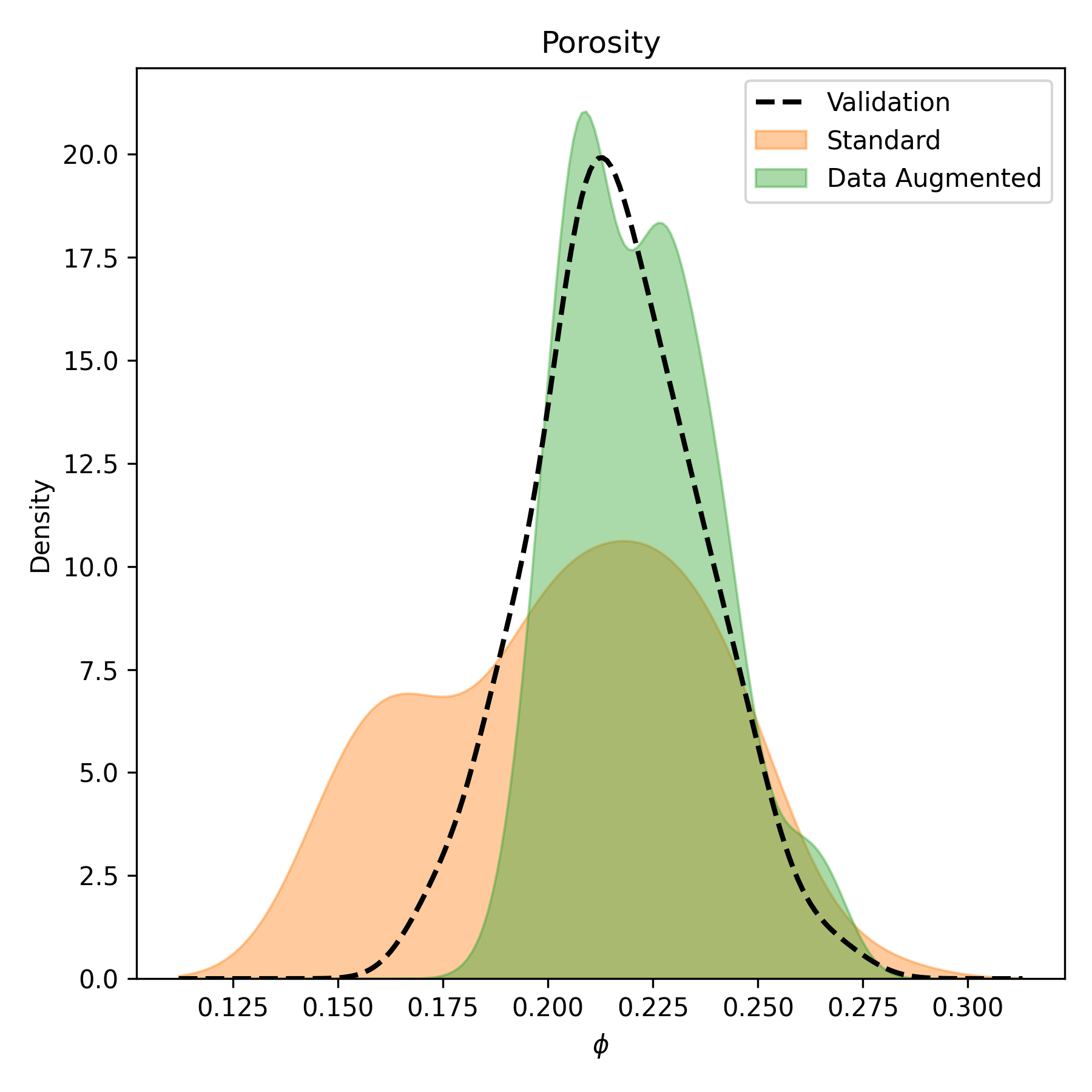}
        \caption{}
        \label{fig:bentheimer_256cd_a}
    \end{subfigure}
    \begin{subfigure}[t]{0.48\textwidth}
        \includegraphics[width=\textwidth]{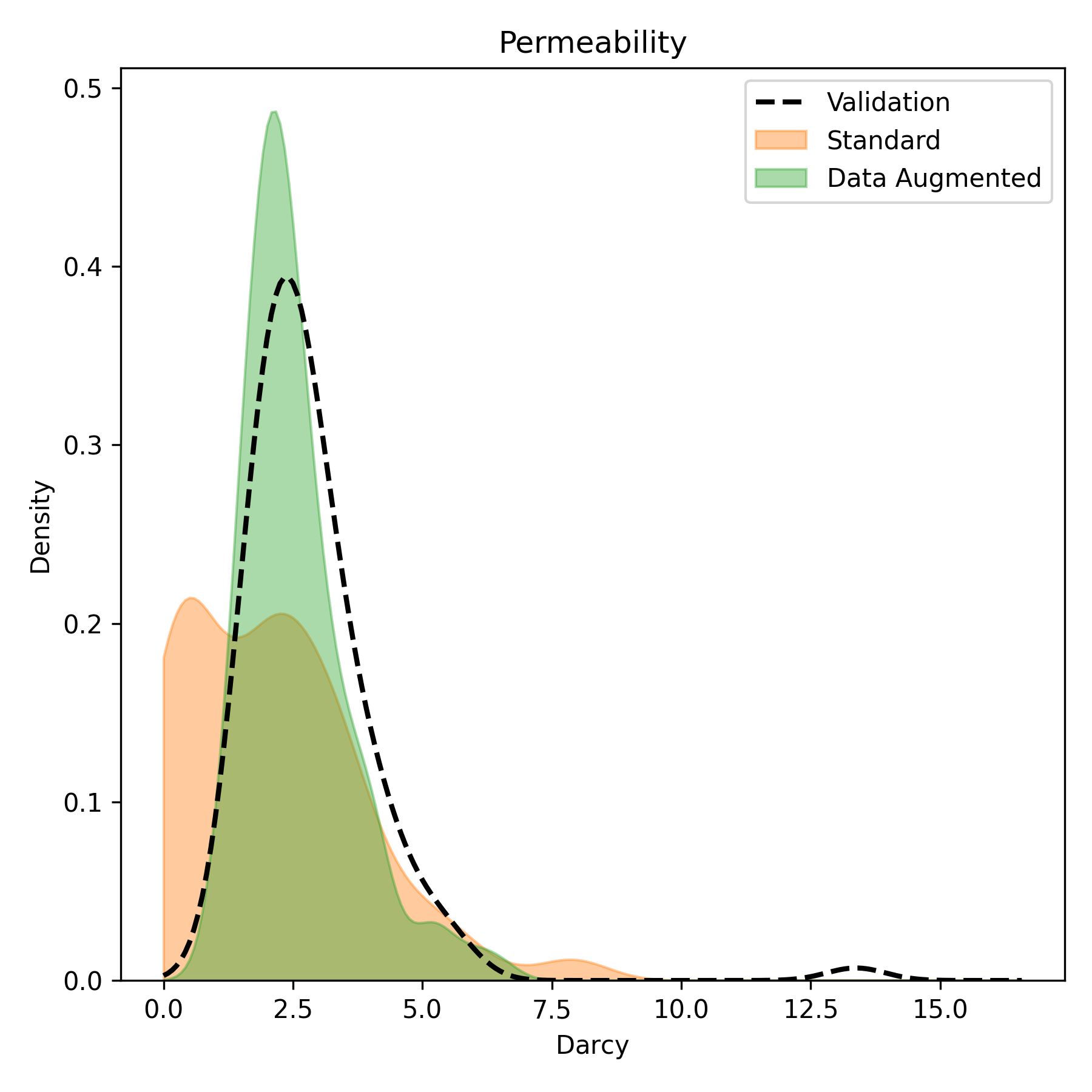}
        \caption{}
        \label{fig:bentheimer_256cd_b}
    \end{subfigure}
    
    \begin{subfigure}[t]{0.48\textwidth}
        \includegraphics[width=\textwidth]{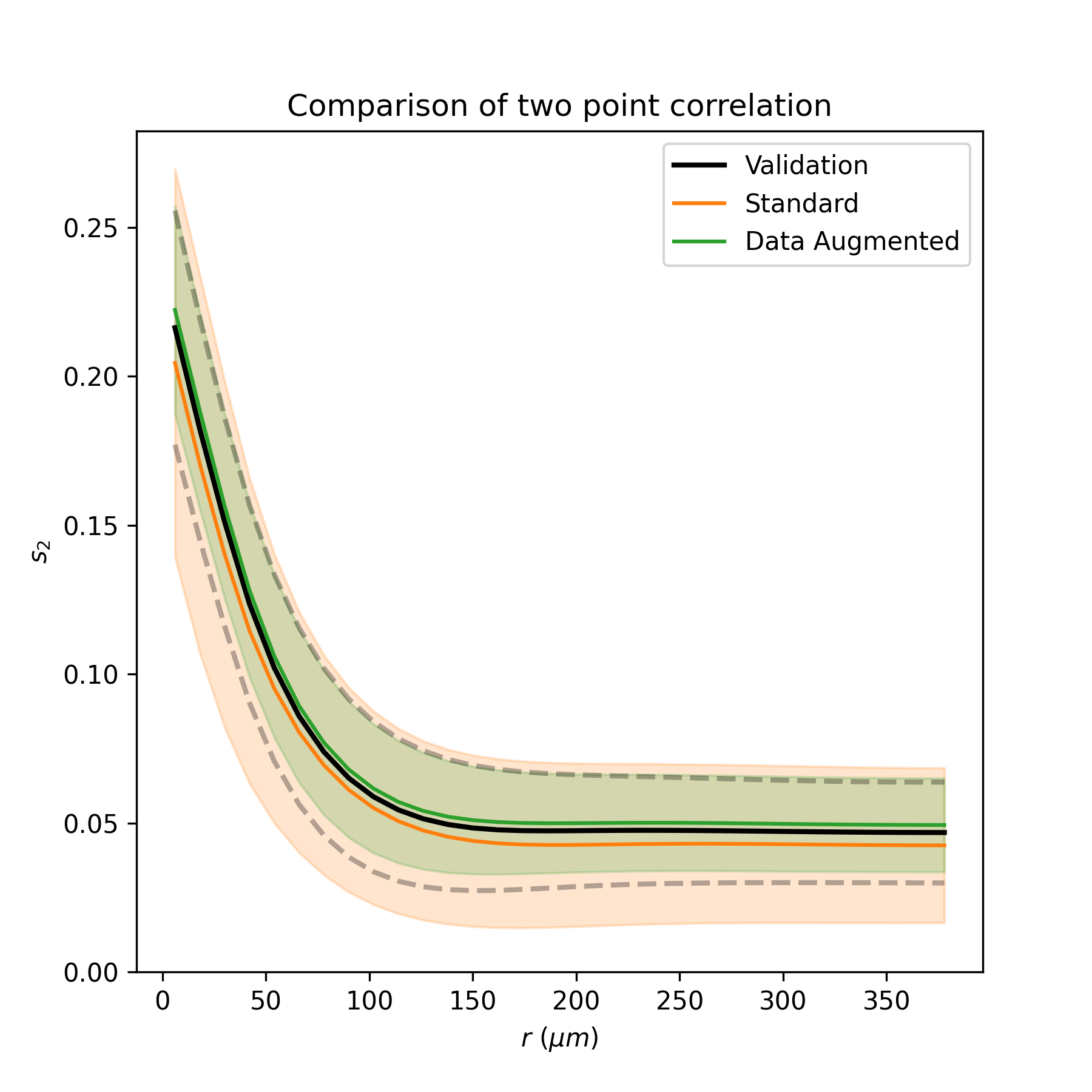}
        \caption{}
        \label{fig:bentheimer_256cd_c}
    \end{subfigure}
    \begin{subfigure}[t]{0.48\textwidth}
        \includegraphics[width=\textwidth]{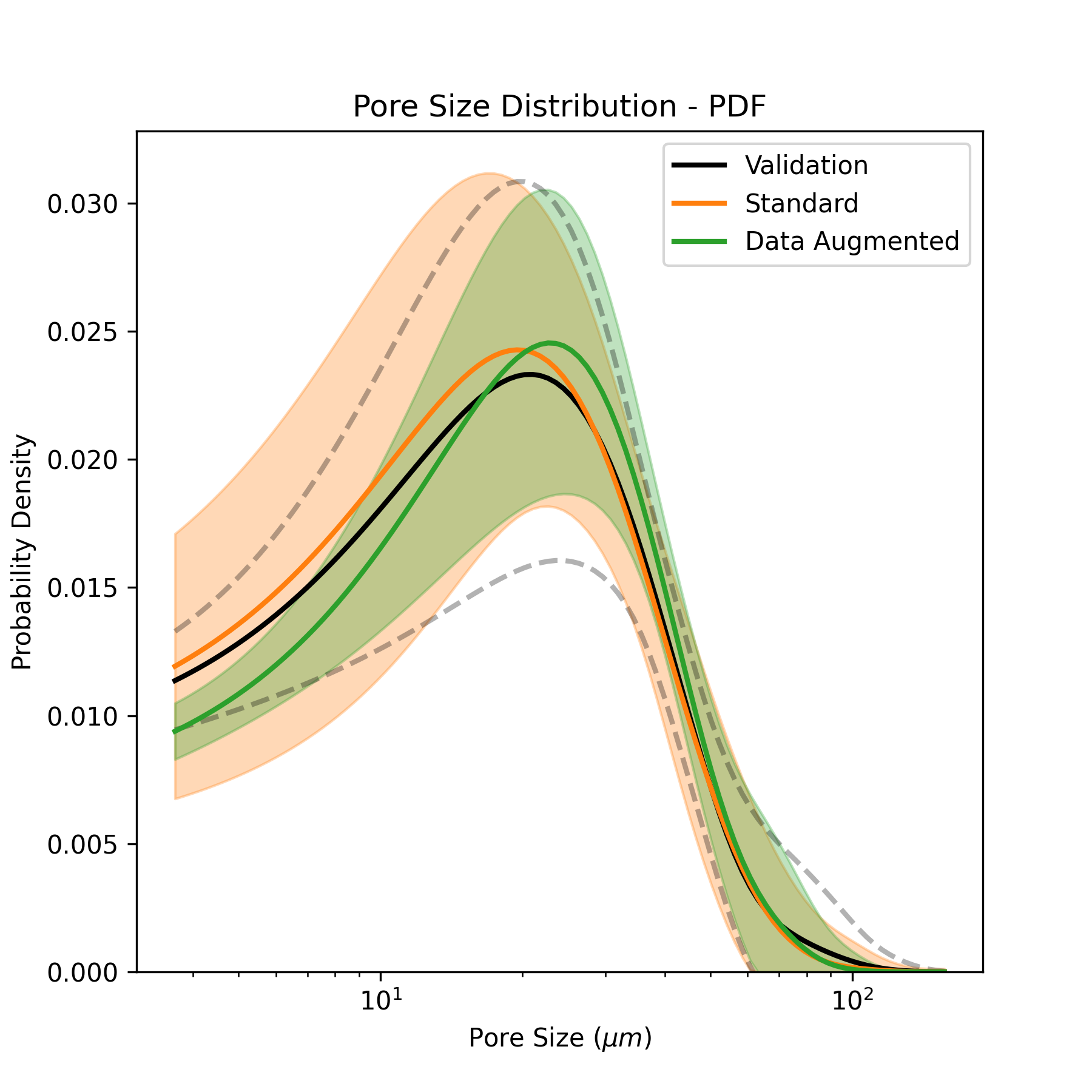}
        \caption{}
        \label{fig:bentheimer_256cd_d}
    \end{subfigure}
    \caption{Statistics for porosity-controlled generated Bentheimer volumes of size $256^3$, with and without data augmentation.}
    \label{fig:bentheimer_256_por_guided_no_data_aug}
\end{figure}

We show results for porosity-controlled and TPC-controlled samples of $256^3$ volumes in Figure \ref{fig:all_rocks_256c_comparison}, where sampling is performed by randomly choosing a volume from the training data and extracting the statistic considered. These are the results for a condition model trained using data augmentation as described in the previous section, which was found to be the most robust and reliable. Controlled sampling with a conditional model trained without data augmentation was unable to eliminate samples with spurious porosity and permeability, as can be seen in Figure \ref{fig:bentheimer_256_por_guided_no_data_aug} for Bentheimer $256^3$ volumes.

\begin{figure}[H]
   \centering
   \begin{subfigure}[t]{0.48\textwidth}
       \centering
       \includegraphics[width=\textwidth]{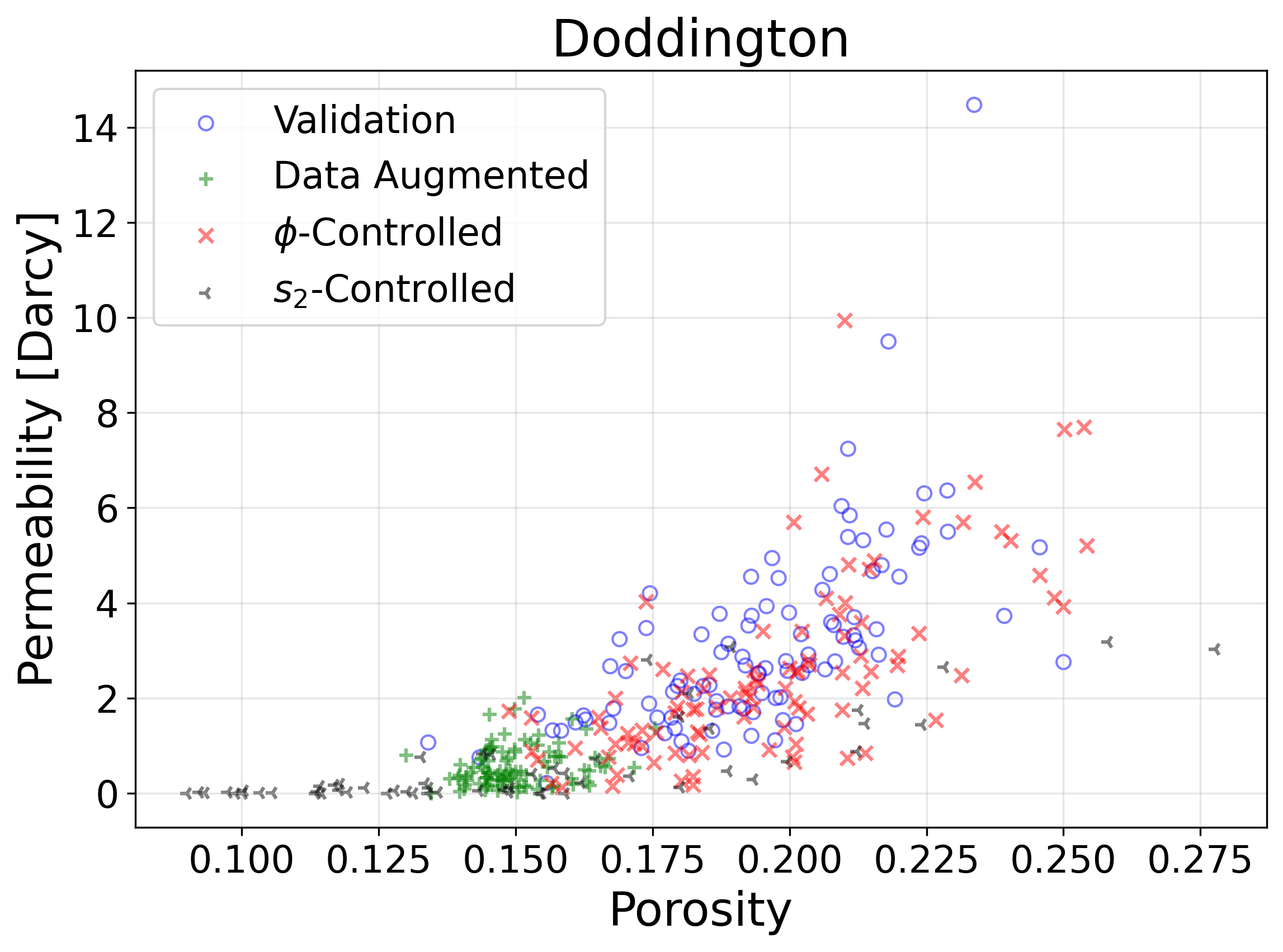}
       \caption{}
       \label{fig:scatter_dodd}
   \end{subfigure}%
   \hfill
   \begin{subfigure}[t]{0.48\textwidth}
       \centering
       \includegraphics[width=\textwidth]{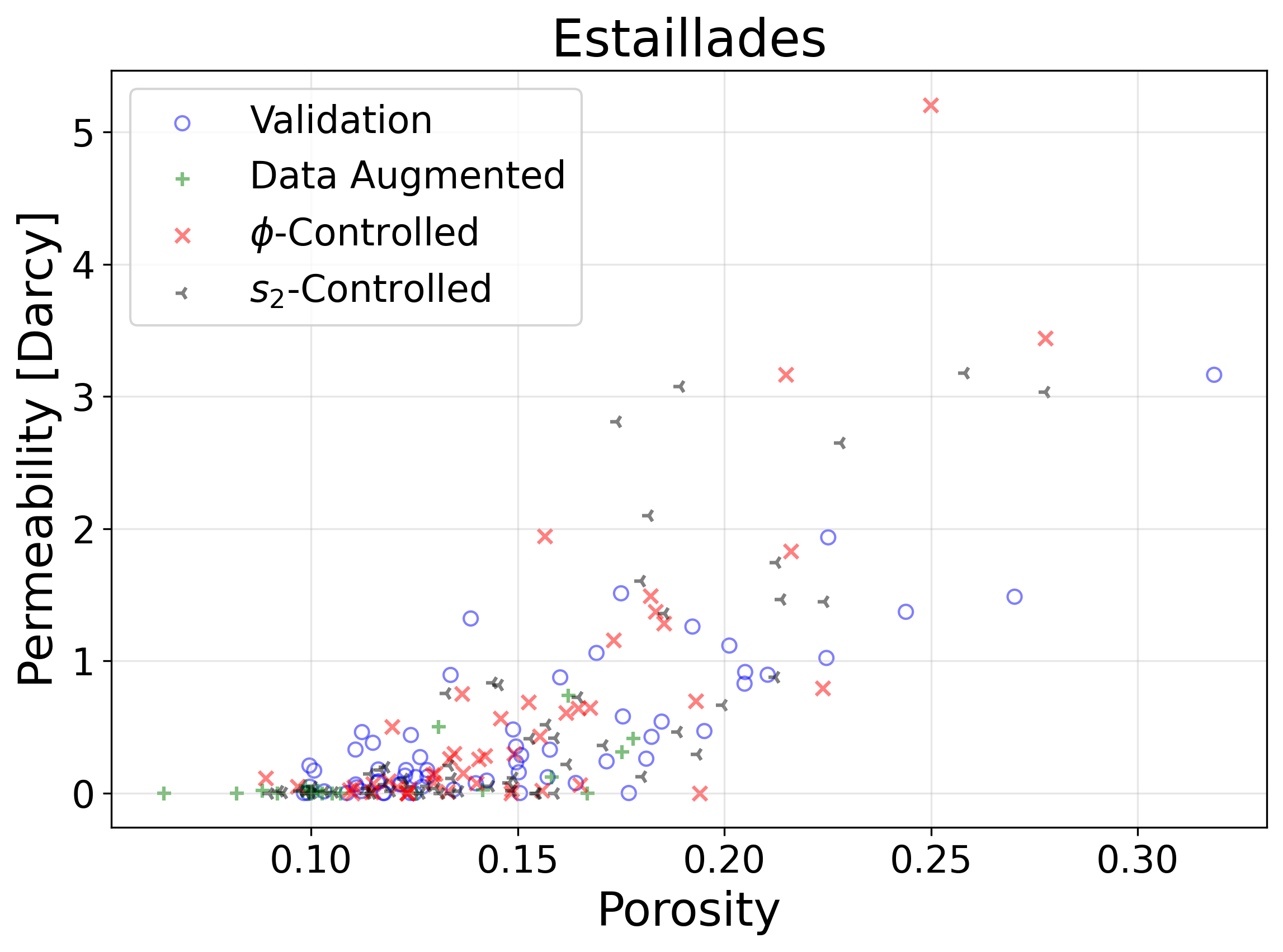}
       \caption{}
       \label{fig:scatter_est}
   \end{subfigure}
   
   \begin{subfigure}[t]{0.48\textwidth}
       \centering
       \includegraphics[width=\textwidth]{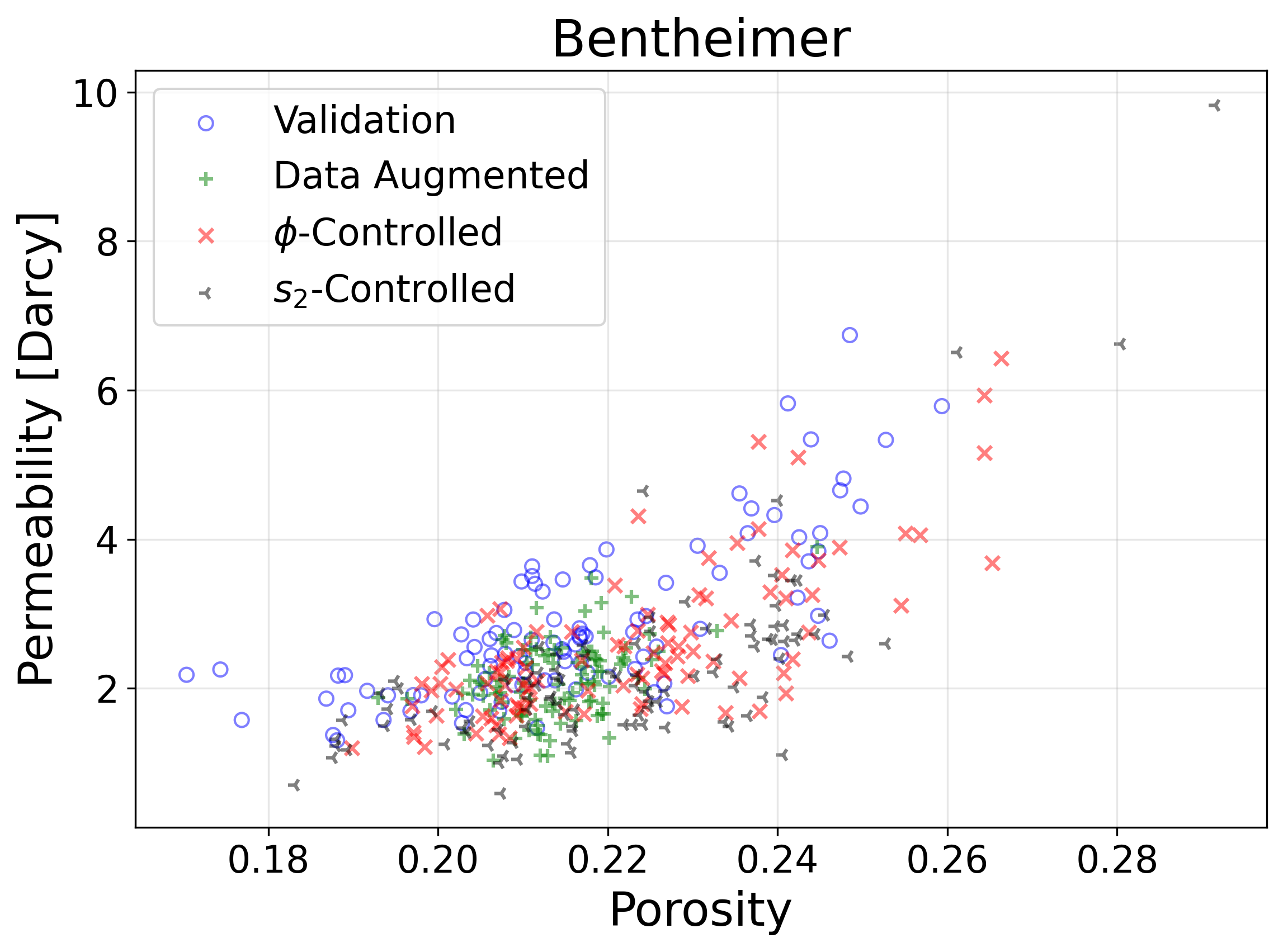}
       \caption{}
       \label{fig:scatter_bent}
   \end{subfigure}%
   \hfill
   \begin{subfigure}[t]{0.48\textwidth}
       \centering
       \includegraphics[width=\textwidth]{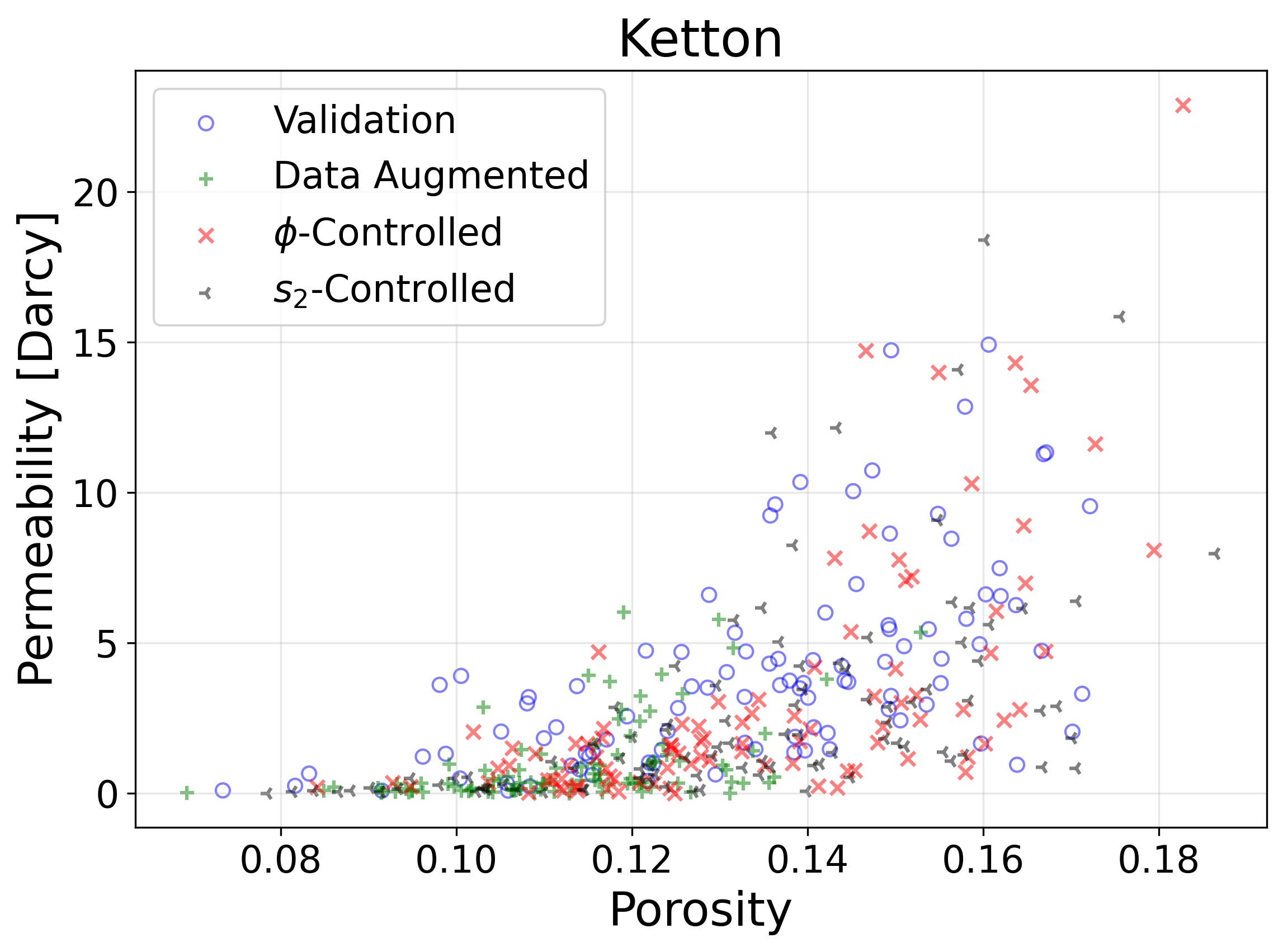}
       \caption{}
       \label{fig:scatter_ket}
   \end{subfigure}
   \caption{Effective porosity versus permeability scatter plots for unconditional (data augmented), porosity controlled and validation samples.}
   \label{fig:scatter_permeability_porosity}
\end{figure}

\begin{figure}[H]
    \centering
    
    % Bentheimer
    \begin{subfigure}{0.8\textwidth}
        \centering
        \includegraphics[width=\textwidth]{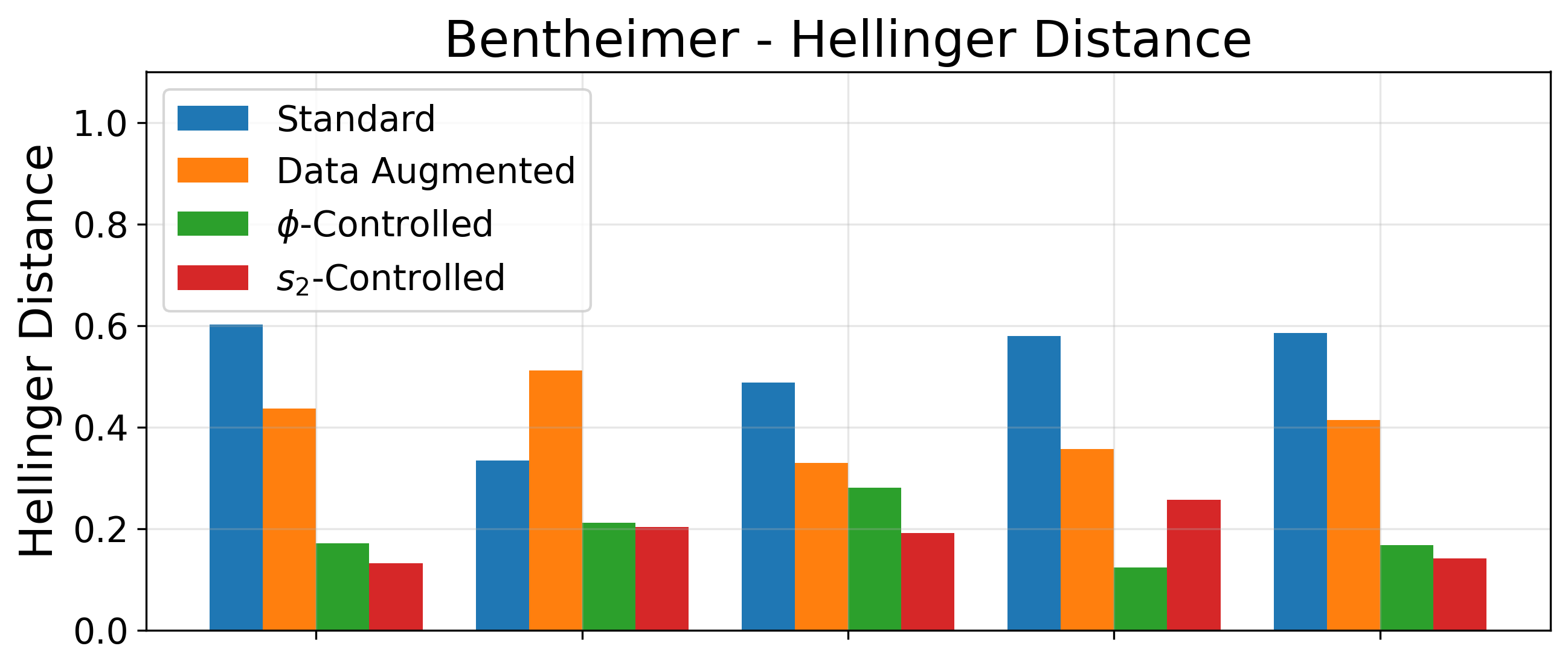}
    \end{subfigure}
    
    % Doddington
    \begin{subfigure}{0.8\textwidth}
        \centering
        \includegraphics[width=\textwidth]{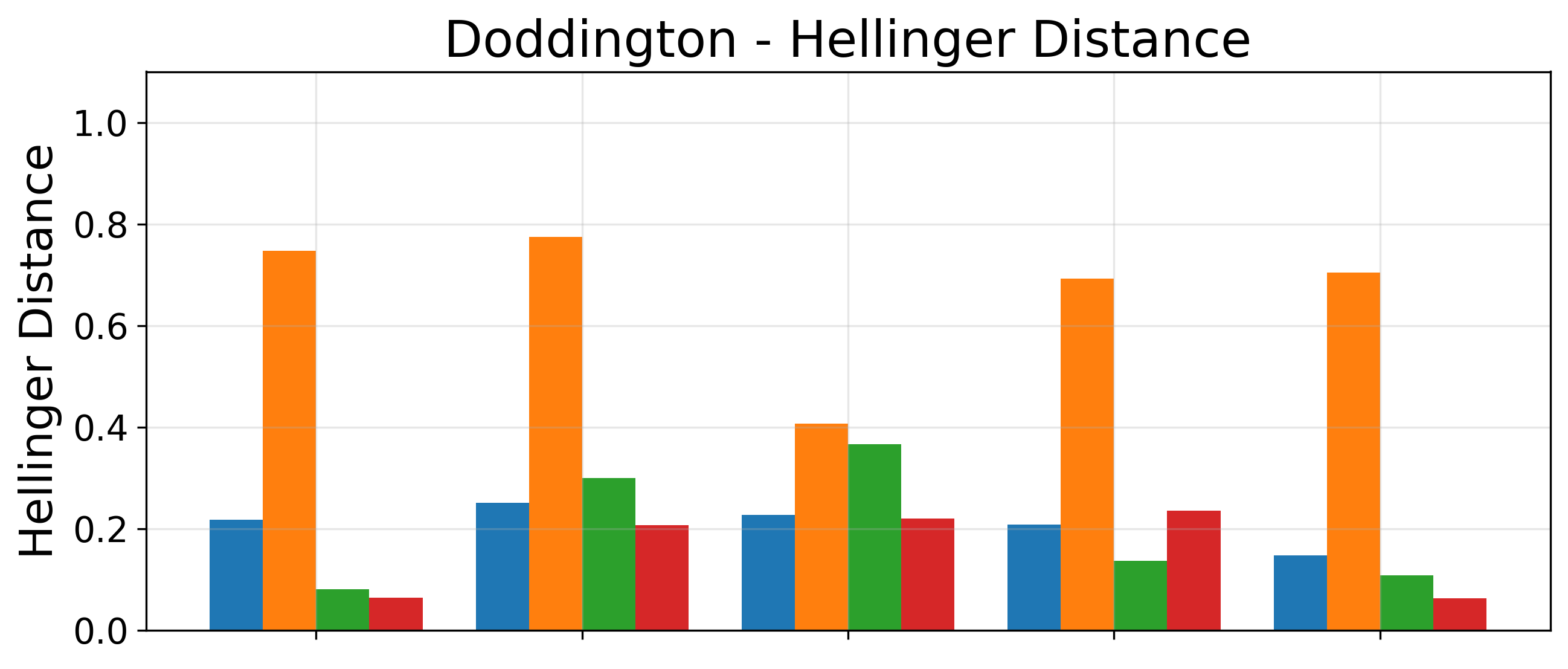}
    \end{subfigure}
    
    % Estaillades
    \begin{subfigure}{0.8\textwidth}
        \centering
        \includegraphics[width=\textwidth]{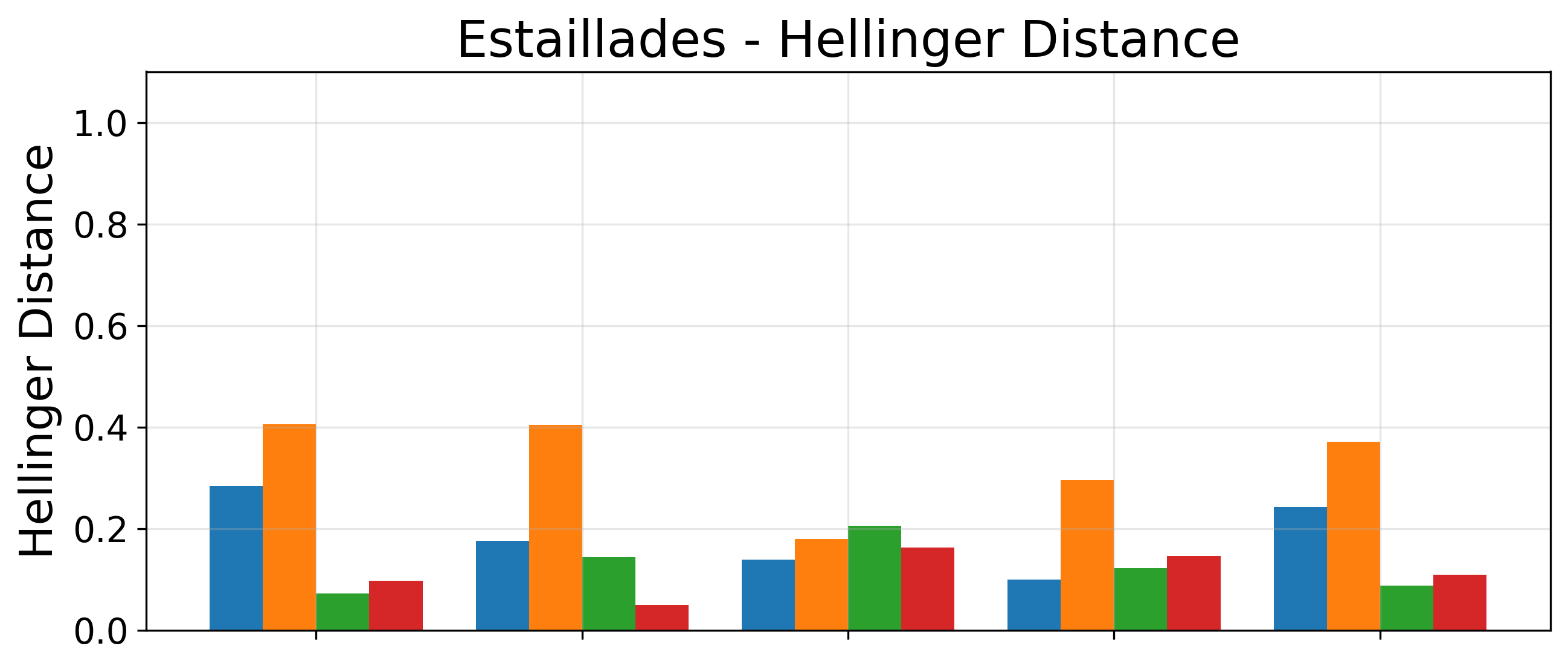}
    \end{subfigure}
    
    % Ketton
    \begin{subfigure}{0.8\textwidth}
        \centering
        \includegraphics[width=\textwidth]{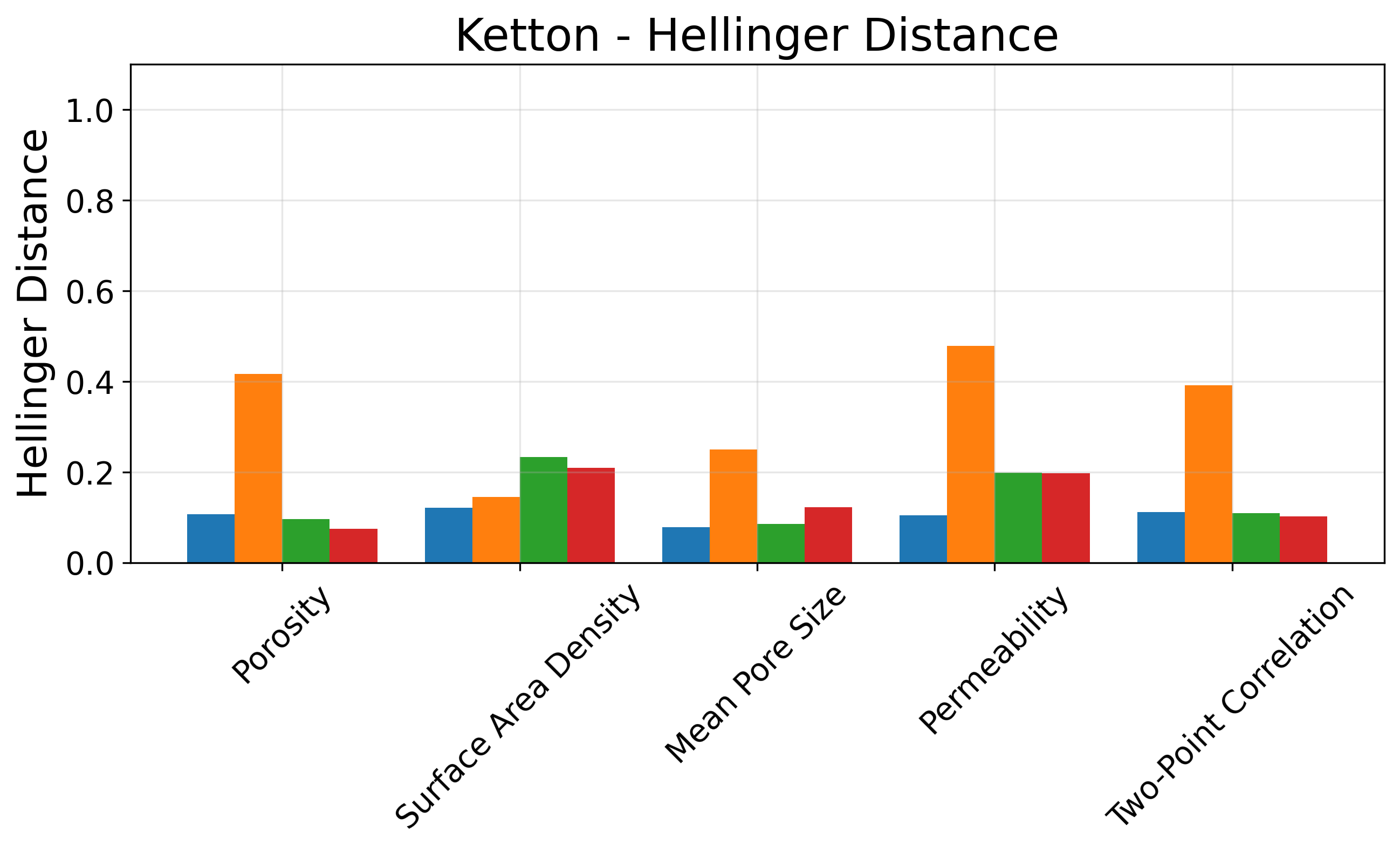}
    \end{subfigure}
    
    \caption{Hellinger distance of the statistics for $256^3$ samples for different sample generation techniques.}
    \label{fig:hellinger_mre}
\end{figure}

Remarkably, controlling only on porosity is sufficient to essentially eliminate mode collapse in all considered statistics, and we can also recover the correct relation between permeability\footnote{See  \ref{app:permeability_calculation} for details on how we calculate permeability.} and porosity\footnote{As discussed in \ref{app:porosity_calculation}, in all samples considered porosity and effective porosity are essentially equal, and thus we use the terms porosity and effective porosity interchangeably.}, as shown in Figure \ref{fig:scatter_permeability_porosity}. Controlling also on the TPC, which increases the training time by $2.5$x, improves the generated distribution of some but not all statistics. A concise comparison between all $256^3$ models can be seen in Figure \ref{fig:hellinger_mre}.

\begin{figure}[H]
    \centering
    % Bentheimer
    \begin{subfigure}[t]{0.15\textwidth}
        \includegraphics[width=\textwidth]{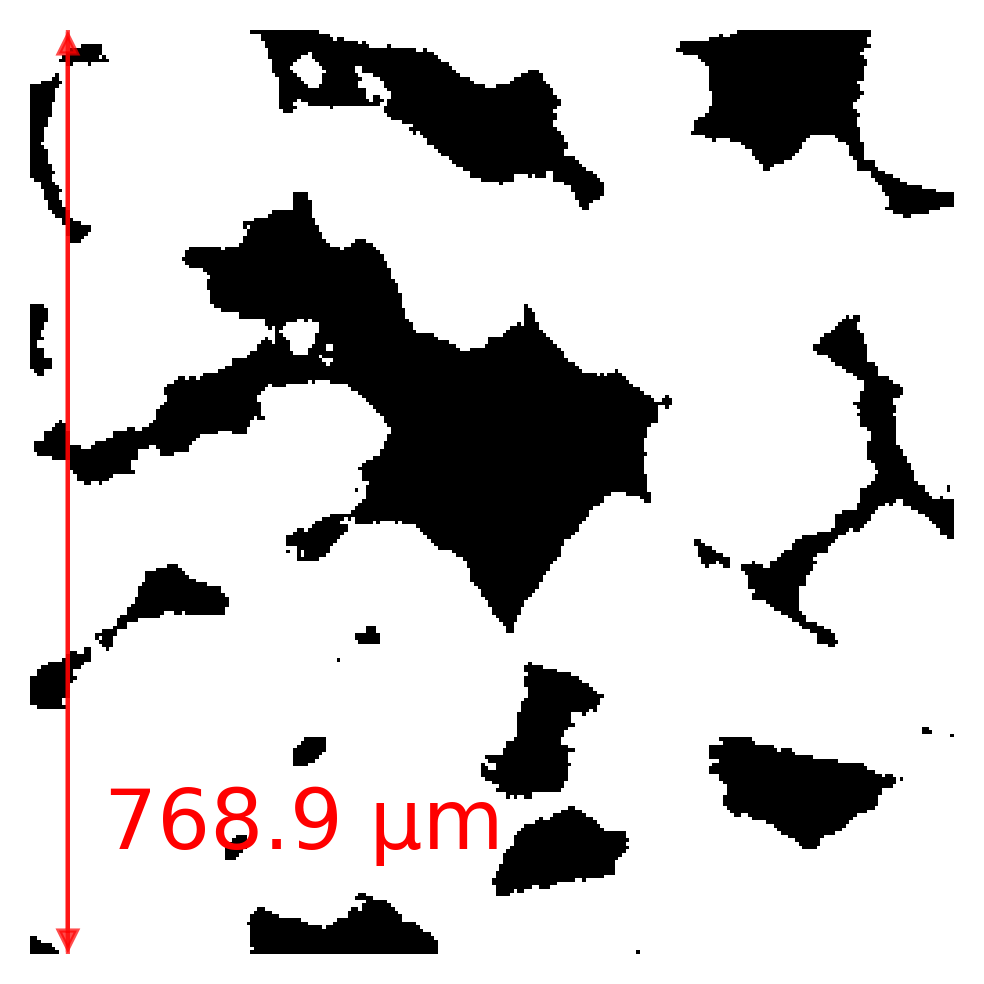}
        \label{fig:bentheimer_valid_0}
    \end{subfigure}
    \begin{subfigure}[t]{0.15\textwidth}
        \includegraphics[width=\textwidth]{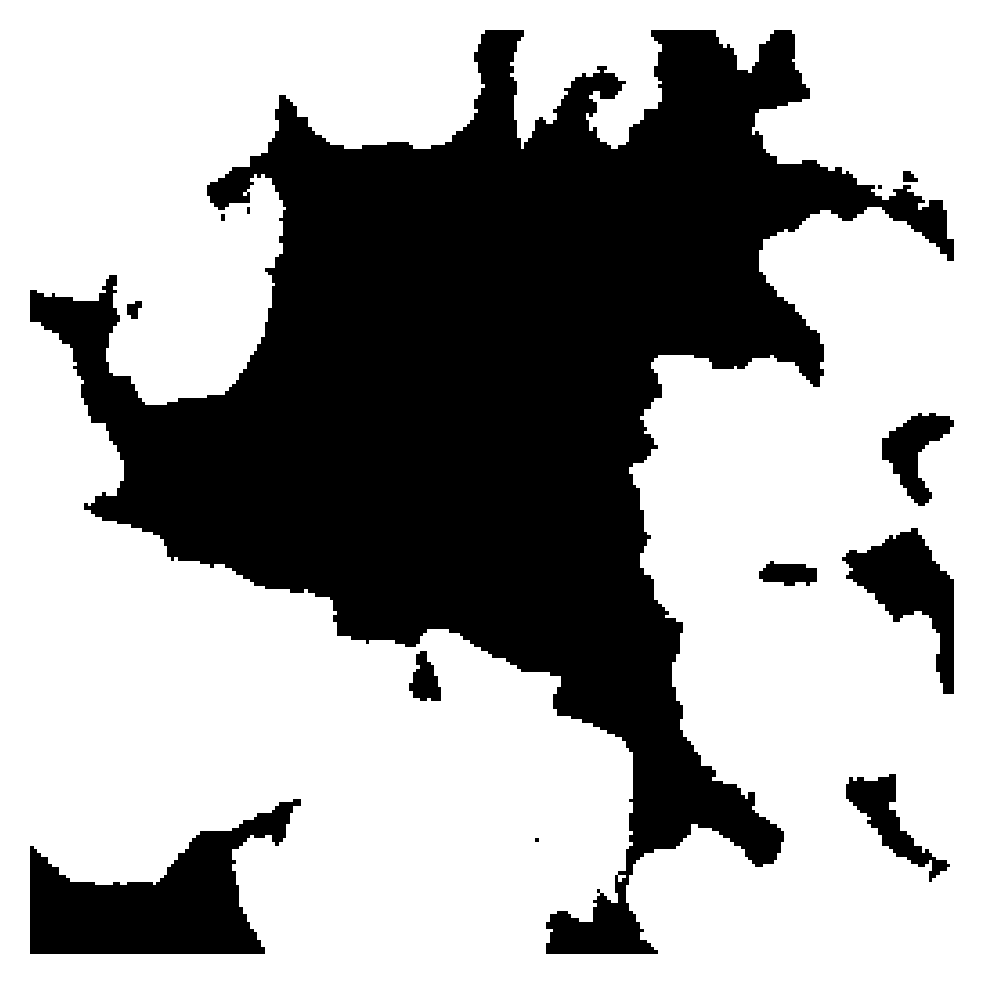}
        \label{fig:bentheimer_valid_127}
    \end{subfigure}
    \begin{subfigure}[t]{0.15\textwidth}
        \includegraphics[width=\textwidth]{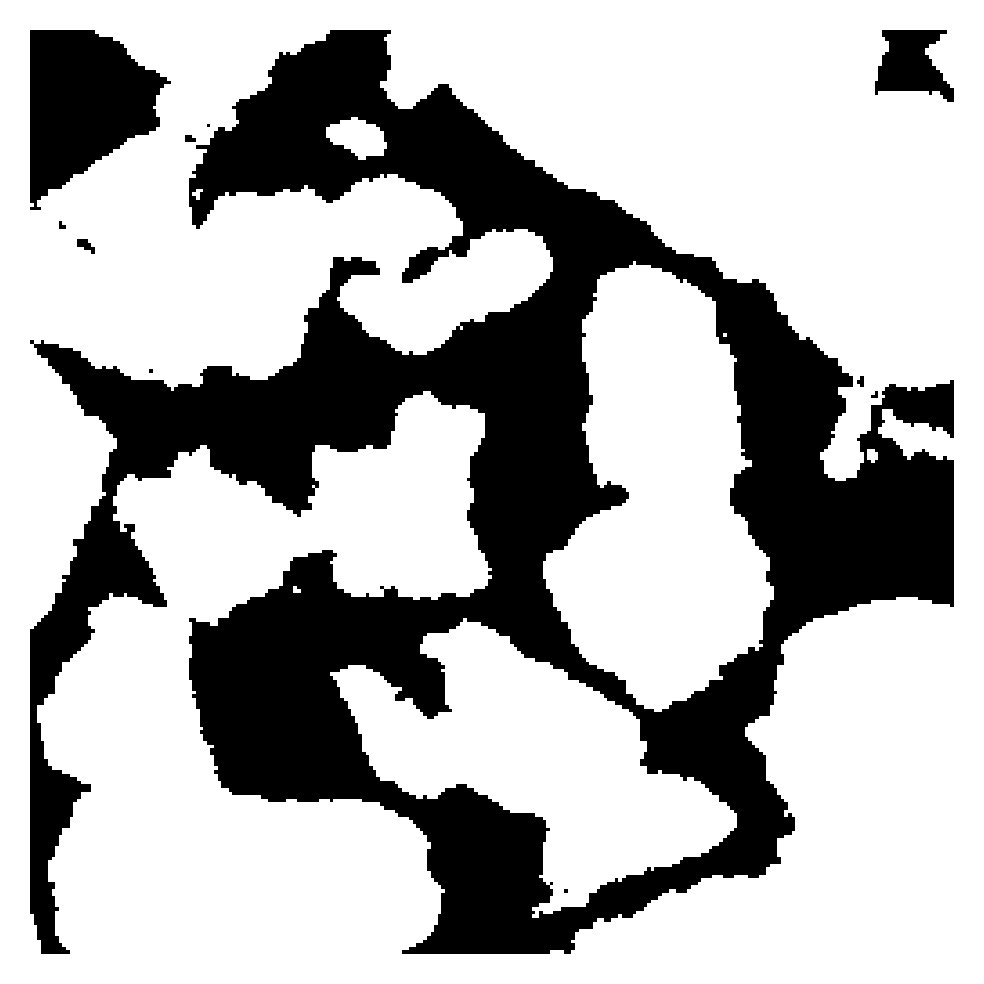}
        \label{fig:bentheimer_valid_255}
    \end{subfigure}
    \vline
    \begin{subfigure}[t]{0.15\textwidth}
        \includegraphics[width=\textwidth]{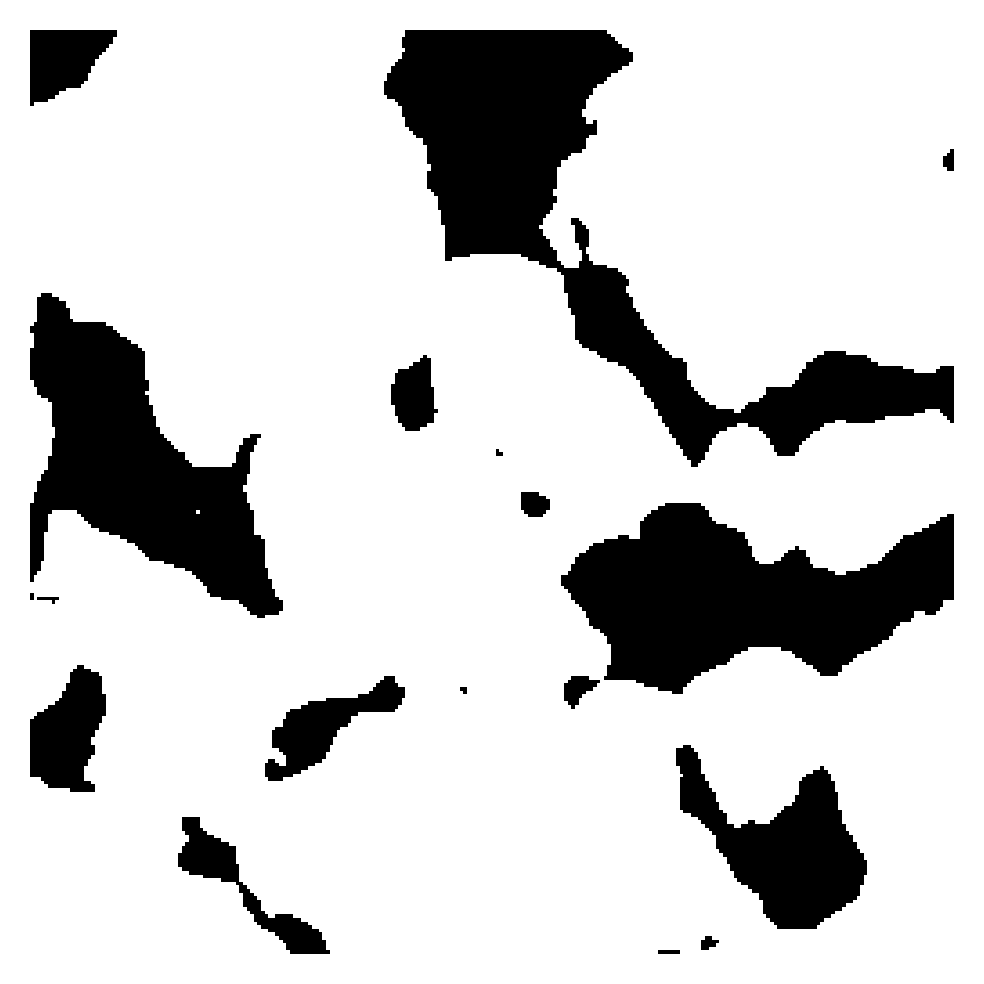}
        \label{fig:bentheimer_gen_0}
    \end{subfigure}
    \begin{subfigure}[t]{0.15\textwidth}
        \includegraphics[width=\textwidth]{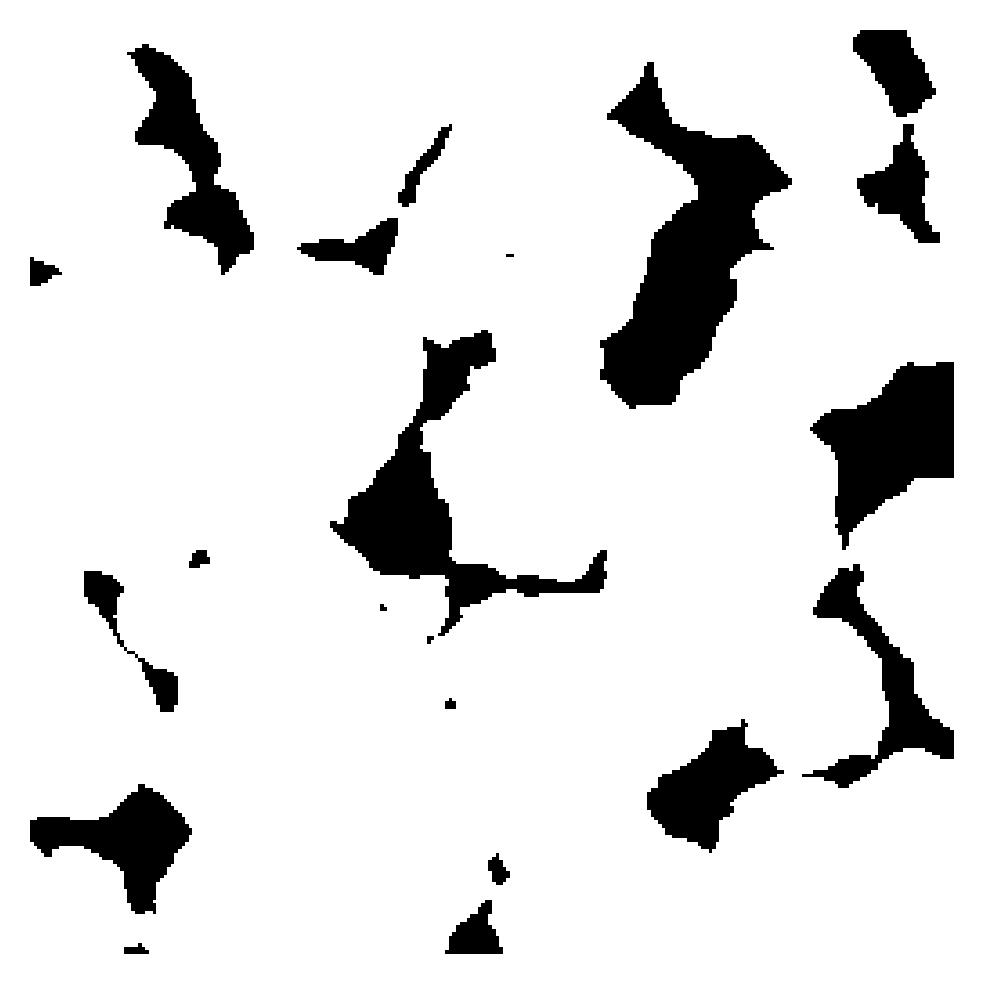}
        \label{fig:bentheimer_gen_127}
    \end{subfigure}
    \begin{subfigure}[t]{0.15\textwidth}
        \includegraphics[width=\textwidth]{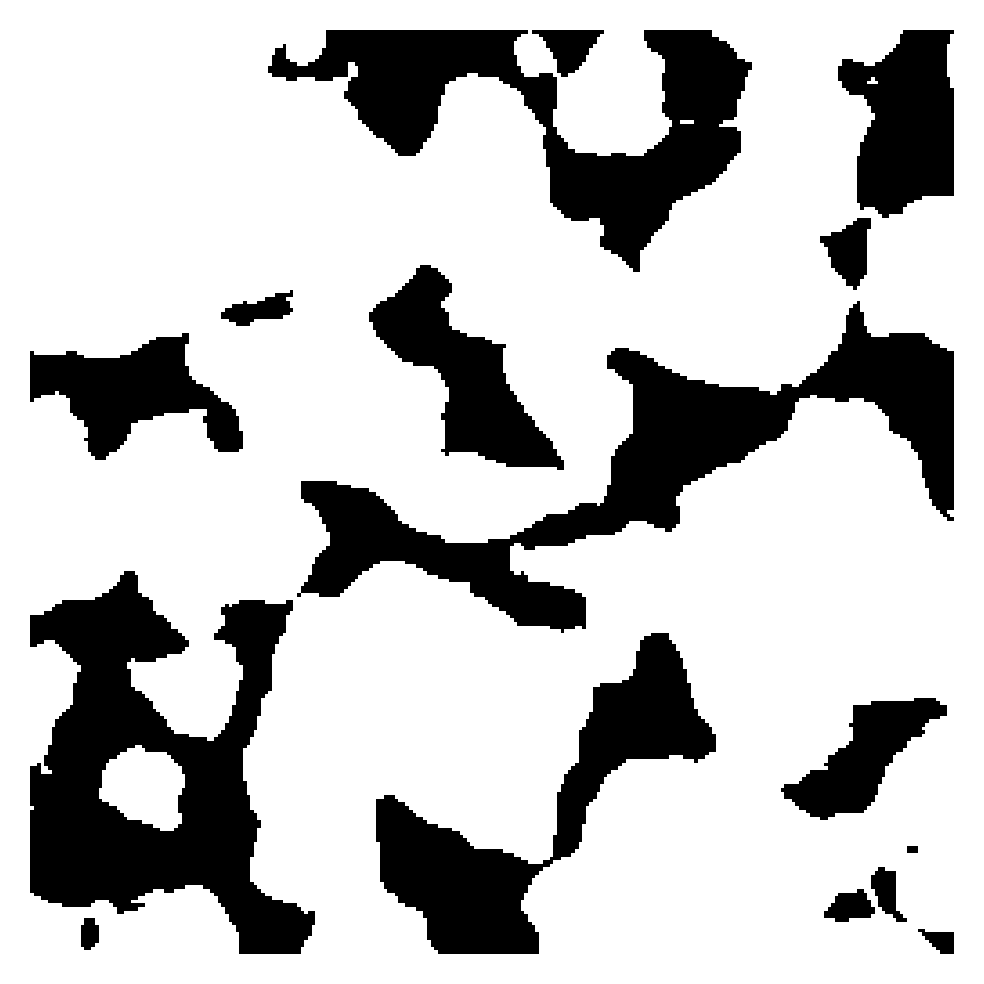}
        \label{fig:bentheimer_gen_255}
    \end{subfigure}

    \hrule

    % Doddington
    \begin{subfigure}[t]{0.15\textwidth}
        \includegraphics[width=\textwidth]{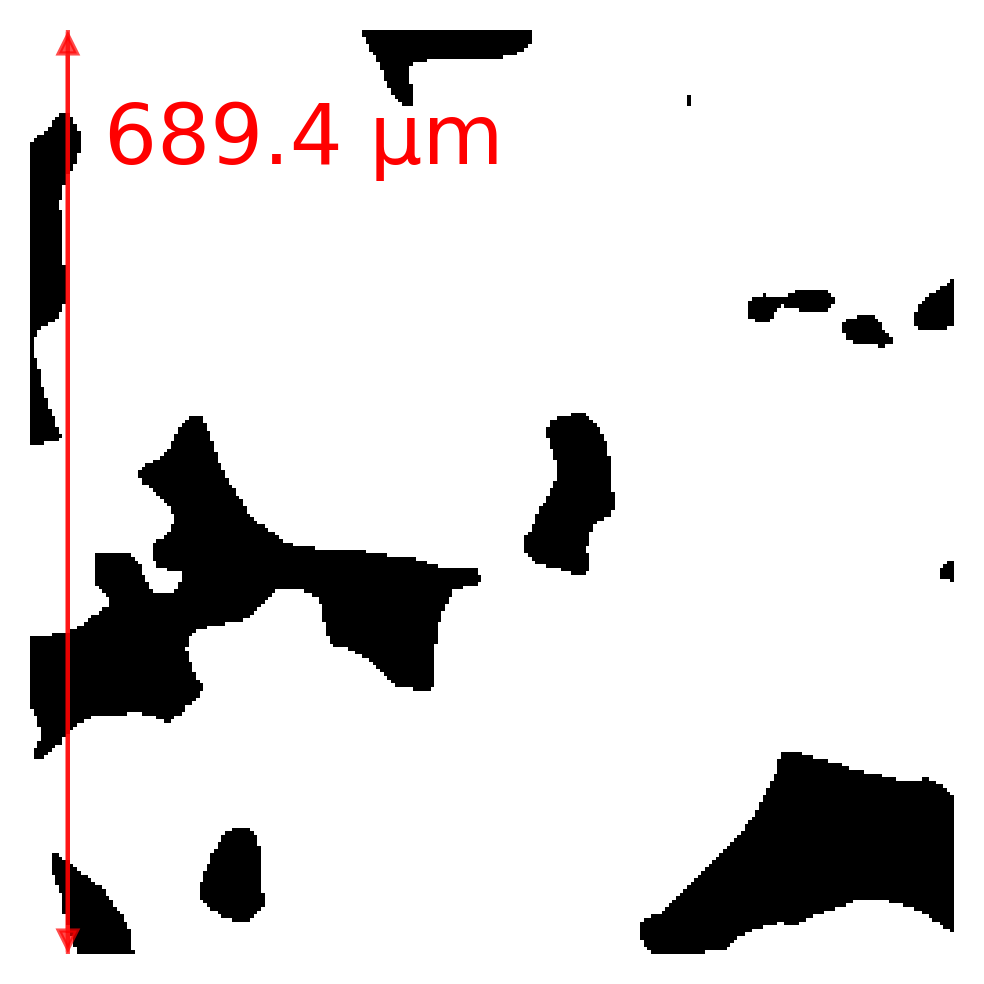}
        \label{fig:doddington_valid_0}
    \end{subfigure}
    \begin{subfigure}[t]{0.15\textwidth}
        \includegraphics[width=\textwidth]{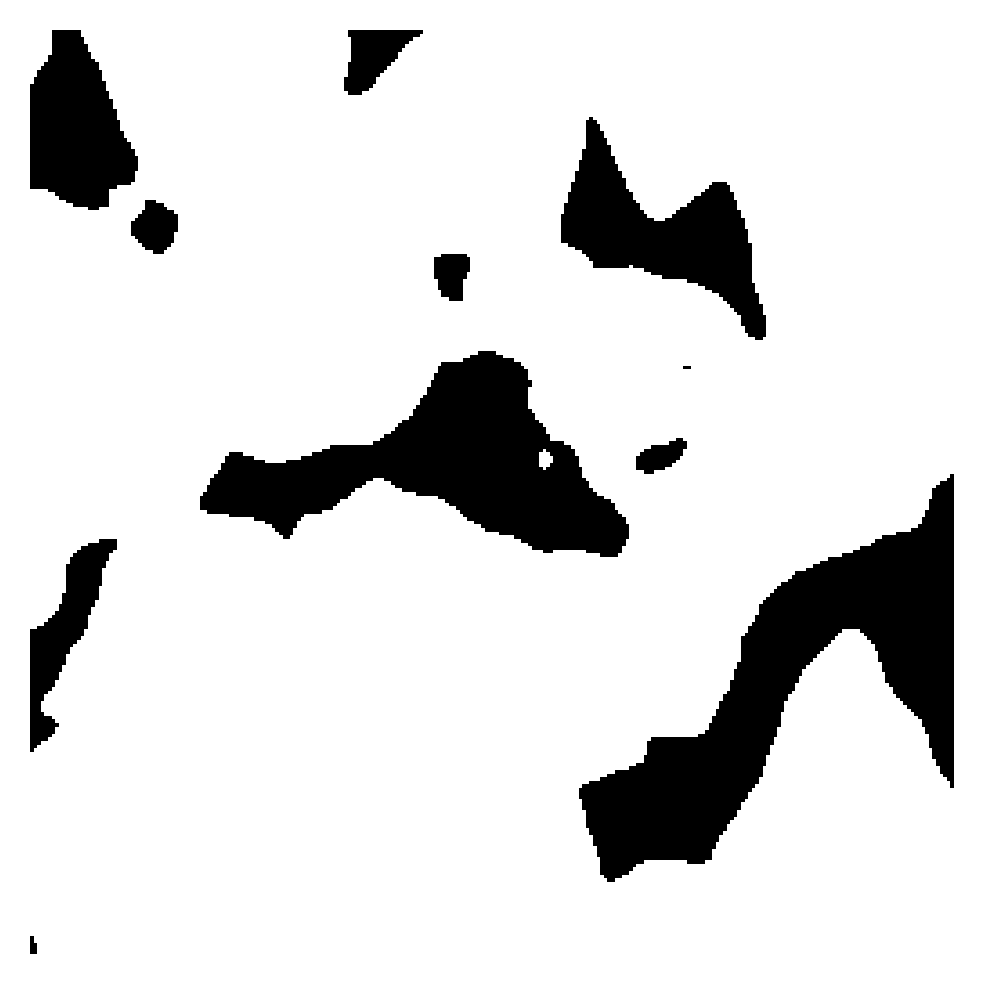}
        \label{fig:doddington_valid_127}
    \end{subfigure}
    \begin{subfigure}[t]{0.15\textwidth}
        \includegraphics[width=\textwidth]{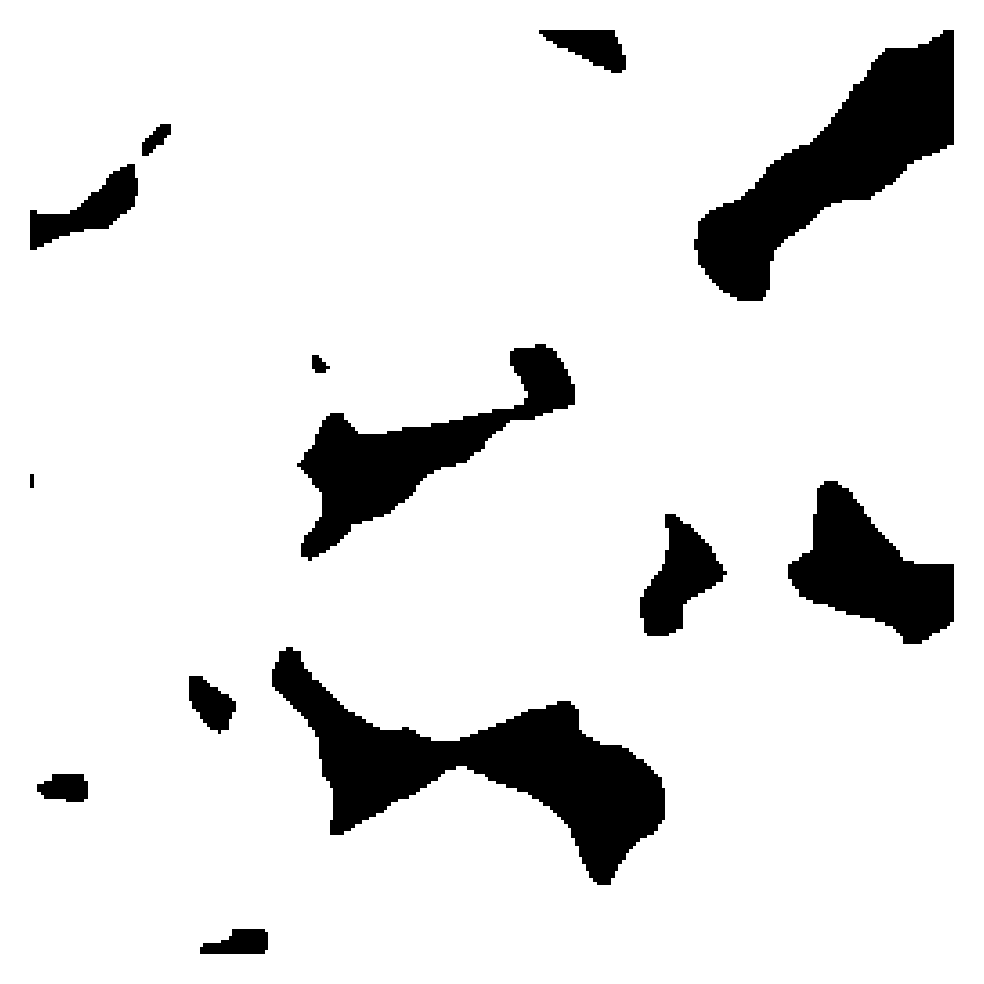}
        \label{fig:doddington_valid_255}
    \end{subfigure}
    \vline
    \begin{subfigure}[t]{0.15\textwidth}
        \includegraphics[width=\textwidth]{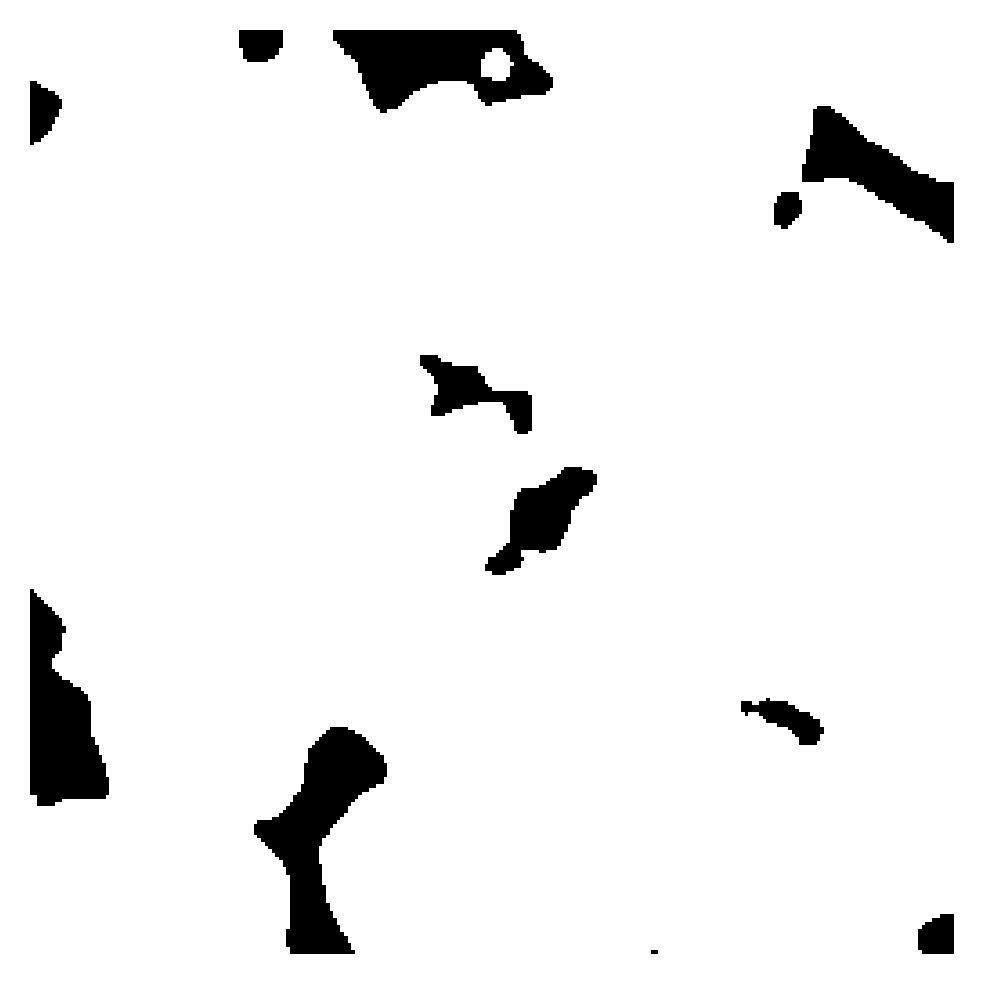}
        \label{fig:doddington_gen_0}
    \end{subfigure}
    \begin{subfigure}[t]{0.15\textwidth}
        \includegraphics[width=\textwidth]{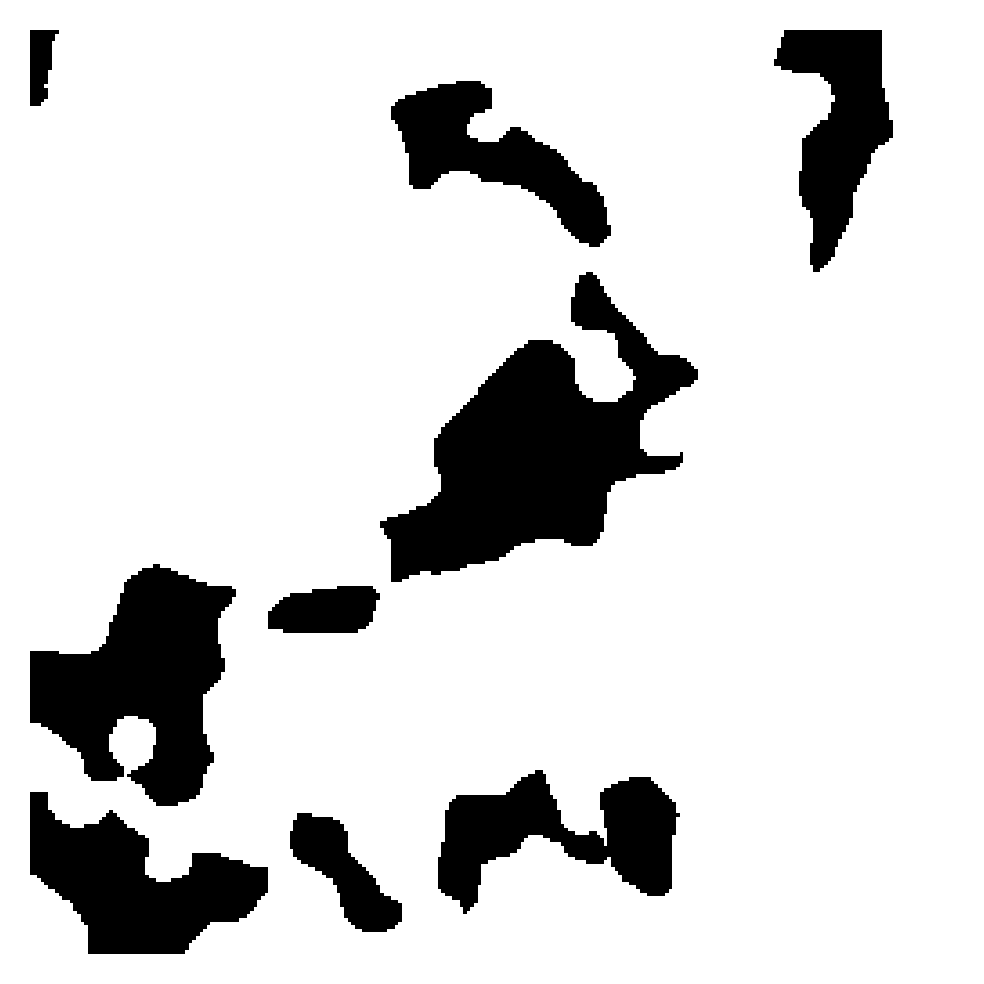}
        \label{fig:doddington_gen_127}
    \end{subfigure}
    \begin{subfigure}[t]{0.15\textwidth}
        \includegraphics[width=\textwidth]{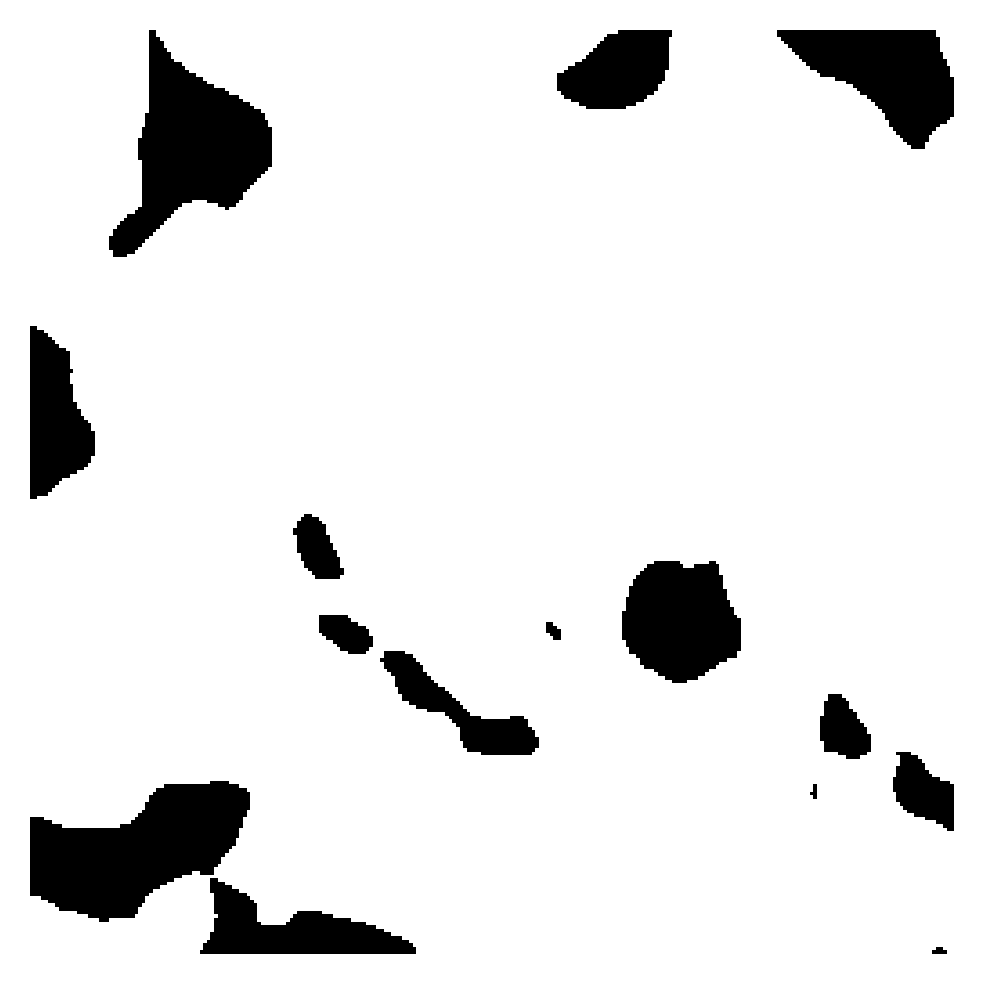}
        \label{fig:doddington_gen_255}
    \end{subfigure}

    \hrule

    % Estaillades
    \begin{subfigure}[t]{0.15\textwidth}
        \includegraphics[width=\textwidth]{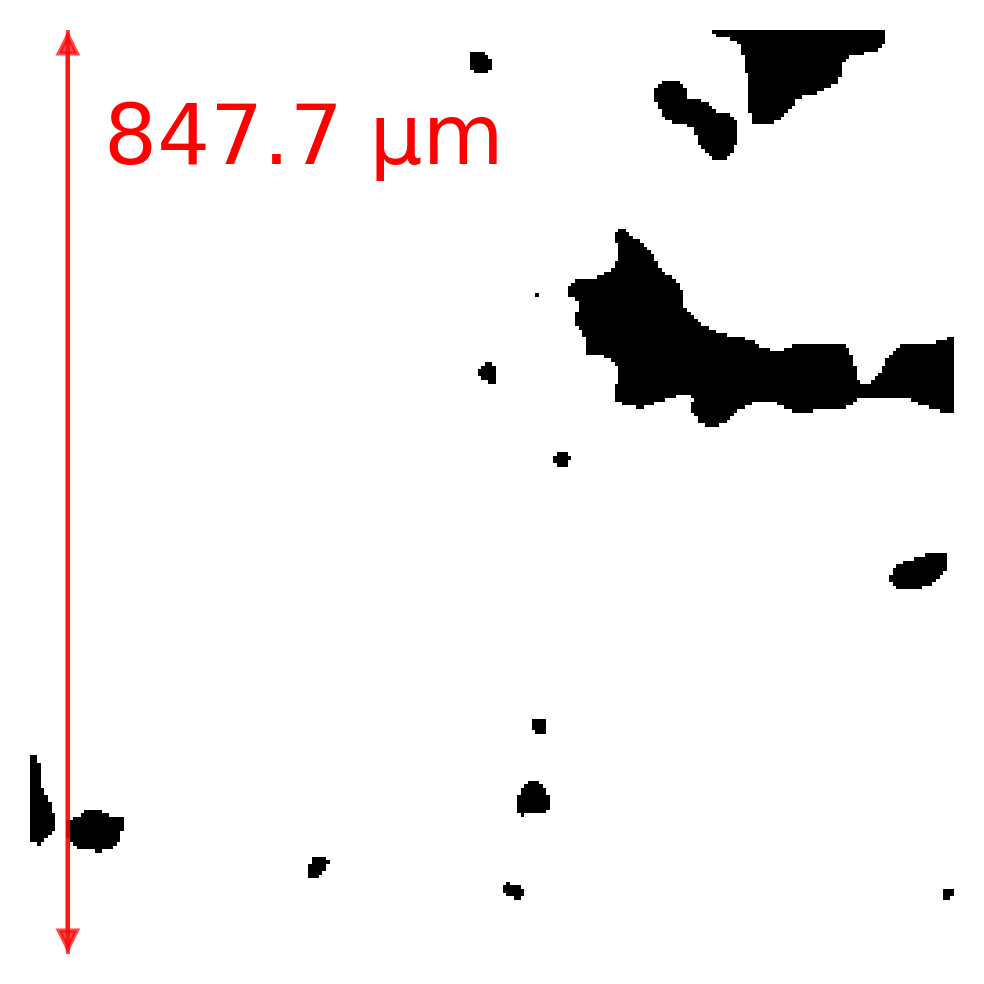}
        \label{fig:estaillades_valid_0}
    \end{subfigure}
    \begin{subfigure}[t]{0.15\textwidth}
        \includegraphics[width=\textwidth]{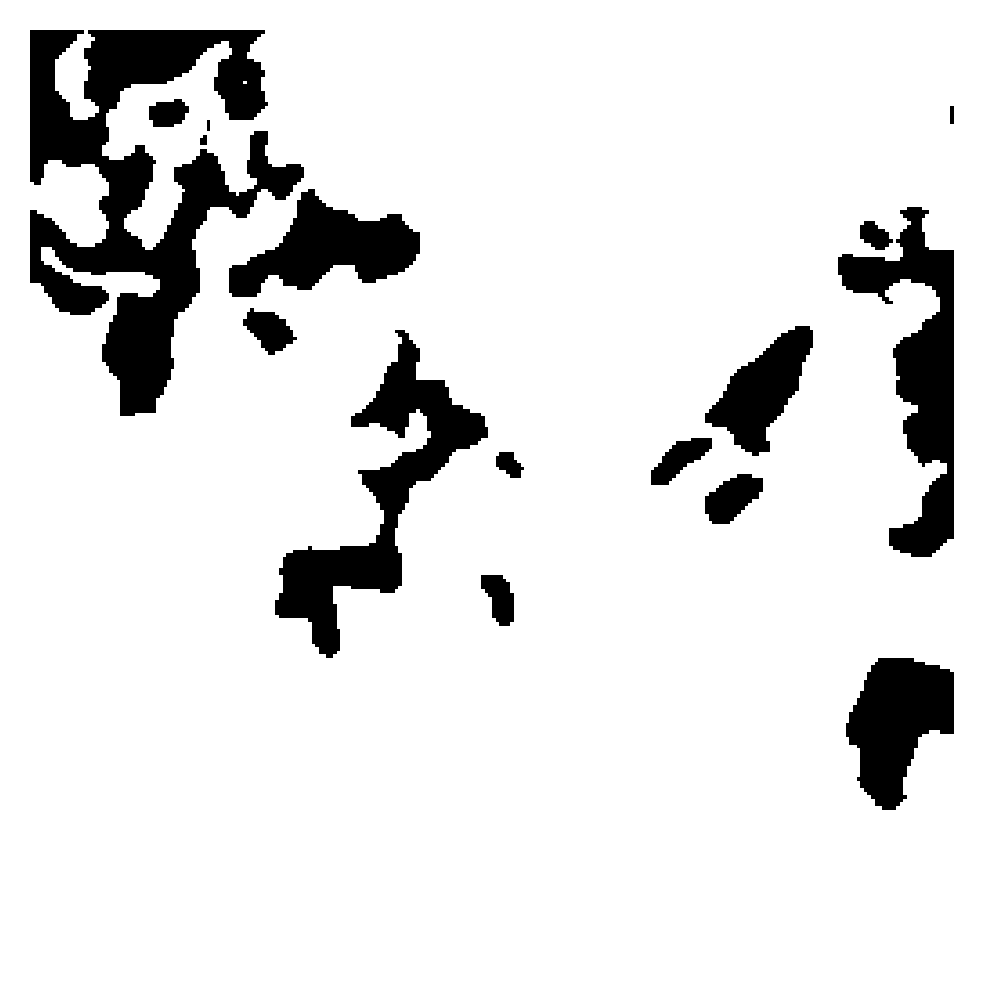}
        \label{fig:estaillades_valid_127}
    \end{subfigure}
    \begin{subfigure}[t]{0.15\textwidth}
        \includegraphics[width=\textwidth]{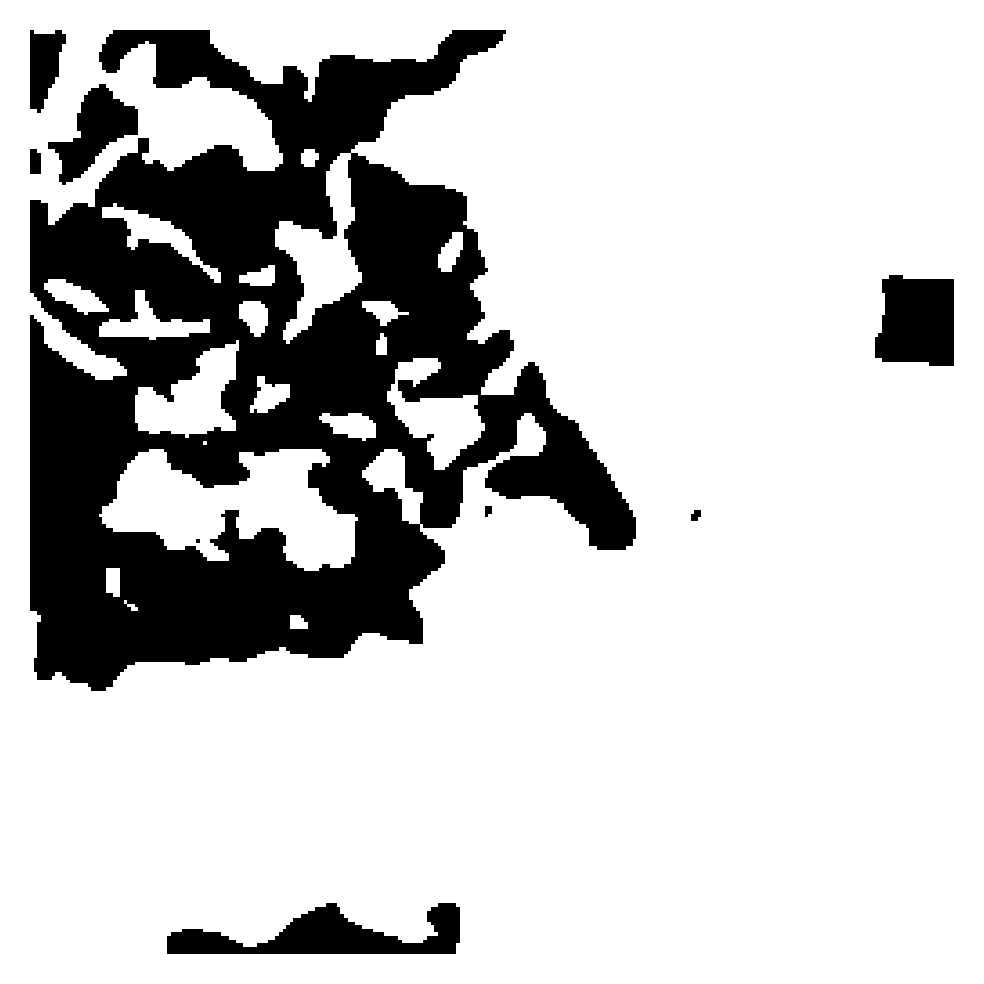}
        \label{fig:estaillades_valid_255}
    \end{subfigure}
    \vline
    \begin{subfigure}[t]{0.15\textwidth}
        \includegraphics[width=\textwidth]{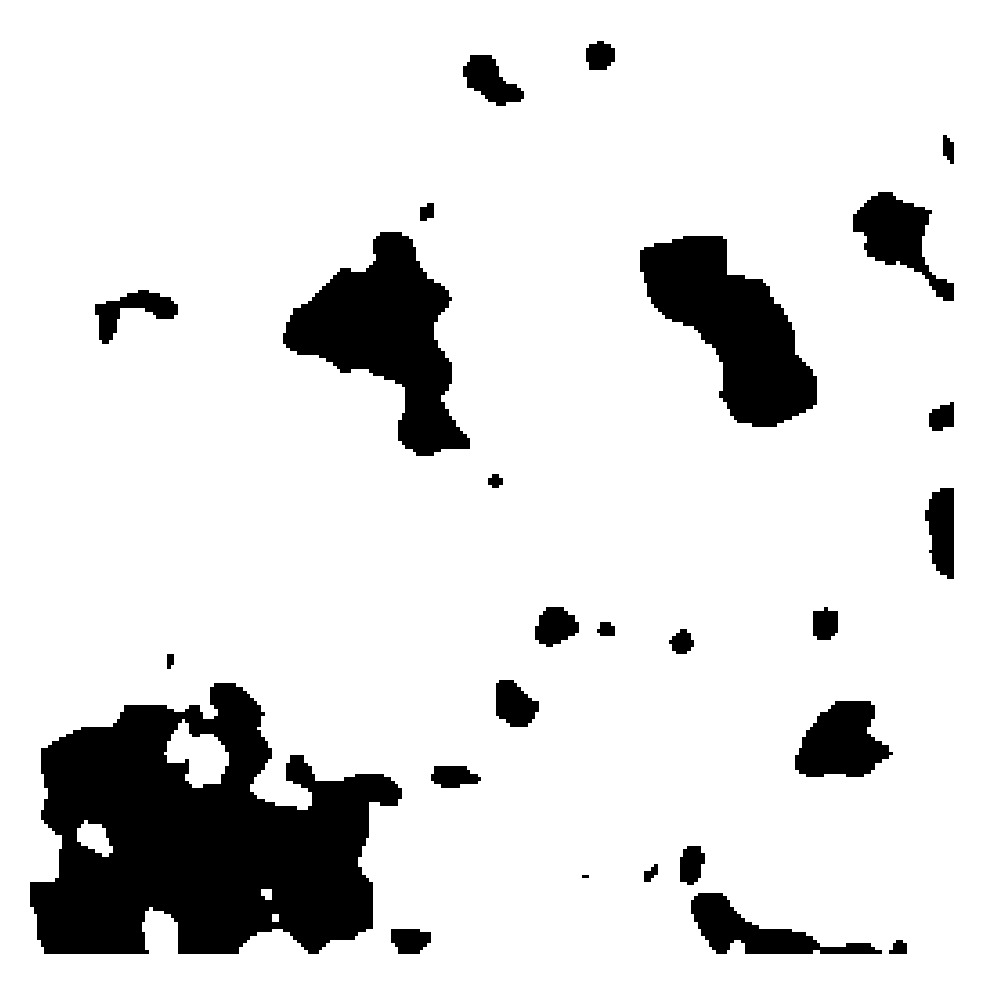}
        \label{fig:estaillades_gen_0}
    \end{subfigure}
    \begin{subfigure}[t]{0.15\textwidth}
        \includegraphics[width=\textwidth]{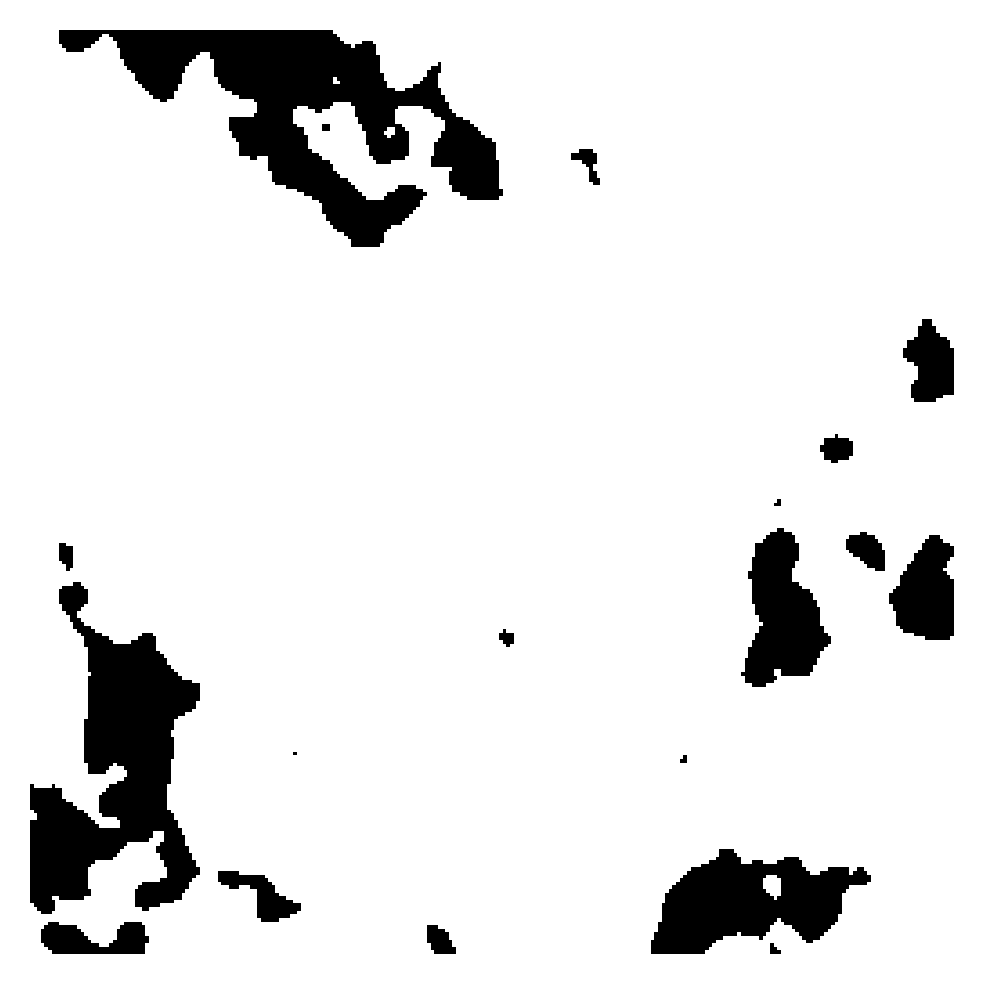}
        \label{fig:estaillades_gen_127}
    \end{subfigure}
    \begin{subfigure}[t]{0.15\textwidth}
        \includegraphics[width=\textwidth]{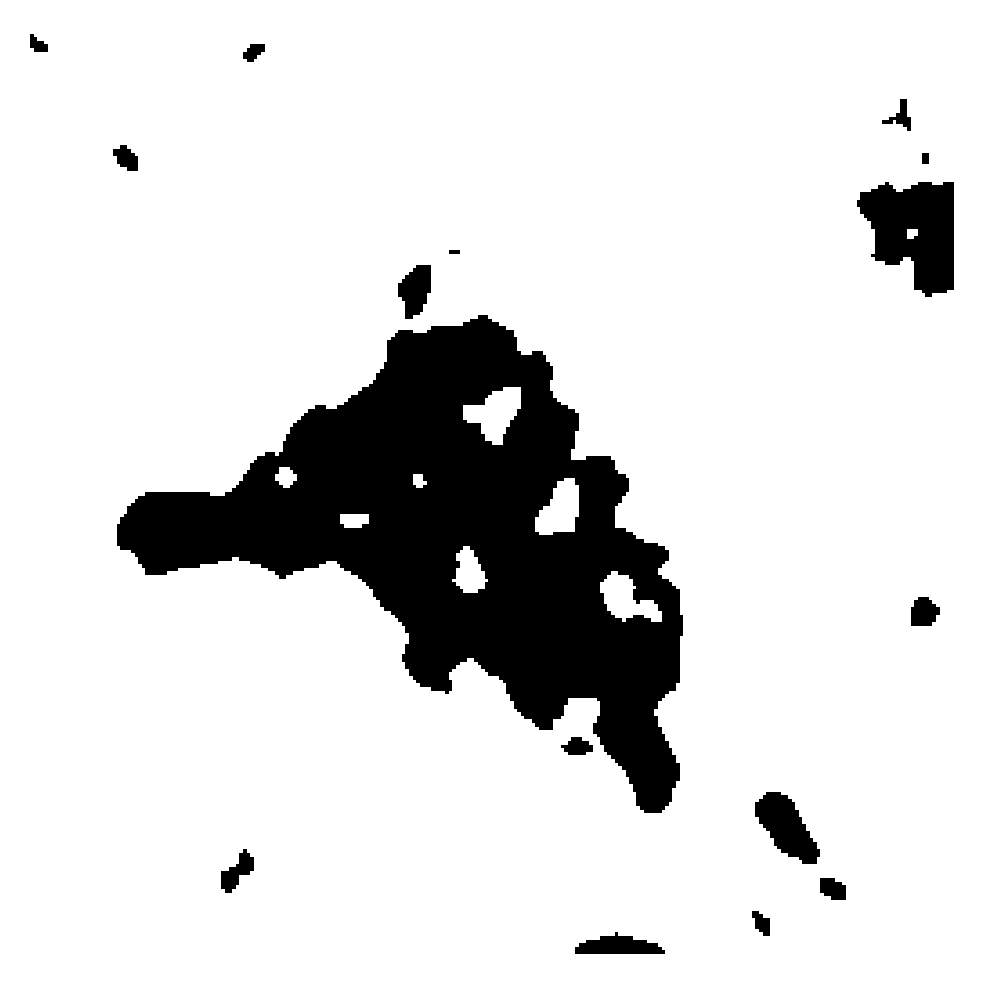}
        \label{fig:estaillades_gen_255}
    \end{subfigure}

    \hrule

    % Ketton
    \begin{subfigure}[t]{0.15\textwidth}
        \includegraphics[width=\textwidth]{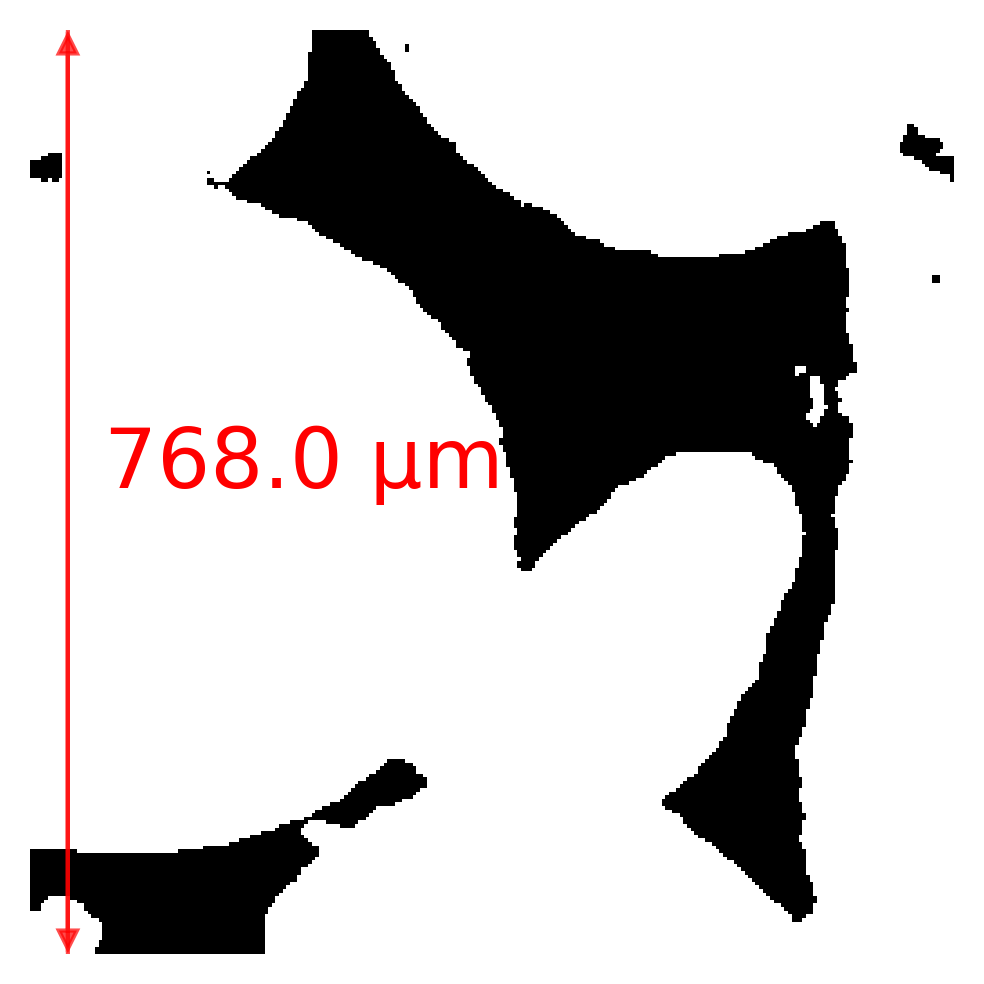}
        \label{fig:ketton_valid_0}
    \end{subfigure}
    \begin{subfigure}[t]{0.15\textwidth}
        \includegraphics[width=\textwidth]{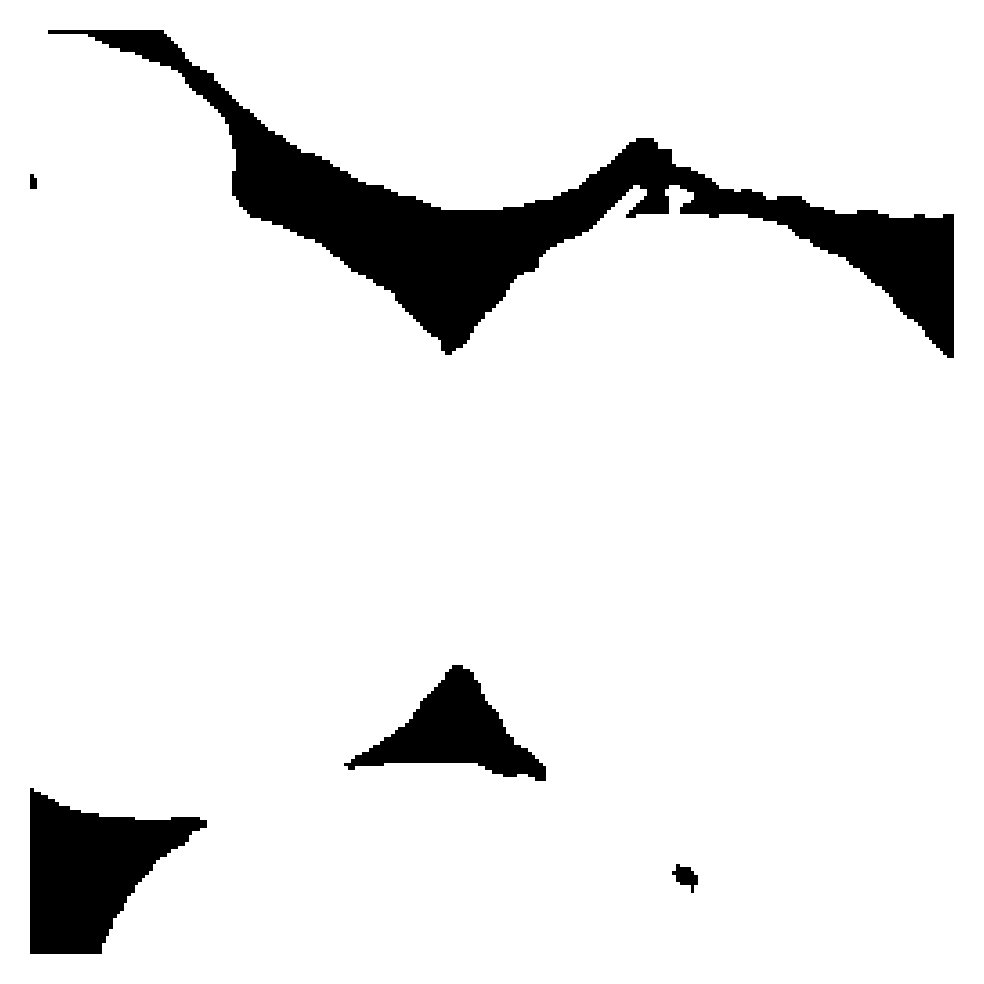}
        \label{fig:ketton_valid_127}
    \end{subfigure}
    \begin{subfigure}[t]{0.15\textwidth}
        \includegraphics[width=\textwidth]{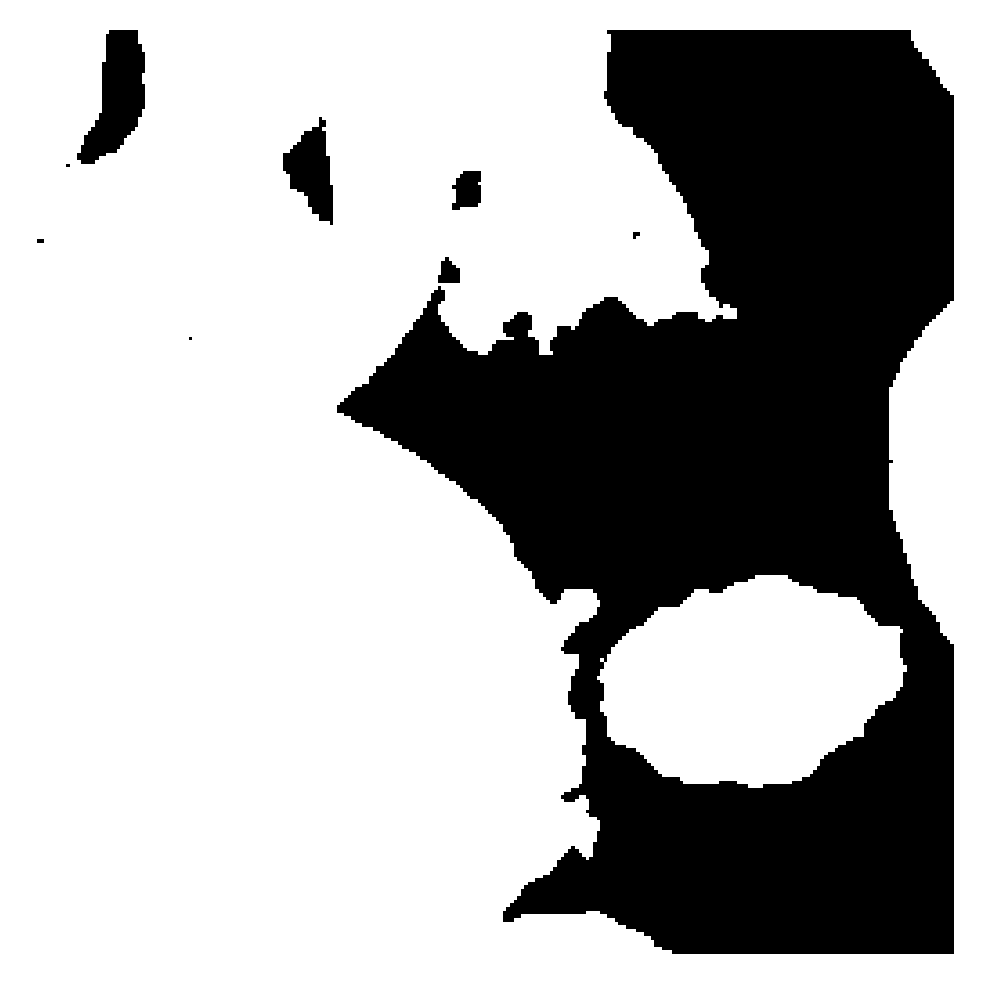}
        \label{fig:ketton_valid_255}
    \end{subfigure}
    \vline
    \begin{subfigure}[t]{0.15\textwidth}
        \includegraphics[width=\textwidth]{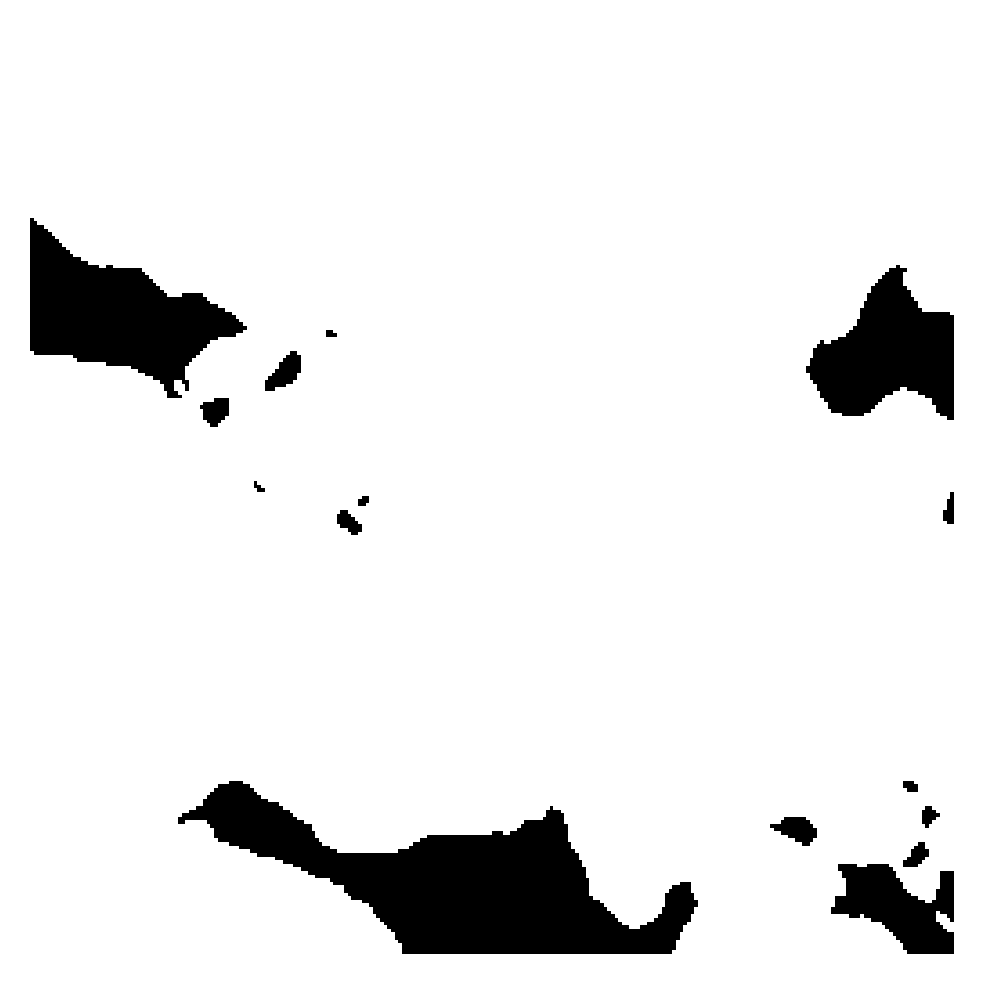}
        \label{fig:ketton_gen_0}
    \end{subfigure}
    \begin{subfigure}[t]{0.15\textwidth}
        \includegraphics[width=\textwidth]{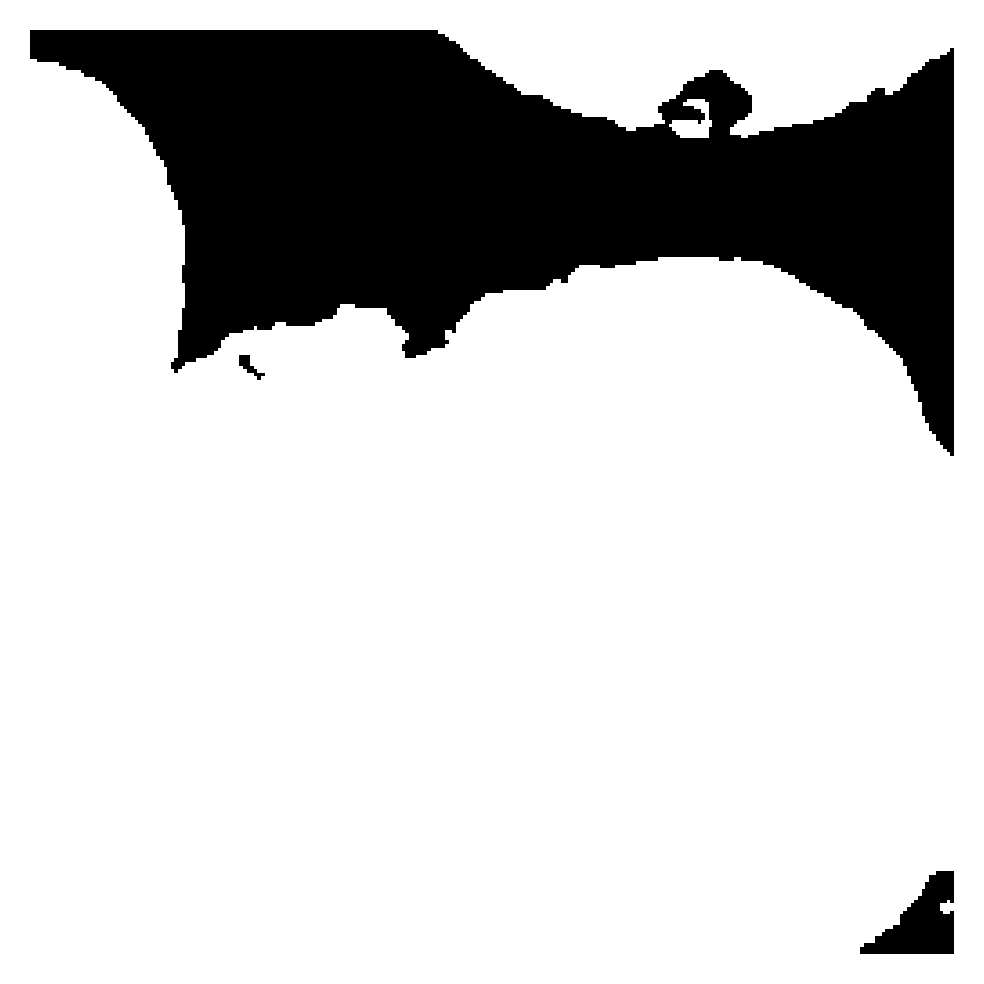}
        \label{fig:ketton_gen_127}
    \end{subfigure}
    \begin{subfigure}[t]{0.15\textwidth}
        \includegraphics[width=\textwidth]{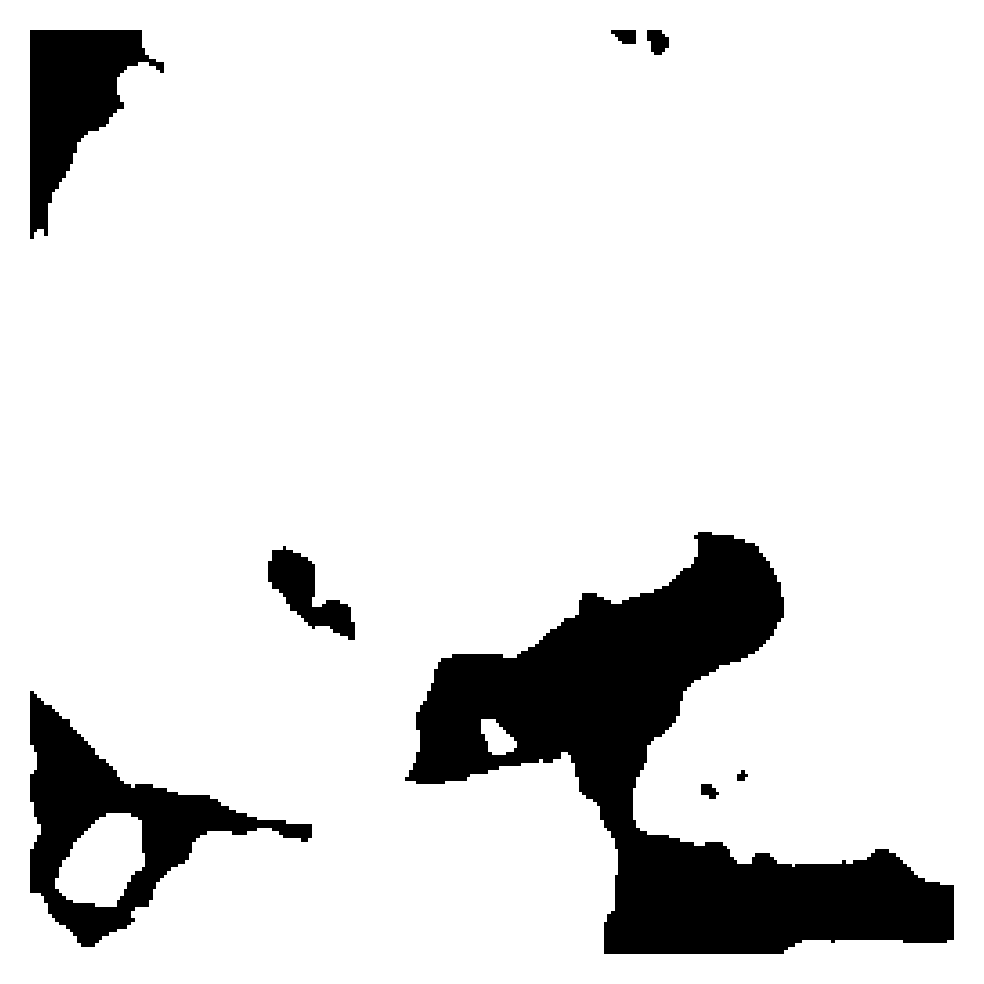}
        \label{fig:ketton_gen_255}
    \end{subfigure}
    
    \caption{Comparison of 2D slices from original validation samples (left) and generated samples (right) at different depths (bottom, middle, top) for different rock types. From top to bottom: Bentheimer, Doddington, Estaillades, and Ketton sandstones. Pore space is in black and solid space in white. The vertical line separates validation data from generated samples.}
    \label{fig:slice_comparison}
\end{figure}

As a final, qualitative, metric, we refer to Figure \ref{fig:pore_spaces} for the pore space of individual samples of $256^3$ volumes, both of generated (through porosity control) and validation samples. We also refer to Figure \ref{fig:slice_comparison} for comparing the slices of the same samples.

% \begin{figure}[H]
%     \centering
    
%     \begin{subfigure}{\textwidth}
%         \centering
%         \includegraphics[width=0.8\textwidth]{figs/hellinger-stats.png}
%     \end{subfigure}
    
%     \begin{subfigure}{\textwidth}
%         \centering
%         \includegraphics[width=0.8\textwidth]{figs/mre-stats.png}
%     \end{subfigure}
    
%     \caption{Hellinger distance and MRE of $256^3$ samples for different sample generation techniques.}
%     \label{fig:hellinger_mre}
% \end{figure}

The pipeline, which combines data augmentation and controlled sampling, effectively addresses the issues encountered during generation and yields robust results. For most rock types, unconditional sampling (without control or data augmentation) already delivers satisfactory outcomes; however, Bentheimer still exhibits spurious outputs in both fully unconditional and controlled models unless data augmentation is employed, as shown in Figure \ref{fig:bentheimer_256_por_guided_no_data_aug}. Because such spurious results can be problematic in practical applications, we recommend using the complete pipeline—data augmentation plus controlled generation—as a general rule to ensure robust and accurate digital rock reconstructions.

% Additionally, we have explored an unsupervised spectral-based filtering technique to automatically detect failed generations by analyzing their frequency-domain signatures. Some preliminary results of this approach are provided in \ref{spectral filtering}. While these early findings are promising, our current pipeline already offers a reliable framework for high-quality sample generation, and spectral-based filtering may serve as an optional enhancement for future refinements.

\section{Conclusion}

In this work, we presented a novel approach to 3D porous media reconstruction using latent diffusion models within the EDM framework. Our method effectively addresses two fundamental challenges in geoscience: the high dimensionality of 3D microstructures and the need for accurate pore-scale geometry capture. By operating in a learned latent space, we achieve significant computational efficiency while preserving critical morphological features. Our experiments demonstrate superior performance in visual realism, structural metrics, and generative diversity compared to existing methods.

The flexibility of latent diffusion models enables the natural integration of domain-specific knowledge through network conditioning, including physics-informed constraints and multi-scale representations. In particular, the conditioning capabilities allow the novel sampling technique of controlled unconditional sampling. By controlling on readily available statistics such as porosity and two-point-correlation, we can obtain a more complete coverage of the data distribution, when combining it with a simple data augmentation procedure.

This opens several promising research directions: coupling with traditional simulation tools to improve flow and transport predictions, leveraging high-performance computing for large-volume generation, and incorporating grayscale reconstruction capabilities. Furthermore, conditioning on multifidelity 2D image databases — spanning micro-CT, SEM, and petrographic analysis — could enhance reconstruction fidelity by integrating complementary data sources. As research progresses to encompass larger datasets, diverse rock types, and comprehensive validation metrics, we anticipate that these models will become an essential tool in computational geosciences for 3D porous media analysis.

\section{Data and Code availability}\label{datacode availability}

The main dataset used in this work is the 4-rock database from Imperial College London \footnote{Found at \url{https://www.imperial.ac.uk/earth-science/research/research-groups/pore-scale-modelling/micro-ct-images-and-networks/}}. Additionally, we trained the autoencoder on Berea Sandstone data from the Eleven Sandstones database at Digital Rocks Portal\footnote{Found at \url{https://www.digitalrocksportal.org/projects/317}}, which is made available under the ODC Attribution License.

Our code consists of two main packages, DiffSci and PoreGen, which can be found at \url{https://github.com/Lacadame/DiffSci} and \url{https://github.com/Lacadame/PoreGen}. DiffSci is a general package for the exploration of diffusion models in the EDM framework, while PoreGen contains the specific features designed for its use in porous media modeling.

\section*{Acknowledgements}

The authors gratefully acknowledge the financial support provided by ExxonMobil Exploração Brasil Ltda. and Agência Nacional do Petróleo, Gás Natural e Biocombustíveis (ANP) through grant no. 23789-1. This research was also supported by Coordenação de Aperfeiçoamento de Pessoal de Nível Superior (CAPES). We are particularly thankful to Claudio Verdun and Akshay Pimpalkar for their foundational contributions during the early stages of this project. We also thank Professor Martin Blunt and The Imperial College Consortium on Pore-Scale Modelling and Imaging for making the micro-CT images of the Bentheimer, Doddington, Estaillades and Ketton stone samples freely available. We also extend our sincere appreciation to all members of the Diff-Twins Project at the Universidade Federal do Rio de Janeiro (UFRJ) for their valuable contributions throughout its development. The opinions expressed in this publication are solely those of the authors and do not necessarily reflect the views of the supporting organizations.

\bibliographystyle{abbrv}
\bibliography{references_diff}

\appendix
\section{Evaluation statistics}\label{section:evaluation_statistics}

To evaluate our results, we used the following statistics to compare the generative model distribution with the real distribution in Section \ref{sec:results}.

In the following, we consider a continuous cubic domain $\mathcal{X} \in \mathbb{R}^3$, and our pore space to be a function $X:\mathcal{X} \to \{0, 1\}$, where $f(x) = 0$ if $x$ is a solid, and $f(x)=1$ if it is a pore. We describe our statistics in this continuous domain, and the application to a volume of size $[L, W, H]$ is done through domain discretization. We calculate all the described statistics using PoreSpy\citep{gostick2019porespy}.

\subsection{Porosity}\label{app:porosity_calculation}
This is the pore fraction of our images, a fundamental metric for geology because it defines the volumetric fraction in our rock that can potentially be filled with fluids and, therefore, how useful the rock could be as a reservoir.

\begin{figure}[htbp]
    \centering
    \begin{subfigure}[t]{0.48\textwidth}
        \includegraphics[width=\textwidth]{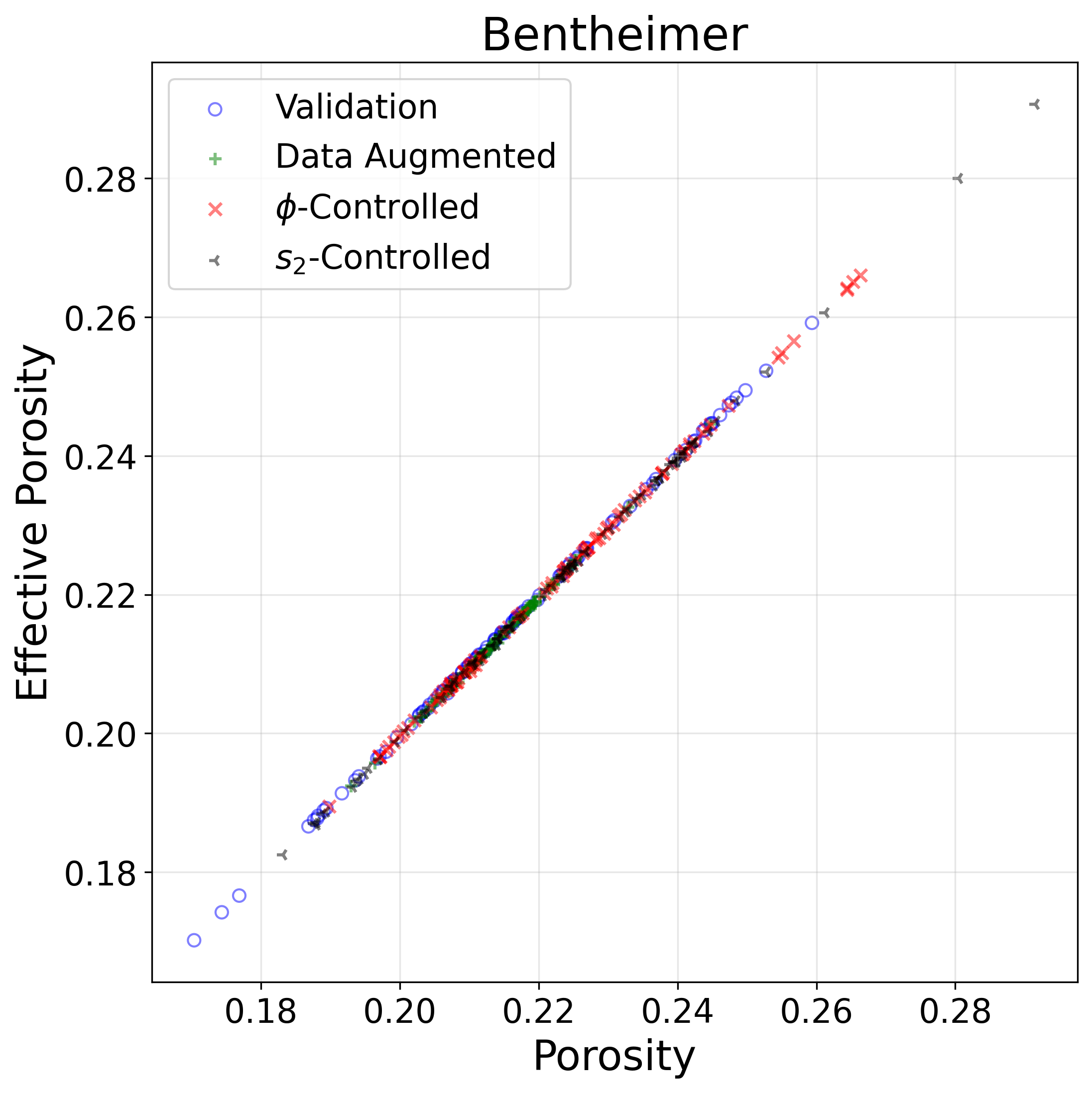}
        \caption{}
        \label{fig:scatter_bentheimer}
    \end{subfigure}
    \begin{subfigure}[t]{0.48\textwidth}
        \includegraphics[width=\textwidth]{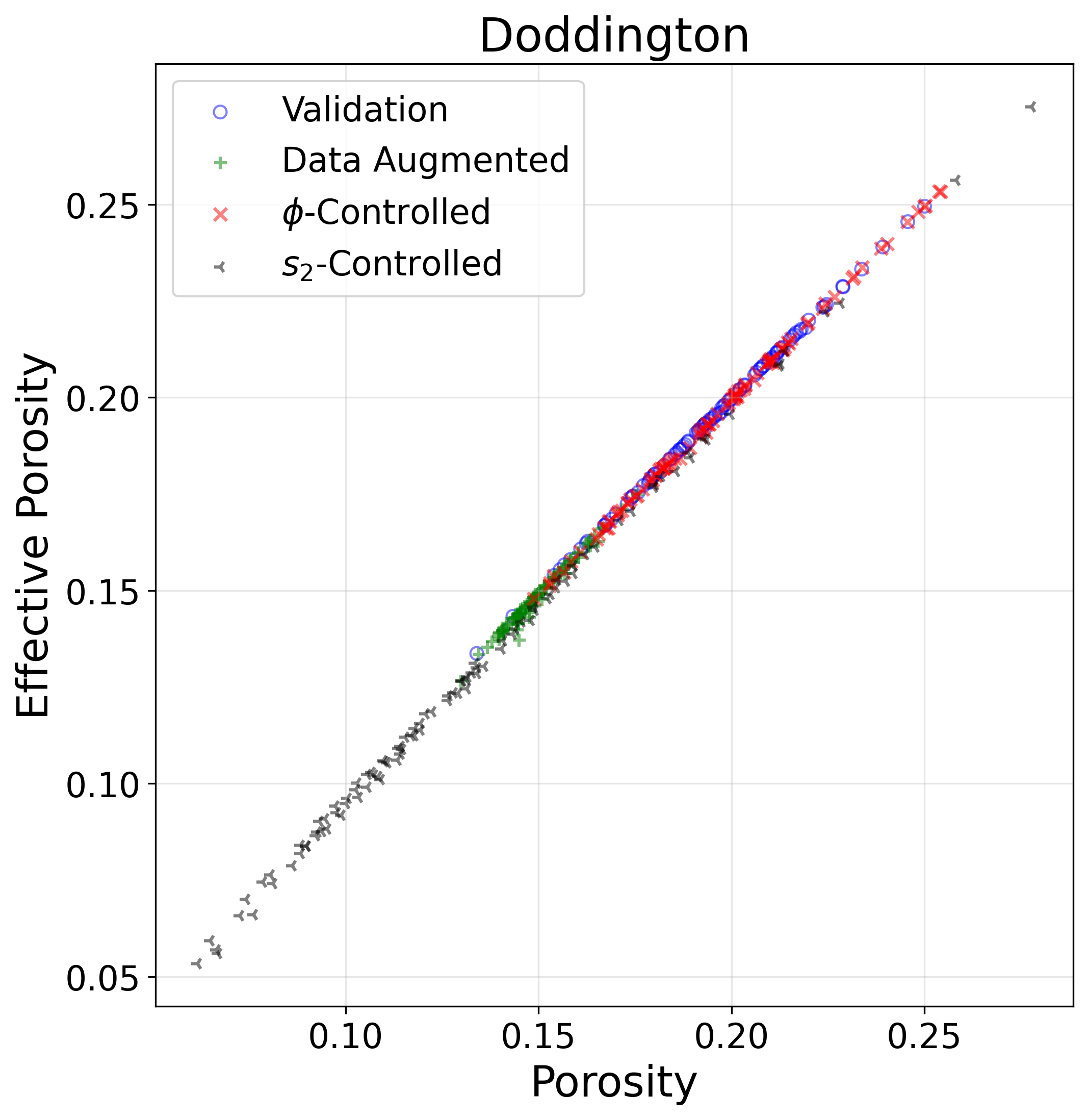}
        \caption{}
        \label{fig:scatter_doddington}
    \end{subfigure}
    \begin{subfigure}[t]{0.48\textwidth}
        \includegraphics[width=\textwidth]{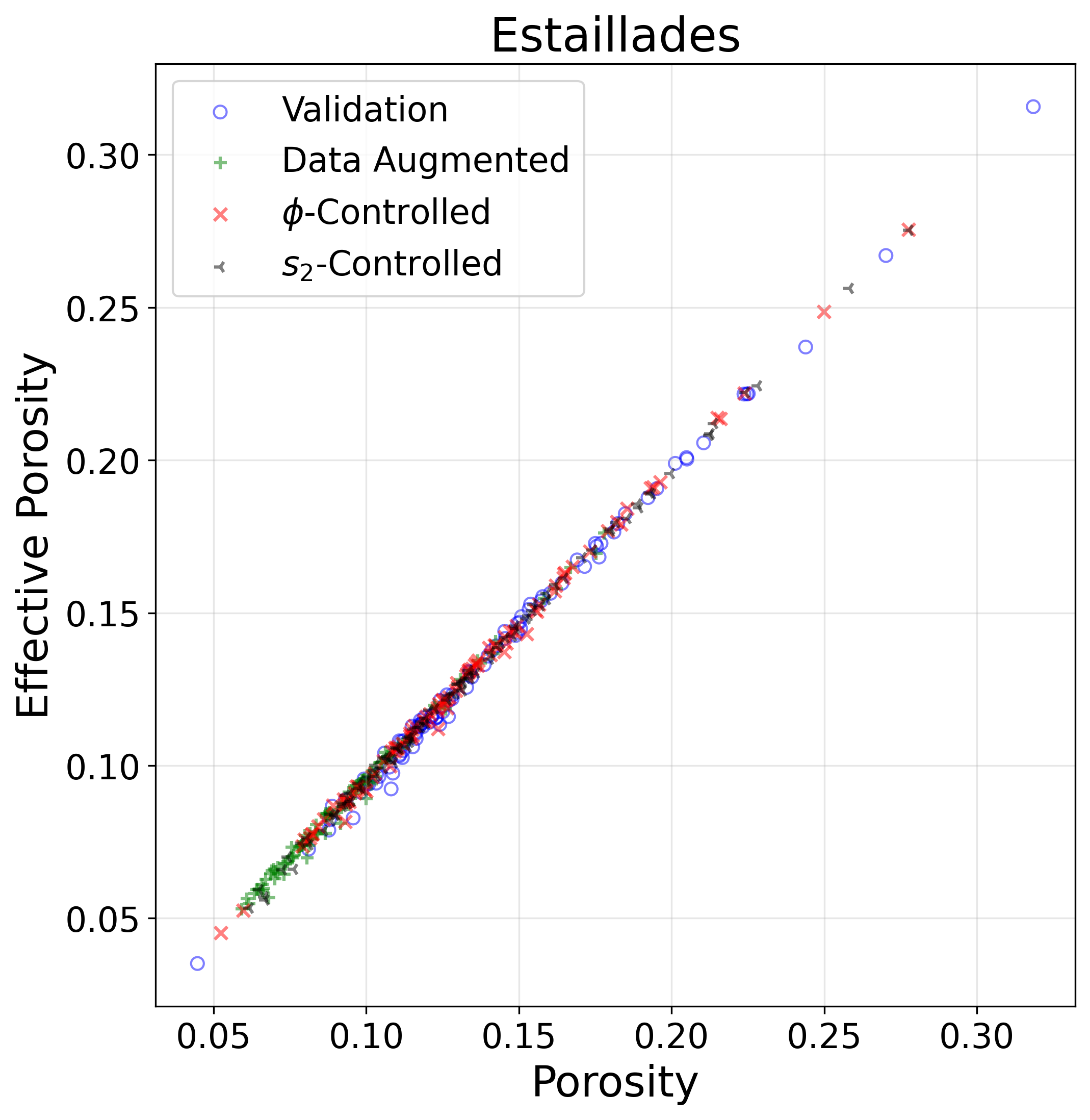}
        \caption{}
        \label{fig:scatter_estaillades}
    \end{subfigure}
    \begin{subfigure}[t]{0.48\textwidth}
        \includegraphics[width=\textwidth]{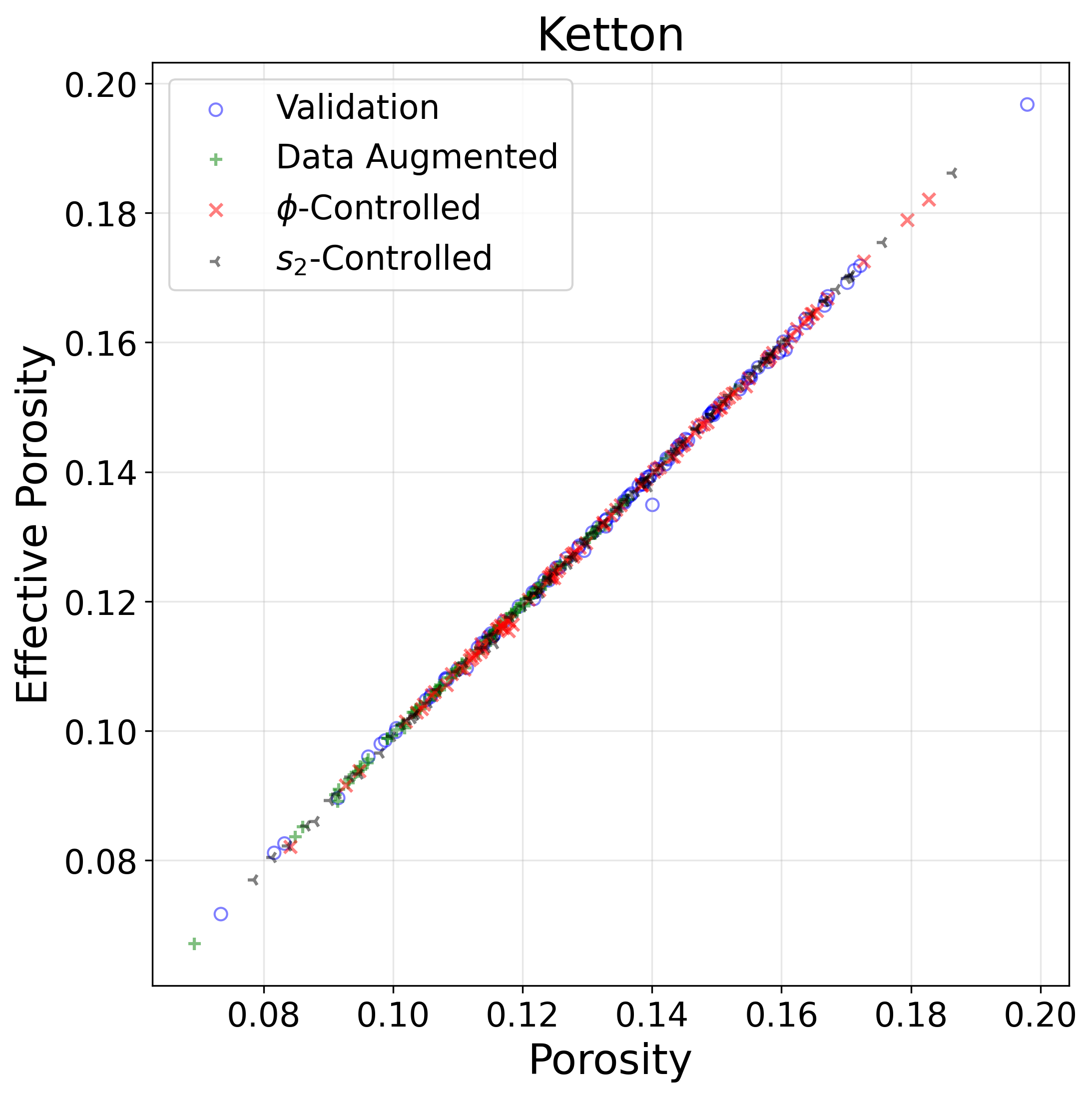}
        \caption{}
        \label{fig:scatter_ketton}
    \end{subfigure}
    \caption{Porosity scatter plots for different rock types: (a) Bentheimer, (b) Doddington, (c) Estaillades, and (d) Ketton sandstones.}
    \label{fig:scatter_porosity_effective}
\end{figure}

Formally, we can define the porosity $\phi(X)$ as follows:
\begin{equation}
    \phi(X) = \frac{1}{\operatorname{vol}(\mathcal{X})} \int_{\mathcal{X}} X(x) dx.
\end{equation}
There is also the associated concept of effective porosity, which is the pore space connected with the pore surface. In our samples, the effective porosity and porosity were essentially the same, as shown in \ref{fig:scatter_porosity_effective}, for both validation and generated samples. Therefore, we use porosity and effective porosity interchangeably in this work.

\subsection{Two-point correlation function}
Two-point correlation functions are widely used to study not only porous media but heterogeneous materials in general\citep{Jiao2007twopoint}. The two-point correlation function $\rho_f(y_1, y_2;X_f)$ is defined as
\begin{equation}
    \begin{split}
    & \rho_f(y_1, y_2;X_f) := \int_{[0, L]^2} X_f(x_1 + y_1, x_2 + y_2) X_f(x_1, x_2) I(x_1 + y_1, x_2 + y_2) dx_1 dx_2 \\
    & I(z_1, z_2) := \begin{cases}
        1 &  (z_1, z_2) \in [0, L]^2 \\
        0 & \text{ otherwise}.
    \end{cases}
    \end{split}
\end{equation}

For an isotropic domain, it makes sense to use instead a radial two-point correlation $\rho_f(r;X_f)$, defined as
\begin{equation}
    \rho_f(r;X_f) = \frac{1}{2 \pi} \int_{0}^{2\pi} \rho_f(r \cos \theta, r \sin \theta ) d \theta.
\end{equation}
We assume isotropy in our considered domains, and will from now on refer to the radial two-point correlation simply as the two-point correlation.

The two-point correlation measures the probability that both points $\mathbf{x}$ and $\mathbf{x} + r$ are pores. In particular, this implies that
\begin{equation}
    \begin{split}
        & \rho_f(0;X_f) = \phi \\
        & L \gg r > \xi \implies \rho_f(r;X_f) \approx \phi^2,
    \end{split}
\end{equation}
where $\xi$ is the correlation length, which is a characteristic length such that $\mathbf{x}$ and $\mathbf{x} + \xi$ are independent.

% Thus, a normalized two-point correlation $\hat{\rho}_f(r;X_f)$ can be defined as 
% \begin{equation}
%     \hat{\rho}_f(r;X_f) := \frac{\rho_f(r, X_f) - \phi^2}{\phi - \phi^2}.
% \end{equation}

\subsection{Pore size distribution}

The third metric we consider is the local thickness pore size distribution, which is defined through the following steps:

\begin{itemize}
    \item For each pore space $x \in X^{-1}(1)$, we define $B_0(x, r_x)$ to be the ball centered on $x$ with the largest radius such that $B_0(x, r_x)$ only contains pore space. This creates a set $\mathcal{B} = \{B_0(x, r_x); x \in X^{-1}(1)\}$.
    \item Then, for each $x \in X^{-1}(1)$, we choose the larger ball $B_0(x', r_{x'})\in\mathcal{B}$ such that $x \in B_0(x', r_{x'})$. The pore size at $x$ is defined as the radius $r_{x'}$ of this ball.
\end{itemize}
This defines a function $\operatorname{psd}:X^{-1}(1) \to \mathbb{R}^+$ that associates each pore space with a pore size, which we call the pore size distribution (PSD). The mean pore size is given by the mean $\frac{1}{\operatorname{vol}(X^{-1}(1))} \int_{X^{-1}(1)} \operatorname{psd}(x) dx$.

For plotting multiple PSD curves (one for each sample), we deploy a kernel density estimate using the SciPy\citep{2020SciPy-NMeth} package \textit{stats.gaussian\_kde}.

\subsection{Surface area density}
We can define the surface area of $X$ as the interface $\partial X := \overline{f^{-1}(1)} \cap \overline{f^{-1}(0)}$ between the pore space and the solid space. Then, assuming enough regularity of $\partial X$, the surface area density is given by
\begin{equation}
    \frac{1}{\operatorname{vol}(X)} \int_{\partial X} dS.
\end{equation}

\subsection{Permeability}\label{app:permeability_calculation}
In the vast majority of porous rocks at a macroscopic scale, the volumetric flow rate $Q$ of a fluid with dynamic viscosity $\mu$, passing through a rock with cross-sectional area $A$ and length $L$, with a pressure drop $\Delta p$ through its length, is given by Darcy's law\citep{whitaker1986flow}
\begin{equation}
    Q = \frac{k A}{\mu L} \Delta p,
\end{equation}
where $k$ is the \textit{permeability} of the rock, depending both on the fluid properties and the pore space geometry of the rock.

In this work, we calculate the permeability of each sample through the axes $x$, $y$, and $z$ for water, using a pore network model \citep{xiong2016porenetworkreview}, as implemented in OpenPNM \citep{gostick2016openpnm}. The pore network was extracted using the SNOW algorithm \citep{gostick2017versatile}, which combines watershed segmentation with network extraction to identify pore bodies and throats. The network model uses pyramidal and cuboidal elements for calculating hydraulic conductance, and disconnected pore clusters are removed to ensure network connectivity. Permeability calculations are performed by imposing a unit pressure differential across each axis. The final permeability value is reported in Darcy units, calculated as the geometric mean of the directional permeabilities under the assumption of approximate rock isotropy at the sample scale, as validated in \ref{appendix:isotropy_proof}.

\section{Evaluation metrics}\label{section:statistical_comparison}

For a scalar statistic $x$, with a generated distribution having probability density $p(x)$ and mean $\mu_p$, and a validation distribution having probability density $q(x)$ and mean $\mu_q$, we use two metrics to compare these distributions, the Hellinger distance and the mean relative error (MRE).

The Hellinger distance is defined as
\begin{equation}
    H(p||q)^2(x) = \frac{1}{2} \int_{-\infty}^{\infty}\left(\sqrt{p(x)} - \sqrt{q(x)}\right)^2 dx.
\end{equation}
In our work, we estimate the densities $p(x)$ and $q(x)$ using a kernel density estimate, the same one used in plotting the pore size distribution.

The mean relative error is defined as
\begin{equation}
    \operatorname{MRE}(p, q) = \frac{\left| \mu_q - \mu_p \right|}{\left|\mu_q\right|}.
\end{equation}

For the two-point correlation, we define the Hellinger distance and the MRE as the mean of these respective metrics at each evaluation point $r$ in a grid, since $\operatorname{TPC}(r)$ is a scalar.

\begin{figure}[H]
    \centering
    
    % Bentheimer
    \begin{subfigure}{0.8\textwidth}
        \centering
        \includegraphics[width=\textwidth]{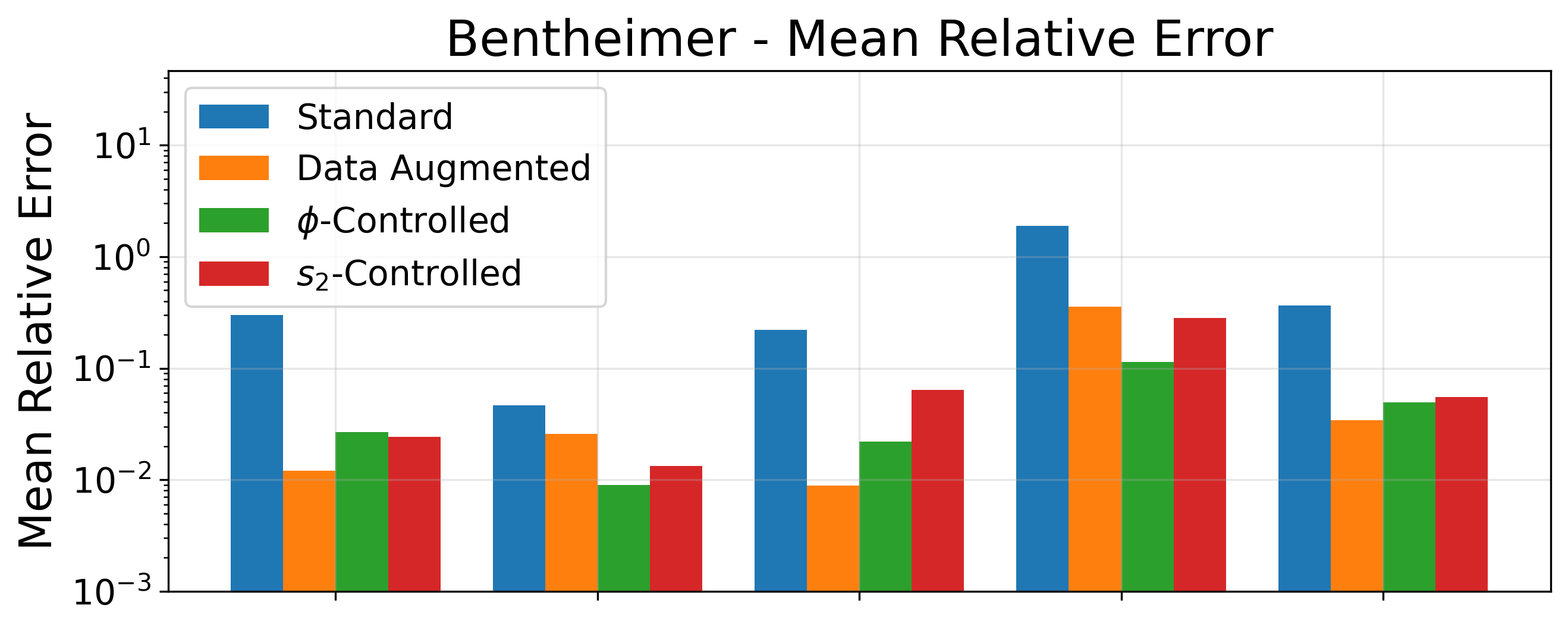}
    \end{subfigure}
    
    % Doddington
    \begin{subfigure}{0.8\textwidth}
        \centering
        \includegraphics[width=\textwidth]{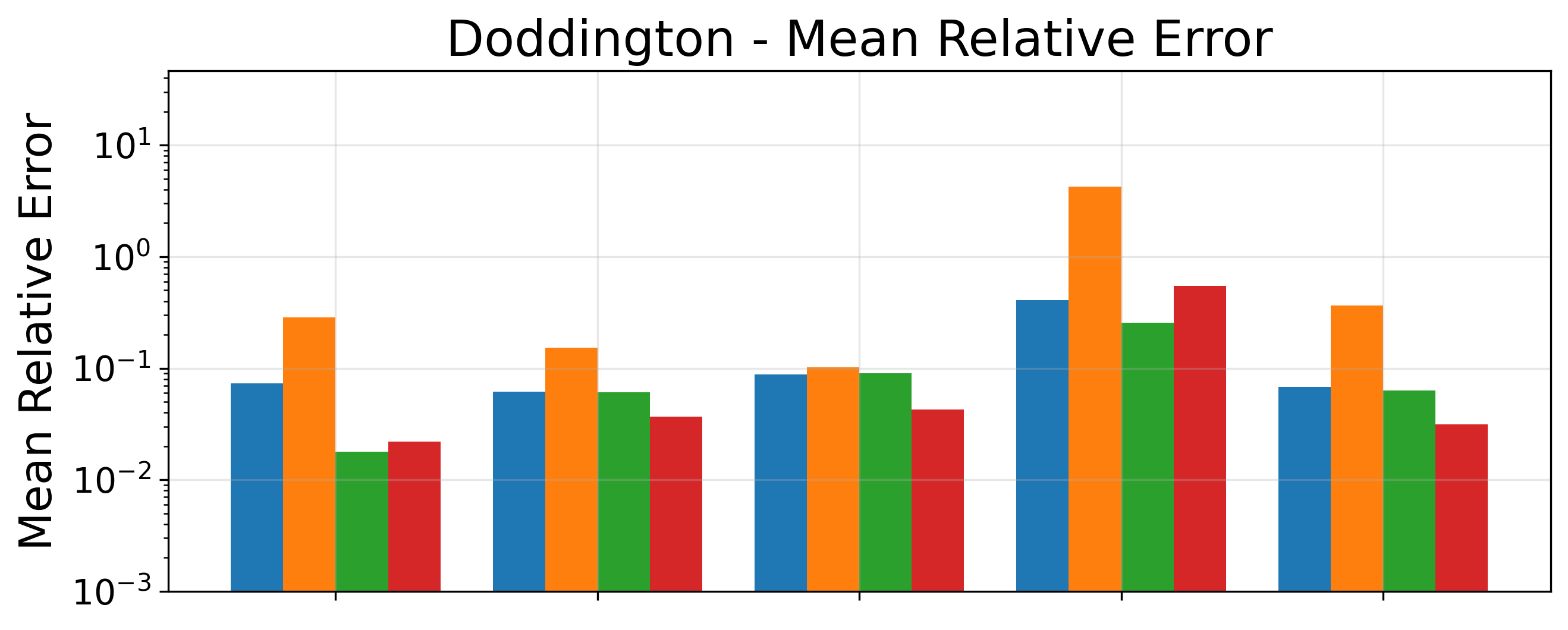}
    \end{subfigure}
    
    % Estaillades
    \begin{subfigure}{0.8\textwidth}
        \centering
        \includegraphics[width=\textwidth]{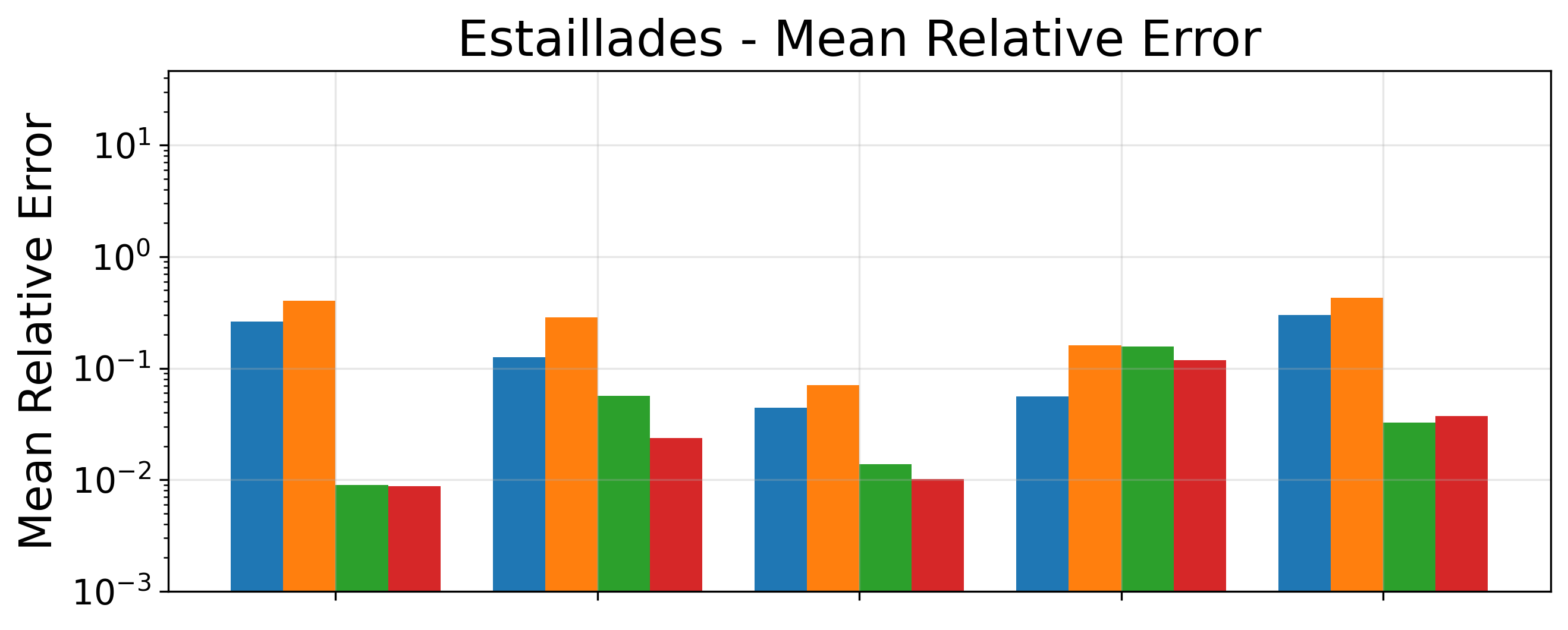}
    \end{subfigure}
    
    % Ketton
    \begin{subfigure}{0.8\textwidth}
        \centering
        \includegraphics[width=\textwidth]{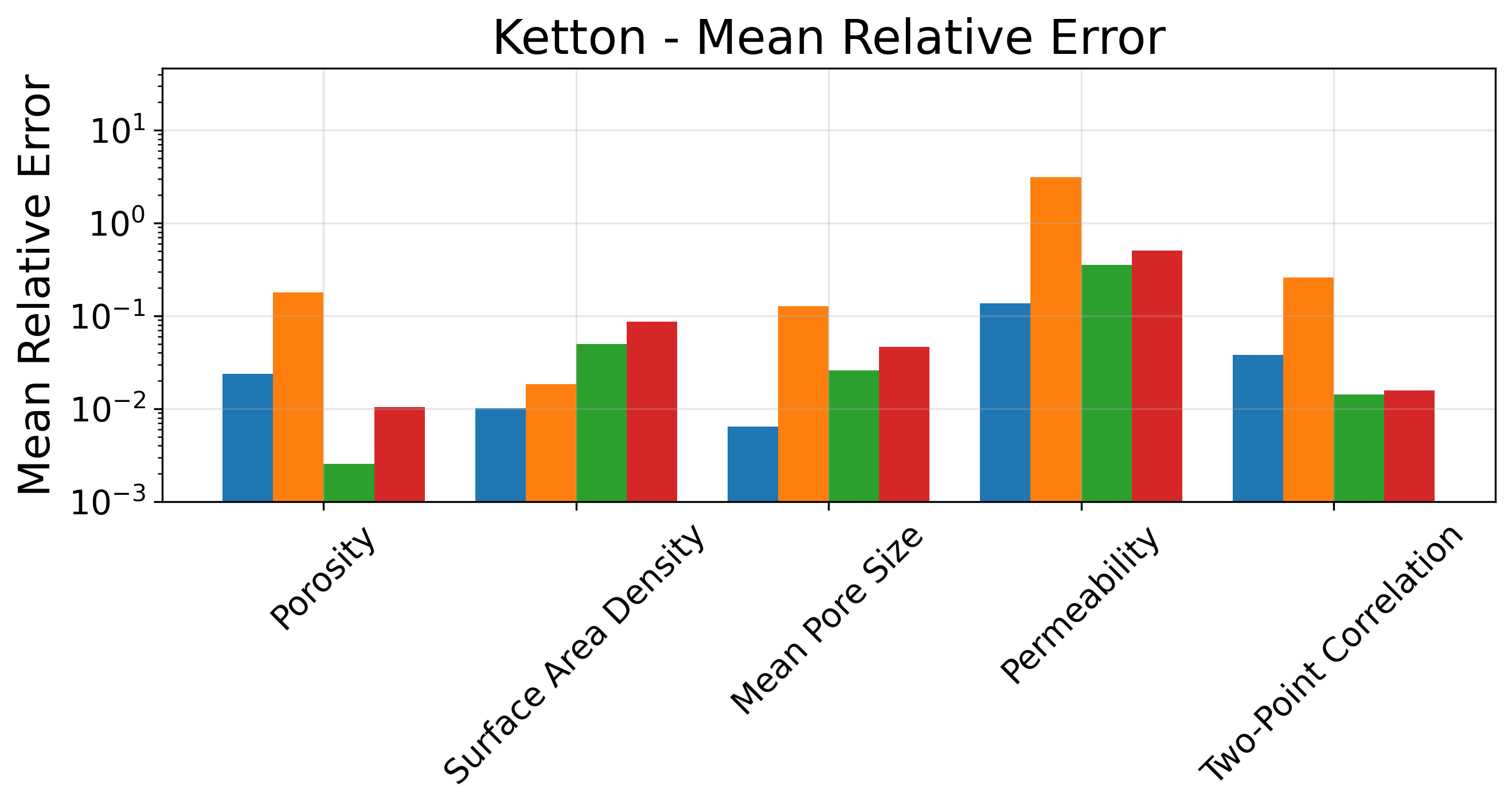}
    \end{subfigure}
    
    \caption{Mean relative error of the statistics for $256^3$ samples for different sample generation techniques.}
    \label{fig:mre_stats}
\end{figure}

\section{Architecture}\label{architecturaldetails}

\subsection{PUNet architectural details}
Each ResNet block in the outer scheme of figure \ref{fig:diffmodelouter} consists of a modified ResNet that processes both spatial and temporal information:
\begin{itemize}
\item Input normalization using Group Layer Normalization
\item First convolution layer (kernel size 3)
\item Time embedding addition, where the time embeddings pass through a three-layer MLP (4× dimension expansion) with SiLU activations and is reshaped to $B \times C_{out} \times 1 \times 1 \times 1$, adding it to the spatial features
\item Second normalization using Group RMS Normalization
\item Second convolution with dropout (p=0.1)
\item A residual connection from the input
\end{itemize}

Referring to Figure \ref{fig:diffmodelouter}, at the bottom of our UNet, the ResNetBlock consists of 6 ResNets, and 2 ResNets in each block in the upward and downward path. The input is projected through a convolutional layer to a channel dimension of 64. Time values are projected from scalar values into high-dimensional embeddings through a Gaussian Fourier projection to an embedding dimension of 64. The Downsampling and upsampling blocks consist of a convolutional layer followed by a downsampling (upsampling) of a factor of 2.
 
\subsection{Conditional embeddings}

We deploy our conditional models, both for porosity and two-point correlation, through embedding modules, whose outputs are summed to the time embedding of the PUNet, as shown schematically in Figure \ref{fig:diffmodelouter}. We describe the details of each embedding module.

\subsubsection{Porosity}

For embedding the porosity, we project the scalar value into high-dimensional embeddings through a Gaussian Fourier projection to an embedding dimension of 64, followed by a three-layer MLP, with hidden layers of dimension 256, using SiLU as an activation function.

\subsubsection{Two-point correlation}

For embedding the two-point correlation functions, we pass it through a combination of positional and Gaussian embeddings, followed by a transformer encoder. The network consists of two main components:

\begin{itemize}
    \item The embedder, which sums a positional encoding of the TPC arguments (the distances) with a Gaussian Fourier projection of the correlation values, both to an embedding dimension of 64. We use a scale parameter of 30.0 for the Gaussian projections.
    \item A standard transformer encoder with two layers, 4 attention heads for each layer, and an expansion factor of 4 for the feed-forward neural network expansion. We process the output through a mean pooling across the sequence dimension.    
\end{itemize}

\section{Training details}\label{app: training details}

In our training, each epoch uses 34560 data volumes, randomly sampled from the training volume described in \ref{datasection}. When features are used, they are extracted on the fly from each training sample. We use AdamW as the optimizer, with a constant learning rate after an initial warm-up phase. The final model is chosen by the lowest validation loss, calculated at the end of each epoch from 3840 validation samples, obtained from the validation volume described in \ref{datasection}.

All training was performed on an NVIDIA DGX A100, using 6 A100 GPU cores. The typical training time for the unconditional and porosity-conditional latent models is 3 hours ($64^3$ volumes), 10 hours ($128^3$ volumes), 36 hours ($256^3$ volumes), and 72 hours for the TPC-conditional on $256^3$ volumes. For the autoencoder, it is 5 hours, and for pixel-space models in $64^3$ volumes, 10 hours.

\section{Isotropy testing}\label{appendix:isotropy_proof}
We test the isotropy of rock samples in our work by comparing the two-point correlation (TPC) function calculated in different ways:
\begin{enumerate}
    \item The TPC function of the entire 3D volume (the reference)
    \item The TPC functions along the x, y, and z axes (directional slices)
\end{enumerate}

For an infinite volume of a completely isotropic medium, the TPC function should be identical regardless of direction. The Mean Relative Error (MRE) quantifies the deviation between the directional TPC functions and the full volumetric TPC function.

For each sample:
\begin{enumerate}
    \item We compute the TPC function for the full 3D volume as our reference.
    \item We compute the TPC functions along the x, y, and z axes.
    \item For each direction (x, y, z), we calculate the relative error at each distance r:
    \begin{equation}
        \operatorname{RE}(r) = \frac{|\rho_{\text{direction}}(r) - \rho(r)|}{\rho(r)} \times 100\%
    \end{equation}
    \item The Mean Relative Error for each direction is then
    \begin{equation}
        \operatorname{MRE}_{\text{direction}} = \frac{1}{N} \sum_{r} \operatorname{RE}(r),
    \end{equation}
    where N is the number of distance points.
\end{enumerate}

We calculate the MRE values of 16 samples of volume $256^3$ for each of our stones, in the $x$, $y$, and $z$ directions. In the following table, we show the averaged MRE values for each stone.
% The low resulting MRE values confirm that our rock samples exhibit good isotropy, validating our assumption that the TPC functions calculated along any direction are reasonable estimators of the full volumetric TPC function, as well as the fact that we can perform data augmentation by flipping on the $x$, $y$, and $z$ directions.

\begin{table}[H]
\centering
\caption{Stone-averaged Mean Relative Error (MRE) by axis (\%)}
\begin{tabular}{lccc}
\hline
\textbf{Stone} & \textbf{x-axis} & \textbf{y-axis} & \textbf{z-axis} \\
\hline
Bentheimer & 2.89\% & 1.53\% & 1.72\% \\
Doddington & 3.23\% & 2.56\% & 2.87\% \\
Estaillades & 6.11\% & 4.64\% & 4.60\% \\
Ketton & 8.01\% & 5.29\% & 6.31\% \\
\hline
\end{tabular}
\label{tab:isotropy_summary}
\end{table}

Note that even a completely isotropic material could exhibit nonzero MRE values for a finite volume. The fact that the higher values are observed for Ketton are consistent with this observation, since this rock has the largest grain size and, therefore, should have a larger field of view. With this in mind, we consider that the values in Table \ref{tab:isotropy_summary} support the isotropy assumption, which is used to justify using slice TPC as a proxy for volumetric TPC, and performing data augmentation by flipping subvolumes in the $x$, $y$, and $z$ directions.

\section{Additional experimental results}\label{app: all experiments}

\begin{figure}[htbp]
    \centering
    \begin{subfigure}[t]{0.95\textwidth}
        \centering
        \includegraphics[width=\textwidth]{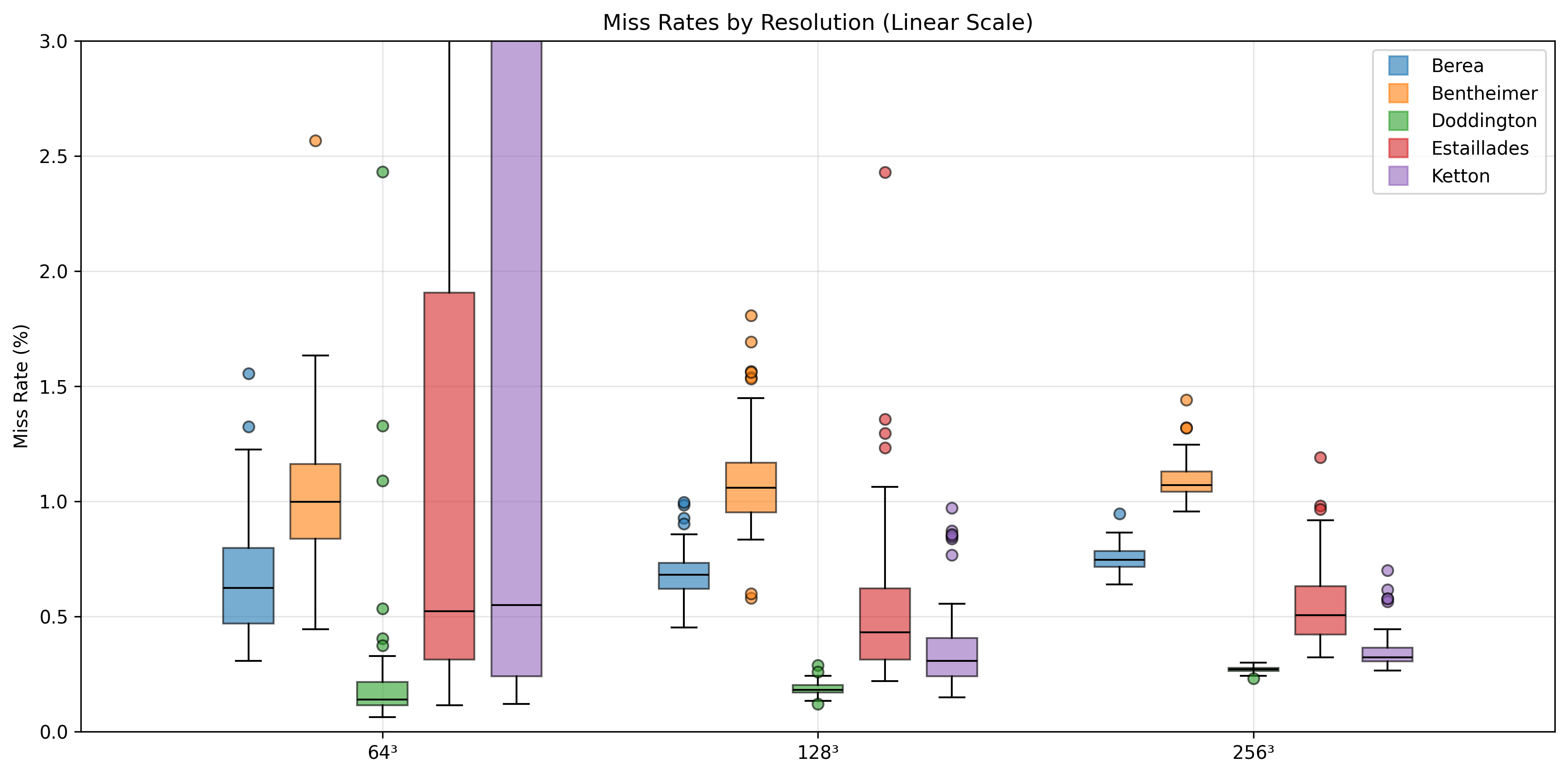}
        \caption{Boxplots of reconstruction error up to 3\% for different rock types and volume sizes.}
        \label{fig:miss_rates_berea}
    \end{subfigure}
    
    \begin{subfigure}[t]{0.95\textwidth}
        \centering
        \includegraphics[width=\textwidth]{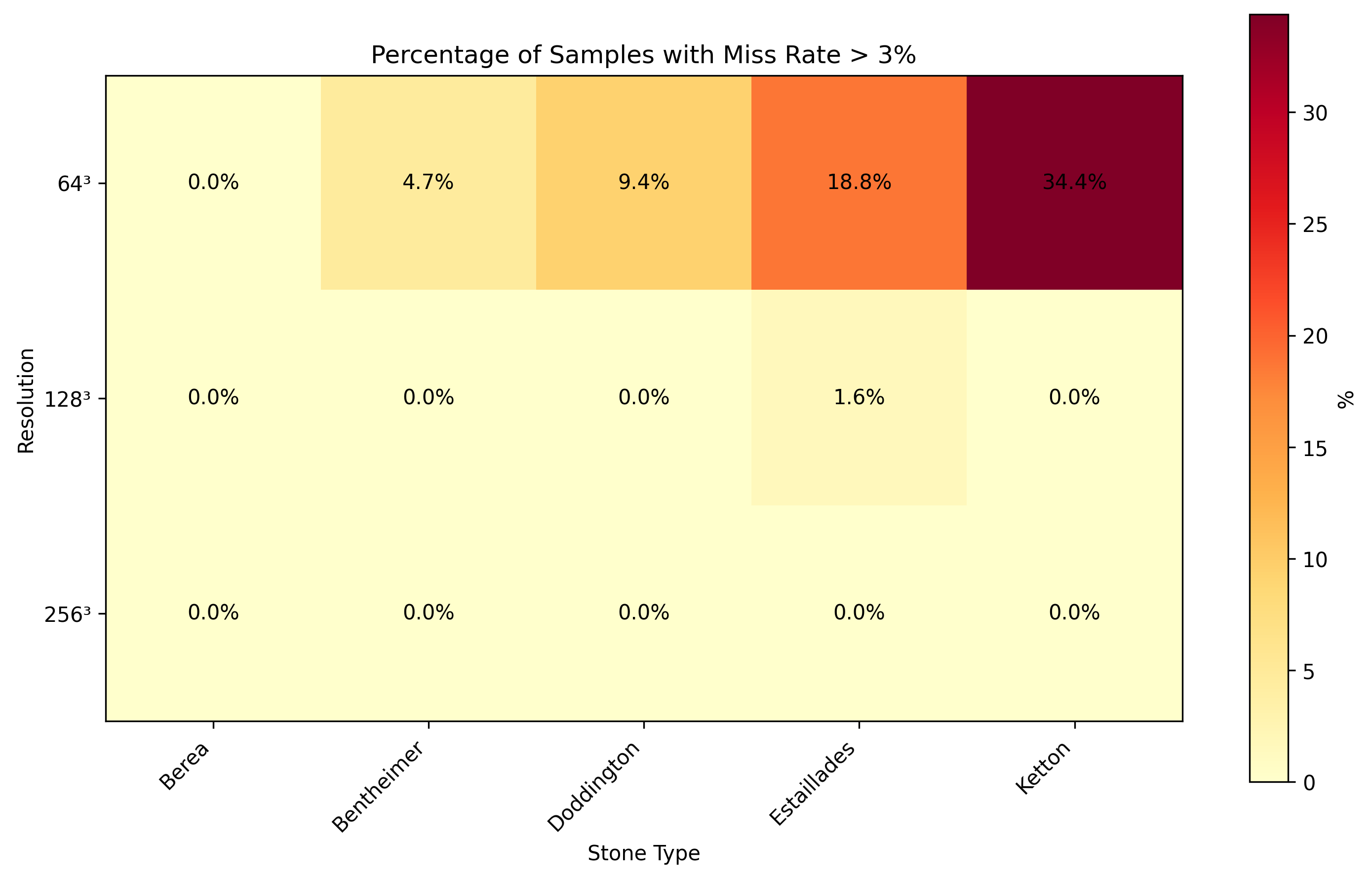}
        \caption{Percentage of reconstructions with error rates exceeding 3\%.}
        \label{fig:miss_table_berea}
    \end{subfigure}
    \caption{Analysis of autoencoder reconstruction error for different rocks, for an autoencoder trained only on Berea sandstone $64^3$ volumes.}
    \label{fig: berea reconstruction_analysis}
\end{figure}

Figure \ref{fig: berea reconstruction_analysis} shows reconstruction errors for an autoencoder trained on $64^3$ volumes extracted from a $1000^3$ cube of another rock, Berea Sandstone. See \ref{datacode availability} for data availability. Remarkably, the reconstruction miss rates are very similar to those for an autoencoder trained in all other four rock types, as can be seen comparing Figure  \ref{fig: berea reconstruction_analysis} with Figure \ref{fig:reconstruction_analysis}.

These comparable reconstruction miss rates across different rock types suggest that the autoencoder is capturing a universal representation of rock structures in its latent space, transcending the specific characteristics of individual rock formations. This finding points to fundamental structural patterns shared across diverse rock types, which may allow more generalized approaches to digital rock analysis.

% Berea autoencoder errors

\begin{figure}[H]
    \centering
    % Bentheimer
    \begin{subfigure}[t]{0.15\textwidth}
        \includegraphics[width=\textwidth]{figs/0005-figures/Bentheimer_256_cond/stats_kde_porosity_27012025.png}
        \caption{}
        \label{fig:app_bentheimer_256c_a}
    \end{subfigure}
    \begin{subfigure}[t]{0.15\textwidth}
        \includegraphics[width=\textwidth]{figs/0005-figures/Bentheimer_256_cond/stats_kde_permeability_27012025.png}
        \caption{}
        \label{fig:app_bentheimer_256c_b}
    \end{subfigure}
    \begin{subfigure}[t]{0.15\textwidth}
        \includegraphics[width=\textwidth]{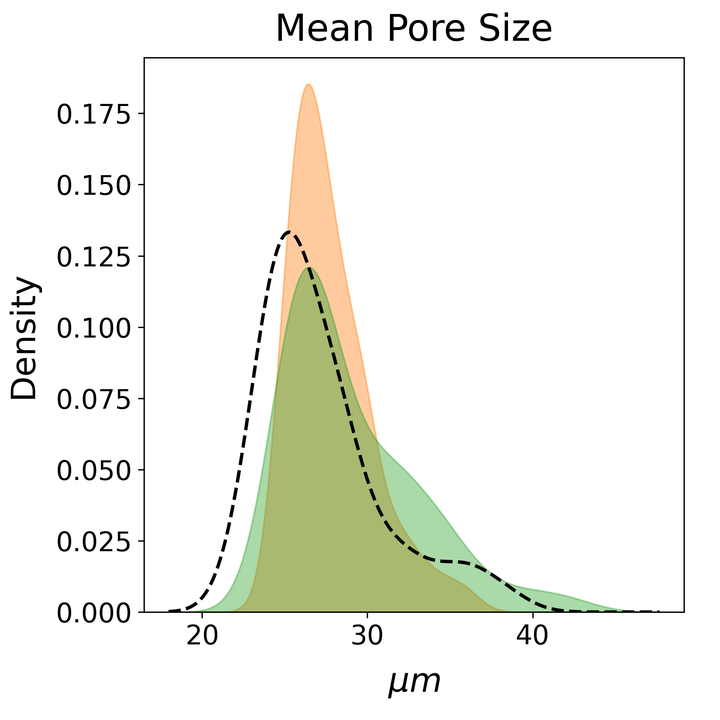}
        \caption{}
        \label{fig:app_bentheimer_256c_c}
    \end{subfigure}
    \begin{subfigure}[t]{0.15\textwidth}
        \includegraphics[width=\textwidth]{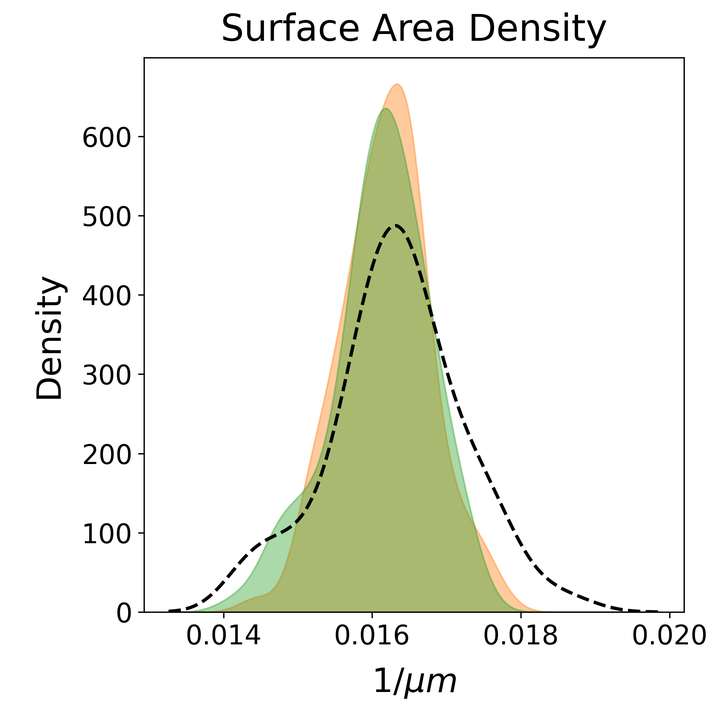}
        \caption{}
        \label{fig:app_bentheimer_256c_d}
    \end{subfigure}
    \begin{subfigure}[t]{0.15\textwidth}
        \includegraphics[width=\textwidth]{figs/0005-figures/Bentheimer_256_cond/tpc_comparison_27012025.png}
        \caption{}
        \label{fig:app_bentheimer_256c_e}
    \end{subfigure}
    \begin{subfigure}[t]{0.15\textwidth}
        \includegraphics[width=\textwidth]{figs/0005-figures/Bentheimer_256_cond/psd_comparison_27012025.png}
        \caption{}
        \label{fig:app_bentheimer_256c_f}
    \end{subfigure}
    
    % Doddington
    \begin{subfigure}[t]{0.15\textwidth}
        \includegraphics[width=\textwidth]{figs/0005-figures/Doddington_256_cond/stats_kde_porosity_27012025.png}
        \caption{}
        \label{fig:app_doddington_256c_a}
    \end{subfigure}
    \begin{subfigure}[t]{0.15\textwidth}
        \includegraphics[width=\textwidth]{figs/0005-figures/Doddington_256_cond/stats_kde_permeability_27012025.png}
        \caption{}
        \label{fig:app_doddington_256c_b}
    \end{subfigure}
    \begin{subfigure}[t]{0.15\textwidth}
        \includegraphics[width=\textwidth]{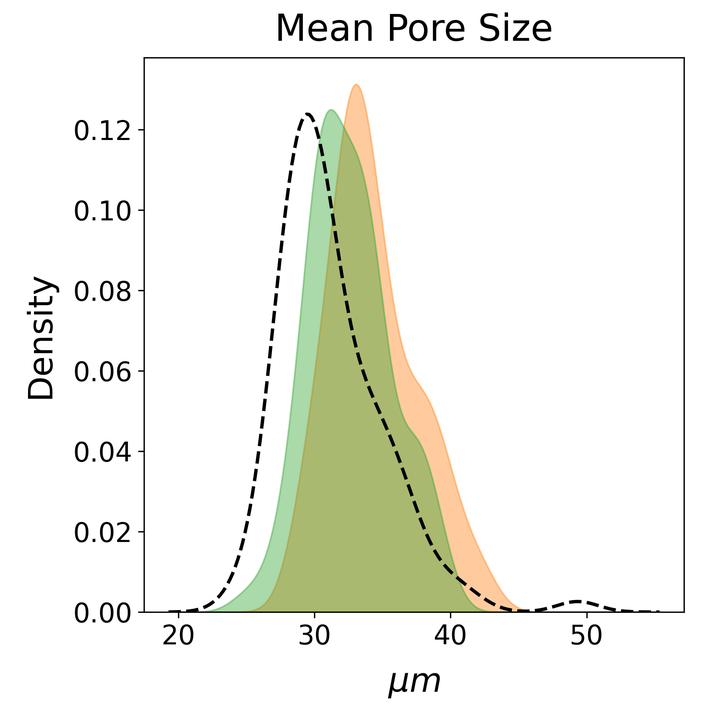}
        \caption{}
        \label{fig:app_doddington_256c_c}
    \end{subfigure}
    \begin{subfigure}[t]{0.15\textwidth}
        \includegraphics[width=\textwidth]{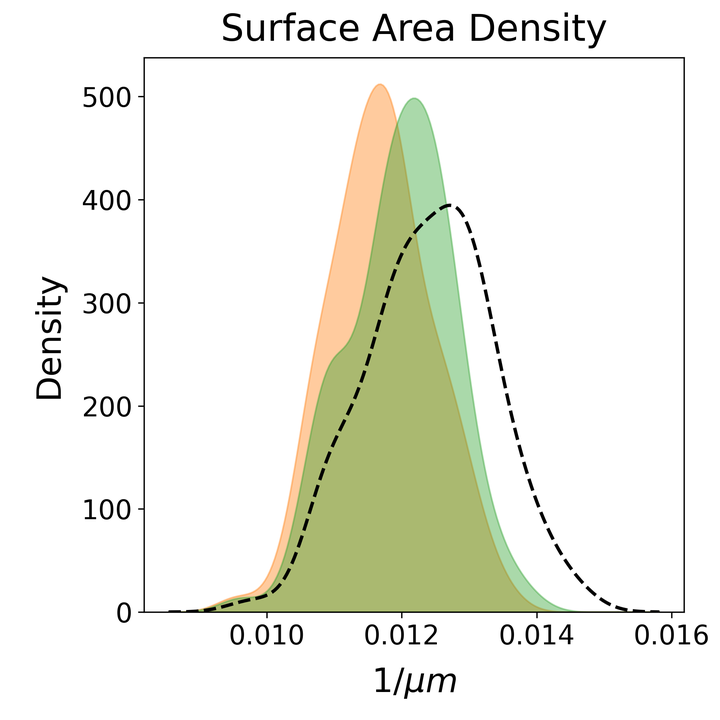}
        \caption{}
        \label{fig:app_doddington_256c_d}
    \end{subfigure}
    \begin{subfigure}[t]{0.15\textwidth}
        \includegraphics[width=\textwidth]{figs/0005-figures/Doddington_256_cond/tpc_comparison_27012025.png}
        \caption{}
        \label{fig:app_doddington_256c_e}
    \end{subfigure}
    \begin{subfigure}[t]{0.15\textwidth}
        \includegraphics[width=\textwidth]{figs/0005-figures/Doddington_256_cond/psd_comparison_27012025.png}
        \caption{}
        \label{fig:app_doddington_256c_f}
    \end{subfigure}
    
    % Estaillades
    \begin{subfigure}[t]{0.15\textwidth}
        \includegraphics[width=\textwidth]{figs/0005-figures/Estaillades_256_cond/stats_kde_porosity_27012025.png}
        \caption{}
        \label{fig:app_estaillades_256c_a}
    \end{subfigure}
    \begin{subfigure}[t]{0.15\textwidth}
        \includegraphics[width=\textwidth]{figs/0005-figures/Estaillades_256_cond/stats_kde_permeability_27012025.png}
        \caption{}
        \label{fig:app_estaillades_256c_b}
    \end{subfigure}
    \begin{subfigure}[t]{0.15\textwidth}
        \includegraphics[width=\textwidth]{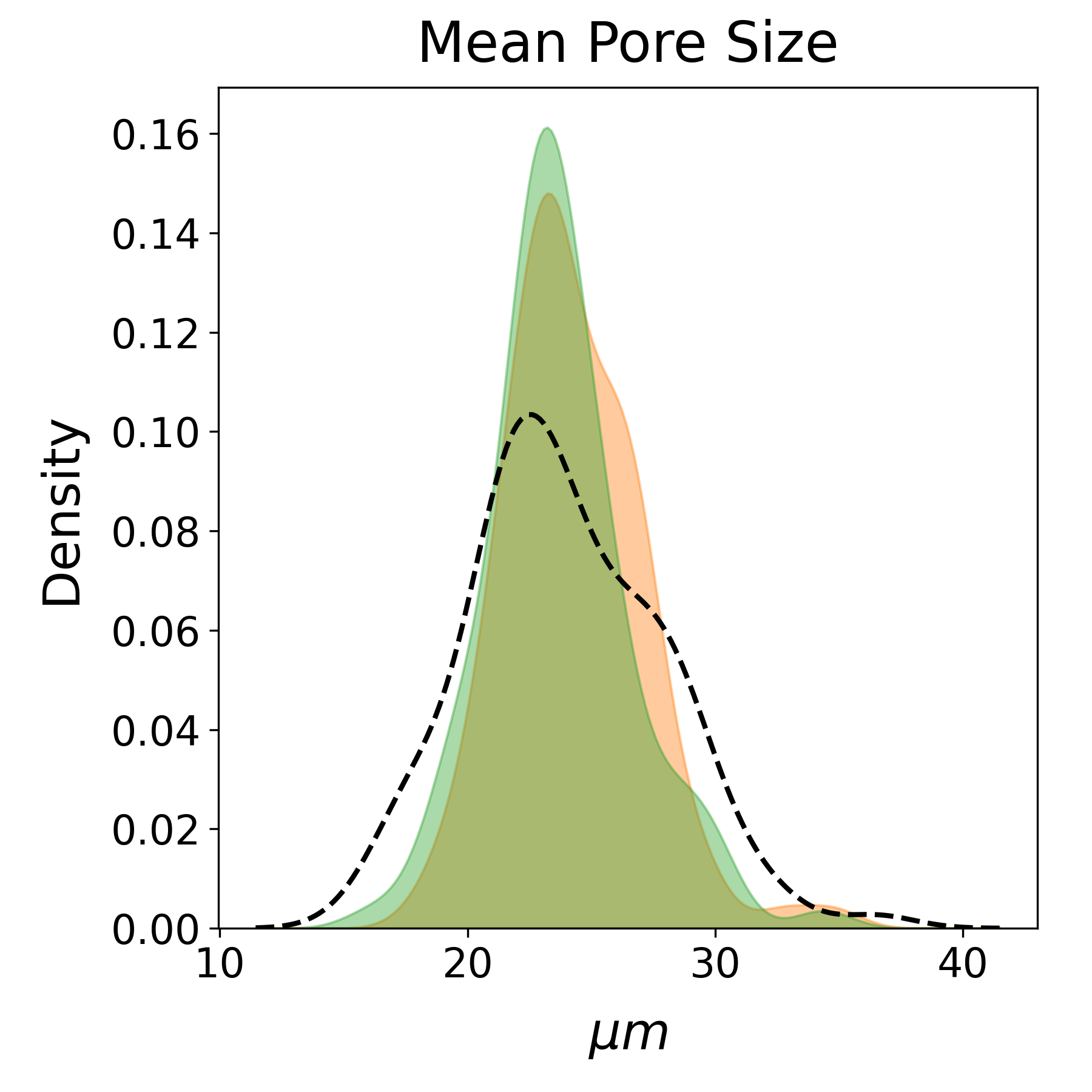}
        \caption{}
        \label{fig:app_estaillades_256c_c}
    \end{subfigure}
    \begin{subfigure}[t]{0.15\textwidth}
        \includegraphics[width=\textwidth]{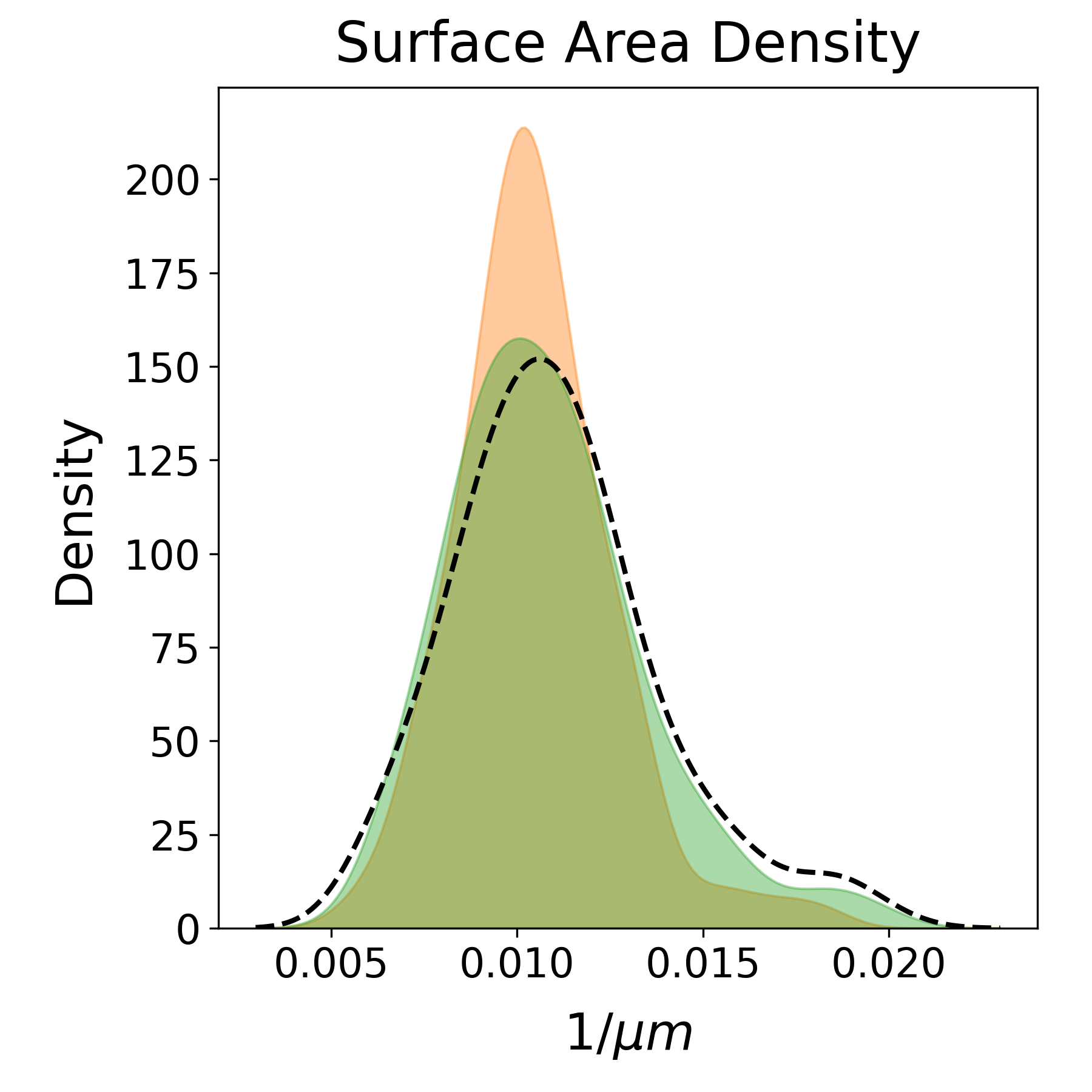}
        \caption{}
        \label{fig:app_estaillades_256c_d}
    \end{subfigure}
    \begin{subfigure}[t]{0.15\textwidth}
        \includegraphics[width=\textwidth]{figs/0005-figures/Estaillades_256_cond/tpc_comparison_27012025.png}
        \caption{}
        \label{fig:app_estaillades_256c_e}
    \end{subfigure}
    \begin{subfigure}[t]{0.15\textwidth}
        \includegraphics[width=\textwidth]{figs/0005-figures/Estaillades_256_cond/psd_comparison_27012025.png}
        \caption{}
        \label{fig:app_estaillades_256c_f}
    \end{subfigure}
    
    % Ketton
    \begin{subfigure}[t]{0.15\textwidth}
        \includegraphics[width=\textwidth]{figs/0005-figures/Ketton_256_cond/stats_kde_porosity_27012025.png}
        \caption{}
        \label{fig:app_ketton_256c_a}
    \end{subfigure}
    \begin{subfigure}[t]{0.15\textwidth}
        \includegraphics[width=\textwidth]{figs/0005-figures/Ketton_256_cond/stats_kde_permeability_27012025.png}
        \caption{}
        \label{fig:app_ketton_256c_b}
    \end{subfigure}
    \begin{subfigure}[t]{0.15\textwidth}
        \includegraphics[width=\textwidth]{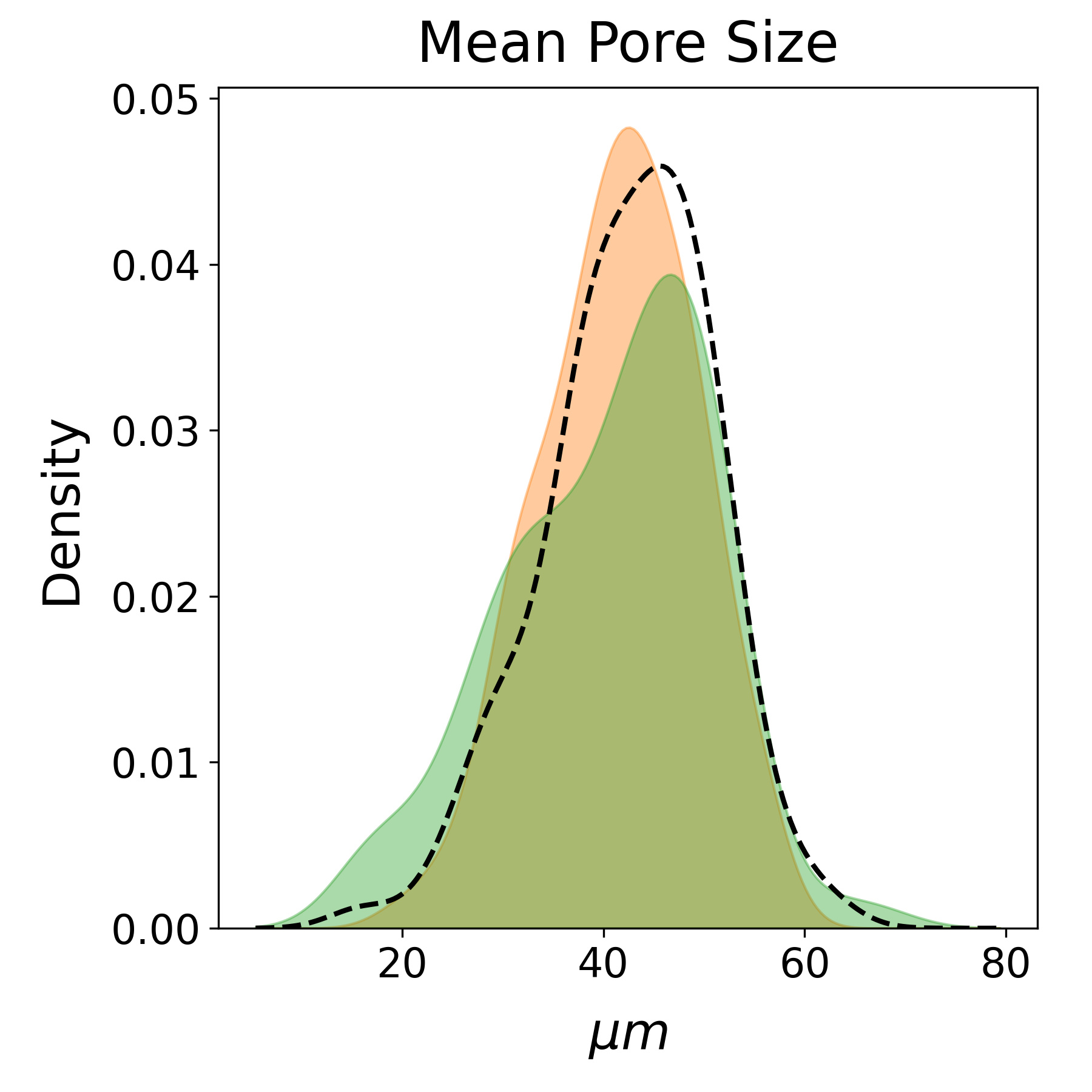}
        \caption{}
        \label{fig:app_ketton_256c_c}
    \end{subfigure}
    \begin{subfigure}[t]{0.15\textwidth}
        \includegraphics[width=\textwidth]{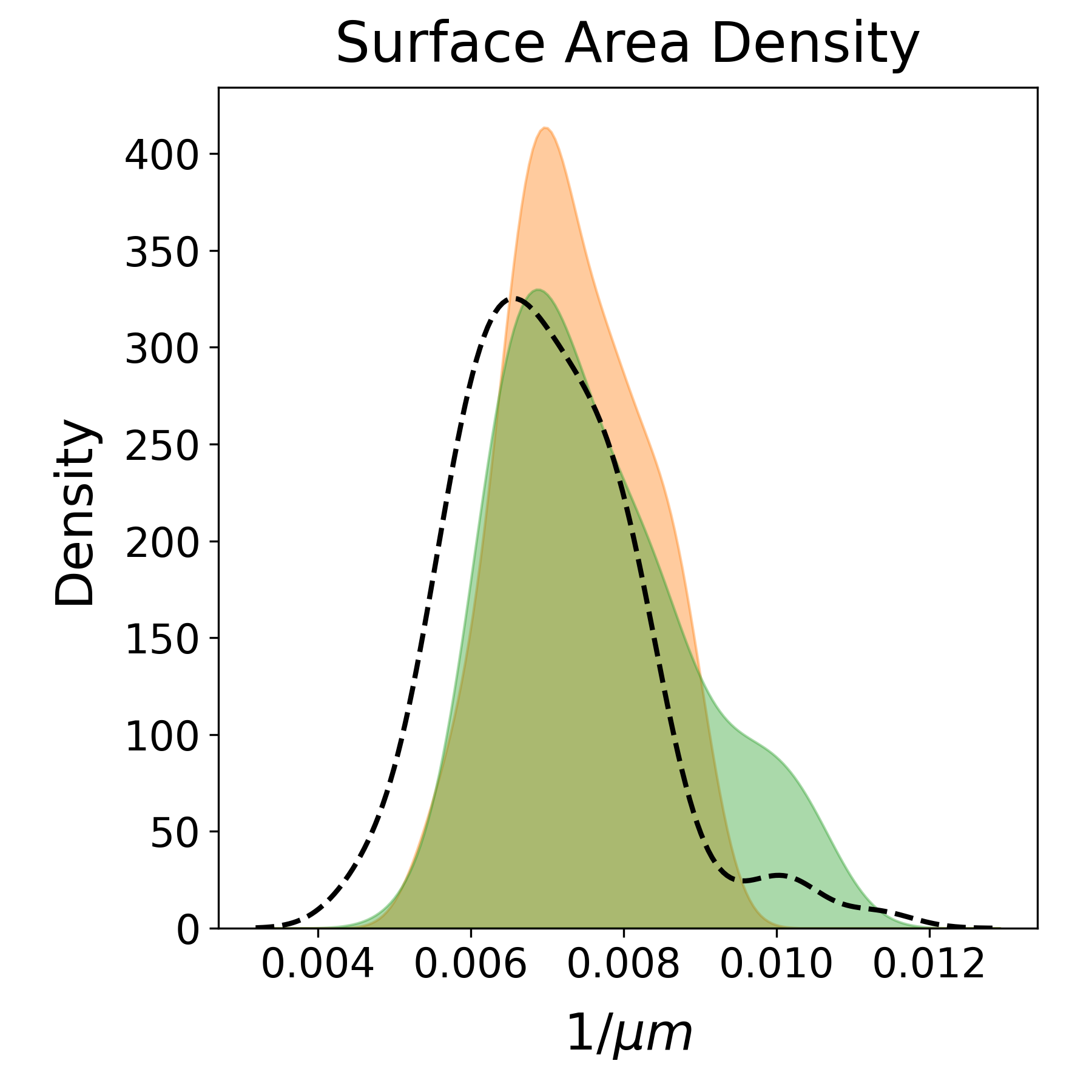}
        \caption{}
        \label{fig:app_ketton_256c_d}
    \end{subfigure}
    \begin{subfigure}[t]{0.15\textwidth}
        \includegraphics[width=\textwidth]{figs/0005-figures/Ketton_256_cond/tpc_comparison_27012025.png}
        \caption{}
        \label{fig:app_ketton_256c_e}
    \end{subfigure}
    \begin{subfigure}[t]{0.15\textwidth}
        \includegraphics[width=\textwidth]{figs/0005-figures/Ketton_256_cond/psd_comparison_27012025.png}
        \caption{}
        \label{fig:app_ketton_256c_f}
    \end{subfigure}
    
    \caption{Full statistical properties comparison for control generated 256$^3$ volume samples across different rock types. Top to bottom: (a-f) Bentheimer, (g-l) Doddington, (m-r) Estaillades, and (s-x) Ketton sandstones. Each row shows (from left to right): porosity distribution, permeability distribution, mean pore size distribution, surface area density distribution, two-point correlation function, and pore size distribution.}
    \label{fig:appendix_all_rocks_256c_comparison}
\end{figure}

\begin{figure}[H]
    \centering
    % Bentheimer
    \begin{subfigure}[t]{0.15\textwidth}
        \includegraphics[width=\textwidth]{figs/0005-figures/Bentheimer_256_noncond/stats_kde_porosity_27012025.png}
        \caption{}
        \label{fig:app_bentheimer_256nc_a}
    \end{subfigure}
    \begin{subfigure}[t]{0.15\textwidth}
        \includegraphics[width=\textwidth]{figs/0005-figures/Bentheimer_256_noncond/stats_kde_permeability_27012025.png}
        \caption{}
        \label{fig:app_bentheimer_256nc_b}
    \end{subfigure}
    \begin{subfigure}[t]{0.15\textwidth}
        \includegraphics[width=\textwidth]{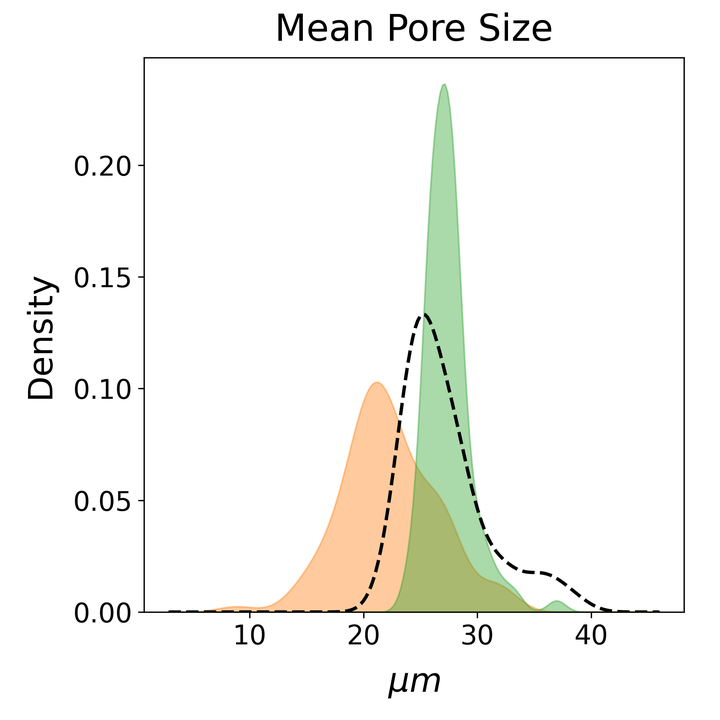}
        \caption{}
        \label{fig:app_bentheimer_256nc_c}
    \end{subfigure}
    \begin{subfigure}[t]{0.15\textwidth}
        \includegraphics[width=\textwidth]{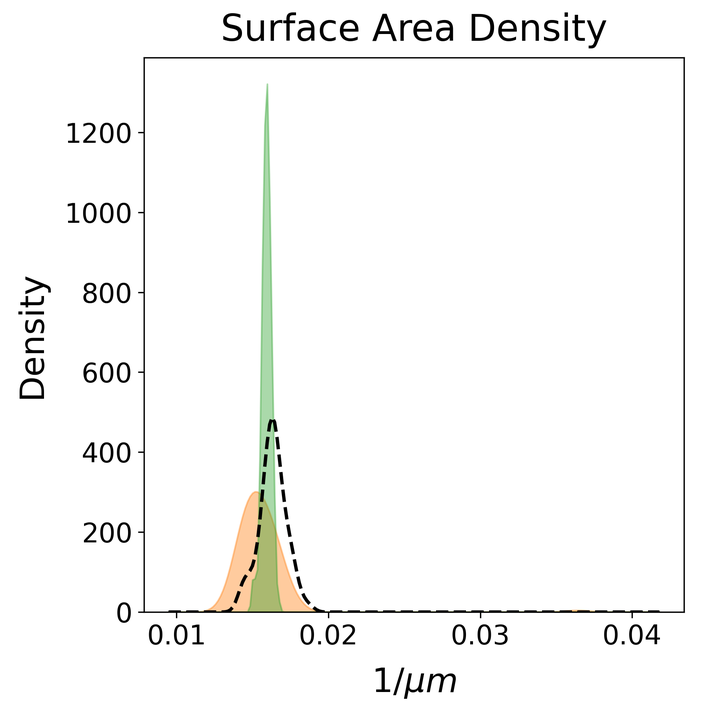}
        \caption{}
        \label{fig:app_bentheimer_256nc_d}
    \end{subfigure}
    \begin{subfigure}[t]{0.15\textwidth}
        \includegraphics[width=\textwidth]{figs/0005-figures/Bentheimer_256_noncond/tpc_comparison_27012025.png}
        \caption{}
        \label{fig:app_bentheimer_256nc_e}
    \end{subfigure}
    \begin{subfigure}[t]{0.15\textwidth}
        \includegraphics[width=\textwidth]{figs/0005-figures/Bentheimer_256_noncond/psd_comparison_27012025.png}
        \caption{}
        \label{fig:app_bentheimer_256nc_f}
    \end{subfigure}
    
    % Doddington
    \begin{subfigure}[t]{0.15\textwidth}
        \includegraphics[width=\textwidth]{figs/0005-figures/Doddington_256_noncond/stats_kde_porosity_27012025.png}
        \caption{}
        \label{fig:app_doddington_256nc_a}
    \end{subfigure}
    \begin{subfigure}[t]{0.15\textwidth}
        \includegraphics[width=\textwidth]{figs/0005-figures/Doddington_256_noncond/stats_kde_permeability_27012025.png}
        \caption{}
        \label{fig:app_doddington_256nc_b}
    \end{subfigure}
    \begin{subfigure}[t]{0.15\textwidth}
        \includegraphics[width=\textwidth]{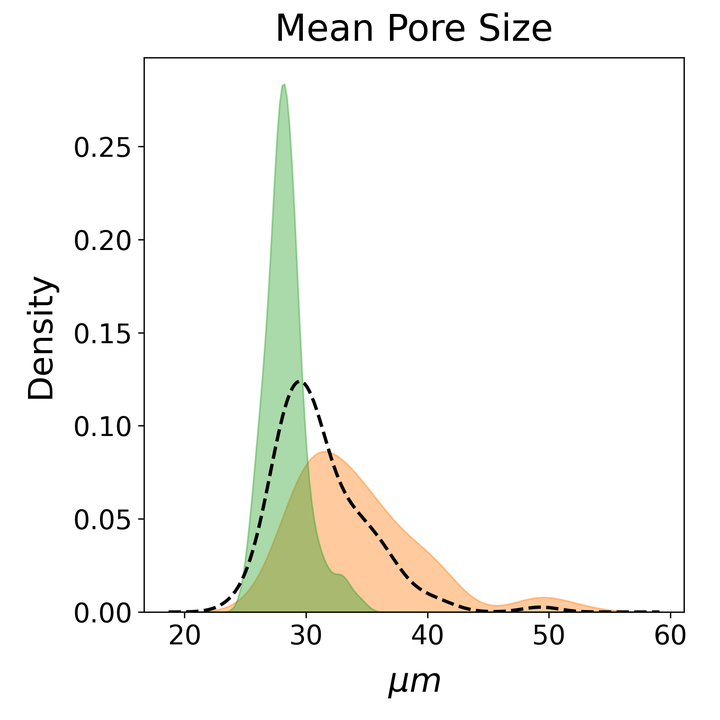}
        \caption{}
        \label{fig:app_doddington_256nc_c}
    \end{subfigure}
    \begin{subfigure}[t]{0.15\textwidth}
        \includegraphics[width=\textwidth]{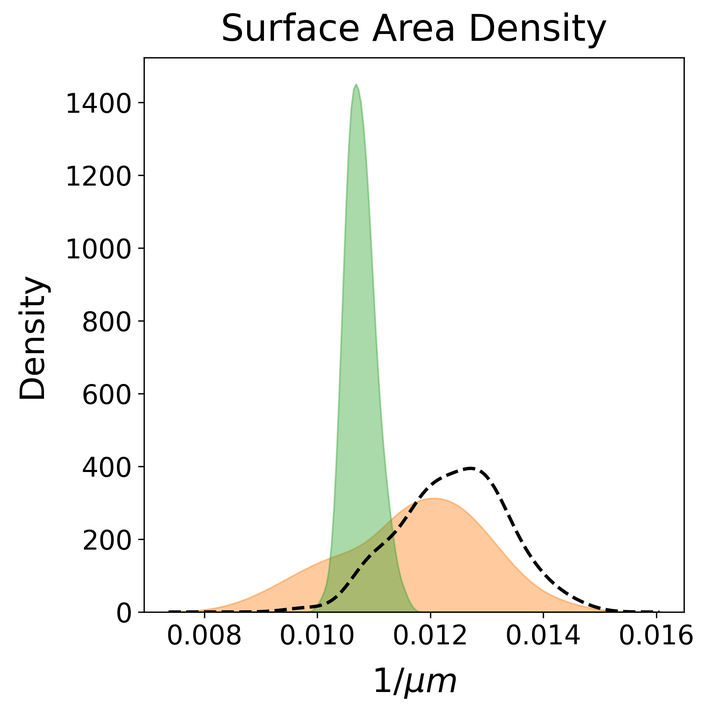}
        \caption{}
        \label{fig:app_doddington_256nc_d}
    \end{subfigure}
    \begin{subfigure}[t]{0.15\textwidth}
        \includegraphics[width=\textwidth]{figs/0005-figures/Doddington_256_noncond/tpc_comparison_27012025.png}
        \caption{}
        \label{fig:app_doddington_256nc_e}
    \end{subfigure}
    \begin{subfigure}[t]{0.15\textwidth}
        \includegraphics[width=\textwidth]{figs/0005-figures/Doddington_256_noncond/psd_comparison_27012025.png}
        \caption{}
        \label{fig:app_doddington_256nc_f}
    \end{subfigure}
    
    % Estaillades
    \begin{subfigure}[t]{0.15\textwidth}
        \includegraphics[width=\textwidth]{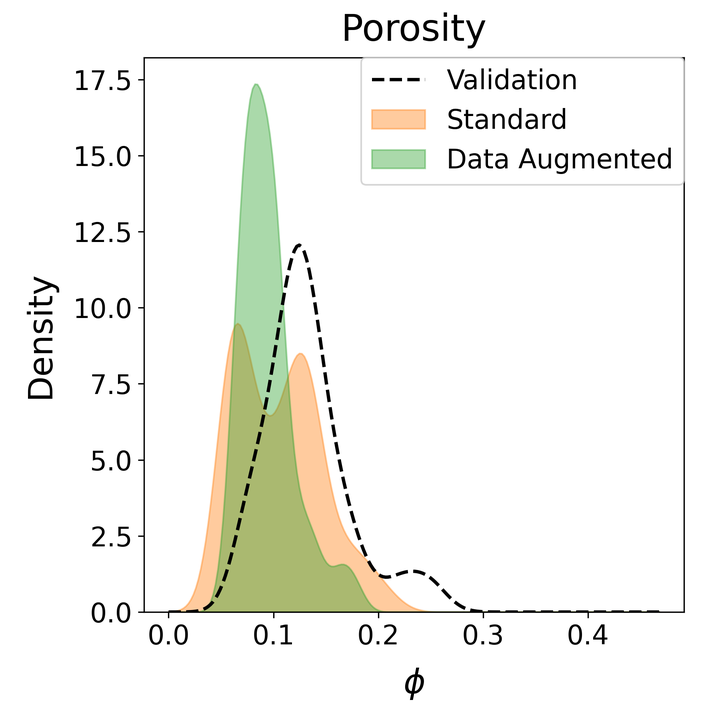}
        \caption{}
        \label{fig:app_estaillades_256nc_a}
    \end{subfigure}
    \begin{subfigure}[t]{0.15\textwidth}
        \includegraphics[width=\textwidth]{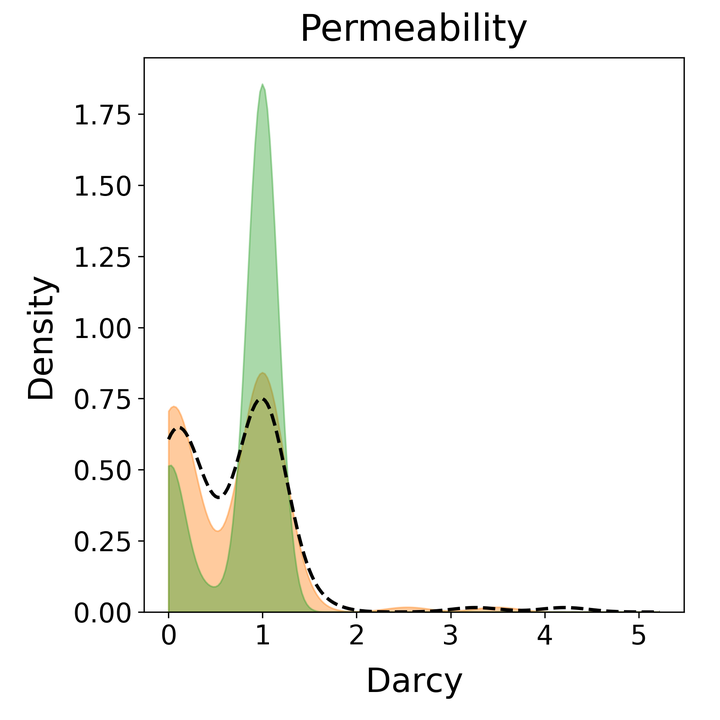}
        \caption{}
        \label{fig:app_estaillades_256nc_b}
    \end{subfigure}
    \begin{subfigure}[t]{0.15\textwidth}
        \includegraphics[width=\textwidth]{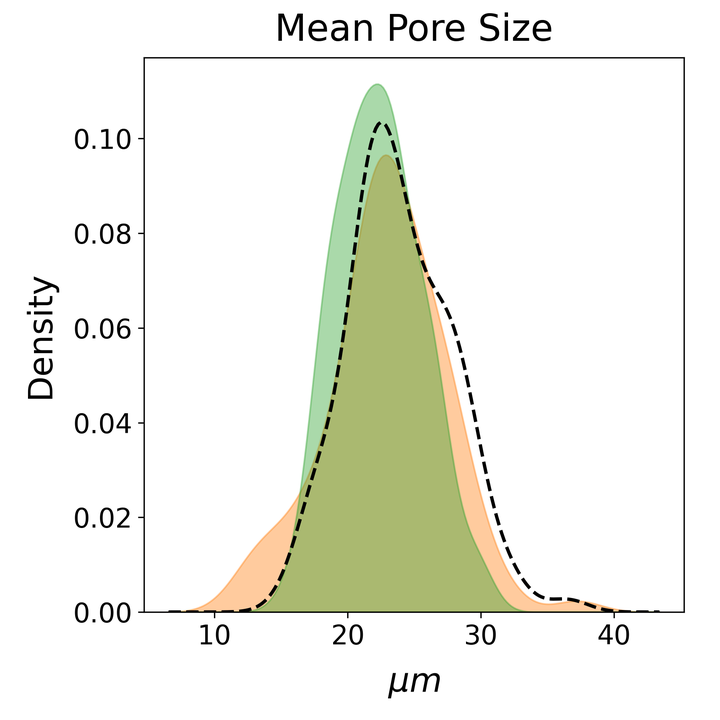}
        \caption{}
        \label{fig:app_estaillades_256nc_c}
    \end{subfigure}
    \begin{subfigure}[t]{0.15\textwidth}
        \includegraphics[width=\textwidth]{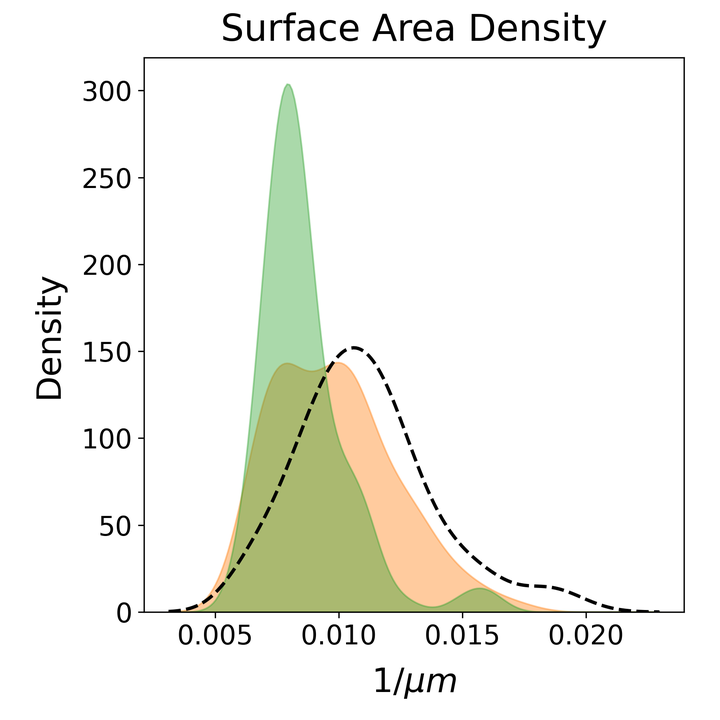}
        \caption{}
        \label{fig:app_estaillades_256nc_d}
    \end{subfigure}
    \begin{subfigure}[t]{0.15\textwidth}
        \includegraphics[width=\textwidth]{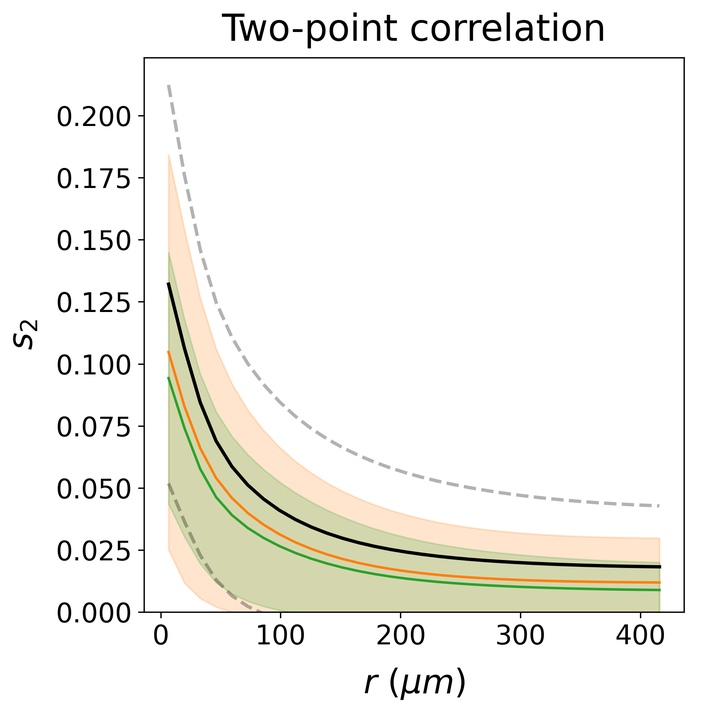}
        \caption{}
        \label{fig:app_estaillades_256nc_e}
    \end{subfigure}
    \begin{subfigure}[t]{0.15\textwidth}
        \includegraphics[width=\textwidth]{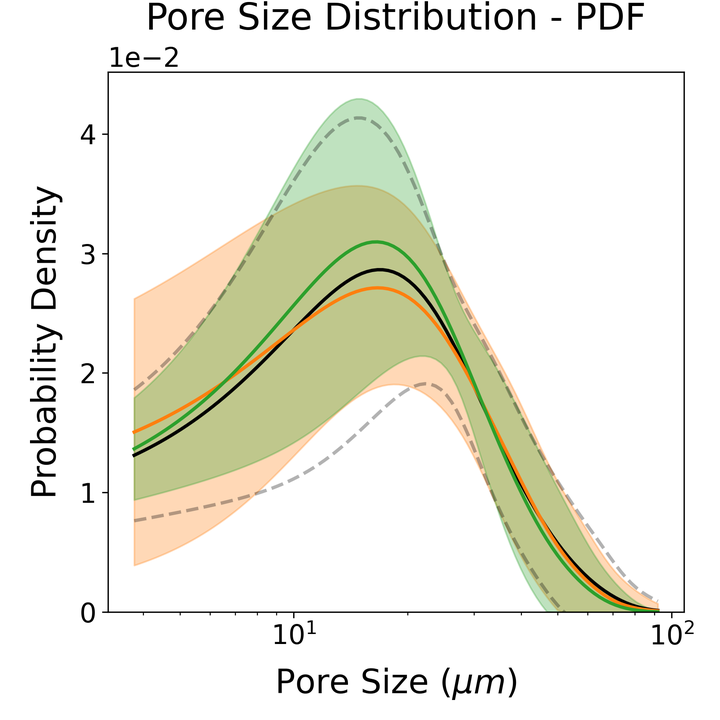}
        \caption{}
        \label{fig:app_estaillades_256nc_f}
    \end{subfigure}
    
    % Ketton
    \begin{subfigure}[t]{0.15\textwidth}
        \includegraphics[width=\textwidth]{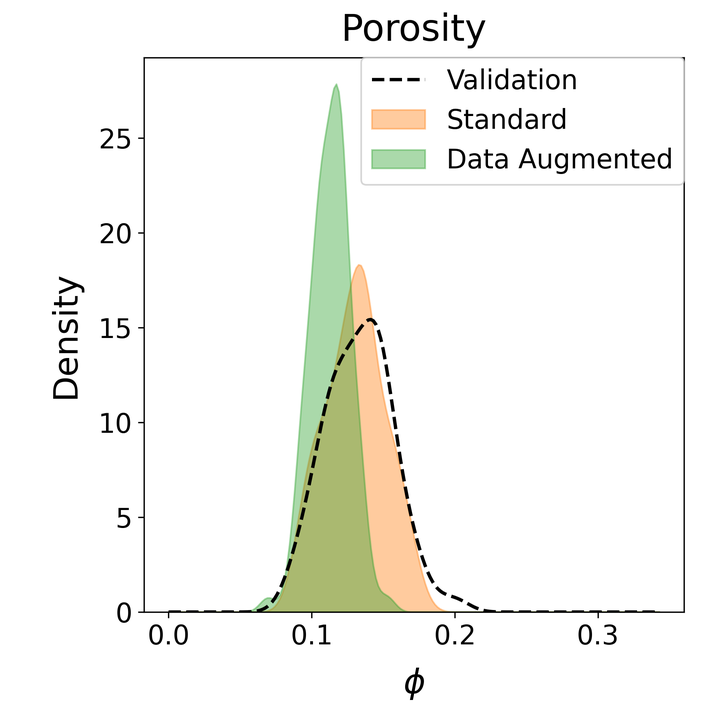}
        \caption{}
        \label{fig:app_ketton_256nc_a}
    \end{subfigure}
    \begin{subfigure}[t]{0.15\textwidth}
        \includegraphics[width=\textwidth]{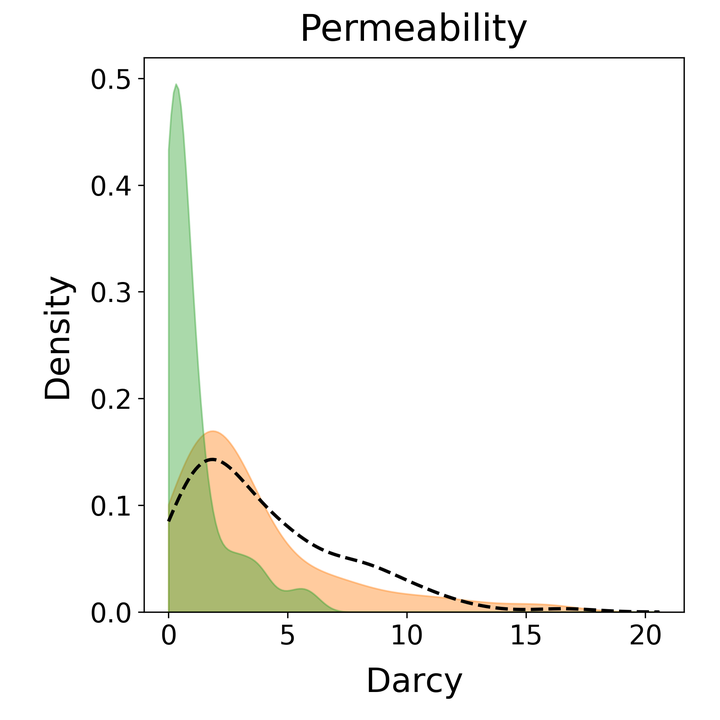}
        \caption{}
        \label{fig:app_ketton_256nc_b}
    \end{subfigure}
    \begin{subfigure}[t]{0.15\textwidth}
        \includegraphics[width=\textwidth]{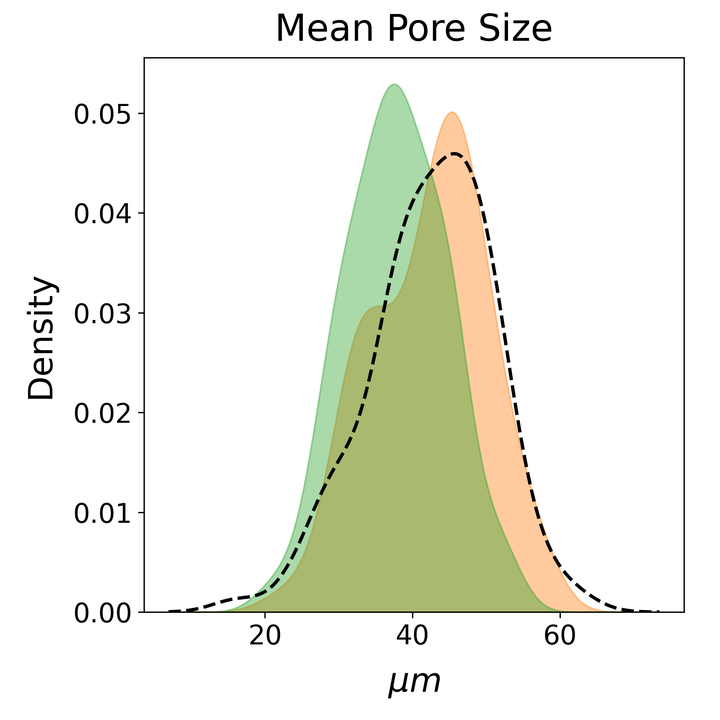}
        \caption{}
        \label{fig:app_ketton_256nc_c}
    \end{subfigure}
    \begin{subfigure}[t]{0.15\textwidth}
        \includegraphics[width=\textwidth]{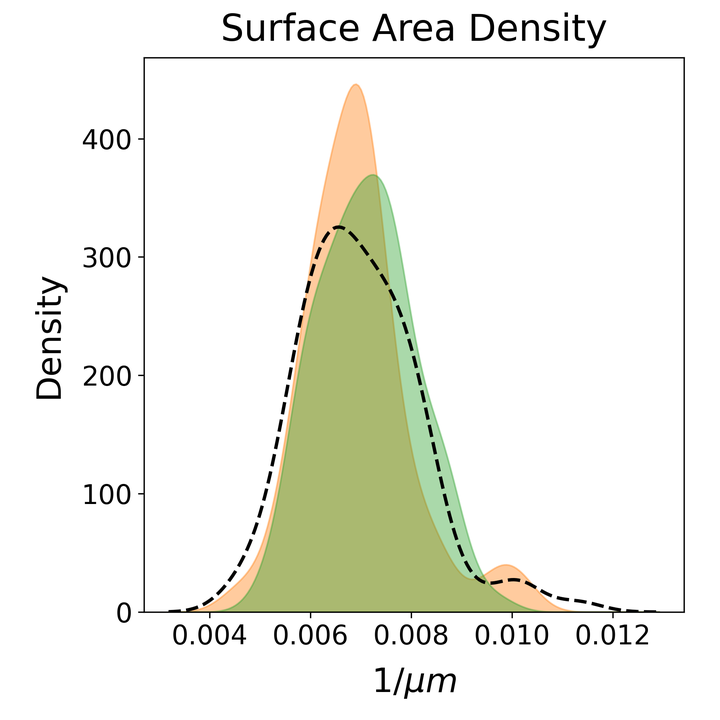}
        \caption{}
        \label{fig:app_ketton_256nc_d}
    \end{subfigure}
    \begin{subfigure}[t]{0.15\textwidth}
        \includegraphics[width=\textwidth]{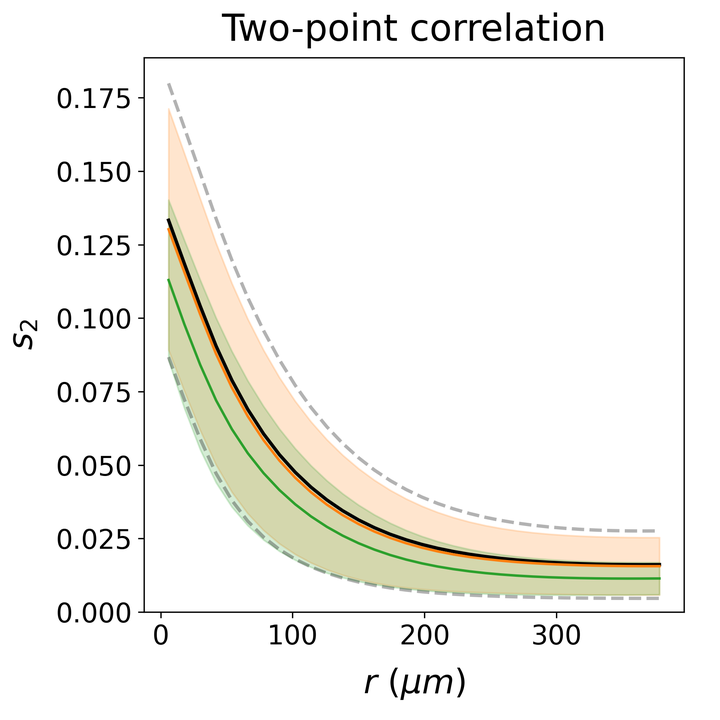}
        \caption{}
        \label{fig:app_ketton_256nc_e}
    \end{subfigure}
    \begin{subfigure}[t]{0.15\textwidth}
        \includegraphics[width=\textwidth]{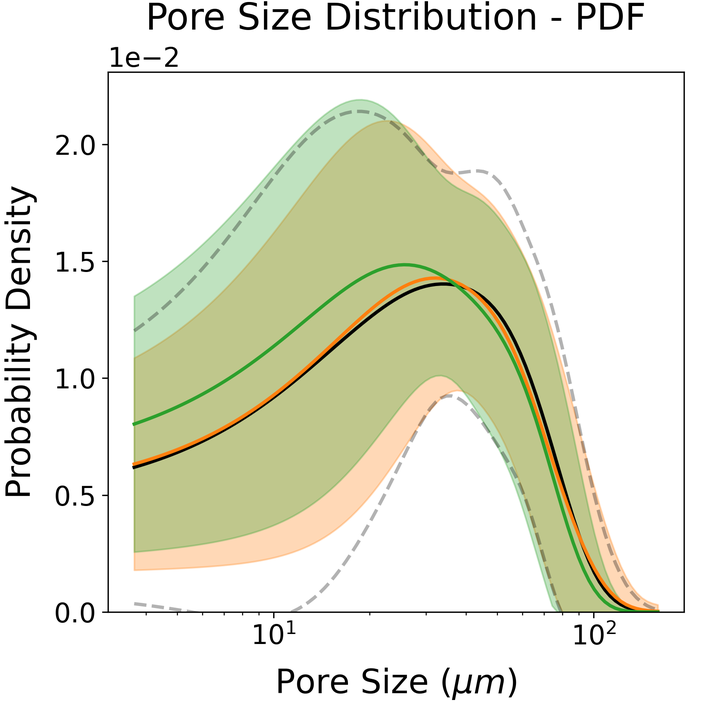}
        \caption{}
        \label{fig:app_ketton_256nc_f}
    \end{subfigure}
    
    \caption{Full statistical properties comparison for unconditional generated 256$^3$ volume samples across different rock types. Top to bottom: (a-f) Bentheimer, (g-l) Doddington, (m-r) Estaillades, and (s-x) Ketton sandstones. Each row shows (from left to right): porosity distribution, permeability distribution, mean pore size distribution, surface area density distribution, two-point correlation function, and pore size distribution.}
    \label{fig:appendix_all_rocks_256nc_comparison}
\end{figure}

\begin{figure}[H]
    \centering
    % Bentheimer
    \begin{subfigure}[t]{0.15\textwidth}
        \includegraphics[width=\textwidth]{figs/0005-figures/Bentheimer_128/stats_kde_porosity_27012025.png}
        \caption{}
        \label{fig:app_bentheimer_128_a}
    \end{subfigure}
    \begin{subfigure}[t]{0.15\textwidth}
        \includegraphics[width=\textwidth]{figs/0005-figures/Bentheimer_128/stats_kde_permeability_27012025.png}
        \caption{}
        \label{fig:app_bentheimer_128_b}
    \end{subfigure}
    \begin{subfigure}[t]{0.15\textwidth}
        \includegraphics[width=\textwidth]{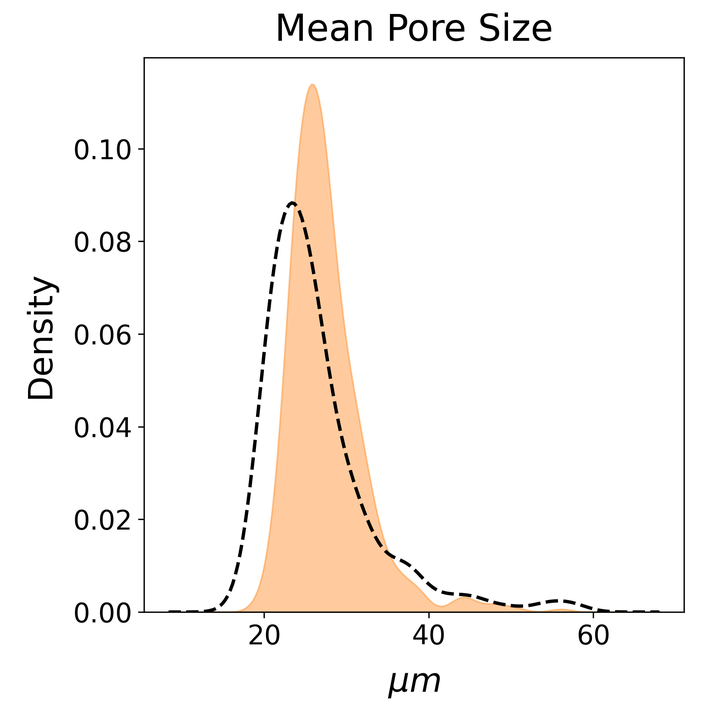}
        \caption{}
        \label{fig:app_bentheimer_128_c}
    \end{subfigure}
    \begin{subfigure}[t]{0.15\textwidth}
        \includegraphics[width=\textwidth]{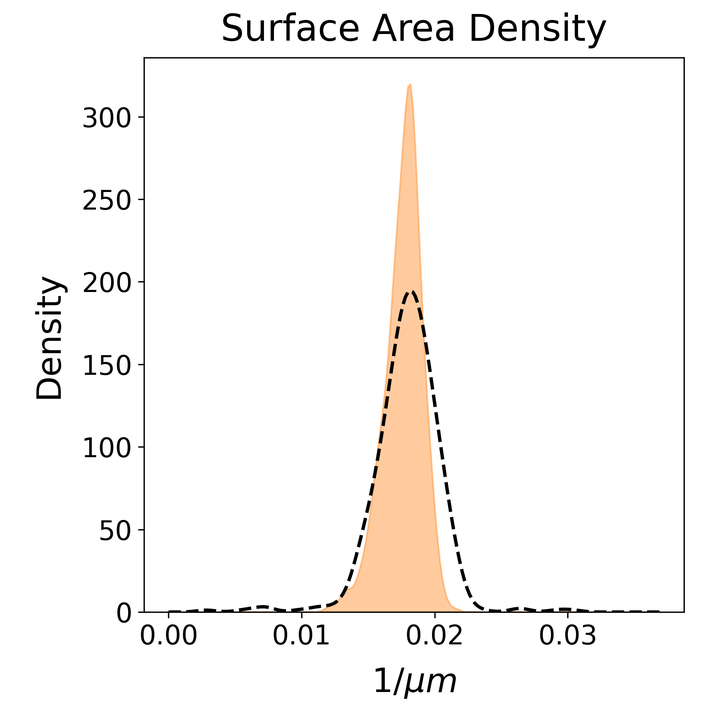}
        \caption{}
        \label{fig:app_bentheimer_128_d}
    \end{subfigure}
    \begin{subfigure}[t]{0.15\textwidth}
        \includegraphics[width=\textwidth]{figs/0005-figures/Bentheimer_128/tpc_comparison_27012025.png}
        \caption{}
        \label{fig:app_bentheimer_128_e}
    \end{subfigure}
    \begin{subfigure}[t]{0.15\textwidth}
        \includegraphics[width=\textwidth]{figs/0005-figures/Bentheimer_128/psd_comparison_27012025.png}
        \caption{}
        \label{fig:app_bentheimer_128_f}
    \end{subfigure}
    
    % Doddington
    \begin{subfigure}[t]{0.15\textwidth}
        \includegraphics[width=\textwidth]{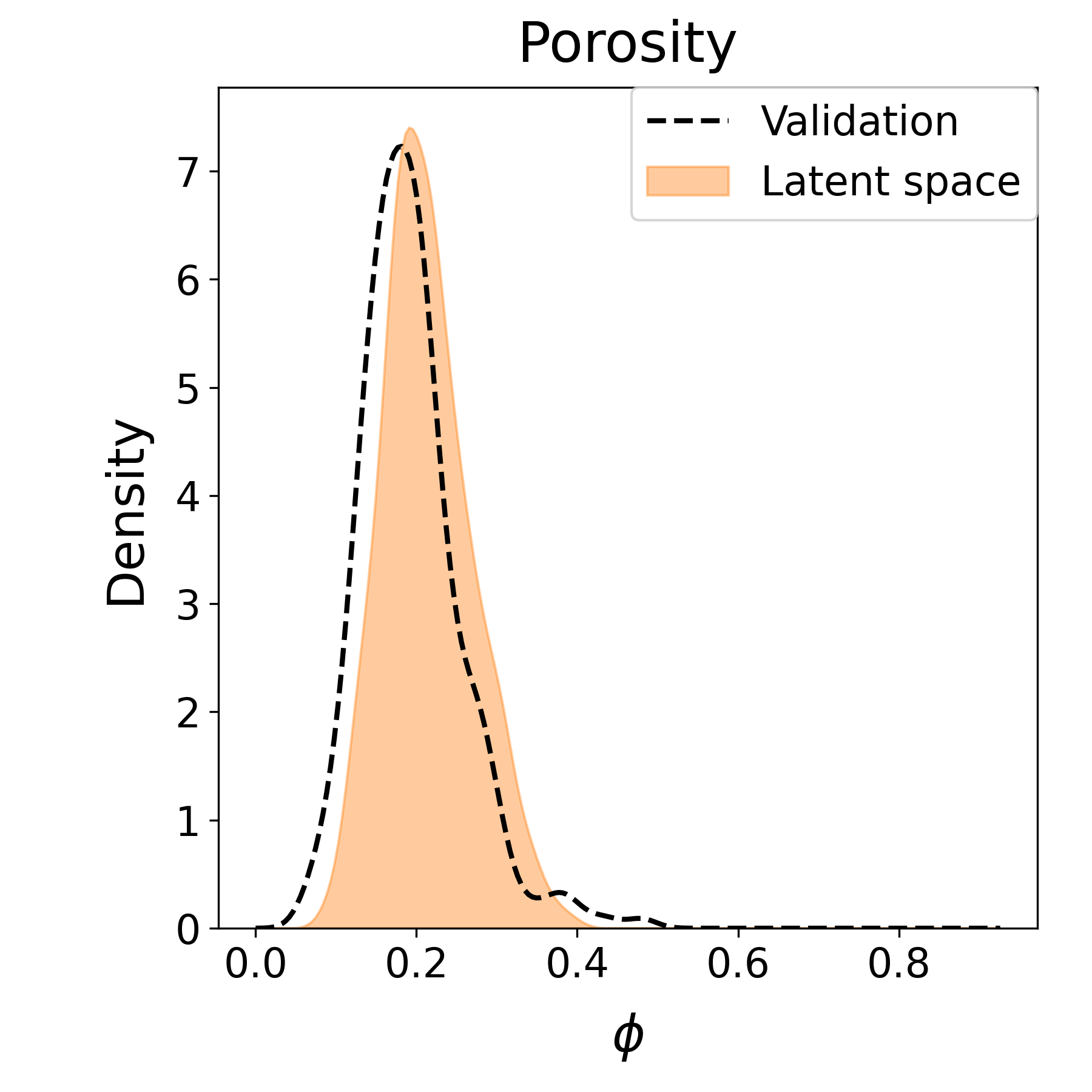}
        \caption{}
        \label{fig:app_doddington_128_a}
    \end{subfigure}
    \begin{subfigure}[t]{0.15\textwidth}
        \includegraphics[width=\textwidth]{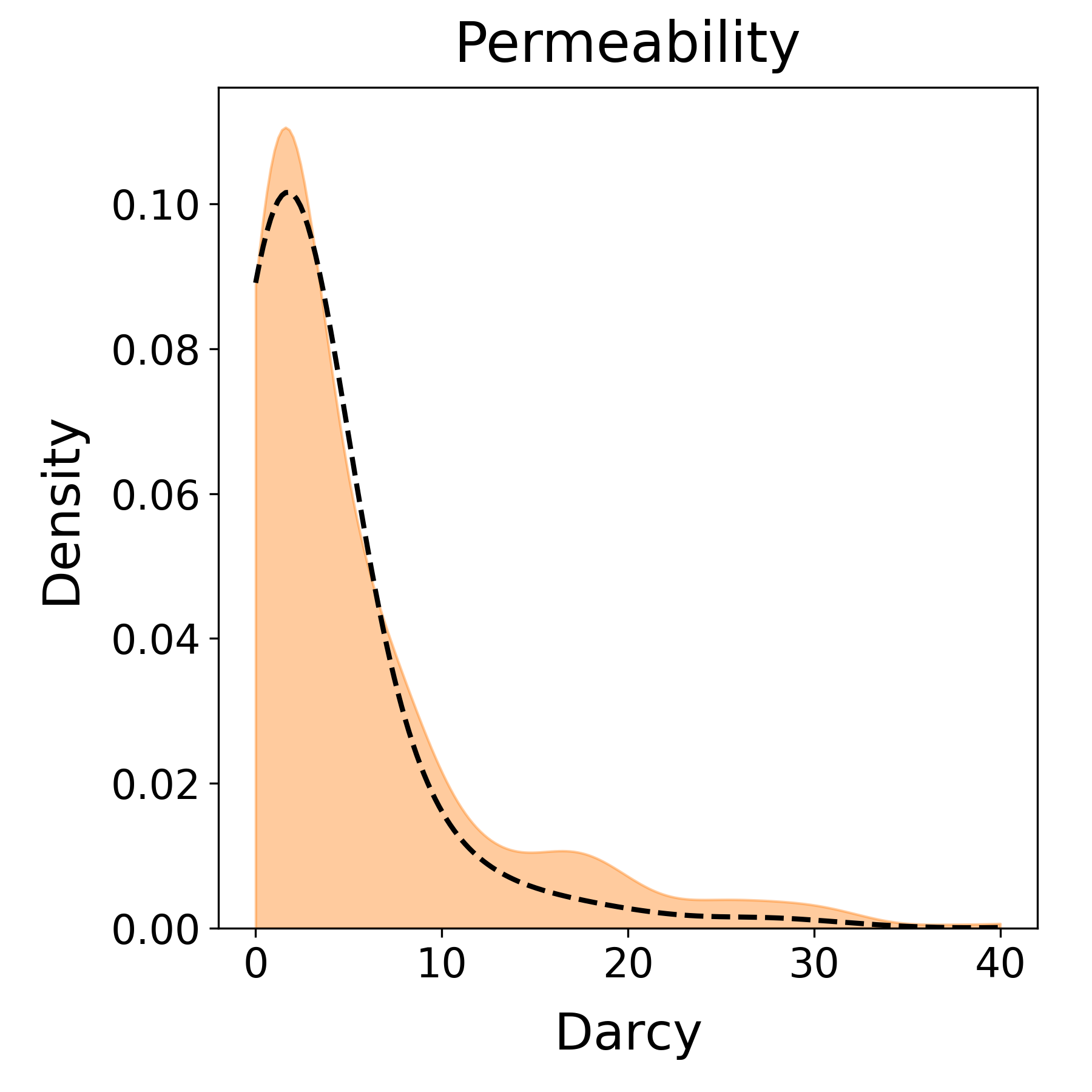}
        \caption{}
        \label{fig:app_doddington_128_b}
    \end{subfigure}
    \begin{subfigure}[t]{0.15\textwidth}
        \includegraphics[width=\textwidth]{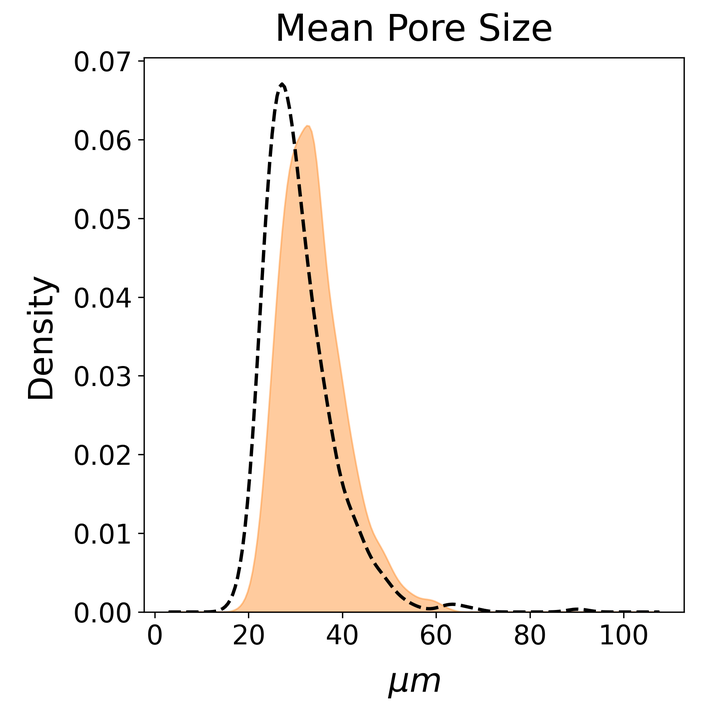}
        \caption{}
        \label{fig:app_doddington_128_c}
    \end{subfigure}
    \begin{subfigure}[t]{0.15\textwidth}
        \includegraphics[width=\textwidth]{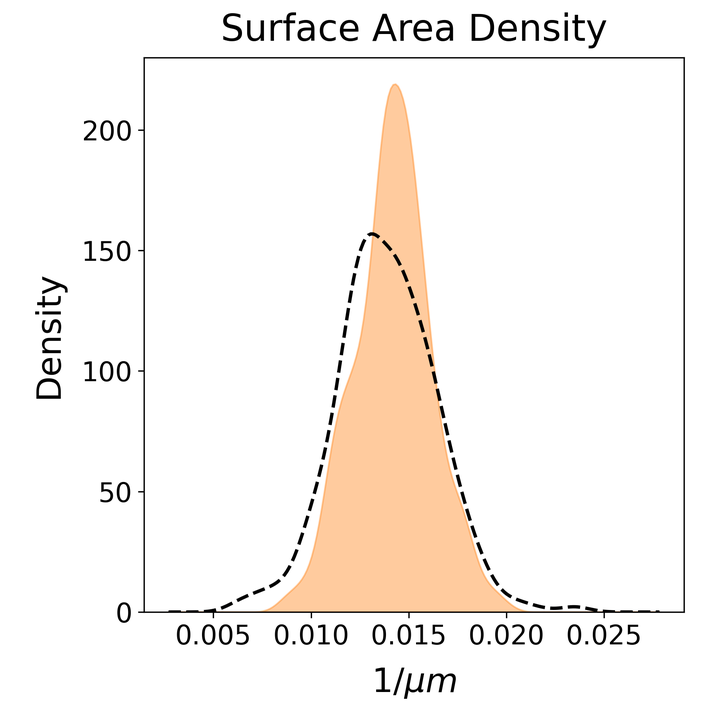}
        \caption{}
        \label{fig:app_doddington_128_d}
    \end{subfigure}
    \begin{subfigure}[t]{0.15\textwidth}
        \includegraphics[width=\textwidth]{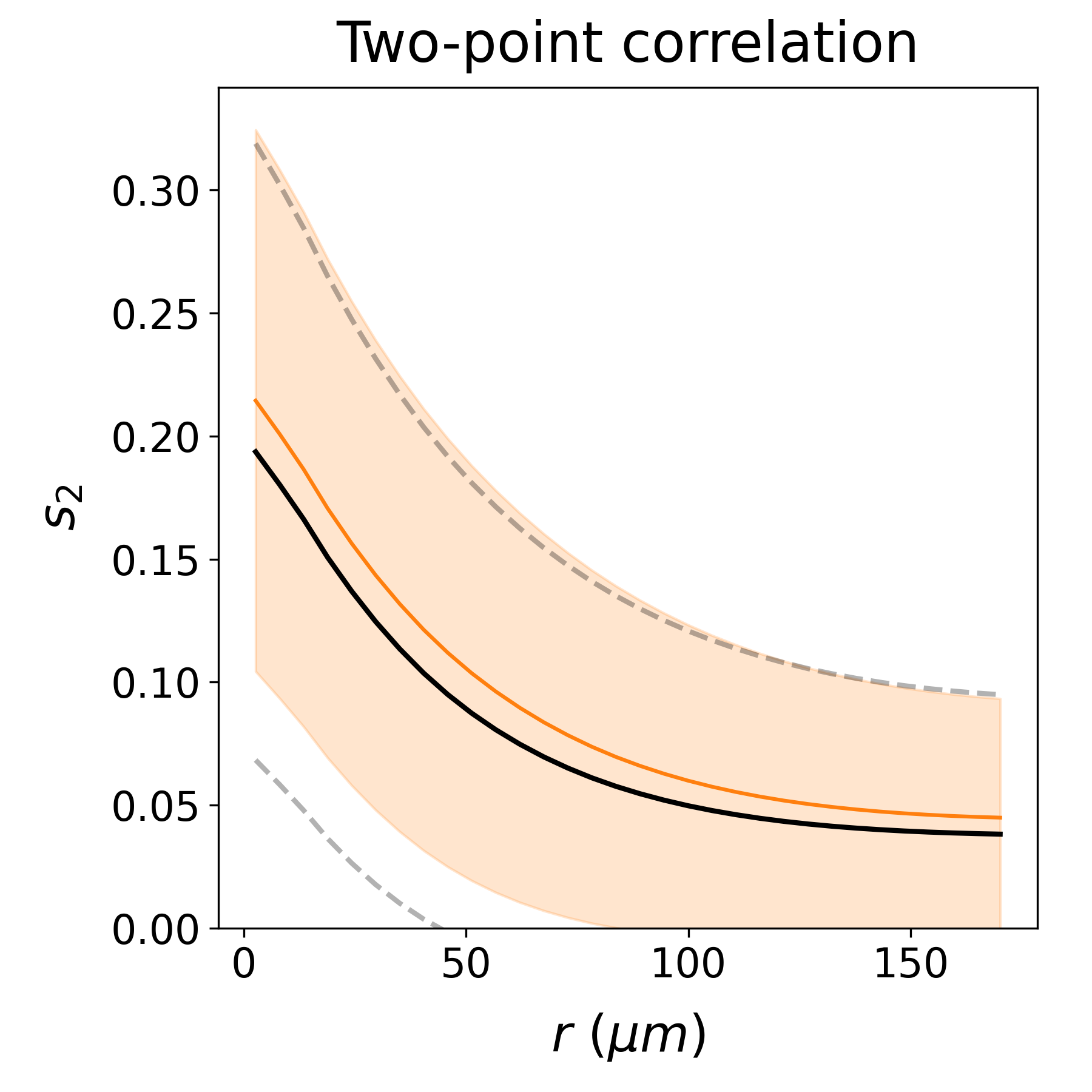}
        \caption{}
        \label{fig:app_doddington_128_e}
    \end{subfigure}
    \begin{subfigure}[t]{0.15\textwidth}
        \includegraphics[width=\textwidth]{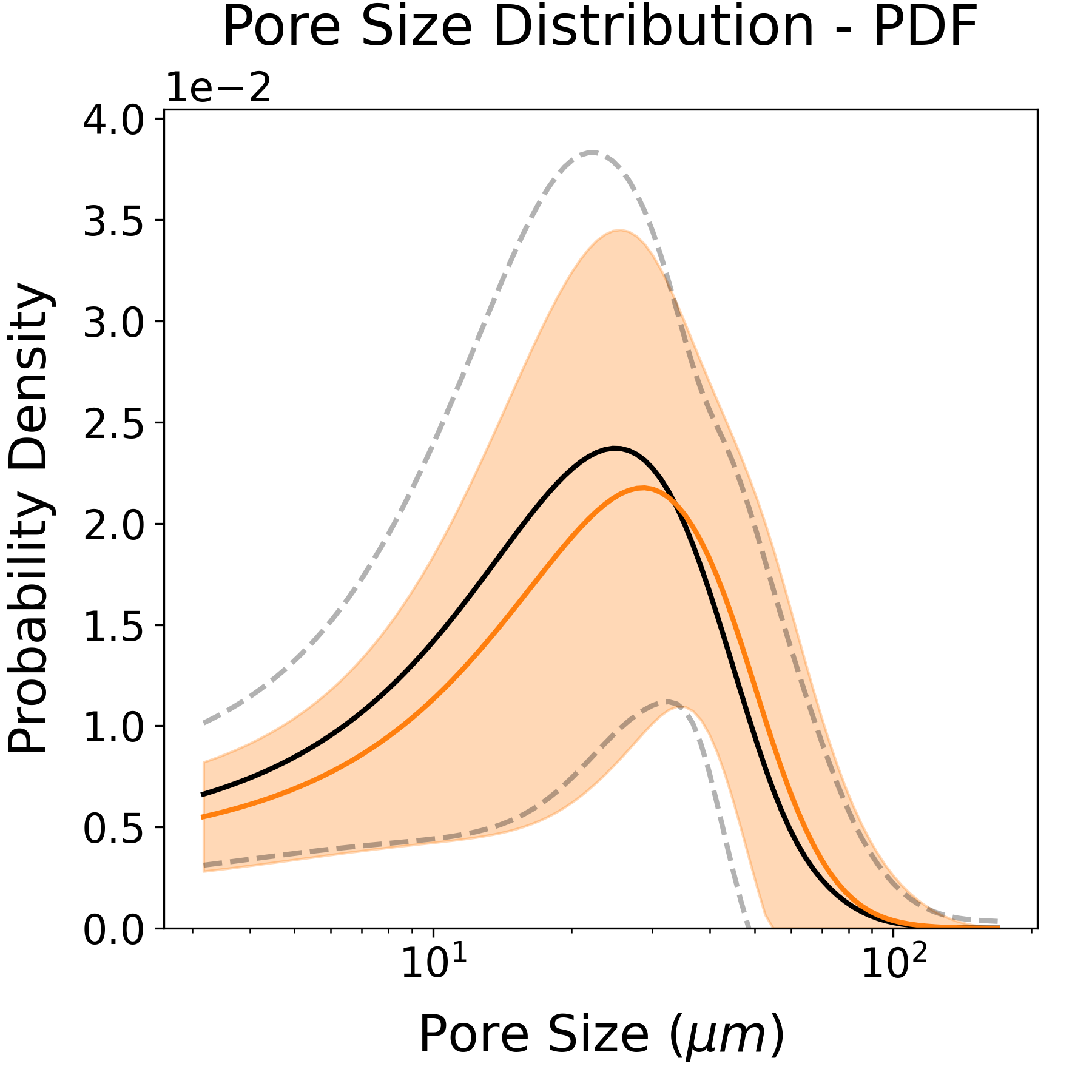}
        \caption{}
        \label{fig:app_doddington_128_f}
    \end{subfigure}
    
    % Estaillades
    \begin{subfigure}[t]{0.15\textwidth}
        \includegraphics[width=\textwidth]{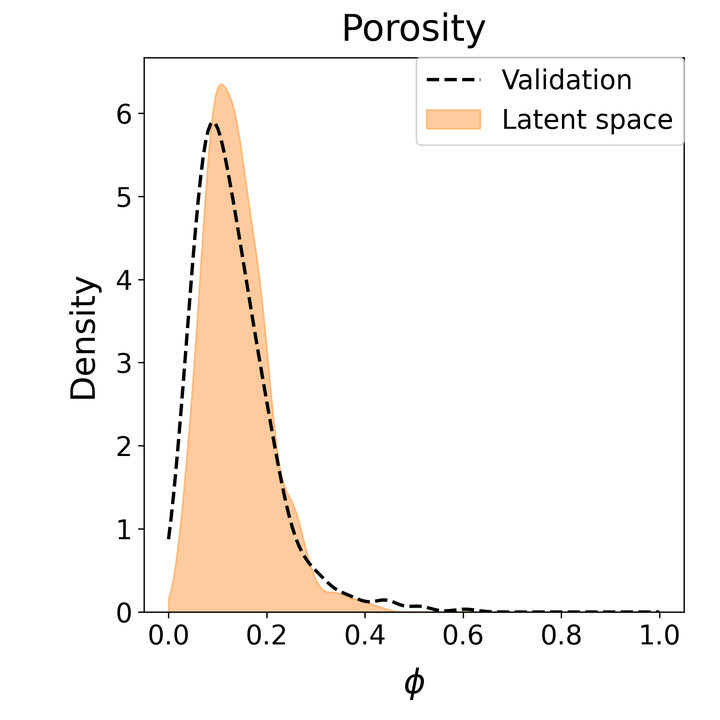}
        \caption{}
        \label{fig:app_estaillades_128_a}
    \end{subfigure}
    \begin{subfigure}[t]{0.15\textwidth}
        \includegraphics[width=\textwidth]{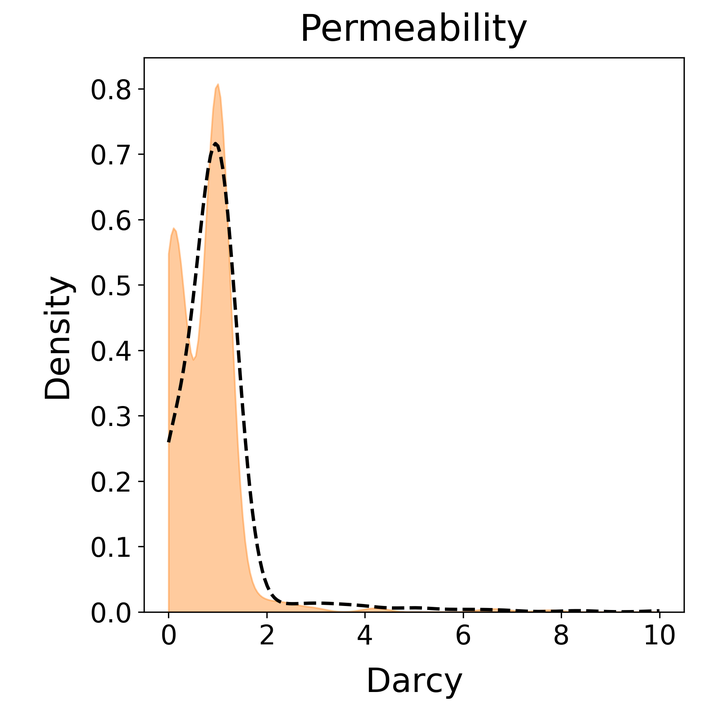}
        \caption{}
        \label{fig:app_estaillades_128_b}
    \end{subfigure}
    \begin{subfigure}[t]{0.15\textwidth}
        \includegraphics[width=\textwidth]{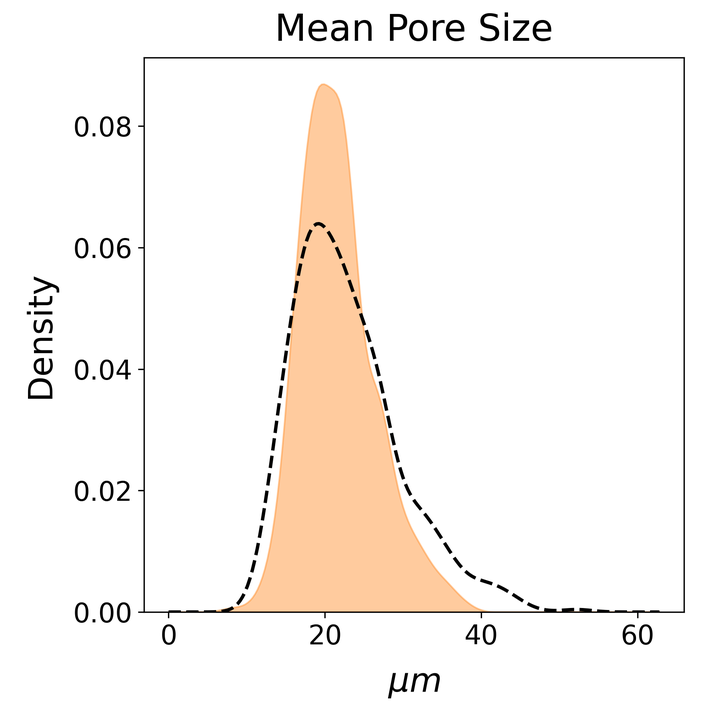}
        \caption{}
        \label{fig:app_estaillades_128_c}
    \end{subfigure}
    \begin{subfigure}[t]{0.15\textwidth}
        \includegraphics[width=\textwidth]{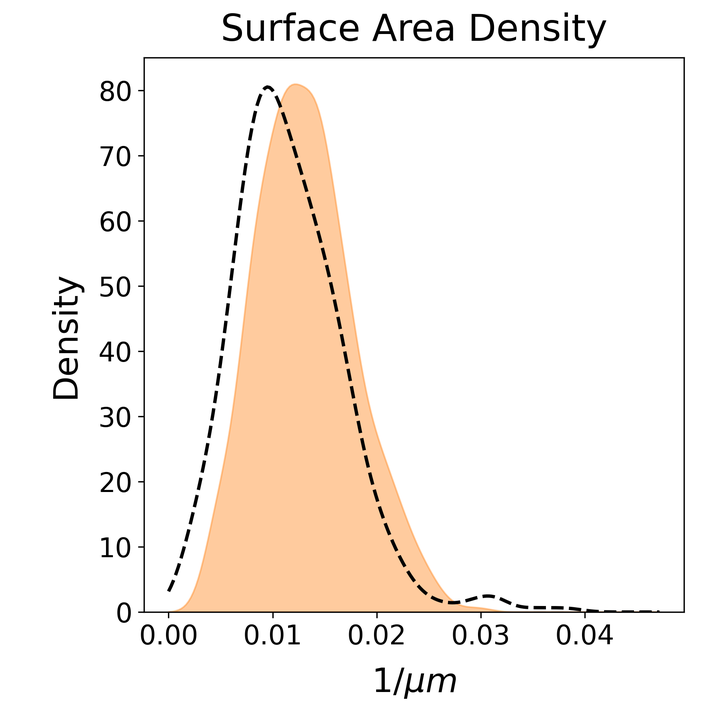}
        \caption{}
        \label{fig:app_estaillades_128_d}
    \end{subfigure}
    \begin{subfigure}[t]{0.15\textwidth}
        \includegraphics[width=\textwidth]{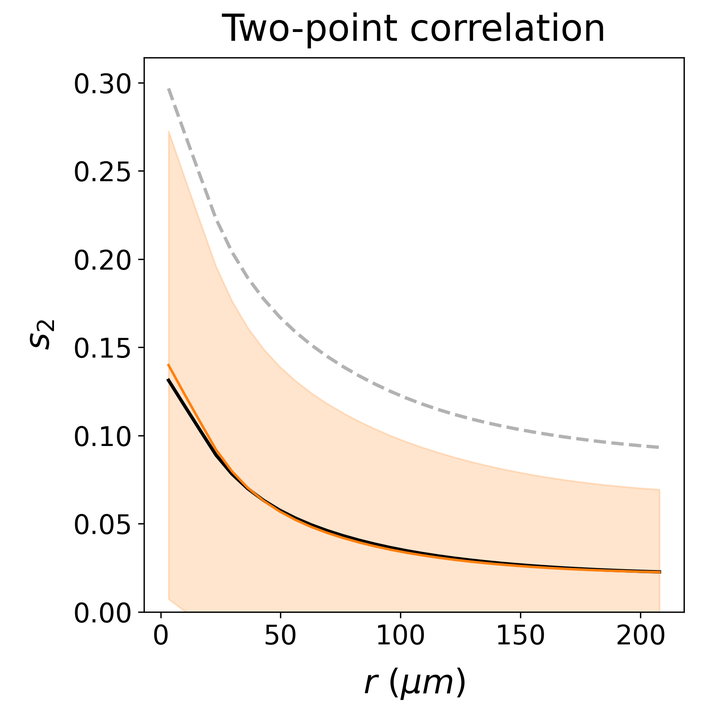}
        \caption{}
        \label{fig:app_estaillades_128_e}
    \end{subfigure}
    \begin{subfigure}[t]{0.15\textwidth}
        \includegraphics[width=\textwidth]{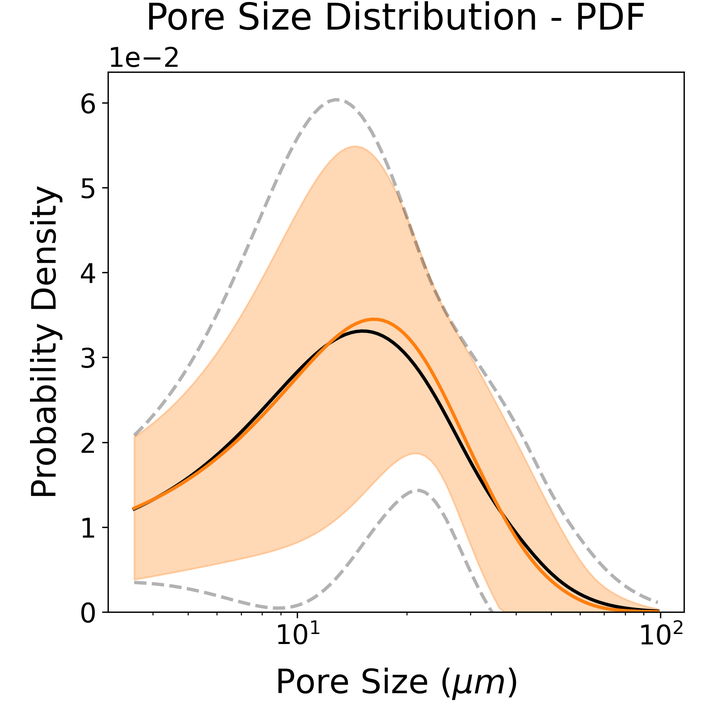}
        \caption{}
        \label{fig:app_estaillades_128_f}
    \end{subfigure}
    
    % Ketton
    \begin{subfigure}[t]{0.15\textwidth}
        \includegraphics[width=\textwidth]{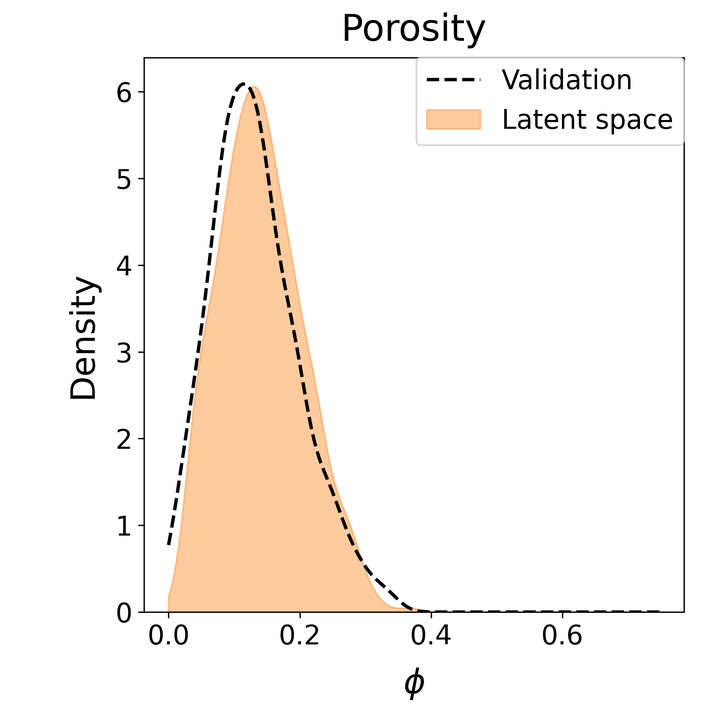}
        \caption{}
        \label{fig:app_ketton_128_a}
    \end{subfigure}
    \begin{subfigure}[t]{0.15\textwidth}
        \includegraphics[width=\textwidth]{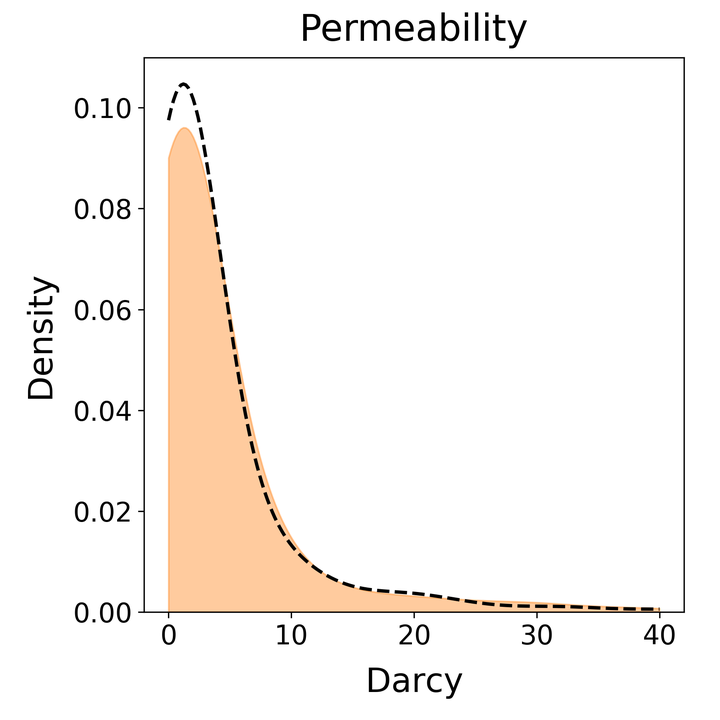}
        \caption{}
        \label{fig:app_ketton_128_b}
    \end{subfigure}
    \begin{subfigure}[t]{0.15\textwidth}
        \includegraphics[width=\textwidth]{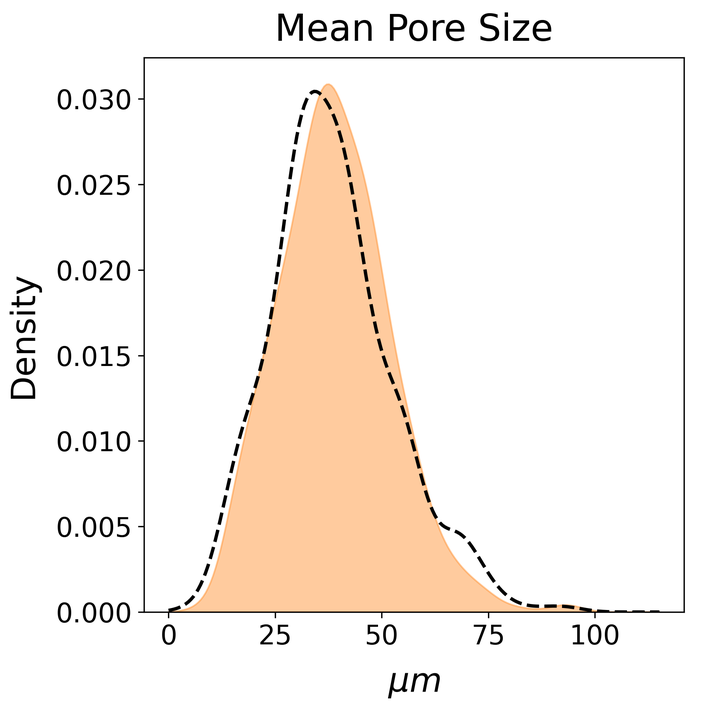}
        \caption{}
        \label{fig:app_ketton_128_c}
    \end{subfigure}
    \begin{subfigure}[t]{0.15\textwidth}
        \includegraphics[width=\textwidth]{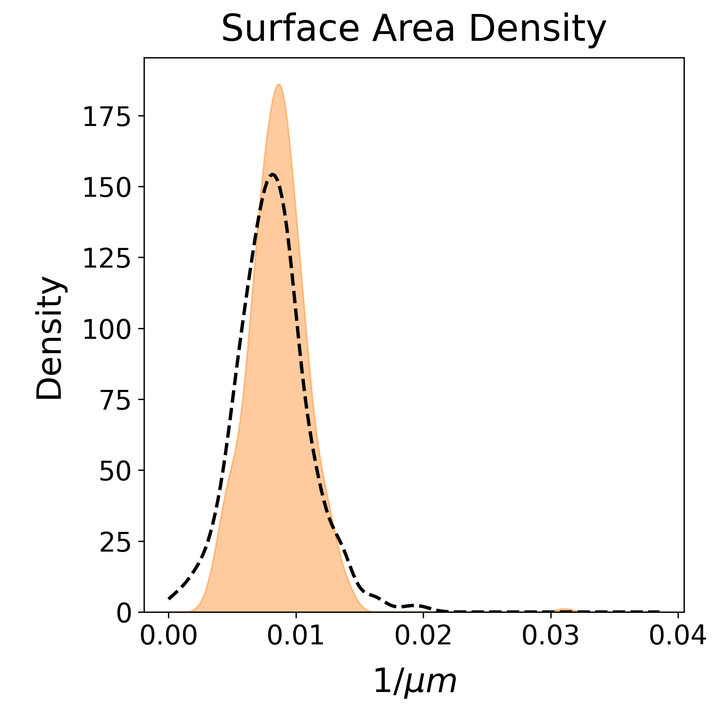}
        \caption{}
        \label{fig:app_ketton_128_d}
    \end{subfigure}
    \begin{subfigure}[t]{0.15\textwidth}
        \includegraphics[width=\textwidth]{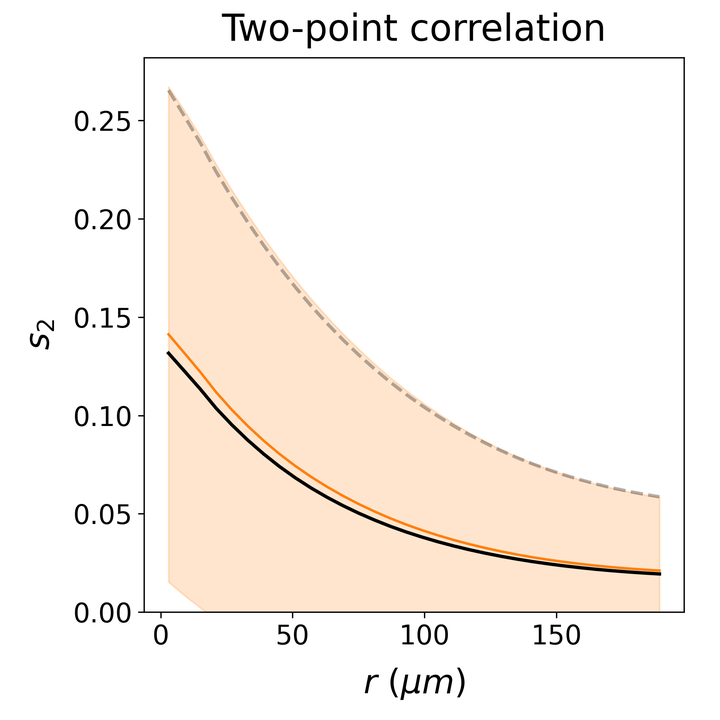}
        \caption{}
        \label{fig:app_ketton_128_e}
    \end{subfigure}
    \begin{subfigure}[t]{0.15\textwidth}
        \includegraphics[width=\textwidth]{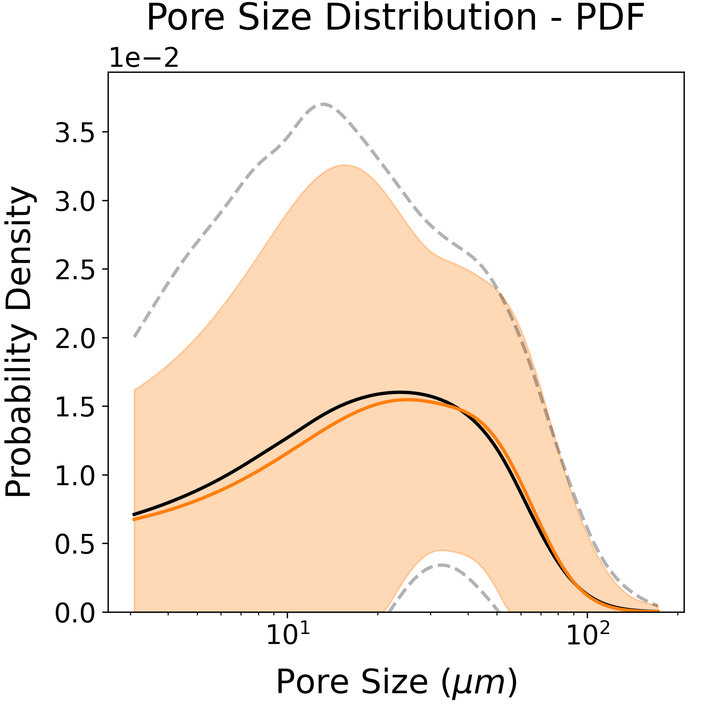}
        \caption{}
        \label{fig:app_ketton_128_f}
    \end{subfigure}
    
    \caption{Full statistical properties comparison for 128$^3$ volume samples across different rock types. Top to bottom: (a-f) Bentheimer, (g-l) Doddington, (m-r) Estaillades, and (s-x) Ketton sandstones. Each row shows (from left to right): porosity distribution, permeability distribution, mean pore size distribution, surface area density distribution, two-point correlation function, and pore size distribution.}
    \label{fig:appendix_all_rocks_128_comparison}
\end{figure}

\begin{figure}[H]
    \centering
    % Bentheimer
    \begin{subfigure}[t]{0.15\textwidth}
        \includegraphics[width=\textwidth]{figs/0005-figures/Bentheimer_64/stats_kde_porosity_27012025.png}
        \caption{}
        \label{fig:app_bentheimer_64_a}
    \end{subfigure}
    \begin{subfigure}[t]{0.15\textwidth}
        \includegraphics[width=\textwidth]{figs/0005-figures/Bentheimer_64/stats_kde_permeability_27012025.png}
        \caption{}
        \label{fig:app_bentheimer_64_b}
    \end{subfigure}
    \begin{subfigure}[t]{0.15\textwidth}
        \includegraphics[width=\textwidth]{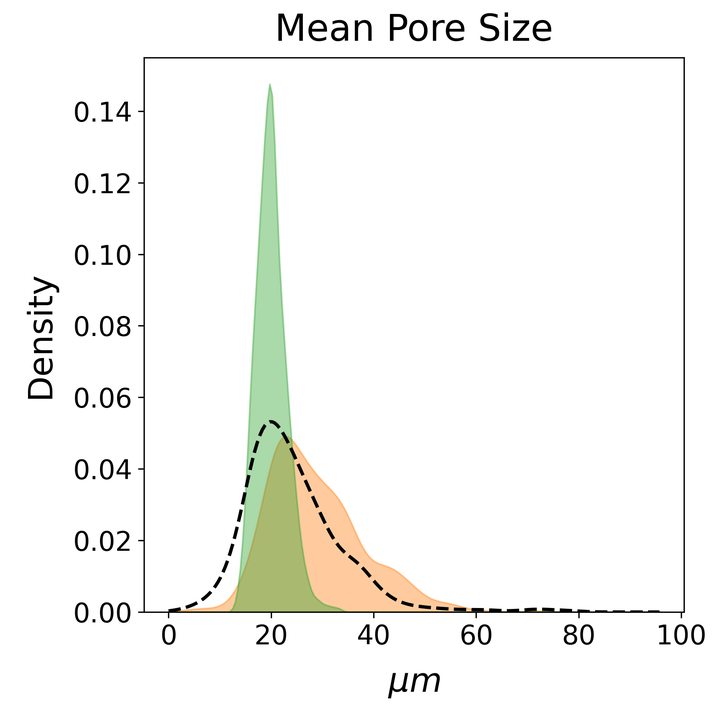}
        \caption{}
        \label{fig:app_bentheimer_64_c}
    \end{subfigure}
    \begin{subfigure}[t]{0.15\textwidth}
        \includegraphics[width=\textwidth]{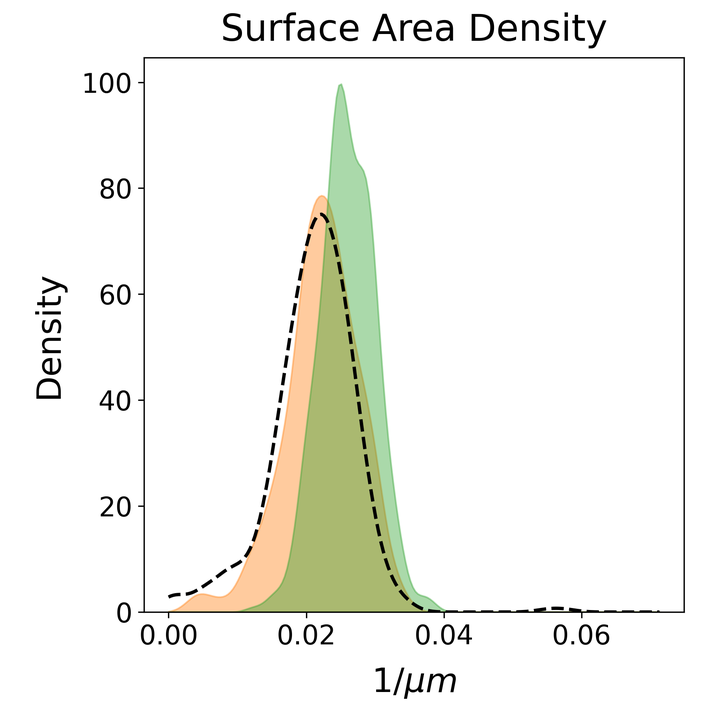}
        \caption{}
        \label{fig:app_bentheimer_64_d}
    \end{subfigure}
    \begin{subfigure}[t]{0.15\textwidth}
        \includegraphics[width=\textwidth]{figs/0005-figures/Bentheimer_64/tpc_comparison_27012025.png}
        \caption{}
        \label{fig:app_bentheimer_64_e}
    \end{subfigure}
    \begin{subfigure}[t]{0.15\textwidth}
        \includegraphics[width=\textwidth]{figs/0005-figures/Bentheimer_64/psd_comparison_27012025.png}
        \caption{}
        \label{fig:app_bentheimer_64_f}
    \end{subfigure}
    
    % Doddington
    \begin{subfigure}[t]{0.15\textwidth}
        \includegraphics[width=\textwidth]{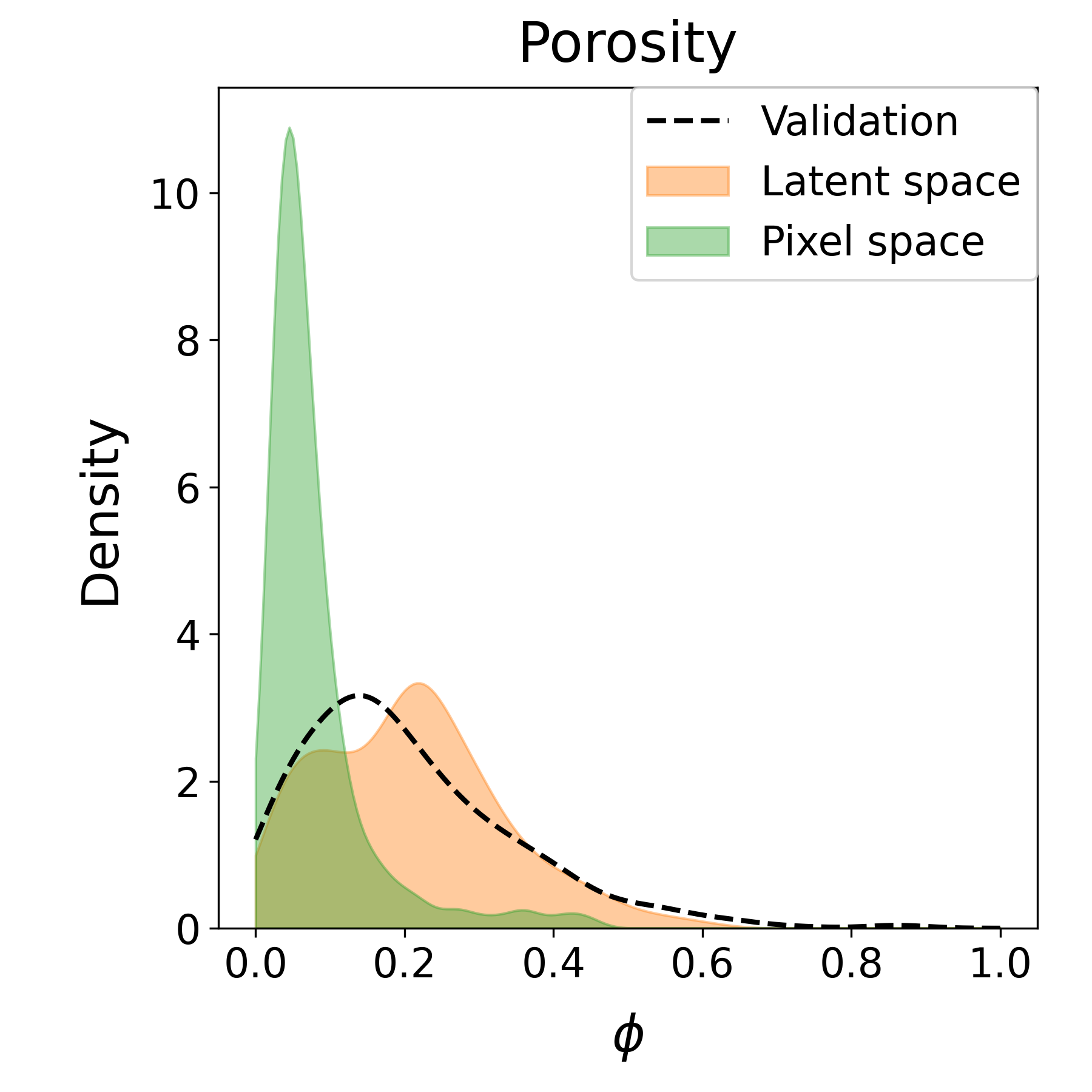}
        \caption{}
        \label{fig:app_doddington_64_a}
    \end{subfigure}
    \begin{subfigure}[t]{0.15\textwidth}
        \includegraphics[width=\textwidth]{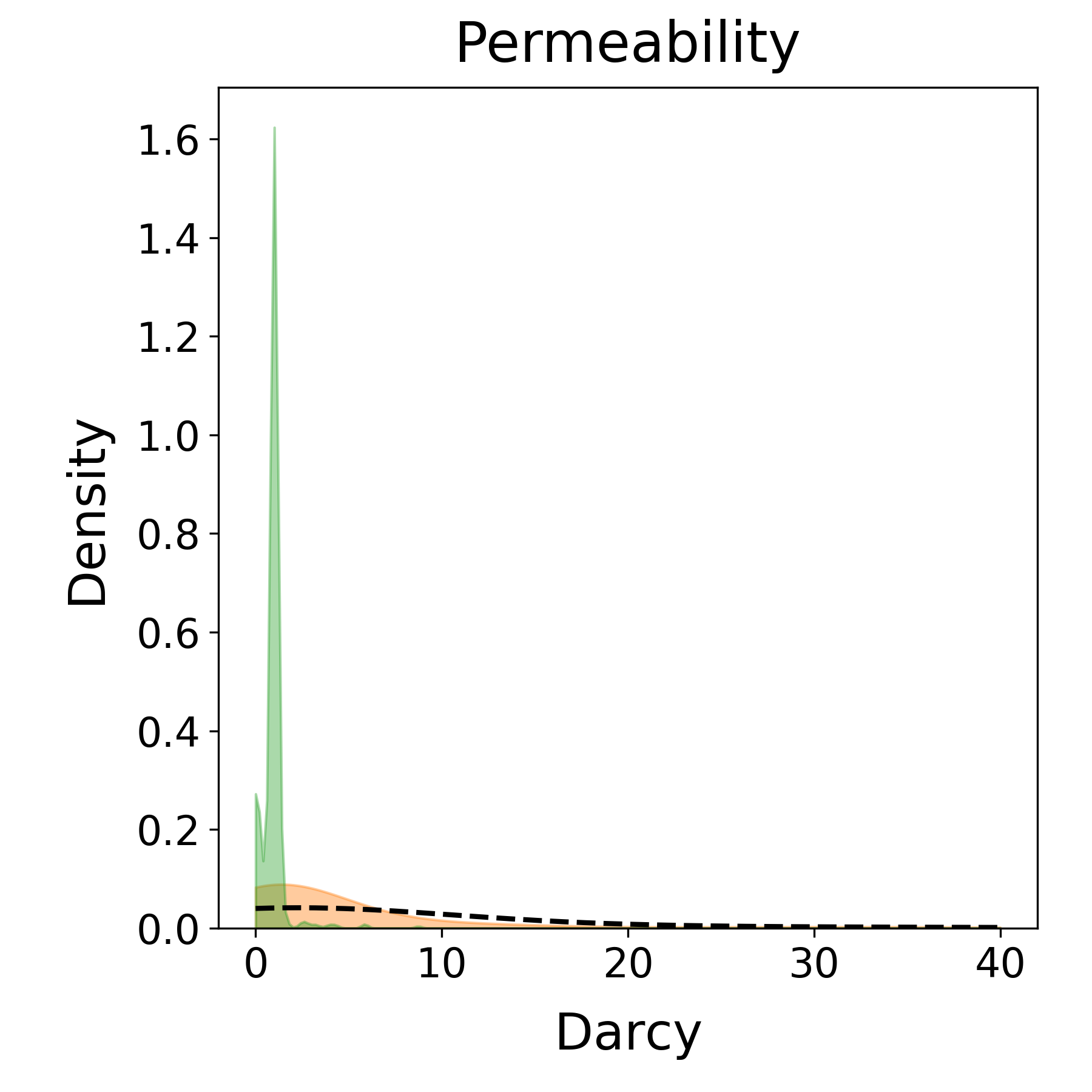}
        \caption{}
        \label{fig:app_doddington_64_b}
    \end{subfigure}
    \begin{subfigure}[t]{0.15\textwidth}
        \includegraphics[width=\textwidth]{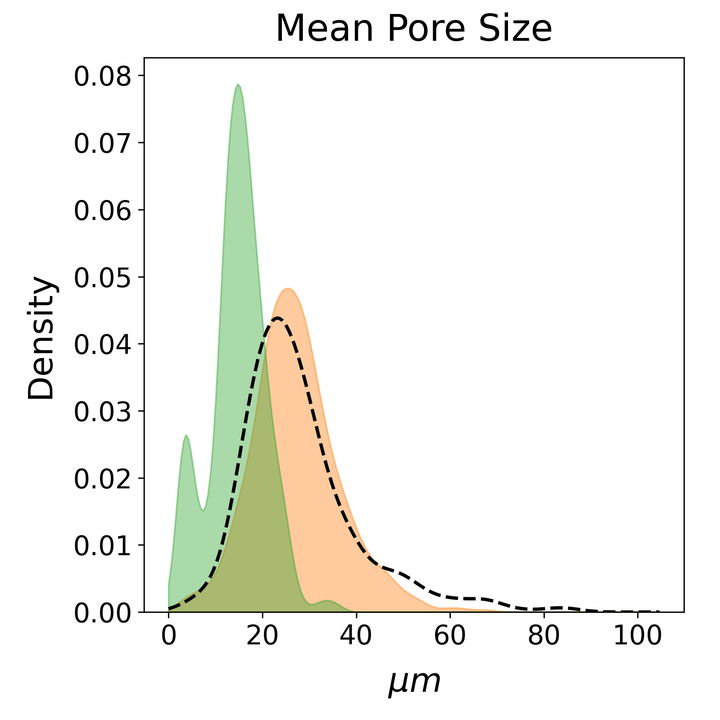}
        \caption{}
        \label{fig:app_doddington_64_c}
    \end{subfigure}
    \begin{subfigure}[t]{0.15\textwidth}
        \includegraphics[width=\textwidth]{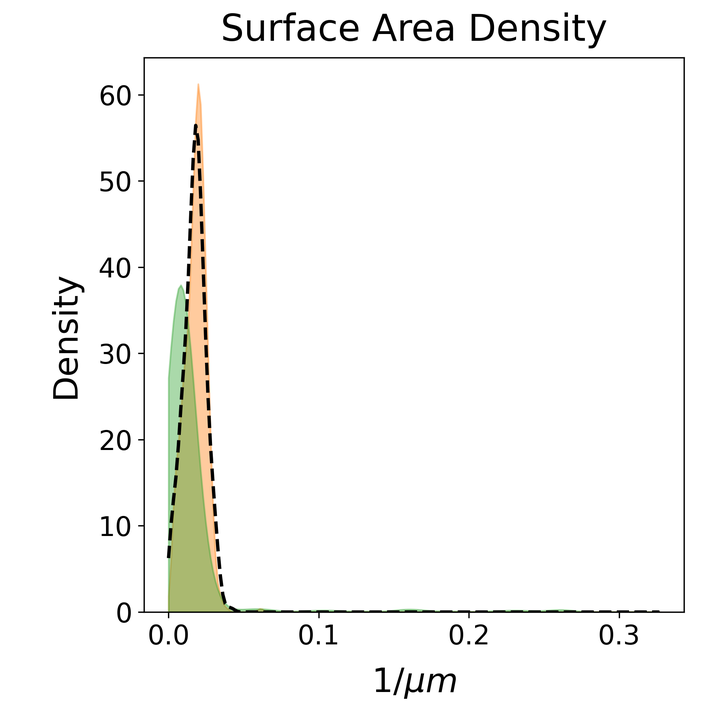}
        \caption{}
        \label{fig:app_doddington_64_d}
    \end{subfigure}
    \begin{subfigure}[t]{0.15\textwidth}
        \includegraphics[width=\textwidth]{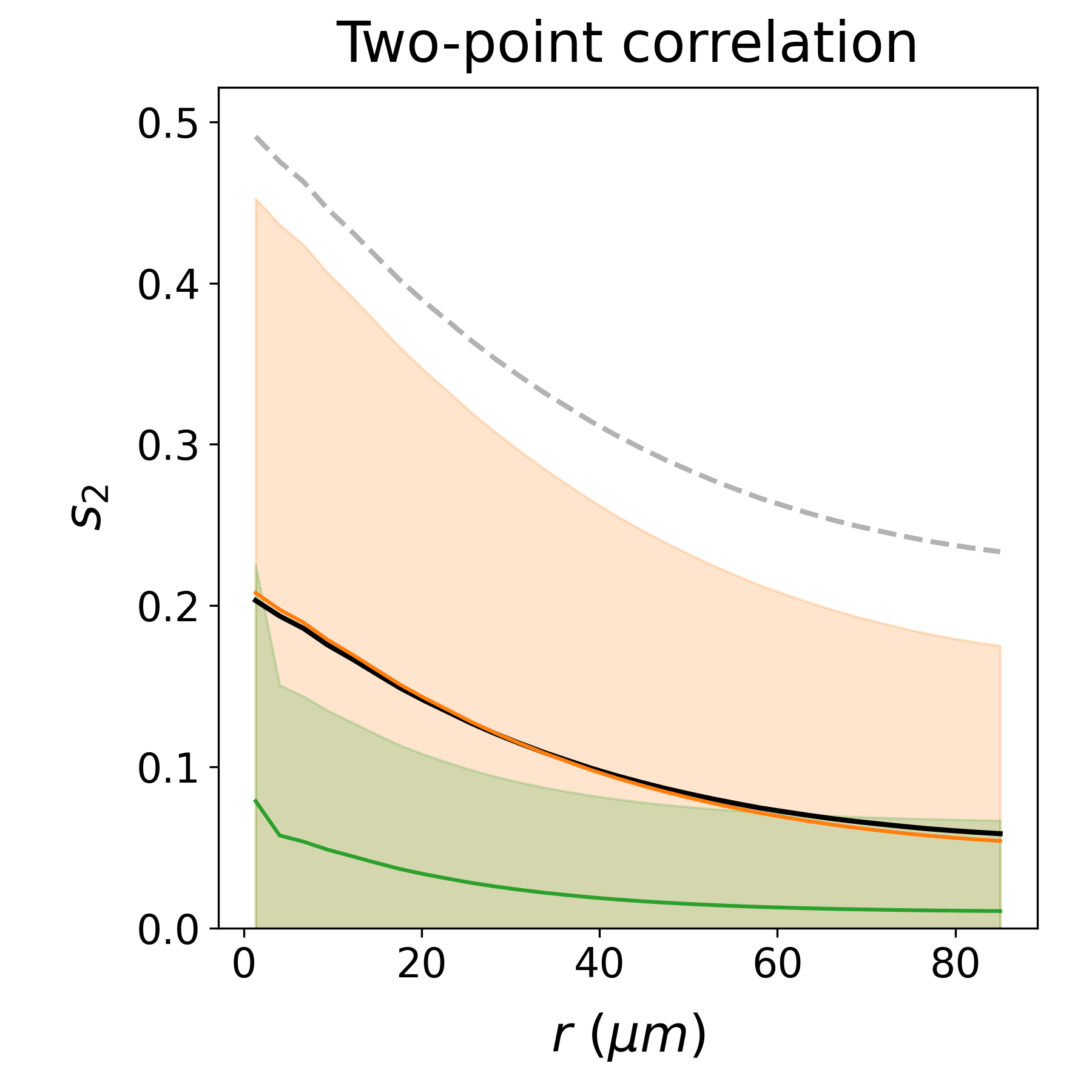}
        \caption{}
        \label{fig:app_doddington_64_e}
    \end{subfigure}
    \begin{subfigure}[t]{0.15\textwidth}
        \includegraphics[width=\textwidth]{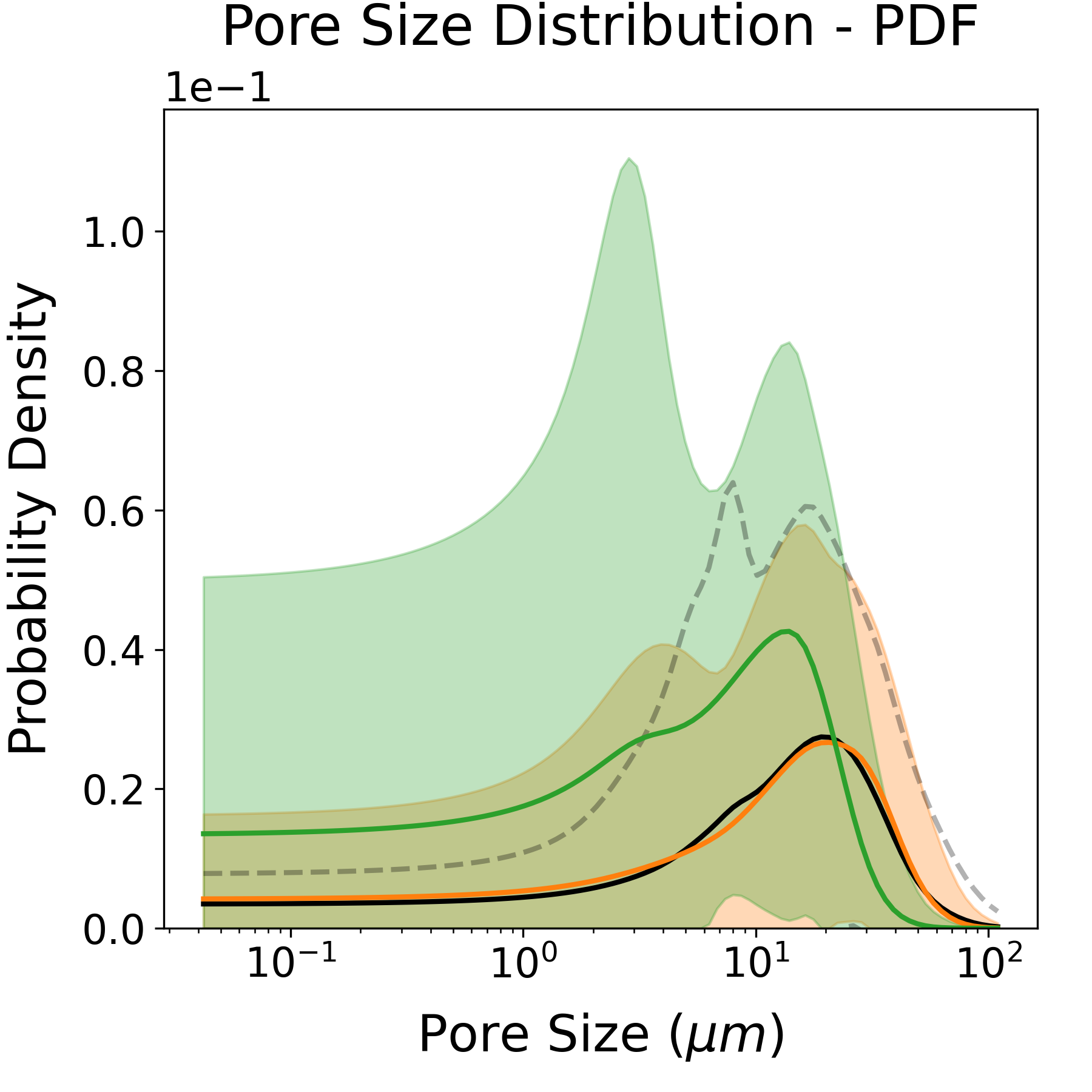}
        \caption{}
        \label{fig:app_doddington_64_f}
    \end{subfigure}
    
    % Estaillades
    \begin{subfigure}[t]{0.15\textwidth}
        \includegraphics[width=\textwidth]{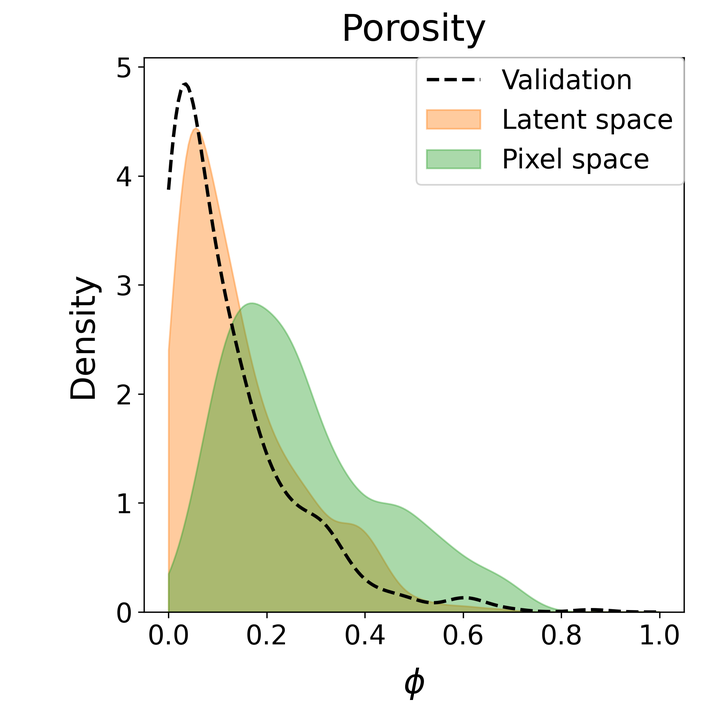}
        \caption{}
        \label{fig:app_estaillades_64_a}
    \end{subfigure}
    \begin{subfigure}[t]{0.15\textwidth}
        \includegraphics[width=\textwidth]{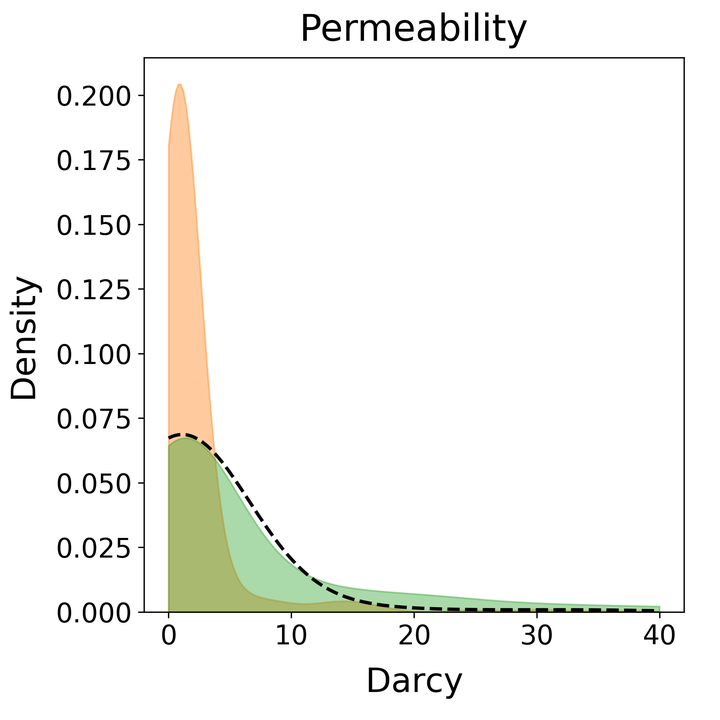}
        \caption{}
        \label{fig:app_estaillades_64_b}
    \end{subfigure}
    \begin{subfigure}[t]{0.15\textwidth}
        \includegraphics[width=\textwidth]{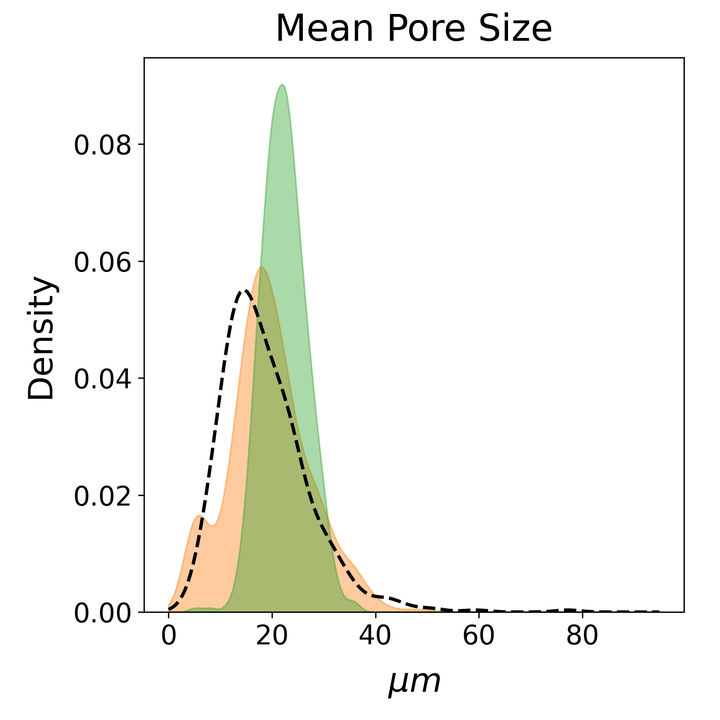}
        \caption{}
        \label{fig:app_estaillades_64_c}
    \end{subfigure}
    \begin{subfigure}[t]{0.15\textwidth}
        \includegraphics[width=\textwidth]{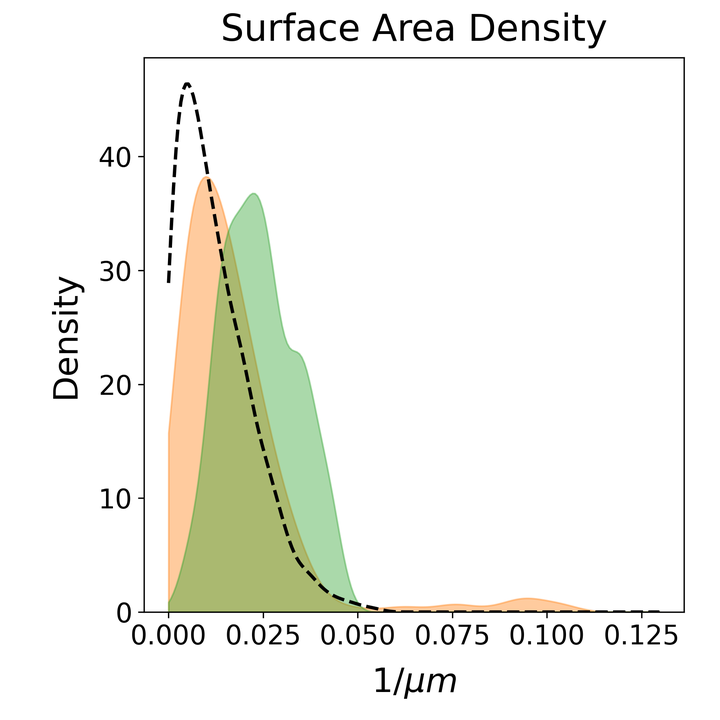}
        \caption{}
        \label{fig:app_estaillades_64_d}
    \end{subfigure}
    \begin{subfigure}[t]{0.15\textwidth}
        \includegraphics[width=\textwidth]{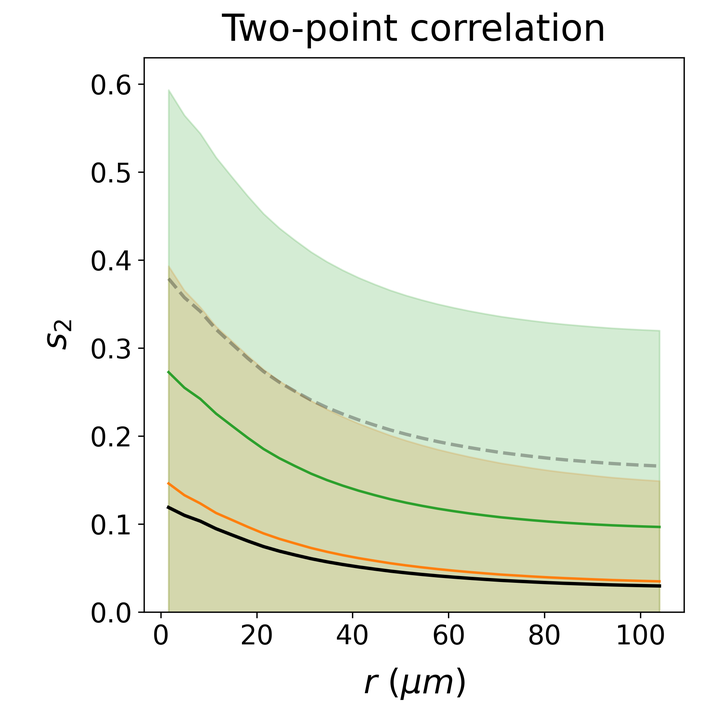}
        \caption{}
        \label{fig:app_estaillades_64_e}
    \end{subfigure}
    \begin{subfigure}[t]{0.15\textwidth}
        \includegraphics[width=\textwidth]{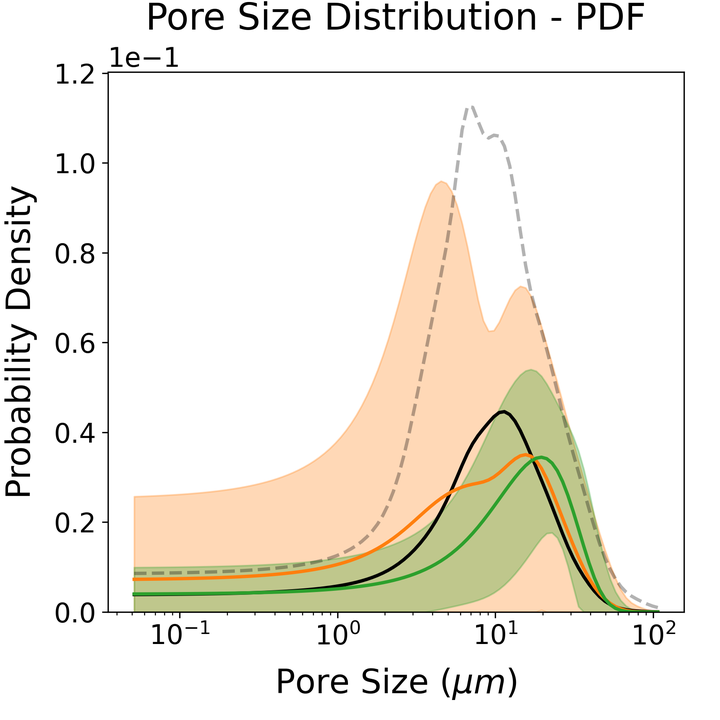}
        \caption{}
        \label{fig:app_estaillades_64_f}
    \end{subfigure}
    
    % Ketton
    \begin{subfigure}[t]{0.15\textwidth}
        \includegraphics[width=\textwidth]{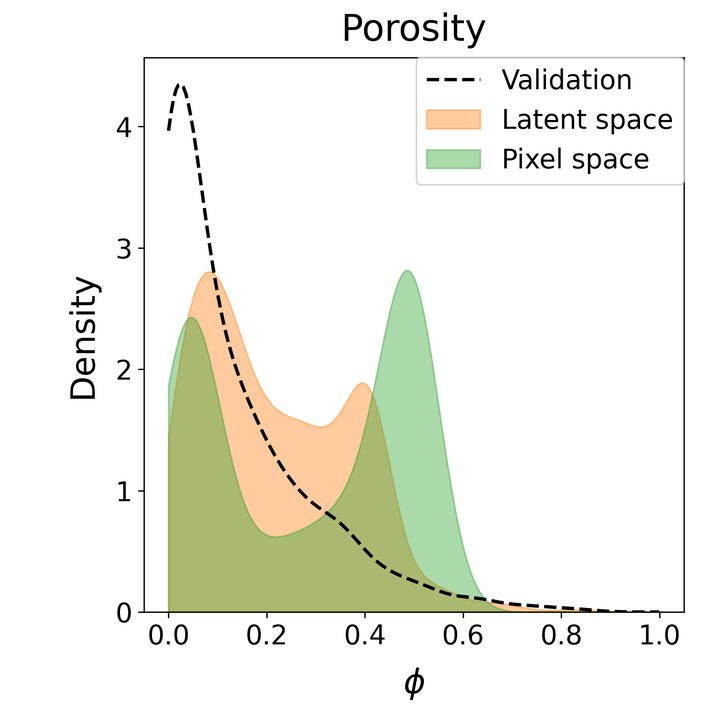}
        \caption{}
        \label{fig:app_ketton_64_a}
    \end{subfigure}
    \begin{subfigure}[t]{0.15\textwidth}
        \includegraphics[width=\textwidth]{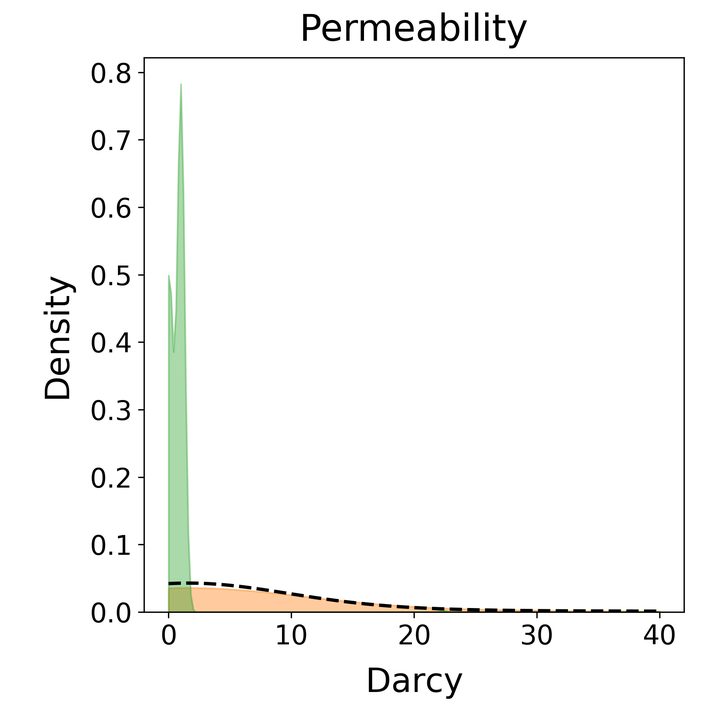}
        \caption{}
        \label{fig:app_ketton_64_b}
    \end{subfigure}
    \begin{subfigure}[t]{0.15\textwidth}
        \includegraphics[width=\textwidth]{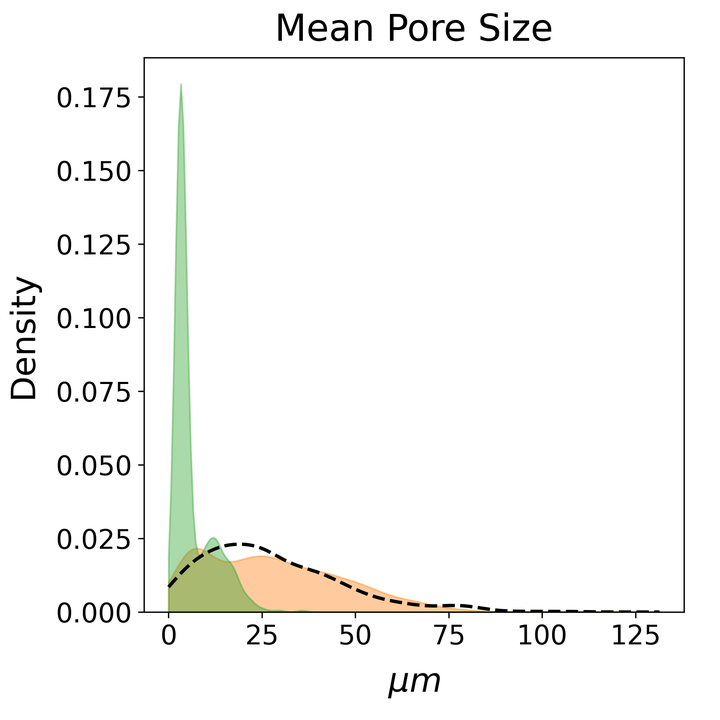}
        \caption{}
        \label{fig:app_ketton_64_c}
    \end{subfigure}
    \begin{subfigure}[t]{0.15\textwidth}
        \includegraphics[width=\textwidth]{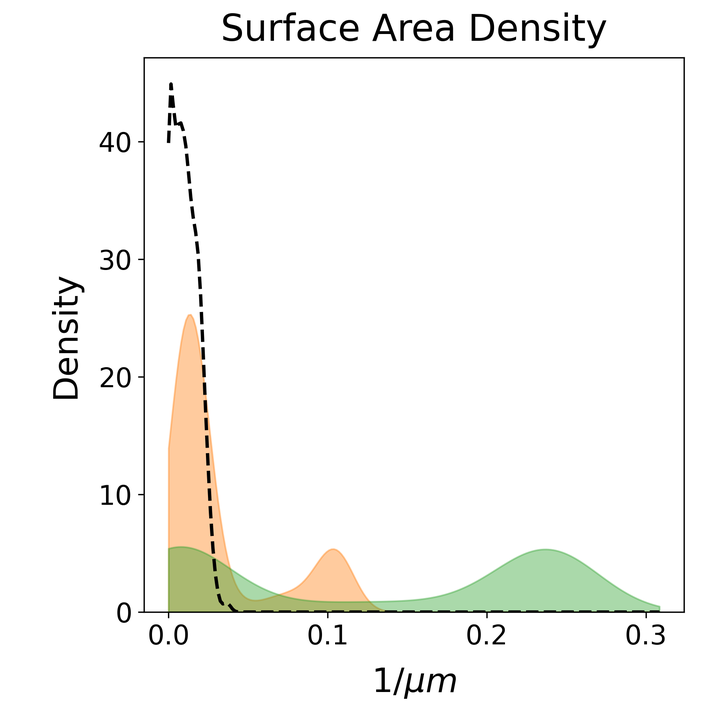}
        \caption{}
        \label{fig:app_ketton_64_d}
    \end{subfigure}
    \begin{subfigure}[t]{0.15\textwidth}
        \includegraphics[width=\textwidth]{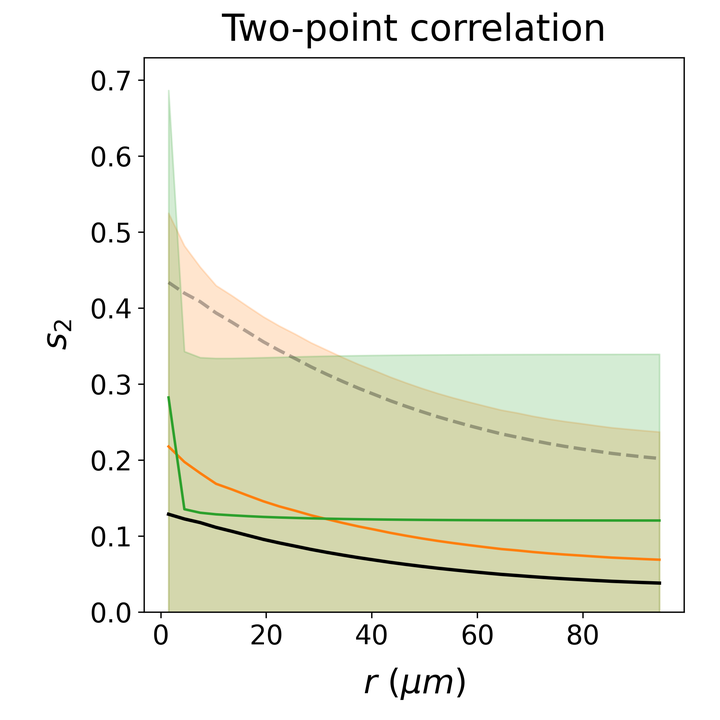}
        \caption{}
        \label{fig:app_ketton_64_e}
    \end{subfigure}
    \begin{subfigure}[t]{0.15\textwidth}
        \includegraphics[width=\textwidth]{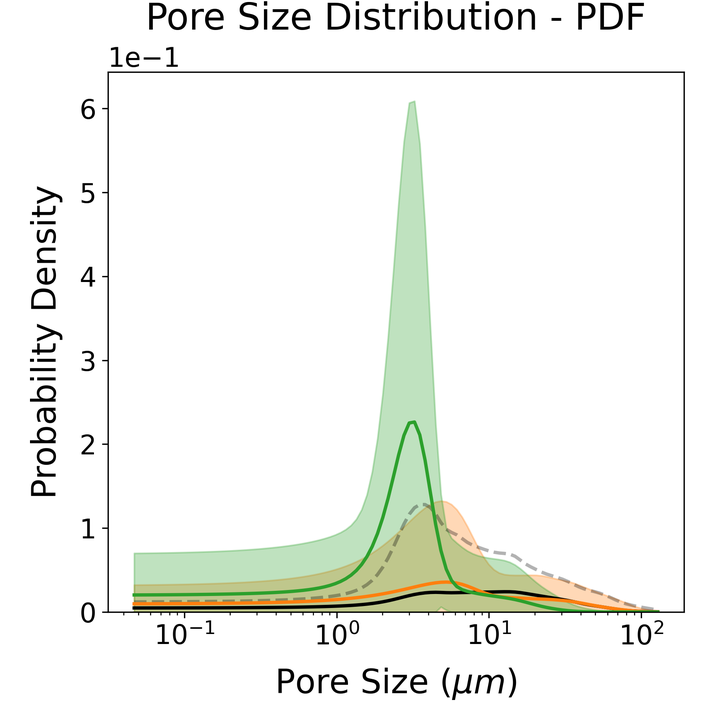}
        \caption{}
        \label{fig:app_ketton_64_f}
    \end{subfigure}
    
    \caption{Full statistical properties comparison for 64$^3$ volume samples across different rock types. Top to bottom: (a-f) Bentheimer, (g-l) Doddington, (m-r) Estaillades, and (s-x) Ketton sandstones. Each row shows (from left to right): porosity distribution, permeability distribution, mean pore size distribution, surface area density distribution, two-point correlation function, and pore size distribution.}
    \label{fig:appendix_all_rocks_64_comparison}
\end{figure}

Figures \ref{fig:appendix_all_rocks_256c_comparison}, \ref{fig:appendix_all_rocks_256nc_comparison}, \ref{fig:appendix_all_rocks_128_comparison}, \ref{fig:appendix_all_rocks_64_comparison} show the full statistical results for every rock and technique we presented in this article.

\end{document}